\documentclass{aa}  

\usepackage{graphicx}
\usepackage{txfonts}
\usepackage[T1]{fontenc}

\usepackage{multirow}

\usepackage{footnote}
\makesavenoteenv{tabular} 
\makesavenoteenv{table}

\usepackage{subfigure}
\usepackage[font=small,labelfont=bf,labelsep=period]{caption}

\usepackage{graphicx}
\usepackage{booktabs}
\usepackage{siunitx} 
\usepackage{amsmath}
\usepackage{amssymb}
\usepackage{pifont}
\newcommand{\cmark}{\ding{51}}%
\newcommand{\xmark}{\ding{55}}%
\usepackage{longtable,booktabs} %
\usepackage[online]{threeparttablex}
\usepackage{lscape}

\def\HII{H~\textsc{ii} }
\def\HI{H~\textsc{i} }

\def\CII{[C~\textsc{ii}] }
\def\CI{[C~\textsc{i}] }
\def\CII{[C~\textsc{ii}] }

\def\tCO{$^{13}$CO(2--1) }

\begin{document} 

\title{ATLASGAL-selected massive clumps in the inner Galaxy. X. Observations of atomic carbon at 492 GHz}

\author{M.-Y. Lee \inst{1,2}\and 
        F. Wyrowski\inst{2}\and 
        K. Menten\inst{2}\and 
        M. Tiwari\inst{3}\and 
        R. G\"usten\inst{2}
       } 

\institute{Korea Astronomy and Space Science Institute, 776 Daedeok-daero, 34055 Daejeon, Republic of Korea\label{1}\and
           Max-Planck-Institut f\"ur Radioastronomie, Auf dem H\"ugel 69, 53121 Bonn, Germany\label{2}\and
           University of Maryland, Department of Astronomy, College Park, MD 20742-2421, USA\label{3}
          }

\date{Received; accepted}

\abstract
{While high-mass stars are key drivers of the evolution of galaxies, 
how they form and interact with the surrounding gas remains not fully understood.
To shed light on this overarching issue, we have been performing a multi-telescope campaign 
to observe carbon species in $\sim$100 massive clumps identified by the ATLASGAL survey (Top100 sample).  
Our targets constitute a representative sample of high-mass star-forming regions 
with a wide range of masses ($\sim$20--10$^{4}$ $M_{\odot}$), bolometric luminosities ($\sim$60--10$^{6}$ $L_{\odot}$), 
and evolutionary stages (70 $\mu$m weak, infrared weak, infrared bright, and \HII region sources).}
{We aim to probe the physical conditions of [C~\textsc{i}]-traced gas in the Top100 sample based on APEX \CI 492 GHz single-pointing observations. 
This is the first of a series of papers presenting results from our \CII and \CI campaign.}
{To determine physical properties such as the temperature, density, and column density, 
we combined the obtained \CI 492 GHz spectra with APEX observations of \CI 809 GHz and $^{13}$CO(2--1), as well as with other multi-wavelength data, 
and employed both local thermodynamic equilibrium (LTE) and non-LTE methods.}
{Our 98 sources are clearly detected in \CI 492 GHz emission, 
and the observed integrated intensities and line widths tend to increase toward evolved stages of star formation. 
In addition to these ``main'' components that are associated with the Top100 sample,  
41 emission and two absorption features are identified by their velocities toward 28 and two lines of sight respectively as ``secondary'' components. 
The secondary components have systematically smaller integrated intensities and line widths than the main components. 
We found that \CI 492 GHz and $^{13}$CO(2--1) are well correlated with 
the $^{13}$CO(2--1)-to-[C~\textsc{i}] 492 GHz integrated intensity ratio varying from 0.2 to 5.3.
In addition, we derived the H$_{2}$-to-[C~\textsc{i}] conversion factor, $X$(C~\textsc{i}), 
by dividing 870 $\mu$m-based H$_{2}$ column densities by the observed \CI 492 GHz integrated intensities 
and found that $X$(C~\textsc{i}) (in units of cm$^{-2}$ (K km s$^{-1}$)$^{-1}$) 
ranges from 2.3 $\times$ 10$^{20}$ to 1.3 $\times$ 10$^{22}$ with a median of 1.7 $\times$ 10$^{21}$. 
In contrast to the strong correlation with $^{13}$CO(2--1), 
\CI 492 GHz has a scattered relation with the 870 $\mu$m-traced molecular gas.
Finally, we performed LTE and non-LTE analyses of the \CI 492 GHz and 809 GHz data for a sub-set of the Top100 sample  
and inferred that \CI emission likely originates from warm (kinetic temperature $\gtrsim$ 60 K), 
optically thin (opacity $<$ 0.5), and highly pressurized (thermal pressure $\sim$ a few (10$^{5}$--10$^{8}$) K cm$^{-3}$) regions.} 
{Our \CI 492 GHz survey demonstrates that \CI 492 GHz is prevalent in the inner Galaxy and 
traces not only massive clumps, but also non-star-forming relatively diffuse gas. 
The strong correlation between \CI 492 GHz and $^{13}$CO(2--1) indicates that they probe similar conditions, 
and the observed variations in the intensity ratio of the two transitions likely reflect local conditions of the interstellar medium. 
The scattered relation between \CI 492 GHz and the 870 $\mu$m-based molecular gas, on the other hand, implies that 
\CI 492 GHz and $^{13}$CO(2--1) probe warm molecular gas that surrounds denser and colder clumps traced by 870 $\mu$m emission.}

\maketitle
%

\section{Introduction}



High-mass stars play a key role in the evolution of galaxies (e.g., \citealt{Kennicutt12}). 
For instance, massive stars inject a substantial amount of radiative and mechanical energy into the surrounding gas 
via ultraviolet (UV) radiation fields, stellar outflows/winds, and supernova explosions. 
This stellar feedback has been argued to be critical for regulating star formation in galaxies 
(e.g., \citealt{Cox81}; \citealt{Silk97}; \citealt{Ostriker10}; \citealt{Hopkins14}). 
In addition, high-mass stars enrich the interstellar medium with heavy elements throughout their lives, 
driving the chemical evolution of galaxies (e.g., \citealt{Matteucci21}). 

Despite their importance, how high-mass stars form and interact with the surrounding gas is not yet fully understood (e.g., \citealt{Motte18}). 
One of the challenges has been that massive star-forming regions tend to be widely separated and located far awary from us  
due to the rarity of high-mass stars. 
Furthermore, high-mass stars reach the main sequence while deeply embedded in their parental molecular clouds, 
preventing us from probing the early stage of star formation with traditional optical and near-infrared (NIR) observations. 
To overcome these difficulties and provide a comprehensive census of high-mass star-forming regions, 
a number of Galactic plane surveys at far-infrared (FIR) and submillimeter (submm) wavelengths have been performed in recent years, 
including the APEX Telescope Large Area Survey of the Galaxy (ATLASGAL; \citealt{Schuller09}). 
ATLASGAL is a 870 $\mu$m survey of the inner Galaxy 
(covering $|l| < 60^{\circ}$ with $|b| < 1.5^{\circ}$ and $280^{\circ} < l < 300^{\circ}$ with $-2^{\circ} < b < 1^{\circ}$)
conducted with the Atacama Pathfinder EXperiment (APEX) 12 m submillimeter telescope in Chile. 
This survey identified $\sim$10$^{4}$ massive clumps  
(e.g., \citealt{Contreras13}; \citealt{Csengeri14}; \citealt{Urquhart14}), 
and the brightest $\sim$100 of them (Top100 hereafter) have been a subject of various follow-up studies, 
e.g., \cite{Giannetti14} (investigating CO depletion and isotopic ratios), 
\cite{Csengeri16} (probing SiO-traced shocked gas), 
\cite{Kim17} (searching for H~\textsc{ii} regions), 
and \cite{Koenig17} (characterizing physical properties based on dust continuum emission). 
More specifically, \cite{Koenig17} constructed dust spectral energy distributions (SEDs) from 8 $\mu$m to 870 $\mu$m 
and measured dust temperatures and fluxes. 
Along with distance information, the authors then estimated clump masses, luminosities, and column densities on $\sim$19$''$ scales. 
These dust-based analyses showed that the Top100 sample could be divided into four groups, i.e., 
70 $\mu$m weak sources (70w), mid-infrared (MIR) weak sources (IRw), MIR-bright sources (IRb), 
and MIR-bright sources associated with radio continuum emission (\HII regions), 
and the dust temperature and bolometric luminosity increase along this sequence, 
implying that the Top100 sample represents different evolutionary stages of high-mass star formation. 

We have been conducting a multi-telescope campaign for observing carbon species in the Top100 clumps 
to address the following questions:  
(1) What are the physical conditions of the interstellar medium leading to high-mass star formation? 
(2) How do the energetics of gas surrounding massive young stellar objects and protostellar objects vary as the sources evolve? 
Carbon is a critical ingredient in the evolution of the interstellar medium (e.g., \citealt{Henning98}). 
In its different forms, i.e., C$^{+}$, C$^{0}$, and CO, it is one of the primary coolants and provides a powerful diagnostic tool 
for examining the physical conditions and energetics of gas. 
For example, [C~\textsc{ii}], [C~\textsc{i}], and low-$J$ CO transitions (upper $J$ $\lesssim$ 4) 
can be used to probe the properties of photodissociation regions, 
while mid- and high-$J$ CO lines can trace shock-heated gas (e.g., \citealt{Pon14}; \citealt{Lee19}). 
Among these carbon phases, atomic carbon has received relatively little attention,  
mainly because \CI fine-structure transitions are typically fainter than low-$J$ CO emission 
and are heavily affected by atmospheric absorption. 
Theoretically, atomic carbon is expected to be abundant at the surfaces of molecular clouds (e.g., \citealt{Langer76}), 
but subsequent observations have shown that \CI emission is widespread throughout the clouds, 
invoking an interest in \CI emission as a tracer of total gas mass
(e.g., \citealt{Frerking89}; \citealt{Schilke95}; \citealt{Kramer08}; \citealt{Oka01}; \citealt{Shimajiri13}). 

This is the first of a series of papers presenting results from our \CII and \CI campaign for the Top100 sources.  
As a pilot study, we here focus on examining the physical conditions of [C~\textsc{i}]-traced gas based on APEX \CI 492 GHz single-pointing observations, 
and the two science questions described above will be addressed with more data in forthcoming papers. 
The organization of this paper is as follows. 
In Sect. \ref{s:obs}, we present APEX \CI and $^{13}$CO(2--1) observations of the Top100 sources   
and describe the determination of line parameters (including central velocities and line widths) via Gaussian fitting. 
The observed properties of \CI 492 GHz are probed across different evolutionary stages of high-mass star formation (Sect. \ref{s:results}) 
and are compared to molecular gas tracers (e.g., $^{13}$CO(2--1) and dust-based H$_{2}$; Sect. \ref{s:CI_molecular}). 
In addition, the \CI 492 GHz and 809 GHz data for a small sub-set of the Top100 sources are analyzed  
to derive the physical conditions of [C~\textsc{i}]-traced gas (Sect. \ref{s:CI_phases}).  
Finally, we discuss our results and summarize main findings in Sect. \ref{s:summary}.

\section{Observations and data reduction}
\label{s:obs} 

\subsection{\CI 492 GHz}
\label{s:obs_CI492}

Single-pointing observations of the \CI $^{3}P_{1}$--$^{3}P_{0}$ transition at 492.16065 GHz were made toward 98 of the Top100 sources  
using the APEX FLASH$^{+}$ (First Light APEX Submillimeter Heterodyne instrument; \citealt{Klein14}) receiver in 2016 (project IDs: M0039-97 and M0013-98). 
Two different off positions were observed in the total power mode to avoid contamination,   
while additional observations were performed in the wobbler mode to accurately measure continuum levels. 
The characteristic of the observed sources, including locations, distances, and classifications, are summarized in Table \ref{t:appendix_table1}.  
The obtained spectra were processed with the GILDAS CLASS\footnote{\url{https://www.iram.fr/IRAMFR/GILDAS}} software 
and were converted into units of main-beam brightness temperature ($T_{\rm MB}$) 
using a forward efficiency ($F_{\rm eff}$) of 0.95 and a main-beam efficiency ($B_{\rm eff}$) of 0.46 based on Mars observations. 
The final spectra on 13$''$ scales were smoothed to a velocity resolution of 0.46 km s$^{-1}$
and have a root-mean-square (rms) noise level of 0.1--0.6 K (median $\sim$ 0.3 K).

\subsection{\CI 809 GHz}
\label{s:obs_CI809} 

Nine of the Top100 sources were mapped in the \CI $^{3}P_{2}$--$^{3}P_{1}$ transition at 809.34197 GHz
using the APEX CHAMP$^{+}$ (Carbon Heterodyne Array of the MPIfR; \citealt{Gusten08}) instrument in 2013 (project ID: M0010-92). 
The mapping was made in the On-The-Fly (OTF) mode with a coverage of 1.6$'$ $\times$ 1.6$'$, 
and the observed sources are marked in Table \ref{t:appendix_table1}. 
The GILDAS CLASS software was employed for data reduction, 
and the processed data cubes were translated into the $T_{\rm MB}$ units 
by adopting $F_{\rm eff}$ = 0.95 and $B_{\rm eff}$ = 0.38 based on Mars observations.
The final data on 8$''$ scales were smoothed to a velocity resolution of 0.70 km s$^{-1}$ 
and have a median rms noise level of $\sim$1 K. 

\subsection{\tCO}
\label{s:obs_CO}

The 98 sources in our \CI 492 GHz survey were also observed in the \tCO transition at 220.39868 GHz (single-pointing) 
using the APEX PI230 receiver in 2016 (project ID: M0024-97). 
Unlike the \CI 492 GHz observations, only one off position was observed for each source in the total power mode.  
The obtained \tCO data were reduced using the GILDAS CLASS software 
and were converted into the $T_{\rm MB}$ units by assuming $F_{\rm eff}$ = 0.95 and $B_{\rm eff}$ = 0.66 (\citealt{Tang18}). 
The final science-ready spectra have a resolution of 30$''$ 
and a rms noise level of 0.04--0.14 K (median $\sim$ 0.06 K) per 0.66 km s$^{-1}$ channel. 

\subsection{Deriving line parameters}
\label{s:line_params} 

While the observed \CI 809 GHz spectra can be approximated as single Gaussians, 
the \CI 492 GHz and \tCO spectra often show complex shapes with multiple components (Appendix \ref{s:appendix_spectra}). 
For our analyses, we identified individual components by fitting Gaussian functions to the final spectra 
and estimated line parameters such as central velocities and full width at half maximum (FWHMs). 
When two components are close to each other, we examined the difference between their central velocities 
and considered them as one component if the difference is smaller than the sum of their FWHMs. 
In addition, we calculated the integrated intensity ($I$([C~\textsc{i}]) or $I$($^{13}$CO)) of each identified component 
by integrating the emission over a velocity range over which the component is clearly detected. 
For \CI 809 GHz, we smoothed the cubes and used the spectra that were extracted at a resolution of 13$''$ 
to compare with the \CI 492 GHz data (Sect. \ref{s:CI_phases}).
The final \CI and \tCO spectra with the fitted Gaussians are shown in Fig. \ref{f:CI_CO_plot_comp},  
and the derived line parameters and integrated intensities are presented in Table \ref{t:appendix_table2}.

\begin{figure}
\centering
\includegraphics[scale=0.6]{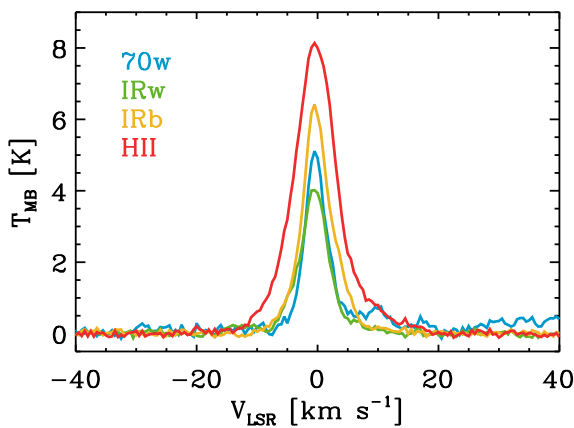} 
\caption{\label{f:stack_CI} Average \CI 492 GHz spectra of the four evolutionary groups
(70w, IRw, IRb, and \HII region groups in blue, green, yellow, and red). 
To derive these spectra, sources with contaminated reference positions were excluded.} 
\end{figure} 

\section{Observational results} 
\label{s:results} 

In this section, we mainly discuss the observed properties of \CI 492 GHz (integrated intensities and FWHMs) 
and how they vary in a diversity of environments. 

\subsection{\CI 492 GHz detection rate} 
\label{s:detection_statistics} 

We regard components whose peak-to-rms ratios are equal to or higher than five as detections. 
With this threshold, we found that all 98 sources are detected in \CI 492 GHz emission. 
The exact breakdown of the sources is as follows: 14, 31, 31, and 22 for the 70w, IRw, IRb, and \HII region groups respectively.  
These different evolutionary groups have comparable median distances ($\sim$4 kpc) and \CI 492 GHz sensitivities ($\sim$0.3 K), 
making comparisons of the groups straightforward. 
In addition to the main components that are directly associated with the Top100 sources,  
secondary components are detected in both emission and absorption (Fig. B.1): 
41 emission components toward 28 lines of sight and two absorption components toward two lines of sight.  
Throughout this paper, we refer to these components as ``diffuse clouds'',  
since they likely trace colder and less dense gas than the main 98 components.

\subsection{Average \CI 492 GHz spectra}
\label{s:stack_CI}

To examine how \CI 492 GHz changes across different environments, 
we first derived average spectra of the four evolutionary groups 
by shifting the observed spectra to a central velocity of 0 km s$^{-1}$ 
based on the source velocities
and stacking them (Fig. \ref{f:stack_CI}). 
For this, 94 sources with clean reference positions were considered
(12, 30, 30, and 22 for the 70w, IRw, IRb, and \HII region groups respectively). 

Fig. \ref{f:stack_CI} shows that the characteristics of \CI 492 GHz emission systematically change  
across the four evolutionary stages in terms of brightness and line width. 
Specifically, the peak temperature increases from 5.8 K to 7.8 K for the IRb and \HII region groups, 
while the 70w spectrum has a slightly higher peak than the IRw spectrum (4.7 K versus 3.8 K). 
On the other hand, the FWHM increases toward more evolved stages: 
4.3, 5.6, 5.9, and 8.4 km s$^{-1}$ for the 70w, IRw, IRb, and \HII region groups. 
The peak temperature and FWHM values here were estimated from Gaussian fitting of the average spectra. 

\begin{figure}
\centering
\includegraphics[scale=0.43]{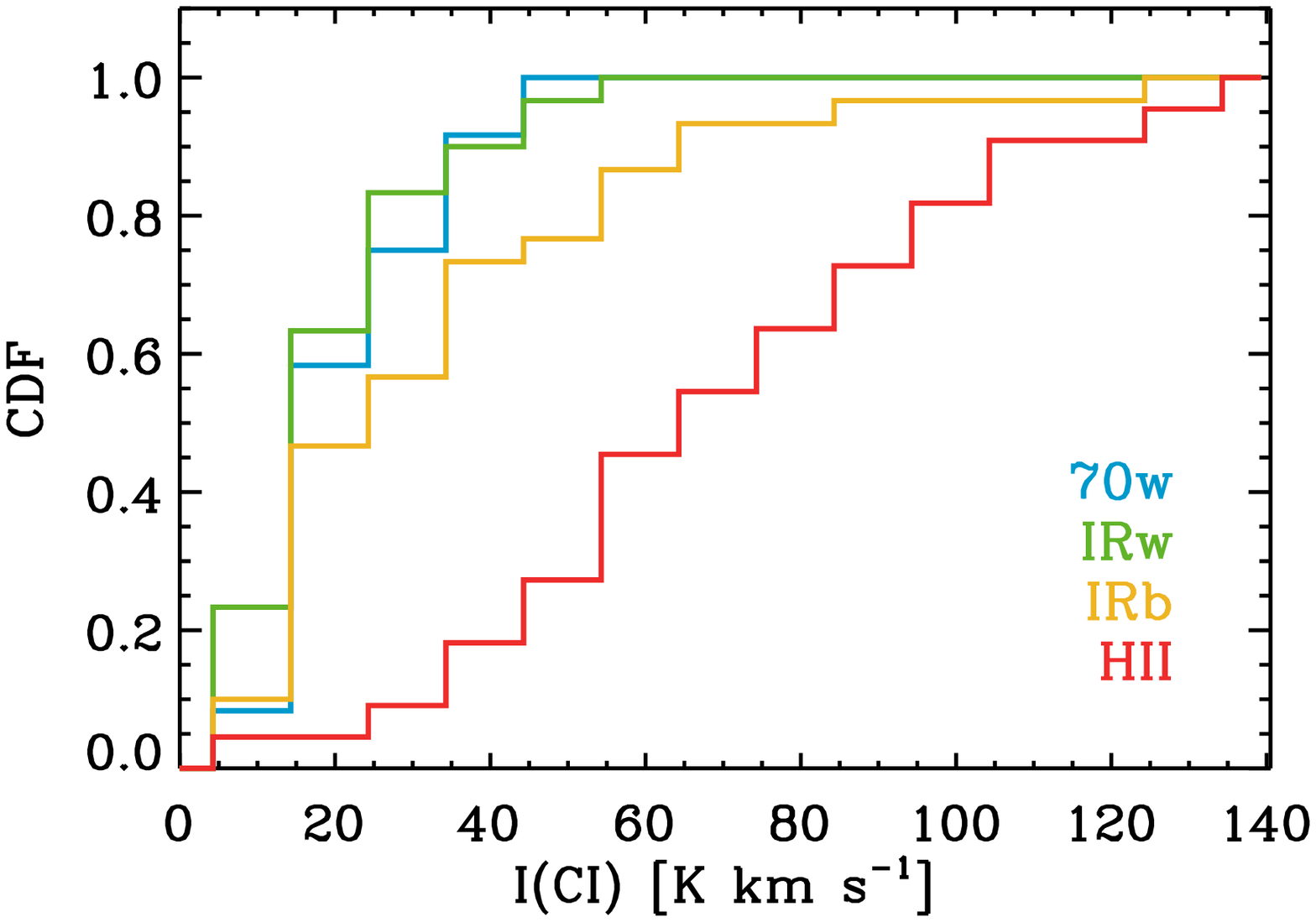}
\includegraphics[scale=0.43]{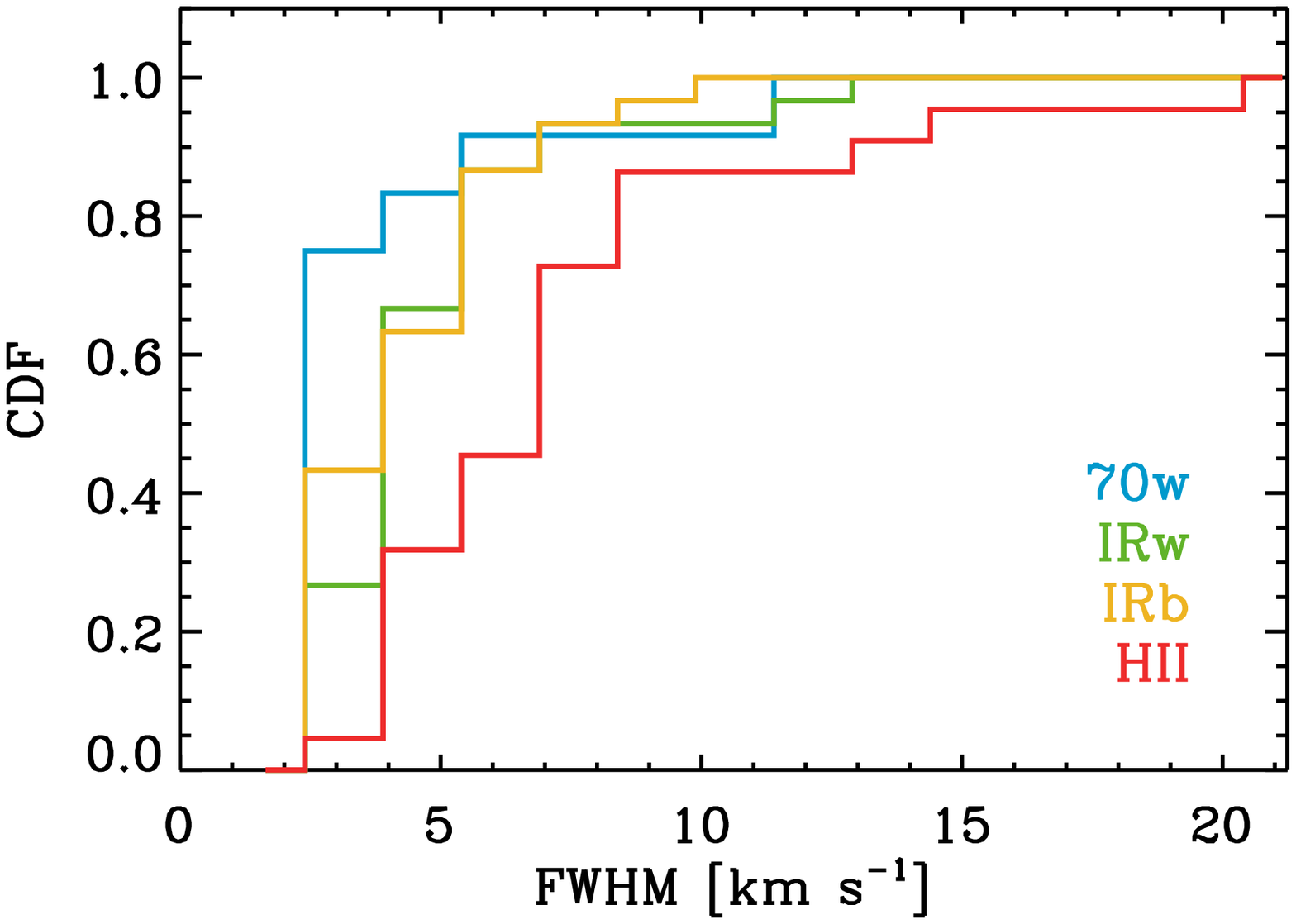}
\caption{\label{f:main_CDF} CDFs of the \CI 492 GHz integrated intensity (\textit{top}) and FWHM (\textit{bottom}). 
For both plots, the 70w, IRw, IRb, and \HII region groups are shown in blue, green, yellow, and red respectively.} 
\end{figure}

\subsection{Integrated intensities and line widths of the main components} 
\label{s:CI_main}

Next we probed how statistically different \CI 492 GHz properties are between the evolutionary stages
by constructing cumulative distribution functions (CDFs) of the integrated intensity and FWHM for the main components (Fig. \ref{f:main_CDF}). 
Like in Sect. \ref{s:stack_CI}, we used the 94 sources with clean reference positions only. 

The top panel of Fig. \ref{f:main_CDF} suggests that 
the 70w, IRw, and IRb groups have comparable distributions of the integrated intensity, 
while the \HII region group has systematically higher values 
(median $I$(C~\textsc{i}) $\sim$ 23, 21, 26, and 67 K km s$^{-1}$ for the 70w, IRw, IRb, and \HII region group respectively). 
This result is supported by two-sided Kolmogorov-Smirnov (K-S) tests\footnote{Throughout our analyses  
two-sided K-S tests were done at a significance level of 5\% when different distributions were compared.
The null hypothesis that the compared two distributions are drawn from the same population is rejected 
if the $D$-statistics is larger than the critical $D$ value (1.36$\sqrt{(m + n)/mn}$)  
where $m$ and $n$ are the number of samples for the compared distributions) and 
the $p$-value is smaller than the significance level of 0.05.}
at a significance level of 5\%, where only the \HII region group is found to be statistically distinct from the rest. 
Similarly, we found that the 70w and \HII region groups have the smallest and largest FWHMs, 
while the IRw and IRb groups have intermediate FWHMs (Fig. \ref{f:main_CDF} bottom):
median FWHM $\sim$ 3.7, 4.6, 4.4, and 7.0 km s$^{-1}$ for the 70w, IRw, IRb, and \HII region group respectively. 
This conclusion again agrees with K-S tests 
where the 70w group is found to be statistically different from the IRw and \HII region groups 
and the \HII region group stands out against the rest. 

\begin{figure}
\centering
\includegraphics[scale=0.43]{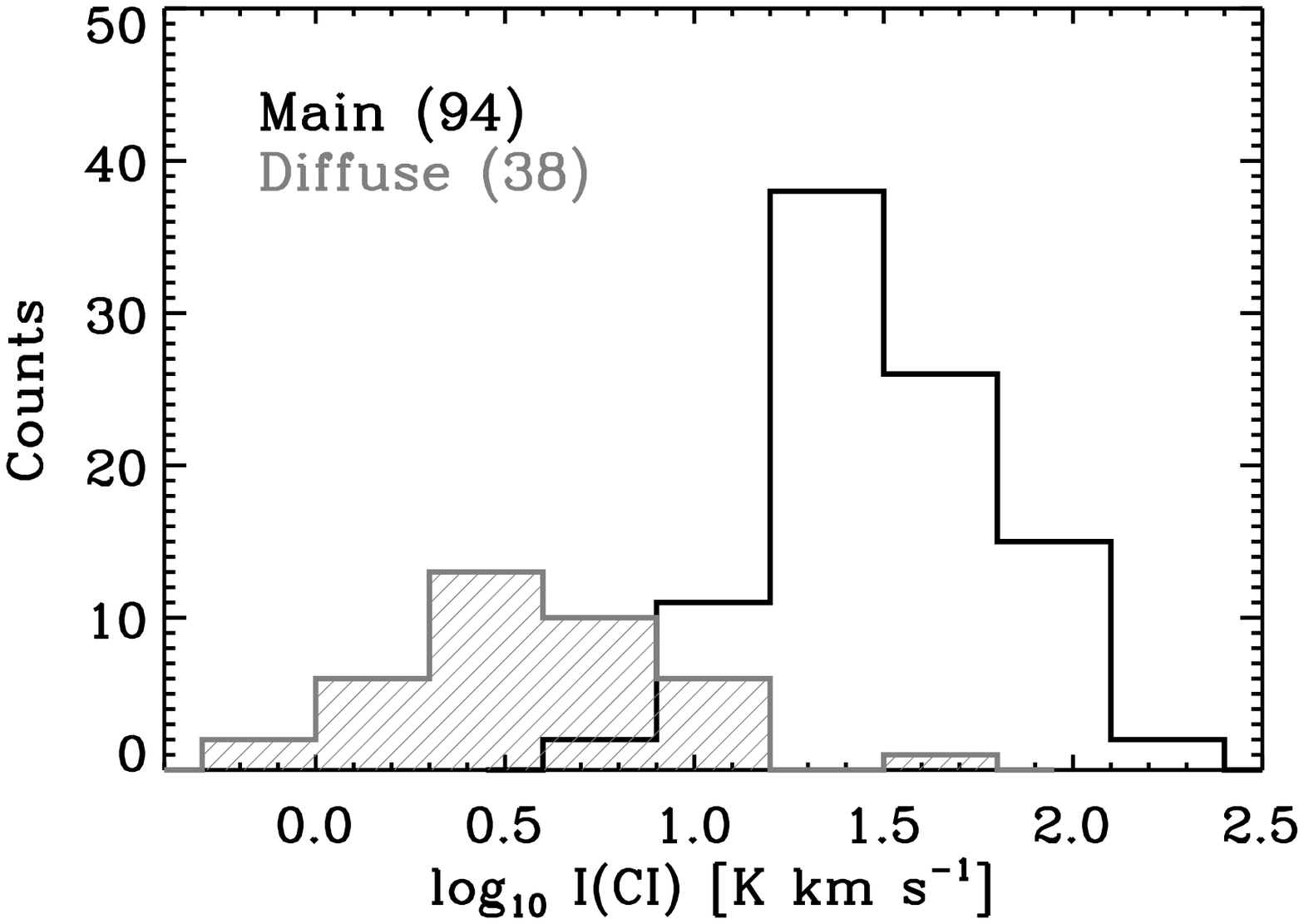}
\includegraphics[scale=0.43]{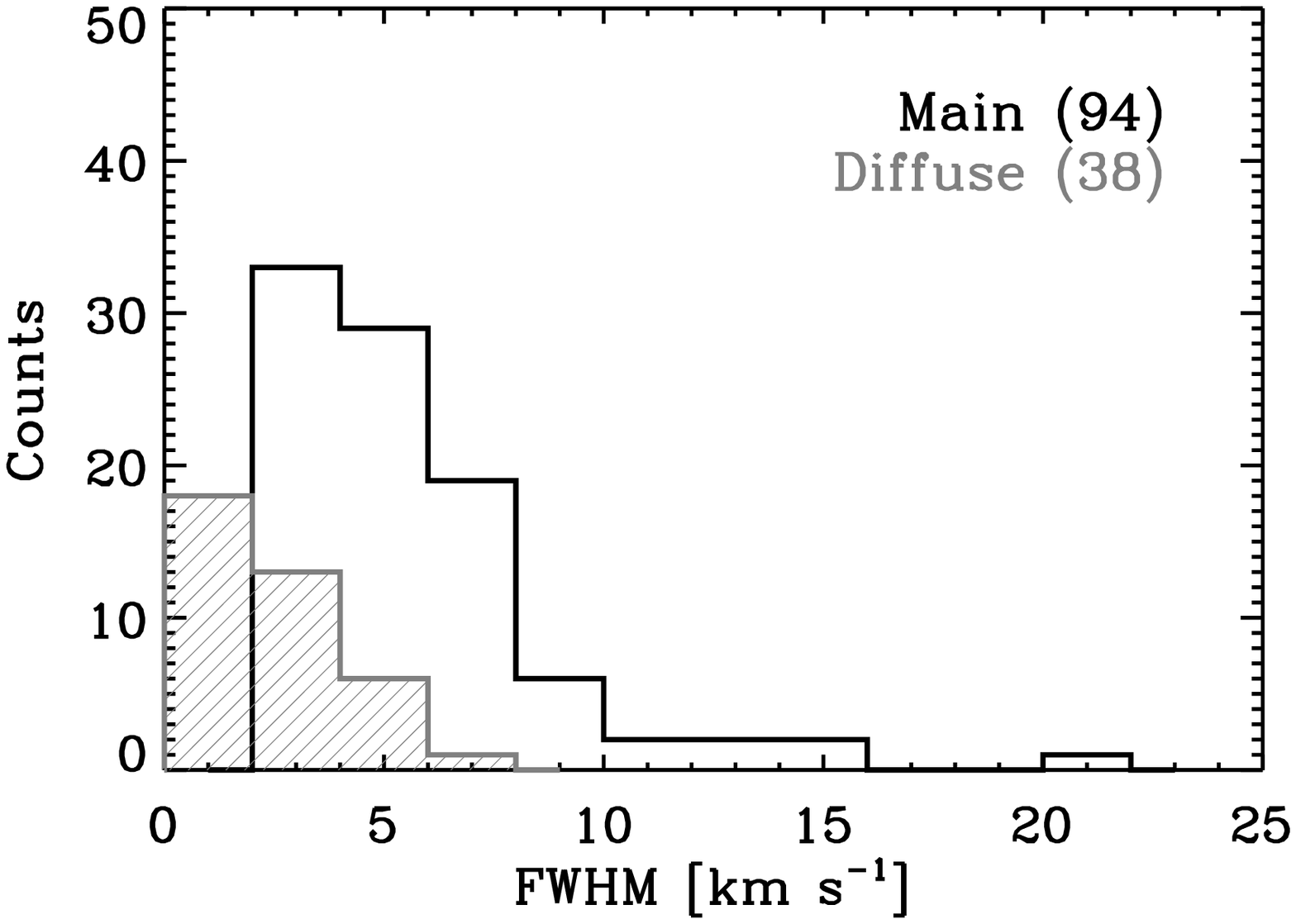}
\caption{\label{f:main_diffuse} Histograms of the \CI 492 GHz integrated intensity (\textit{top}) and FWHM (\textit{bottom}). 
The distributions of the 94 main components and 38 diffuse cloud components are shown in black and gray respectively.} 
\end{figure}

\subsection{Comparison between the main and diffuse cloud components} 
\label{s:main_diffuse}

As noted in Sect. \ref{s:detection_statistics}, 
41 components are detected along 28 lines of sight as secondary emission components. 
In comparison to the main components that are associated with the Top100 sources, 
these secondaries have systematically smaller values of the \CI 492 GHz integrated intensity and line width (Fig. \ref{f:main_diffuse}). 
For example, the median $I$(C~\textsc{i}) is 29 K km s$^{-1}$ and 4 K km s$^{-1}$ for the main and diffuse cloud components respectively, 
while the median FWHM is 5 km s$^{-1}$ and 2 km s$^{-1}$.
For these estimates and the histograms in Fig. \ref{f:main_diffuse}, 
spectra with clean reference positions were used (94 and 38 for the main and diffuse cloud components respectively).

In summary, our APEX survey suggests that \CI 492 GHz is ubiquitous and 
traces not only massive dense clumps, but also non-star-forming relatively diffuse gas. 
The observed integrated intensities and FWHMs 
systematically vary across different environments, 
which is likely driven by local conditions of the interstellar medium.  

\begin{table}
\begin{center}
\caption{\label{t:table6} Types of \CI and $^{13}$CO detections.} 
\begin{tabular}{c c c c} \toprule \toprule
 & Type 0$^{(a)}$ & Type 1$^{(a)}$ & Type 2$^{(a)}$ \\ \midrule 
\CI detection & \cmark & \xmark & \cmark \\ 
$^{13}$CO detection & \cmark & \cmark & \xmark \\ \midrule 
70w & 14 (10) & 0 & 0 \\ 
IRw & 31 (21) & 0 & 0 \\ 
IRb & 31 (21) & 0 & 0 \\ 
\HII region & 22 (22) & 0 & 0 \\
Diffuse$^{(b)}$ & 35 (23) & 45 (42) & 5 (3) \\ \midrule
Total & 133 (97) & 45 (42) & 5 (3) \\ \bottomrule
\end{tabular}
\end{center}
\textbf{Notes.} 
$^{(a)}$ The values in parenthesis are the numbers of detections that have clean reference positions. 
$^{(b)}$ G22.37$+$0.45 shows two components in \CI 492 GHz, 
which are sufficiently close to be considered as one component in $^{13}$CO(2--1). 
When we compared \CI 492 GHz to $^{13}$CO(2--1), 
we therefore used a sum of the two components in \CI 492 GHz for this source. 
Because of this, the total number of diffuse cloud components detected in \CI 492 GHz is 40 in this table, 
as opposed to 41 as we mentioned in Sect. \ref{s:main_diffuse}. 
\end{table}

\section{Relation between \CI 492 GHz and molecular gas tracers} 
\label{s:CI_molecular} 

From which part of a molecular cloud \CI 492 GHz emission arises has long been a subject of debate. 
Considering the small difference between the ionization threshold of C$^{0}$ (11.26 eV) and the dissociation threshold of CO (11.09 eV), 
C$^{0}$ is expected to form a narrow layer between C$^{+}$ and CO, limiting its usefulness as a tracer of total gas mass. 
However, a significant body of research has revealed that this expectation could be wrong and 
C$^{0}$ could exist deep inside molecular clouds 
(e.g., \citealt{Frerking89}; \citealt{Schilke95}; \citealt{Kramer08}). 
For example, the proportionality between the \CI 492 GHz and low-$J$ $^{13}$CO integrated intensities has often been interpreted 
as evidence of the widespread \CI 492 GHz emission throughout molecular clouds 
(e.g., \citealt{Keene97}; \citealt{Oka01}; \citealt{Shimajiri13}; \citealt{Izumi21}). 
In this section, we try to shed light on the origin of \CI 492 GHz emission 
by examining a relation between \CI 492 GHz and molecular gas tracers (including $^{13}$CO(2--1) and 870 $\mu$m-based H$_{2}$).  

\subsection{\CI 492 GHz and $^{13}$CO(2--1)} 
\label{s:CI_CO}

\subsubsection{Classification of \CI and $^{13}$CO detections} 
\label{s:CI_CO_class} 

We here probe the relation between \CI 492 GHz and $^{13}$CO(2--1) in detail. 
First, we started our analysis by classifying the \CI and $^{13}$CO data into three types: 
(1) Type 0 where \CI 492 GHz and $^{13}$CO(2--1) are detected; 
(2) Type 1 where only $^{13}$CO(2--1) is detected; 
(3) Type 2 where only \CI 492 GHz is detected. 
The breakdown of each type is summarized in Table \ref{t:table6}.

Among 139 components that are seen in \CI 492 GHz emission, 
133 show clear $^{13}$CO(2--1) detections, implying a good correspondence between \CI 492 GHz and $^{13}$CO(2--1) in general.  
On the other hand, 45 secondary components are newly detected in $^{13}$CO(2--1) only, 
which is in contrast to a handful number of components seen in \CI 492 GHz only (5). 
This striking difference partially results from the higher sensitivity of the $^{13}$CO(2--1) data (Sect. \ref{s:obs}). 


\begin{figure*}
\centering
\includegraphics[scale=0.475]{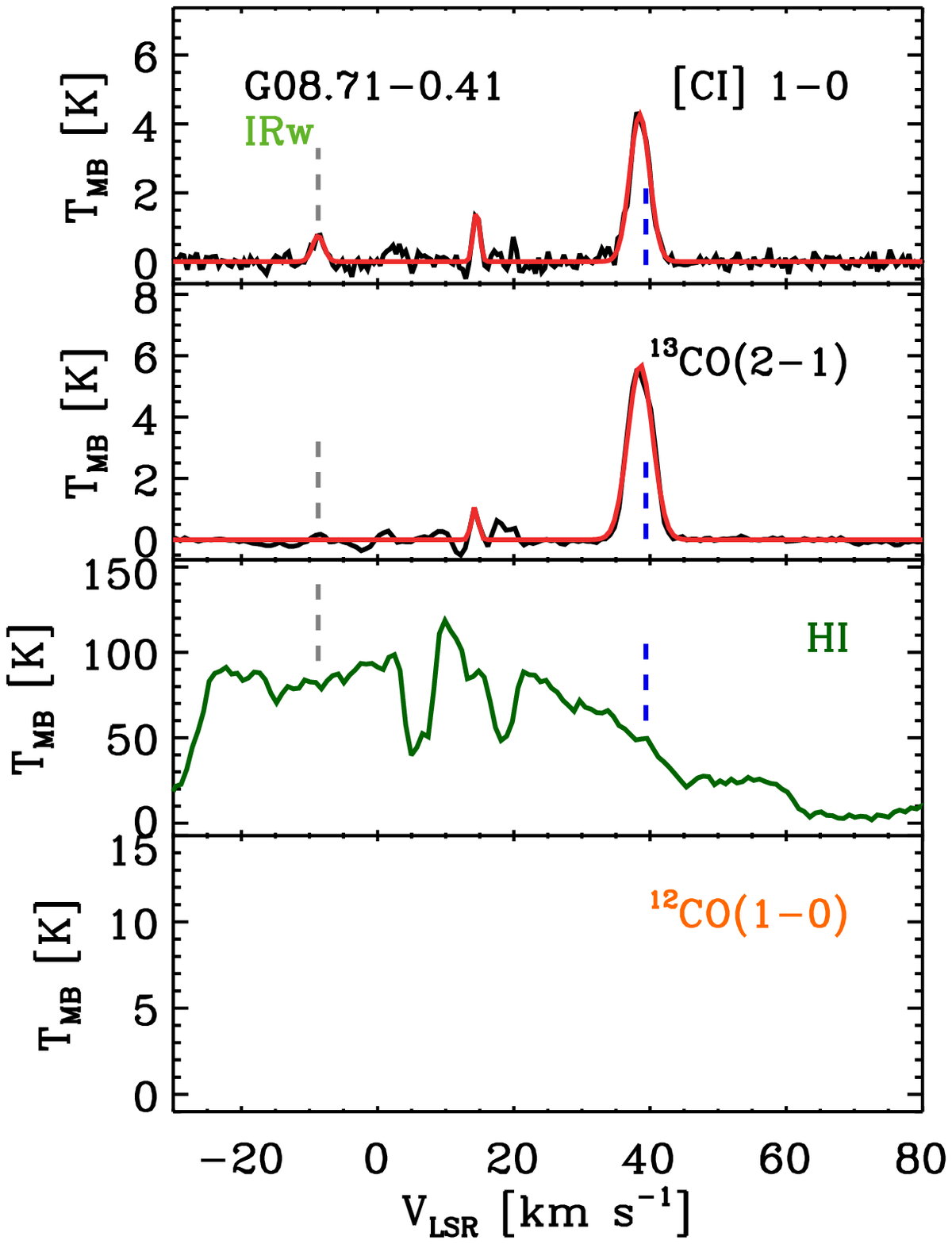} \hspace{0.5cm}
\includegraphics[scale=0.475]{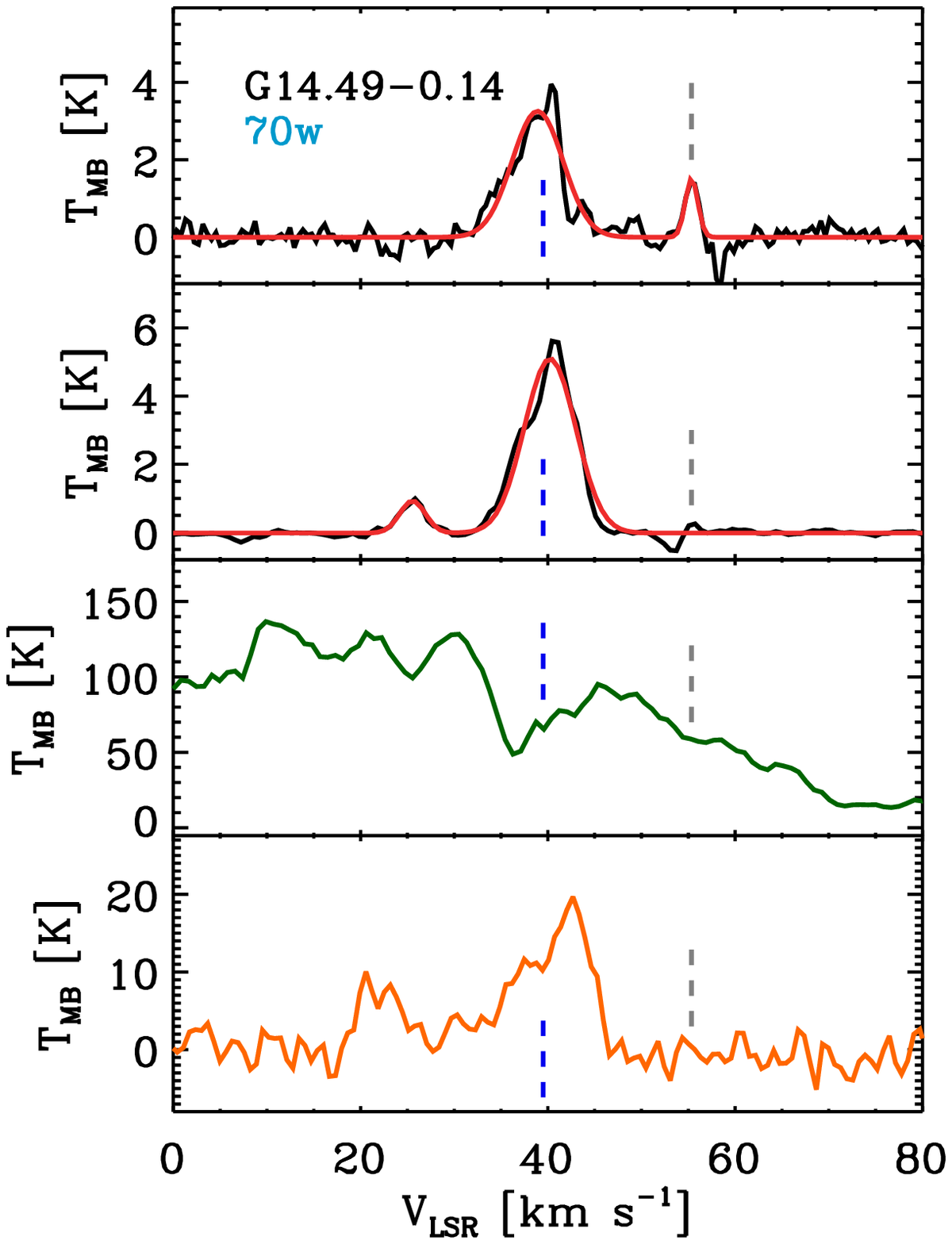} \\ 
\includegraphics[scale=0.475]{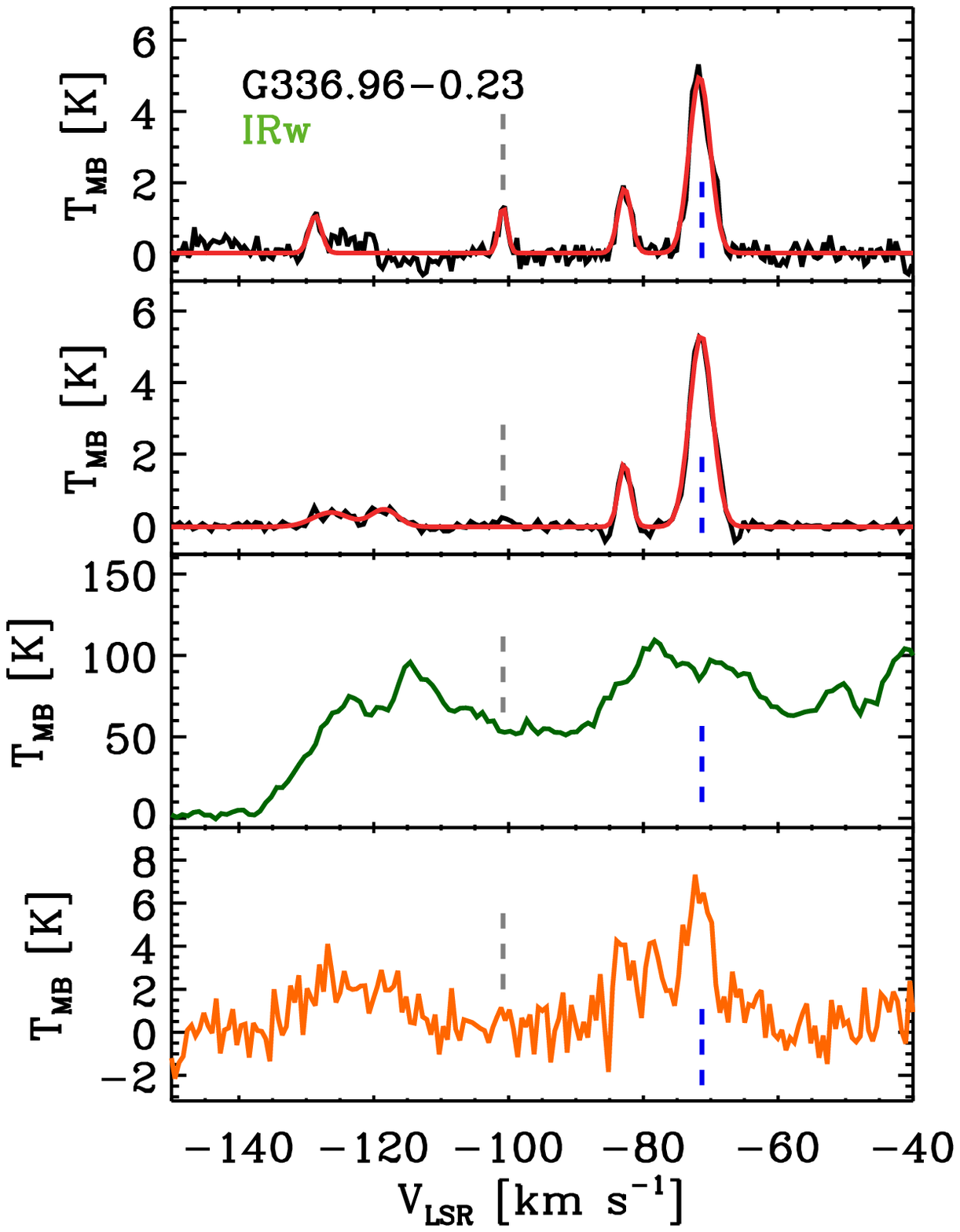}
\includegraphics[scale=0.475]{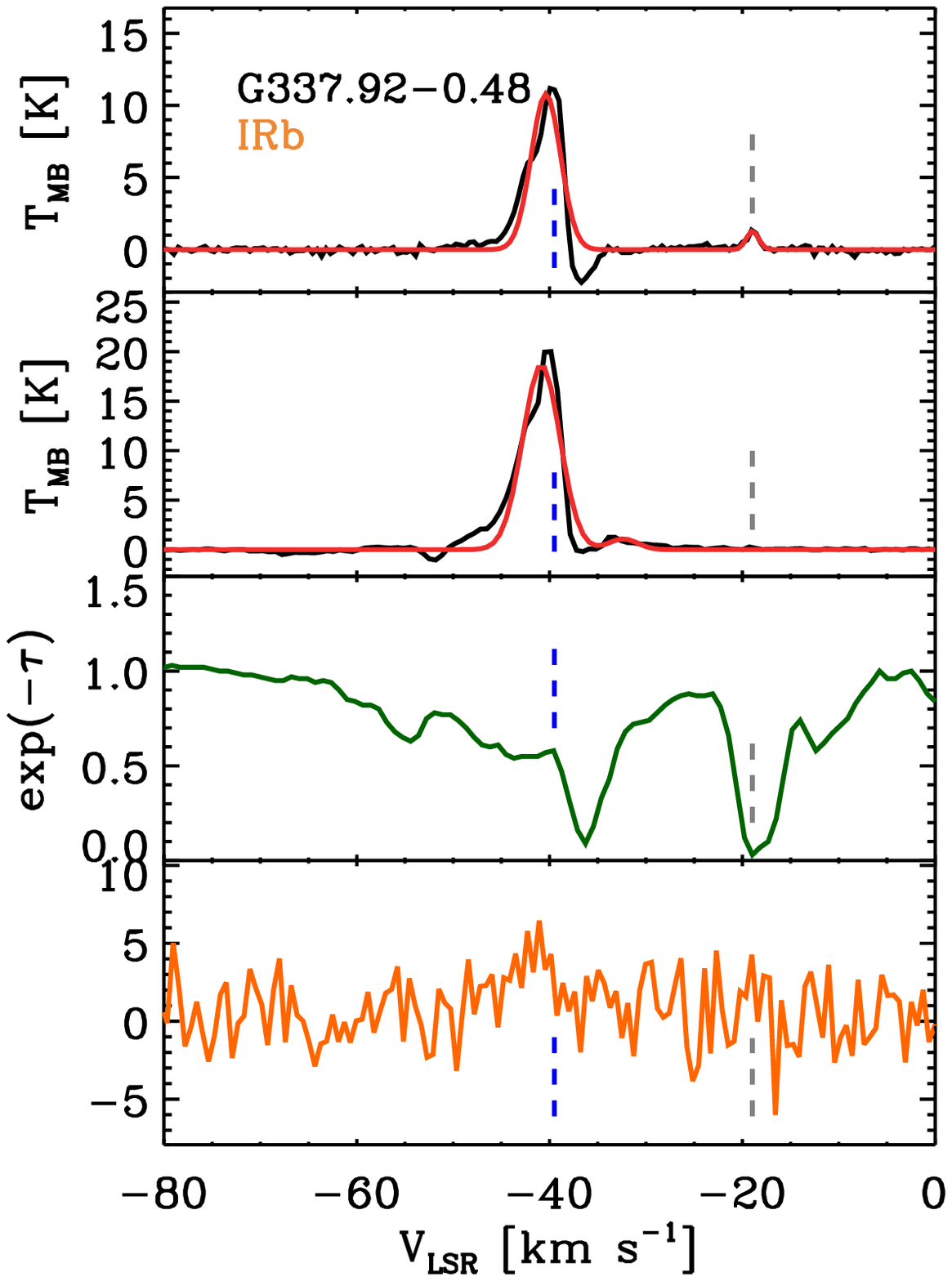}
\includegraphics[scale=0.475]{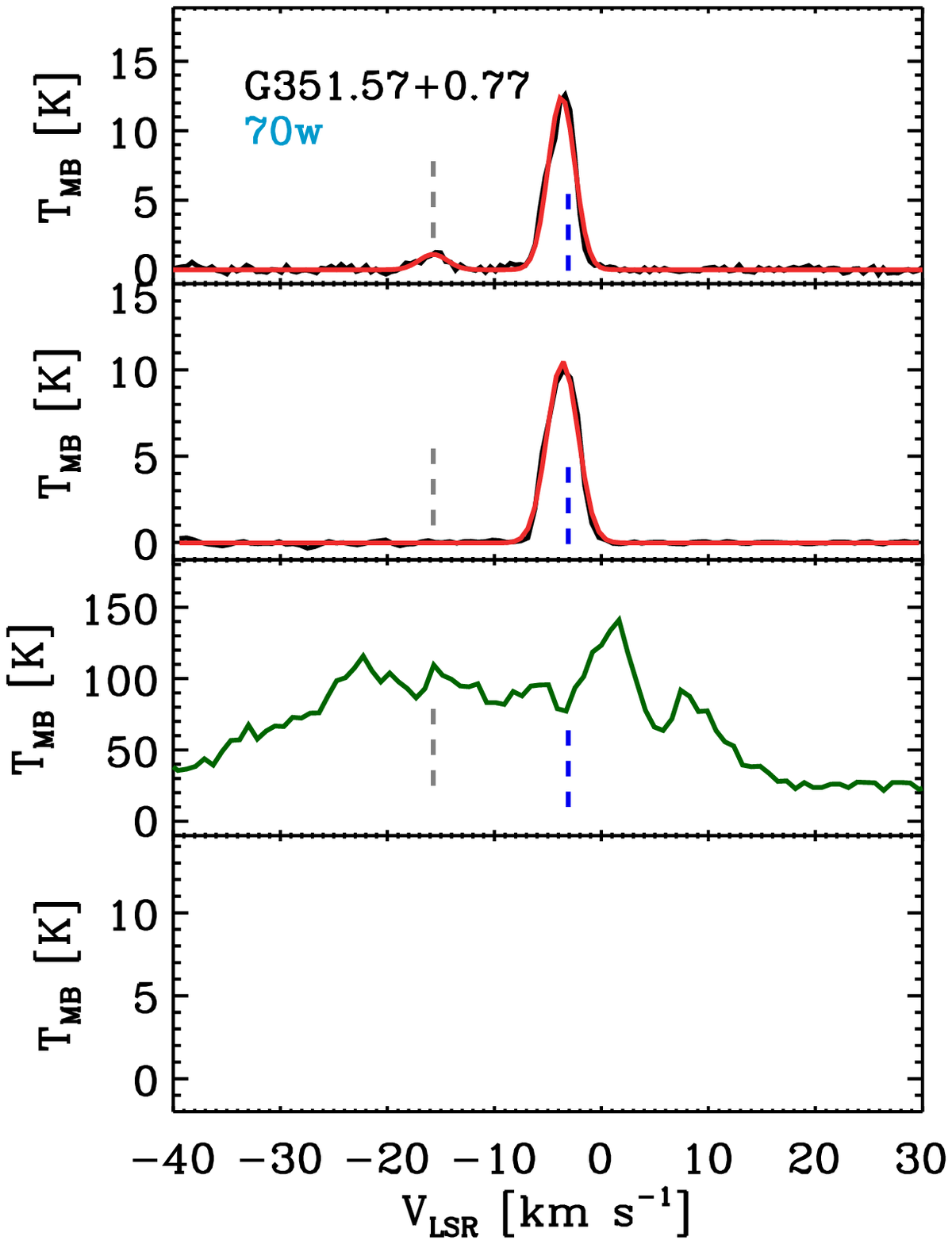} 
\caption{\label{f:CI_CO_dark} Comparison between \CI 492 GHz (black), $^{13}$CO(2--1) (black), 
H~\textsc{i} (green), and $^{12}$CO(1--0) (orange) for five sources. 
For H~\textsc{i}, the absorption spectrum from \cite{Brown14} is shown for G337.92$-$0.48, 
while SGPS data are presented for the remaining sources.   
In addition, $^{12}$CO(1--0) spectra from the FUGIN (G14.49$-$0.14) and Mopra (G336.96$-$0.23 and G337.92$-$0.48) surveys are used. 
Over the \CI 492 GHz and $^{13}$CO(2--1) spectra, the fitted Gaussians are overlaid in red.   
Finally, the source velocities and the secondary components that are seen in \CI 492 GHz 
without corresponding $^{13}$CO(2--1) detections are indicated in blue and gray respectively.}
\end{figure*}

\subsubsection{\CI 492 GHz as a tracer of ``CO-dark'' molecular gas} 
\label{s:CI_CO_dark} 

The \CI 492 GHz components without corresponding $^{13}$CO(2--1) detections 
can be candidates for ``CO-dark'' molecular gas and require an in-depth look.
In Fig. \ref{f:CI_CO_dark}, we compare \CI 492 GHz, $^{13}$CO(2--1), H~\textsc{i}, and $^{12}$CO(1--0) 
for the five sources toward which \CI components without association with $^{13}$CO(2--1) emission are identified (indicated as gray dashed lines). 
For G08.71$-$0.41, G14.49$-$0.14, G336.96$-$0.23, and G351.57$+$0.77, 
H~\textsc{i} is seen in emission, and 2$'$-scale spectra from the Southern Galactic Plane Survey (SGPS; \citealt{McClure-Griffiths05}) are shown. 
On the contrary, absorption features are detected toward G337.92$-$0.48, 
and the H~\textsc{i} absorption spectrum from \cite{Brown14} is presented. 
Finally, for $^{12}$CO(1--0), 20$''$- and 36$''$-scale spectra 
from the FOREST Unbiased Galactic plane Imaging survey with the Nobeyama 45 m telescope (FUGIN; \citealt{Umemoto17}) 
and the Mopra southern Galactic plane survey (\citealt{Braiding18}) are used.  

\begin{figure*}[t]
\centering
\includegraphics[scale=0.45]{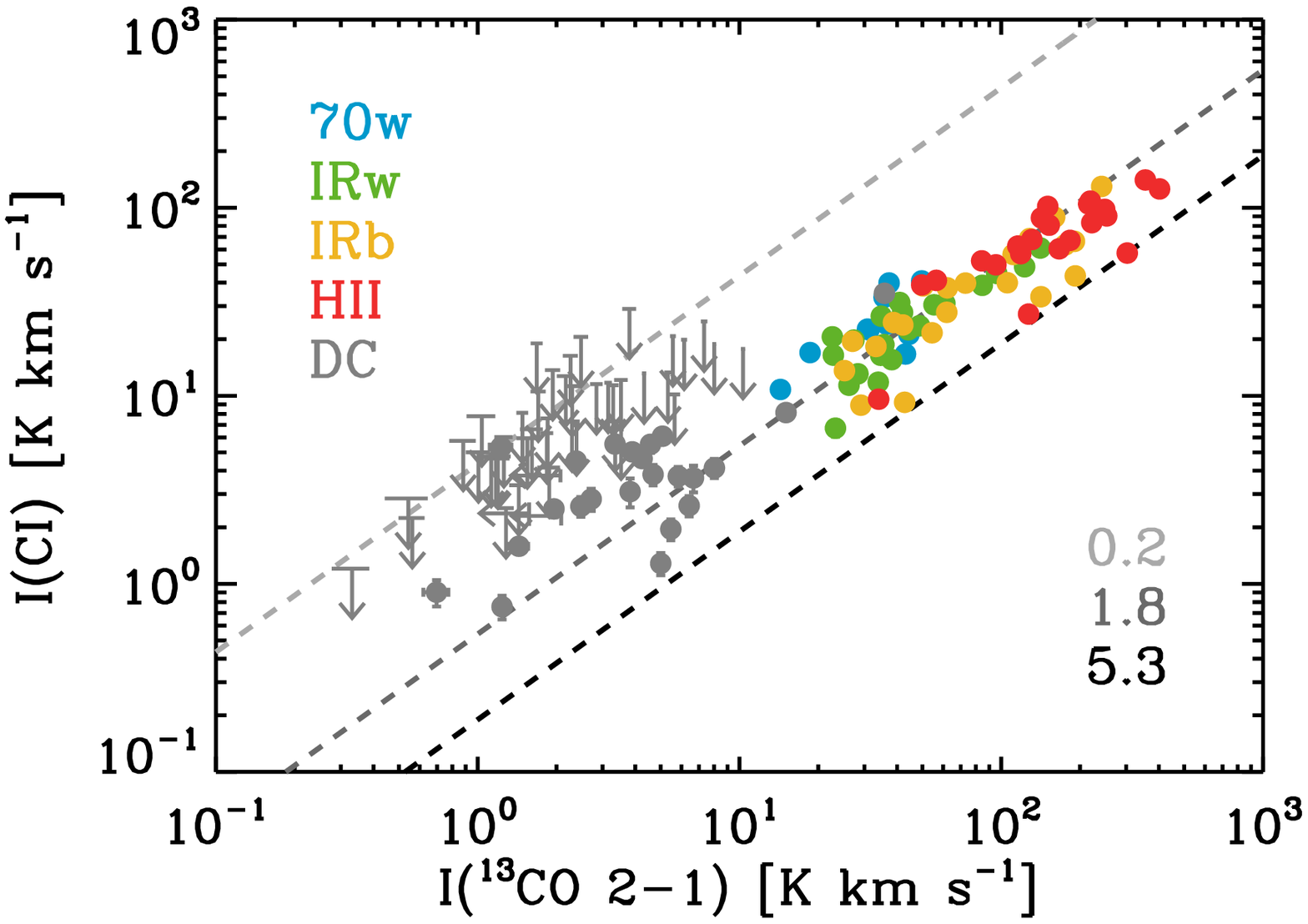}\hspace{0.5cm}
\includegraphics[scale=0.45]{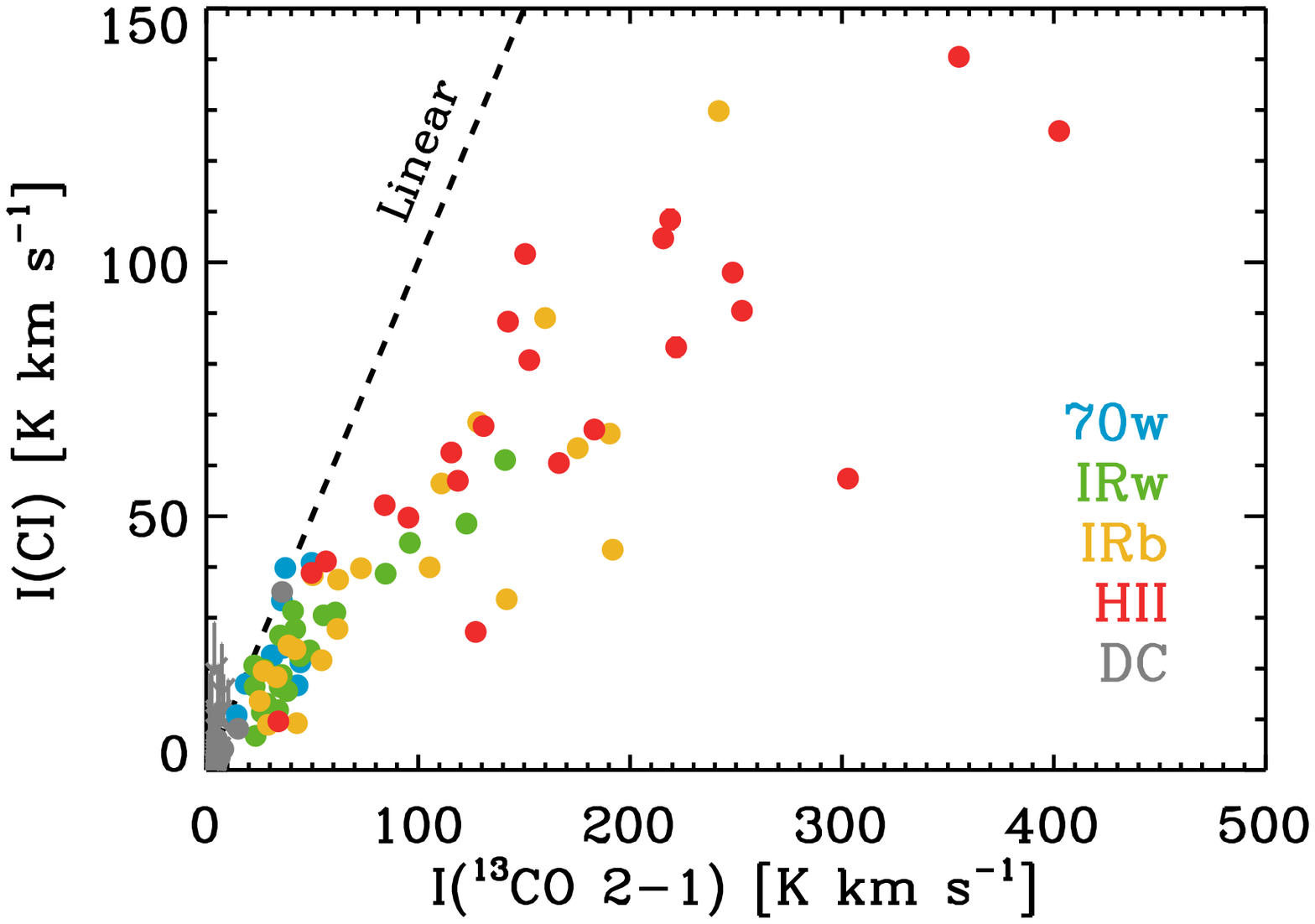}
\caption{\label{f:CI_CO_int} Comparison between the \CI 492 GHz and $^{13}$CO(2--1) integrated intensities. 
All three types (Types 0, 1, and 2; Sect \ref{s:CI_CO_class}) are used here, 
and the 70w, IRw, IRb, \HII region groups, and diffuse cloud components are shown in different colors. 
For the Type 1, upper limits were calculated by $N.C$([C~\textsc{i}]) $\times$ \CI 492 GHz 5$\sigma$ $\times$ $\Delta\varv$([C~\textsc{i}]) 
where $N.C$([C~\textsc{i}]) is the number of [C~\textsc{i}] channels over which $^{13}$CO is detected 
and $\Delta\varv$([C~\textsc{i}]) is the velocity resolution of the [C~\textsc{i}] spectra. 
A similar calculation was done for the Type 2, 
but using the number of $^{13}$CO channels over which [C~\textsc{i}] is detected, 
$^{13}$CO 5$\sigma$, and velocity resolution of the $^{13}$CO spectra instead. 
These derived upper limits are indicated as the arrows. 
\textit{Left}: log scale. 
The measured minimum, median, and maxium $I$($^{13}$CO)-to-$I$([C~\textsc{i}]) ratios (0.2, 1.8, and 5.3) 
are indicated as the dashed lines in different shades of gray color.
\textit{Right}: linear scale. 
A linear one-to-one relation is shown as the black dashed line.}
\end{figure*}

Fig. \ref{f:CI_CO_dark} shows that G08.71$-$0.41 and G14.49$-$0.14 likely suffer from 
bad reference positions in $^{13}$CO(2--1) at the velocities of our interest ($-$9 km s$^{-1}$ and 55 km s$^{-1}$ respectively), 
which means that $^{13}$CO(2--1) could have been detected. 
On the other hand, $^{13}$CO(2--1) is simply too faint at the velocities of our interest for G336.96$-$0.23, G337.92$-$0.48, and G351.57$+$0.77
($-$101 km s$^{-1}$, $-$19 km s$^{-1}$, and $-$16 km s$^{-1}$ respectively), 
implying the $I$($^{13}$CO)-to-$I$([C~\textsc{i}]) ratios lower than 0.7, 0.9, and 0.5 respectively.
For these estimates, $^{13}$CO(2--1) 5$\sigma$ values for the velocity ranges where \CI emission is integrated are considered. 
In general, Fig. \ref{f:CI_CO_dark} shows that H~\textsc{i} is abundant toward the five sources. 
In particular, for G337.92$-$0.48, the \CI component without associaiton with $^{13}$CO(2--1) emission corresponds to the strongest absorption feature, 
implying a substantial contribution from the cold \HI. 
On the contrary, $^{12}$CO(1--0) is not detected at the velocities of our interest for G336.96$-$0.23 and G337.92$-$0.48 (rms of $\sim$1--2 K).  
These results suggest that the \CI 492 GHz secondary components without corresponding $^{13}$CO(2--1) detections have a large amount of atomic gas 
and could trace ``CO-dark'' molecular gas. 
This conclusion, however, is drawn from a couple lines of sight and needs to be confirmed by extensive observations of \CI 492 GHz, 
$^{13}$CO(2--1), and $^{12}$CO(1--0) with matching resolutions and sensitivities. 

\begin{figure}
\centering
\includegraphics[scale=0.45]{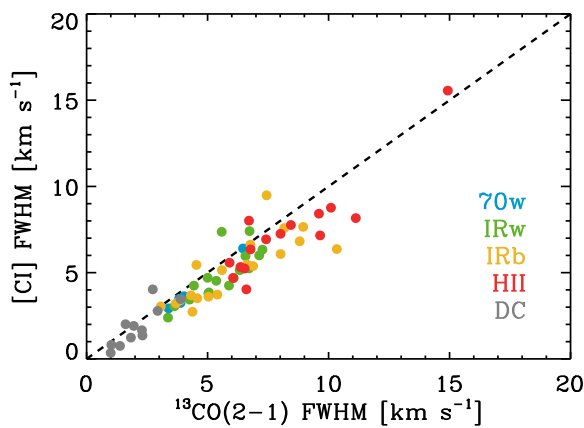}
\caption{\label{f:CI_CO_wid} \CI 492 GHz line width versus $^{13}$CO(2--1) line width. 
Only the Type 0 detections with simple spectra are shown here.}
\end{figure}

\subsubsection{Correlation between the \CI and $^{13}$CO integrated intensities} 
\label{s:CI_CO_int_corr} 

In Fig. \ref{f:CI_CO_int}, we show a comparison between the \CI 492 GHz and $^{13}$CO(2--1) integrated intensities for all three types.
To provide different perspectives, both linear and log versions are presented. 
For this comparison, we excluded 41 detections that suffer from bad reference positions in \CI 492 GHz or $^{13}$CO(2--1)  
and used arrows to mark upper limits for the Types 1 and 2.  

In general, we found that the \CI 492 GHz and $^{13}$CO(2--1) integrated intensities are well correlated. 
For instance, the Type 0 detections have a strong correlation 
between \CI 492 GHz and $^{13}$CO(2--1) with a Spearman rank correlation coefficient of 0.95. 
This suggests that \CI 492 GHz and $^{13}$CO(2--1) arise from similar regions of the interstellar medium, 
which could be supported by comparable line widths of the two transitions (Fig. \ref{f:CI_CO_wid}): 
the median of the absolute difference between the \CI 492 GHz and $^{13}$CO(2--1) line widths is 0.8 km s$^{-1}$. 
Considering a factor of two larger resolution of the $^{13}$CO(2--1) data (13$''$ versus 30$''$), 
this difference in FWHM is surprisingly small. 
Finally, both \CI 492 GHz and $^{13}$CO(2--1) are likely not too optically thick 
(Sect. \ref{s:CI_em_RADEX}; \citealt{Urquhart21}), 
contributing to the good correlation between the two transitions.

\begin{figure}
\centering
\includegraphics[scale=0.45]{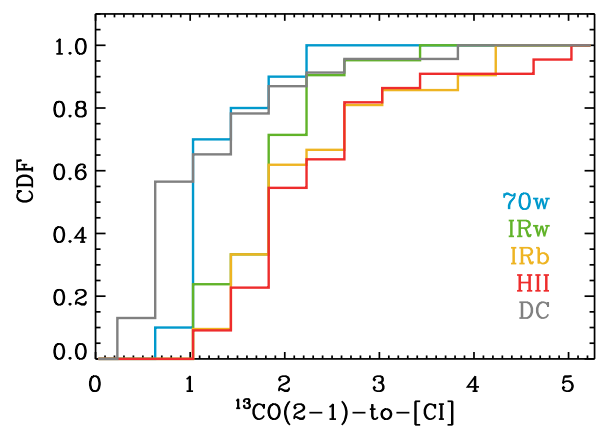}
\caption{\label{f:CI_CO_ratio_cdf} CDFs of the $I$($^{13}$CO)-to-$I$([C~\textsc{i}]) ratio.     
The 70w, IRw, IRb, \HII region groups, and diffuse cloud components are shown in blue, green, yellow, red, and gray. 
For this plot, 97 Type 0 detections with clean off positions were considered.}
\end{figure}

\subsubsection{Ratio of $^{13}$CO to [C~\textsc{i}]} 
\label{s:CO_CI_obs} 

While \CI 492 GHz and $^{13}$CO(2--1) are strongly correlated,  
the ratio of $^{13}$CO(2--1) to \CI 492 GHz does vary, e.g., 
the $I$($^{13}$CO)-to-$I$([C~\textsc{i}]) ratio ranges from 0.2 to 5.3 with a median of 1.8 (Fig. \ref{f:CI_CO_int} left).  
These variations can be seen in Fig. \ref{f:CI_CO_int} (right) as well, 
where the data points deviate from the linear one-to-one relation more significantly 
as the \CI and $^{13}$CO integrated intensities increase beyond $\sim$50 K km s$^{-1}$. 
Considering that $I$([C~\textsc{i}]) increases toward more evolved groups (Sect. \ref{s:CI_main}), 
the systematic deviation from the linear one-to-one relation implies that 
the ratio of $^{13}$CO(2--1) to [C~\textsc{i}] 492 GHz depends on the classification of the sources. 

\begin{figure*}
\centering
\includegraphics[scale=0.45]{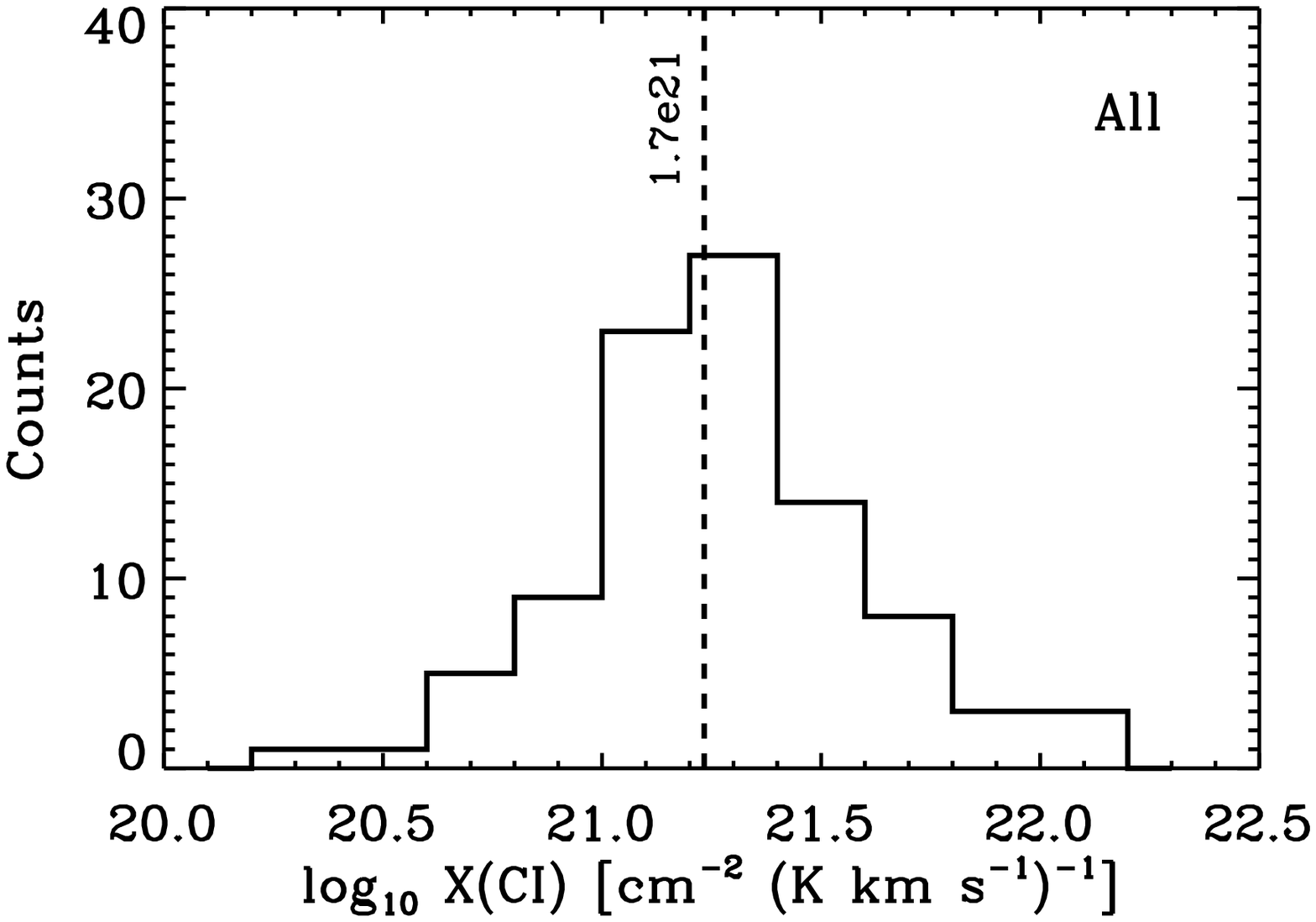}
\includegraphics[scale=0.45]{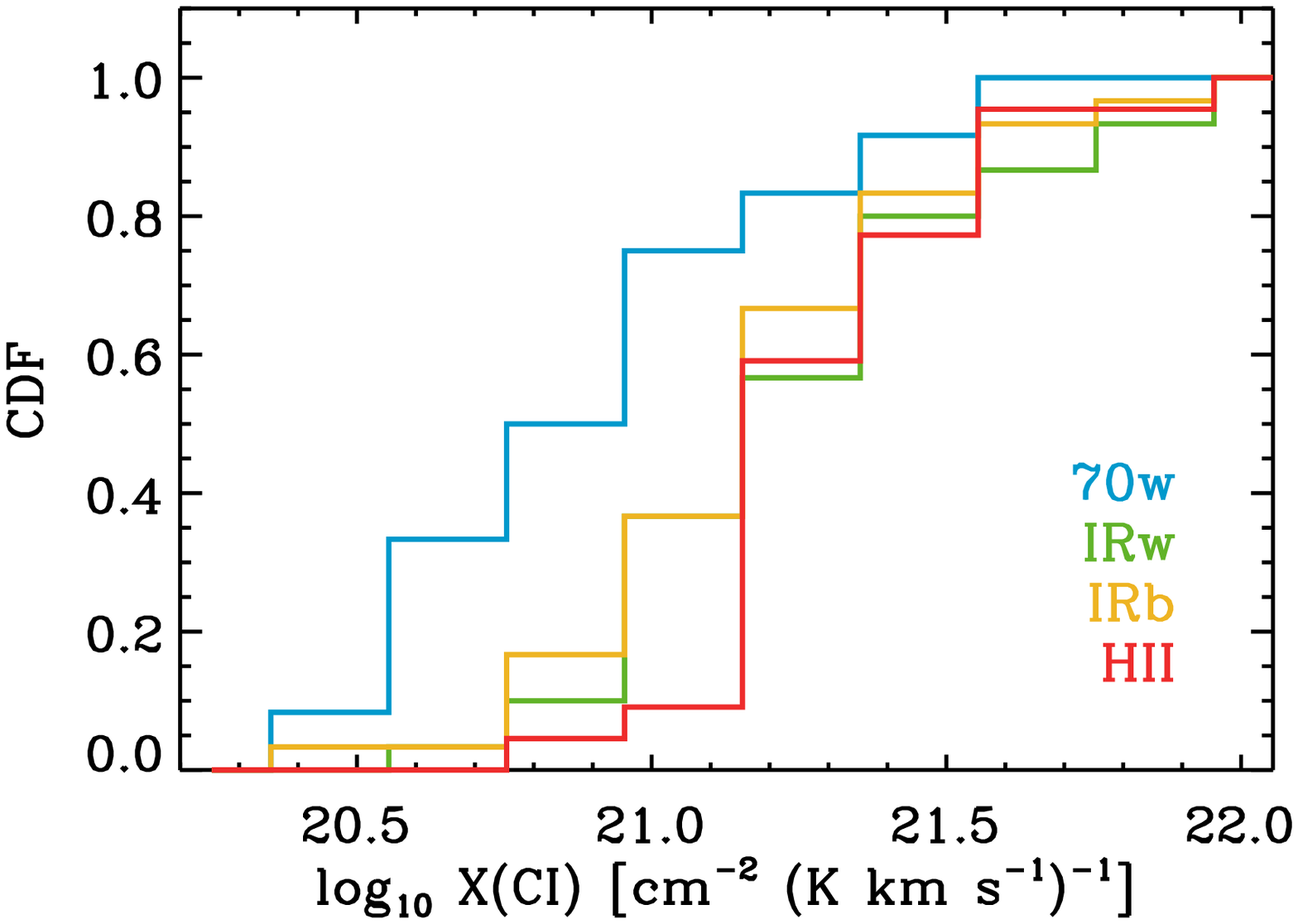} 
\caption{\label{f:XCI_dist} \textit{Left}: $X$(C~\textsc{i}) histogram. 
The median value of 1.7 $\times$ 10$^{21}$ cm$^{-2}$ (K km s$^{-1}$)$^{-1}$ is indicated as the dashed line. 
\textit{Right}: CDFs of the $X$(C~\textsc{i}) factor 
(70w, IRw, IRb, and \HII region sources in blue, green, yellow, and red respectively).}
\end{figure*}

To examine how the ratio of $^{13}$CO(2--1) to [C~\textsc{i}] 492 GHz depends on the classification of the sources, 
we constructed CDFs of $I$($^{13}$CO)/$I$([C~\textsc{i}]) 
for the 70w, IRw, IRb, \HII region, and diffuse cloud groups 
and found that the diffuse cloud and 70w groups have systematically lower ratios (Fig. \ref{f:CI_CO_ratio_cdf}). 
For example, the fraction of the sources with $I$($^{13}$CO)/$I$([C~\textsc{i}]) $<$ 2 are 
87\%, 80\%, 48\%, 57\%, and 41\% for the diffuse cloud, 70w, IRw, IRb, and \HII region groups respectively.
This conclusion is also supported by two-sided K-S tests, 
where the diffuse cloud and 70w groups are found to be drawn from the same distribution, 
while the null hypothesis is rejected at a significance level of 5\%  
when the diffuse cloud components are compared against the IRw/IRb/\HII region groups. 

The observed variations in the $^{13}$CO(2--1)-to-[C~\textsc{i}] 492 GHz integrated intensity ratio 
most likely result from local conditions of gas and 
implies that the Top100 sources are in different stages of carbon phase transition (C$^{+}$-, C$^{0}$-, or CO-dominated). 
The exact stages of carbon phase transition, as well as the underlying physical and energetic conditions, will be examined in forthcoming papers
where observations of important cooling lines ([C~\textsc{ii}], [C~\textsc{i}], [O~\textsc{i}], and CO transitions) for the Top100 sources
are analyzed using state-of-the-art models of photodissociation regions and shocks (e.g., \citealt{LePetit06}; \citealt{Flower15}).

\subsection{\CI 492 GHz and 870 $\mu$m-based H$_{2}$} 
\label{s:XCI}

\subsubsection{$X$(C~\textsc{i}) factor} 
\label{s:XCI_dist} 

In addition to $^{13}$CO(2--1), we examined a relation between \CI 492 GHz and the total H$_{2}$ column density   
by calculating the ratio of $N$(H$_{2}$) to $I$([C~\textsc{i}]) ($X$(C~\textsc{i}) factor).   
For this calculation, we extracted the 870 $\mu$m-based $N$(H$_{2}$) estimates from \cite{Koenig17} 
for the 94 main components whose \CI 492 GHz spectra have clean reference positions   
and scaled them by (13/19)$^{2}$ to compensate the difference in resolution 
(13$''$ and 19$''$ for \CI 492 GHz and H$_{2}$ respectively; this scaling assumes that H$_{2}$ uniformly fills the 19$''$ beam).  
The derived $X$(C~\textsc{i}) ranges from 2.3 $\times$ 10$^{20}$ to 1.3 $\times$ 10$^{22}$ with a median of 1.7 $\times$ 10$^{21}$ 
(in units of cm$^{-2}$ (K km s$^{-1}$)$^{-1}$; Fig. \ref{f:XCI_dist} left),  
and this median value is consistent with recent observations of the massive star-forming region RCW 38 (1.4 $\times$ 10$^{21}$; \citealt{Izumi21}), 
as well as predictions from numerical simulations ($\sim$10$^{21}$; \citealt{Offner14}; \citealt{Glover15}). 

While showing a Gaussian distribution with the well-defined peak of 1.7 $\times$ 10$^{21}$,   
$X$(C~\textsc{i}) has large variations (a factor of 55). 
To examine if there is any trend in these variations, 
we constructed CDFs for the four evolutionary groups (Fig. \ref{f:XCI_dist} right). 
The constructed CDFs show that the \HII region group has more sources with high $X$(C~\textsc{i}) values. 
For example, sources with $X$(C~\textsc{i}) $>$ 1.7 $\times$ 10$^{21}$ constitute 77\% of the \HII region group, 
while accounting for 25\%, 57\%, and 33\% of the 70w, IRw, and IRb groups. 
When examined at a significance level of 5\% for two-sided K-S tests, 
the \HII region group was found to be statistically different from the 70w and IRb groups.

\begin{figure}
\centering
\includegraphics[scale=0.45]{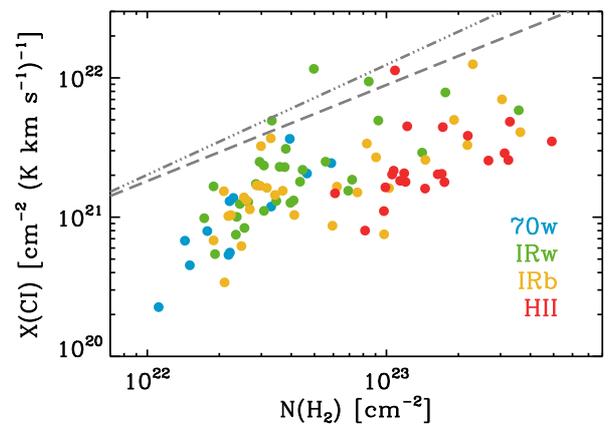} 
\caption{\label{f:XCI_H2} $X$(C~\textsc{i}) versus $N$(H$_{2}$) 
(70w, IRw, IRb, and \HII region sources in blue, green, yellow, and red).
The predicted relations, $X$(C~\textsc{i}) = 8.4 $\times$ 10$^{3}$ $N$(H$_{2}$)$^{0.79}$ ($G_{\rm UV}$ = 1 Draine field) 
and $X$(C~\textsc{i}) = 1.2 $\times$ 10$^{6}$ $N$(H$_{2}$)$^{0.69}$ ($G_{\rm UV}$ = 10 Draine field), by \cite{Offner14} are shown in gray.} 
\end{figure}

\subsubsection{Dependence of $X$(C~\textsc{i}) on $N$(H$_{2}$)}
\label{s:XCI_H2}

Sect. \ref{s:XCI_dist} showed that $N$(H$_{2}$) for a given $I$([C~\textsc{i}]) is not constant 
and varies across the evolutionary groups. 
To better understand the origin of these variations, 
we examined how $X$(C~\textsc{i}) changes with $N$(H$_{2}$) (Fig. \ref{f:XCI_H2}) 
and found that $X$(C~\textsc{i}) has a weak dependence on $N$(H$_{2}$)
(Spearman's rank correlation coefficient of 0.6;
we did not attempt to perform least-squares fitting, since proper fitting is impossible without uncertainties in the H$_{2}$ column density).
The same dependence of $X$(C~\textsc{i}) on the H$_{2}$ column density was also found by \cite{Offner14} and \cite{Izumi21}. 
In particular, \cite{Offner14} performed hydrodynamic simulations of Galactic molecular clouds 
with two different UV radiation fields ($G_{\rm UV}$ = 1 and 10 Draine fields) and 
showed that $X$(C~\textsc{i}) increases with $N$(H$_{2}$) $\gtrsim$ 10$^{22}$ cm$^{-2}$ (gray lines in Fig. \ref{f:XCI_H2}). 
This increase of $X$(C~\textsc{i}) with $N$(H$_{2}$) was interpreted as a result of \CI 492 GHz becoming optically thick 
and less sensitive to high H$_{2}$ column densities. 
In our case, however, \CI 492 GHz is likely not too optically thick (Sect. \ref{s:CI_phases}).
Our result could rather indicate the flattening of C$^{0}$ abundance at high H$_{2}$ column densities,  
which can be verified by a comprehensive excitation study based on extensive \CI 492 GHz and 809 GHz observations. 

\begin{figure}
\centering
\includegraphics[scale=0.45]{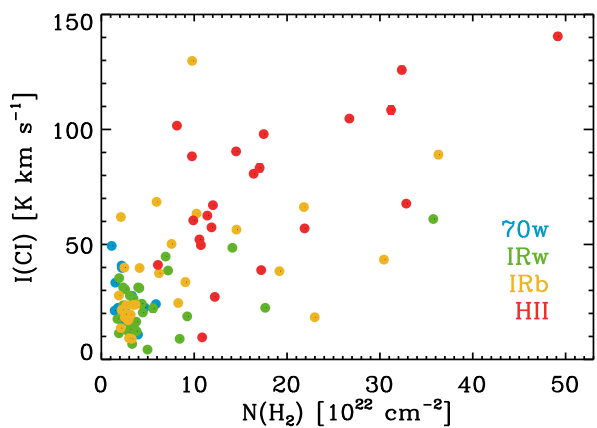} 
\caption{\label{f:CI_H2} $I$([C~\textsc{i}]) versus $N$(H$_{2}$) 
(70w, IRw, IRb, and \HII region sources in blue, green, yellow, and red).}
\end{figure}

\subsubsection{Relation between $I$([C~\textsc{i}]) and $N$(H$_{2}$)}
\label{s:CI_H2}

The observed variations in $X$(C~\textsc{i}) reflect 
a scattered relation between $I$([C~\textsc{i}]) and $N$(H$_{2}$) (Fig. \ref{f:CI_H2}), 
which is in sharp contrast to the tight correlation between $I$([C~\textsc{i}]) and $I$($^{13}$CO) (Fig. \ref{f:CI_CO_int} right).
While bearing in mind possible sources for the scatter 
(e.g., variations in the assumed properties of dust and gas, such as the dust opacity at 870 $\mu$m and dust-to-gas mass ratio), 
we take our result at face value and interpret it as a result of \CI 492 GHz 
being a tracer of moderately dense molecular gas that is more closely associated with $^{13}$CO(2--1). 
Indeed, \cite{Duarte-Cabral21} recently performed a cross-match between the ATLASGAL 870 $\mu$m and SEDIGISM $^{13}$CO(2--1) surveys 
and found that 4824 ATLASGAL clumps are enveloped within 1709 SEDIGISM clouds, 
implying that $^{13}$CO(2--1) probes the molecular gas that surrounds denser and colder clumps traced by 870 $\mu$m.  
Similarly, \cite{Plume00} and \cite{Shimajiri13} found that 
\CI 492 GHz is poorly correlated with CS(1--0) and C$^{18}$O(1--0) (dense molecular gas tracers) in Orion A. 
In general, our finding suggests that \CI 492 GHz would not be a good tracer of total gas mass 
if the cloud mass is dominated by dense and cold molecular gas.


\section{Physical conditions traced by \CI}
\label{s:CI_phases}

Two fine-structure transitions of \CI at 492 GHz and 809 GHz 
are invaluable to constrain the physical properties of gas. 
For instance, in the optically thin regime, 
their flux ratio constrains the excitation temperature ($T_{\rm ex}$) 
as the integrated intensity is proportional to the upper state column density. 
In more general situations, including high optical depth cases, 
the physical properties of gas can be derived through 
escape probability radiative transfer models such as RADEX (\citealt{vanderTak07}).
In this section, we examine the physical conditions of [C~\textsc{i}]-traced gas in a small sub-set of the Top100 sources 
by analyzing the \CI 492 GHz and 809 GHz observations
in the commonly adopted approximation of local thermodynamic equilibrium (LTE). 
Considering that what kind of the interstellar medium \CI traces is not entirely clear, 
we perform a non-LTE calculation as well to evaluate the impact of sub-critical densities.


\subsection{LTE calculation} 
\label{s:CI_em_LTE} 

As mentioned in Sect. \ref{s:obs_CI809}, we observed nine sources in \CI 809 GHz. 
To compare with the \CI 492 GHz data, we smoothed the \CI 809 GHz cubes
and extracted spectra on 13$''$ scales from the pixels that correspond to the observed positions in our \CI 492 GHz survey. 
The \CI 809 GHz emission is detected in four sources, 
and the observed peak temperatures are summarized in Table \ref{t:table4}. 

\begin{figure}
\centering
\includegraphics[scale=0.47]{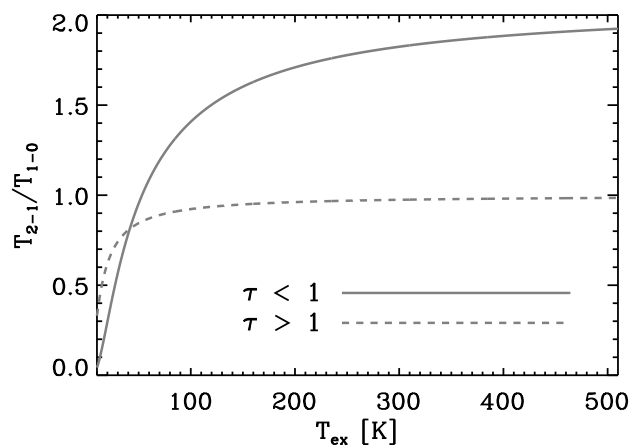} 
\caption{\label{f:CI_ratio_eq} Ratio of the 809 GHz to 492 GHz peak temperature ($T_{2-1}$/$T_{1-0}$) 
as a function of the excitation temperature ($T_{\rm ex}$). 
Optically thin and thick cases are shown in solid and dashed lines.}
\end{figure}

\begin{table*}[t]
\begin{center}
\caption{\label{t:table4} Properties of \CI 492 GHz and 809 GHz.}
\begin{tabular}{c c c c c c c c c c} \toprule \toprule  
Name & Class. & $T_{\rm peak,1-0}$ & $T_{\rm peak,2-1}$ & FWHM$_{1-0}$ & FWHM$_{2-1}$ & $T_{\rm ex}$ & $\tau_{1-0}$ & $\tau_{2-1}$ & $N$(C) \\ 
       &                & (K)                     & (K)                     & (km s$^{-1}$) & (km s$^{-1}$) & (K)          &            &              & (10$^{17}$ cm$^{-2}$) \\ \midrule
G24.63$+$0.17 & IRw & 4.4 & 9.5 $\downarrow$ & 3.0 & $-$ & 510 $\downarrow$ & 0.01 $\uparrow$ & 0.02 $\uparrow$ & 3.1 $\downarrow$ \\ 
G28.56$-$0.24 & IRw & 4.3 & 6.4 $\downarrow$ & 5.4 & $-$ & 120 $\downarrow$ & 0.04 $\uparrow$ & 0.06 $\uparrow$ & 3.5 $\downarrow$ \\ 
G337.17$-$0.03 & IRw & 3.5 & 4.3 $\downarrow$ & 7.4 & $-$ & 72 $\downarrow$ & 0.06 $\uparrow$ & 0.08 $\uparrow$ & 3.5 $\downarrow$ \\ 
G338.07$+$0.01 & 70w & 2.4 & 4.1 $\downarrow$ & 11.4 & $-$ & 196 $\downarrow$ & 0.01 $\uparrow$ & 0.02 $\uparrow$ & 5.5 $\downarrow$ \\ 
G338.78$+$0.48 & 70w & 5.1 $\uparrow$ & 6.7 $\downarrow$ & 4.9 & $-$ & $-$ & $-$ & $-$ & $-$ \\ \midrule 
G344.23$-$0.57 & IRw & 5.3 & 5.6 & 4.3 & 4.6 & 60$_{-6}^{+11}$ & 0.11$_{-0.02}^{+0.03}$ & 0.14$_{-0.01}^{+0.02}$ & 3.3$_{-0.1}^{+0.1}$ \\
G345.00$-$0.23 & \HII region & 1.9 & 3.3 & 3.9 & 6.2 & 186$_{-67}^{+107}$ & 0.01$_{-0.005}^{+0.008}$ & 0.02$_{-0.008}^{+0.009}$ & 1.6$_{-0.1}^{+0.1}$ \\
G345.49$+$0.32 & \HII region & 13.2 & 17.5 & 6.3 & 5.8 & 94$_{-3}^{+4}$ & 0.17$_{-0.01}^{+0.01}$ & 0.26$_{-0.01}^{+0.01}$ & 14.2$_{-0.1}^{+0.1}$ \\ 
G345.51$+$0.35 & IRb & 10.4 & 9.6 & 4.5 & 5.4 & 49$_{-2}^{+2}$ & 0.32$_{-0.02}^{+0.02}$ & 0.35$_{-0.01}^{+0.01}$ & 9.2$_{-0.1}^{+0.1}$ \\ \bottomrule
\end{tabular}
\end{center}
\textbf{Notes.} The columns are as follows. 
Name: ATLASGAL source name; 
Class.: source classification based on \cite{Koenig17}; 
$T_{\rm peak,1-0}$: \CI 492 GHz peak main-beam brightness temperature; 
$T_{\rm peak,2-1}$: \CI 809 GHz peak main-beam brightness temperature; 
FWHM$_{1-0}$: \CI 492 GHz line width; 
FWHM$_{2-1}$: \CI 809 GHz line width; 
$T_{\rm ex}$: excitation temperature; 
$\tau_{1-0}$: \CI 492 GHz opacity; 
$\tau_{2-1}$: \CI 809 GHz opacity; 
$N$(C): \CI 492 GHz-based column density of atomic carbon.
For the five sources where \CI 492 GHz only is detected, 
the upper and lower limits on the relevant parameters are indicated as the downward and upward arrows.
In the case of G338.78$+$0.48, the measured $T_{\rm peak,1-0}$ is most likely a lower limit,  
since its \CI 492 GHz spectrum is affected by a contaminated reference position. 
\end{table*}

Given that the observed sources are dense clumps, 
we could assume LTE conditions and derive physical conditions including  
the excitation temperature, opacities at 492 GHz and 809 GHz ($\tau_{1-0}$ and $\tau_{2-1}$), 
and atomic carbon column density ($N$(C)) (Appendix \ref{s:appendix_mc_LTE}). 
But before the LTE calculation, 
we first explored how the ratio of the 809 GHz to 492 GHz peak main-beam brightenss temperature ($T_{2-1}$/$T_{1-0}$)  
changes with the excitation temperature in two extreme cases: optically thin and thick regimes. 
When $\tau \ll 1$, Eqs. (\ref{e:appendix_mc_LTE_eq1}) and (\ref{e:appendix_mc_LTE_eq5}) suggest that $T_{2-1}$/$T_{1-0}$ can be written as  
\begin{equation}
\label{e:eqn1} 
\frac{T_{2-1}}{T_{1-0}} = 2.1~\textrm{exp}\left(\frac{-23.6}{T_{\rm{ex}}}\right)\frac{\textrm{exp}(23.6/T_{\rm ex}) - 1}{\textrm{exp}(38.9/T_{\rm ex}) - 1}\frac{1 - \textrm{exp}(-38.9/T_{\rm ex})}{1 - \textrm{exp}(-23.6/T_{\rm ex})}. 
\end{equation}

\noindent On the other hand, when $\tau \gg 1$, $T_{2-1}$/$T_{1-0}$ can be written as 
\begin{equation} 
\label{e:eqn2}
\frac{T_{2-1}}{T_{1-0}} = 1.6~\frac{\textrm{exp}(23.6/T_{\rm ex}) - 1}{\textrm{exp}(38.9/T_{\rm ex}) - 1}. 
\end{equation}

\noindent Eqs. (\ref{e:eqn1}) and (\ref{e:eqn2}) are plotted as a function of the excitation temperature in Fig. \ref{f:CI_ratio_eq}.





\begin{figure*}
\centering
\includegraphics[scale=0.52]{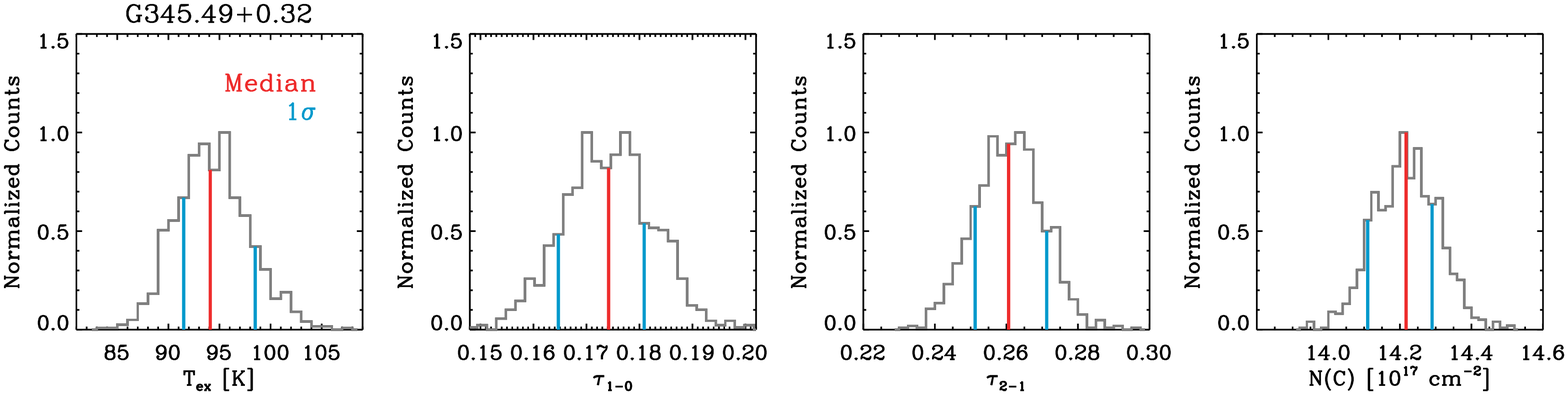} 
\includegraphics[scale=0.52]{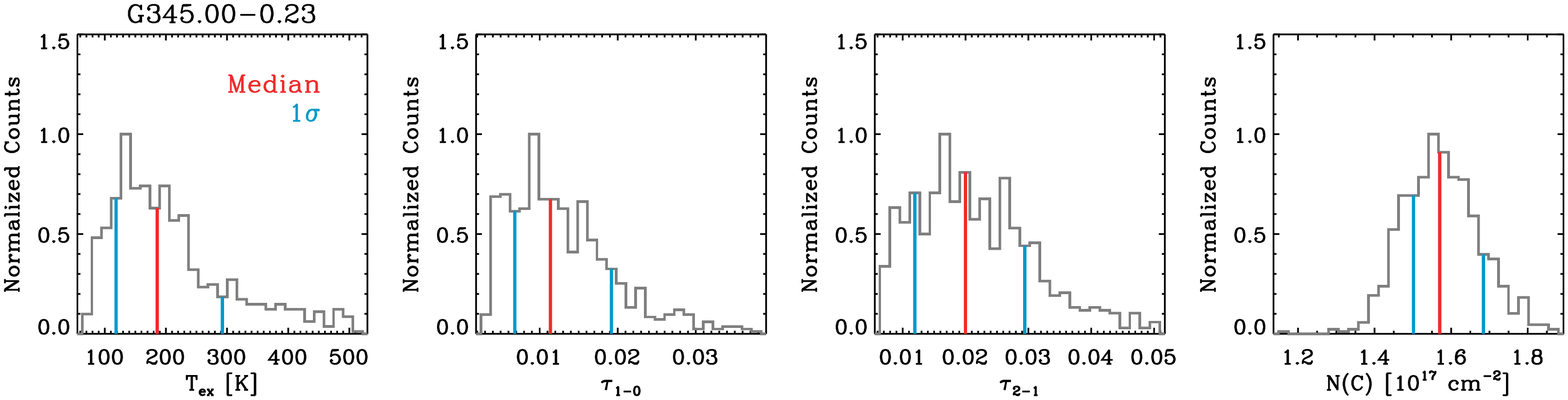} 
\caption{\label{f:MC_hist} Normalized histograms of the MC-based parameters (\textit{top}: G345.49$+$0.32; \textit{bottom}: G345.00$-$0.23).
The median values and 1$\sigma$ boundaries are indicated in red and blue.} 
\end{figure*}

\begin{figure*}
\centering
\includegraphics[scale=0.7]{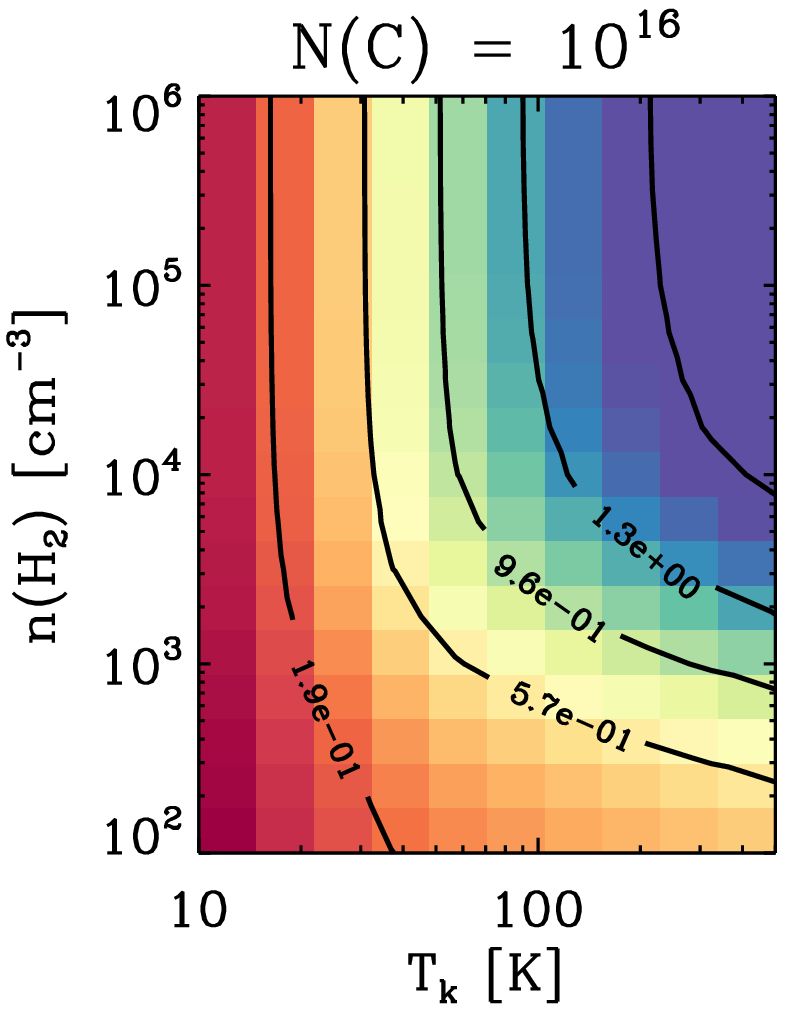}
\includegraphics[scale=0.7]{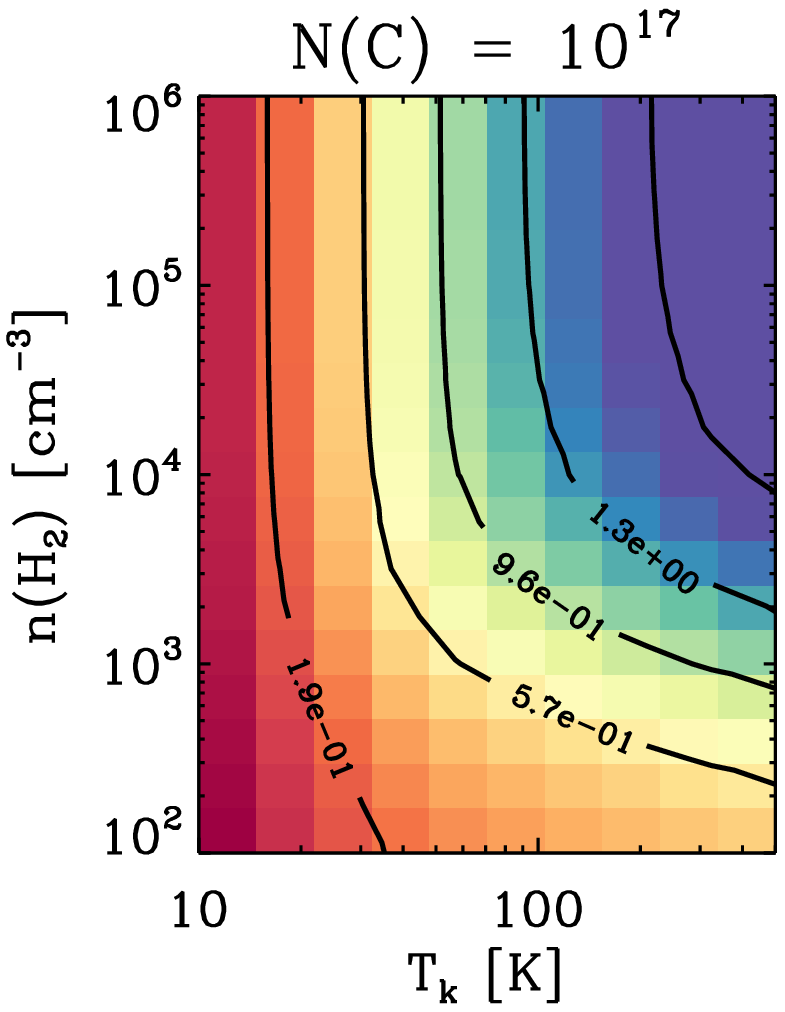}
\includegraphics[scale=0.7]{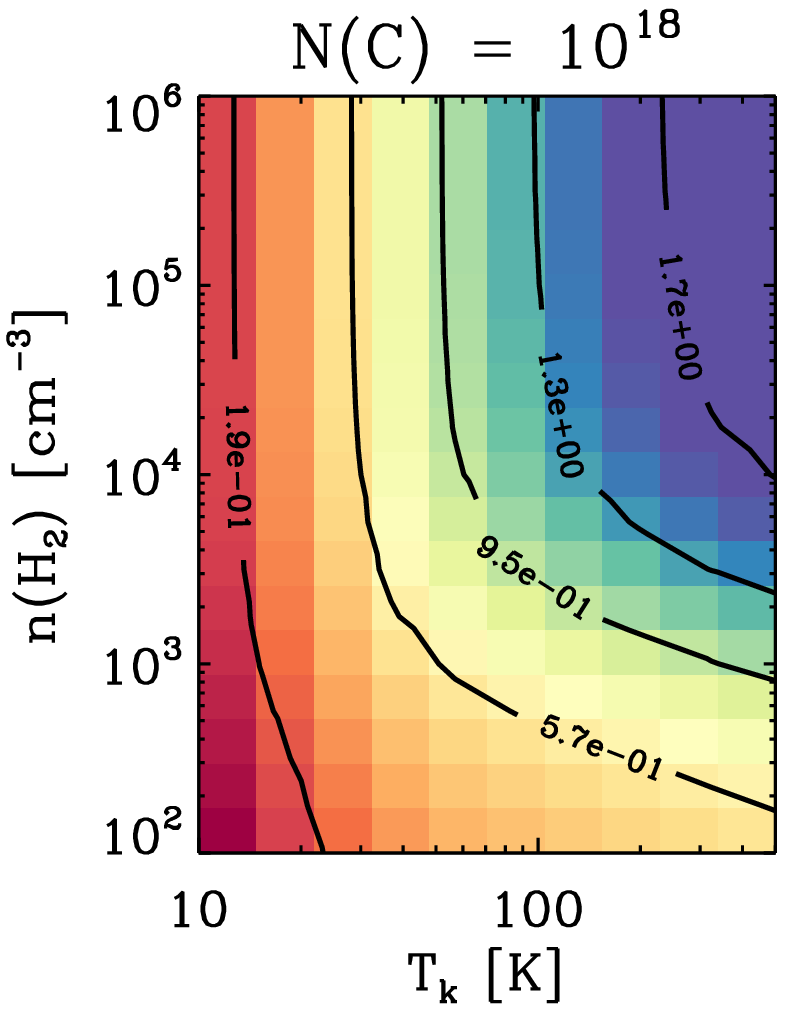}
\caption{\label{f:CI_em_RADEX} Predicted 809 GHz-to-492 GHz temperature ratios for $N$(C) = 10$^{16}$, 10$^{17}$, and 10$^{18}$ cm$^{-2}$ 
(\textit{left, middle, and right} respectively). 
In each plot, the overlaied contours range from 10\% to 90\% of the peak temperature ratio with 20\% steps.} 
\end{figure*}

Fig. \ref{f:CI_ratio_eq} shows interesting limiting behaviors. 
For example, the 809 GHz line would be twice as bright as the 492 GHz line (in temperature units) 
when the [C~\textsc{i}]-emitting gas is warm ($k_{\rm B} T_{\rm ex}$ $\gg$ $h \nu_{2-1}$) and optically thin. 
On the other hand, the 809 GHz line would be about half as bright when the gas is cool ($T_{\rm ex}$ $\sim$ 10--20 K) and optically thick. 
The observed $T_{2-1}$/$T_{1-0}$ ranges from $\sim$0.9 to $\sim$1.7 (Table \ref{t:table4}), 
suggesting that the [C~\textsc{i}]-emitting gas is likely warm and optically thin. 

To estimate $T_{\rm ex}$, $\tau_{1-0}$, $\tau_{2-1}$, $N$(C), and their associated 1$\sigma$ uncertainties, 
we performed Monte Carlo (MC) simulations where the \CI spectra are perturbed 1000 times based on the rms values (Gaussian errors are assumed). 
Each perturbed spectrum is fitted with a Gaussian function to measure the peak temperature, 
and Eqs. (\ref{e:appendix_mc_LTE_eq2}), (\ref{e:appendix_mc_LTE_eq3}), and (\ref{e:appendix_mc_LTE_eq8}) are applied to the measured temperatures 
to derive $T_{\rm ex}$, $\tau_{1-0}$, $\tau_{2-1}$, and $N$(C). 
For our calculation of $N$(C), $\tau_{1-0}$ is used since it has a higher sensitivity than $\tau_{2-1}$.
Finally, the distribution of each parameter is examined to estimate a median value, 
as well as $\pm$34\% boundaries from the median (so that the data points within the boundaries are 68\% of the total number).
These median values and 1$\sigma$ boundaries are provided in Table \ref{t:table4}, 
and the normalized histograms of the derived parameters for G345.49$+$0.32 and G340.00$-$0.23 
are shown in Fig. \ref{f:MC_hist} to demonstrate our calculation. 
We note that G345.00$-$0.23 has asymmetric distributions, leading to a large uncertainty in $T_{\rm ex}$ in particular. 
This is because the measured temperature ratio of 1.7 is on the asymptotic part of the $T_{2-1}$/$T_{1-0}$ versus $T_{\rm ex}$ curve (Fig. \ref{f:CI_ratio_eq}) 
and therefore a small change in the ratio can result in a large variation in $T_{\rm ex}$. 
For the five sources where \CI 809 GHz is not detected, 
the lower and upper limits on the physical parameters are provided in Table \ref{t:table4}.  

\cite{Zmuidzinas88} observed Galactic star-forming regions (OMC 1, NGC 2024, S140, DR 21(OH), W3, M17, and W51) in \CI 492 GHz and 809 GHz 
and performed similar excitation analyses. 
Compared to \cite{Zmuidzinas88} results (32--77 K), our excitation temperatures (49--186 K) are slightly higher. 
In particular, we found a tendency of higher temperatures for more evolved sources (94 K and 186 K for the two \HII region sources), 
though this needs to be verified by extensive \CI 809 GHz observations in the future. 
Our optical depth values are, on the other hand, lower than \cite{Zmuidzinas88}, 
i.e., $\tau_{1-0}$ $\sim$ 0.01--0.32 versus 0.22--1.25.


\begin{table*}[t]
\begin{center}
\caption{\label{t:table5} RADEX parameters for the individual sources.} 
\begin{tabular}{l c c c c} \toprule \toprule 
 & G344.23$-$0.57 & G345.00$-$0.23 & G345.49$+$0.32 & G345.51$+$0.35 \\ \midrule 
$T_{2-1}$/$T_{1-0}$$^{(a)}$ & 1.06 (0.09) & 1.74 (0.25) & 1.33 (0.02) & 0.92 (0.03) \\
$N$(C)$^{(b)}$ (10$^{17}$ cm$^{-2}$) & 3.2 & 1.8 & 10 & 10 \\ \midrule 
\multirow{3}{*}{$T_{\rm k}$$^{(c)}$ (K)} & 57--500 & 136--500 & 136--324 & 57--324 \\ 
                                         & 57 & 324 & 230 & 112 \\ 
                                         & 132 & 140 & 133 & 120 \\ \midrule 
\multirow{3}{*}{$n$(H$_{2}$)$^{(c)}$ (10$^{3}$ cm$^{-3}$)} & 1--10$^{3}$ & 3--10$^{3}$ & 3--10 & 1--10 \\
                                                           & 20 & 100 & 7 & 2 \\ 
                                                           & 300 & 300 & 5 & 4 \\ \midrule 
\multirow{3}{*}{$P/k_{\rm B}$$^{(c)}$ (10$^{5}$ K cm$^{-3}$)} & 3--600 & 20--5 $\times$ 10$^{3}$ & 10--14 & 2--6 \\ 
                                                              & 10 & 200 & 12 & 3 \\ 
                                                              & 200 & 10$^{3}$ & 2 & 1 \\ \midrule 
\multirow{3}{*}{$\tau_{1-0}$$^{(c)}$} & 0.02--0.1 & 0.002--0.02 & 0.03--0.1 & 0.1--0.3 \\ 
                                      & 0.09 & 0.006 & 0.06 & 0.2 \\ 
                                      & 0.03 & 0.005 & 0.04 & 0.1 \\ \midrule 
\multirow{3}{*}{$\tau_{2-1}$$^{(c)}$} & 0.10--0.12 & 0.008--0.03 & 0.20--0.21 & 0.36--0.39 \\ 
                                      & 0.11 & 0.02 & 0.21 & 0.37 \\ 
                                      & 0.01 & 0.01 & 0.01 & 0.01 \\ \midrule 
\multirow{3}{*}{$T_{1-0}$$^{(c)}$} & 4.0--4.4 & 1.9--2.2 & 11.9--12.3 & 11.9--13.3 \\ 
                                    & 4.1 & 2.0 & 12.1 & 12.7 \\ 
                                    & 0.1 & 0.1 & 0.3 & 0.6 \\ \midrule 
\multirow{3}{*}{$T_{2-1}$$^{(c)}$} & 4.0--4.7 & 3.2--3.7 & 15.8--16.4 & 11.0--12.5 \\ 
                                   & 4.2 & 3.5 & 16.1 & 12.0 \\ 
                                   & 0.2 & 0.2 & 0.4 & 0.6 \\ \bottomrule 
\end{tabular}
\end{center}
\textbf{Notes.} 
$^{(a)}$ The observed temperature ratio and its 1$\sigma$ uncertainty (in parentheses). 
$^{(b)}$ The input $N$(C) for RADEX that is closest to the LTE-based $N$(C).  
$^{(c)}$ The RADEX parameters that are found to reproduce the observed temperature ratios within 1$\sigma$ uncertainties 
($P/k_{\rm B}$ = thermal pressure).  
The ranges, median values, and 1$\sigma$ uncertainties are presented in rows. 
\end{table*}

\subsection{Non-LTE consideration}
\label{s:CI_em_RADEX} 

While there is certainly a significant amount of dense gas in the observed massive clumps, 
whether the \CI lines arise from this dense gas or 
from the lower density gas that surrounds the dense clumps is currently not clear. 
Since our previous calculations are based on the LTE assumption, 
it is therefore important to examine the impact of sub-critical densities on the derived parameters.

To this end, we employed RADEX (\citealt{vanderTak07}),  
where the radiative transfer equation for isothermal and homogeneous gas is solved based on the escape probability method
to compute the intensities of atomic and molecular transitions. 
To cover a wide range of physical conditions, 
we created a RADEX model grid with the following parameters: 
kinetic temperature $T_{\rm k}$ = 10--500 K, H$_{2}$ density $n$(H$_{2}$) = 10$^{2}$--10$^{6}$ cm$^{-3}$, 
and C$^{0}$ column density $N$(C) = 10$^{16}$--10$^{18}$ cm$^{-2}$ 
(uniformly sampled in log space with ten, seventeen, and nine points). 
In addition to these key parameters, the line width of 5 km s$^{-1}$ 
(median of the measured FWHM$_{1-0}$; Table \ref{t:table4}) 
and the background radiation of 2.73 K were assumed. 
To illustrate the RADEX modeling results, 
we present the predicted $T_{2-1}$/$T_{1-0}$ for $N$(C) = 10$^{16}$, 10$^{17}$, and 10$^{18}$ cm$^{-2}$ in Fig. \ref{f:CI_em_RADEX}. 


Fig. \ref{f:CI_em_RADEX} illustrates a well-known degeneracy between the temperature and density: 
high density/low temperature and low density/high temperature combinations produce the same line ratio. 
To reproduce the observed $T_{2-1}$/$T_{1-0}$ of 0.9--1.7, 
$T_{\rm k}$ $\gtrsim$ 60 K and $n$(H$_{2}$) $\gtrsim$ 10$^{3}$ cm$^{-3}$ 
are required, mostly ensuring LTE 
(densities lower than $\sim$10$^{4}$ cm$^{-3}$ mainly result in a difference between the kinetic and excitation temperatures). 
To explore the RADEX predictions more closely, we then selected the $N$(C) values that are closest to our LTE-based estimates 
and extracted the physical parameters that reproduce the observed line ratios within 1$\sigma$ uncertainties (Table \ref{t:table5}). 
Our results show that the H$_{2}$ density, kinetic temperature, and thermal pressure are poorly constrained,   
while the optical depths are relatively well constrained and  not too different from our LTE values. 
The large variations in $T_{\rm k}$ and $n$(H$_{2}$) are mostly driven by the density-temperature degeneracy
and hinder us from inferring any systematic variation across the different evolutionary groups. 
On the other hand, the predicted brightness temperatures are comparable to the observed values. 
All these results imply that our LTE-based optical depths and C$^{0}$ column densities are not heavily affected by sub-critical densities 
and \CI emission arises from warm ($T_{\rm k}$ $\gtrsim$ 60 K), optically thin ($\tau$ < 0.5), and 
highly pressurized regions ($P/k_{\rm B}$ $\sim$ a few (10$^{5}$--10$^{8}$) K cm$^{-3}$) in several of the Top100 sources. 

\section{Summary} 
\label{s:summary} 

In this paper, we present APEX observations of \CI 492 GHz toward ATLASGAL-selected massive clumps in the inner Galaxy (Top100 sample).  
Our target sources have been extensively examined in dust continuum and spectral lines 
and constitute a representative sample of high-mass star-forming regions 
covering a significant range of masses ($\sim$20--10$^{4}$ $M_{\odot}$), bolometric luminosities ($\sim$60--10$^{6}$ $L_{\odot}$), 
and evolutionary phases (70 $\mu$m weak, IR-weak, IR-bright, and \HII regions). 
To examine the physical conditions of [C~\textsc{i}]-traced gas in the Top100 sources, 
we combined our APEX \CI 492 GHz spectra with multi-wavelength observations of atomic and molecular gas and obtained the following results.  

\begin{enumerate}
\item All 98 sources are detected in \CI 492 GHz emission. 
In addition, 41 emission components and two absorption components are seen toward 28 and two lines of sight respectively as secondary components. 
These results imply that \CI 492 GHz is prevalent in the inner Galaxy and traces not only massive clumps, but also non-star-forming relatively diffuse gas. 

\item The observed \CI 492 GHz integrated intensities and line widths tend to increase toward evolved stages of star formation. 
In comparison to these main components that are associated with the Top100 sample,  
the secondary emission components have systematically lower \CI 492 GHz intensities and FWHMs. 

\item Among 139 components that are seen in \CI 492 GHz emisison, 133 have clear $^{13}$CO(2--1) detections, 
implying a good correlation between \CI 492 GHz and $^{13}$CO(2--1) in general (Spearman rank correlation coefficient of 0.95). 
The ratio of the $^{13}$CO(2--1) to \CI 492 GHz integrated intensity changes from 0.2 to 5.3 though, 
likely driven by local conditions of the interstellar medium.

\item For the secondary components, 45 are detected in $^{13}$CO(2--1) emission only, 
while five are visible in \CI 492 GHz emission only.
This contrast in the detection statistics is partially due to the difference in sensivitiy for the \CI 492 GHz and $^{13}$CO(2--1) data. 
On the other hand, we found that the \CI 492 GHz diffuse cloud components without corresponding $^{13}$CO(2--1) emission 
contain a significant amount of atomic gas and could trace ``CO-dark'' molecular gas. 
This conclusion, however, is based on few lines of sight
and needs to be verified by sensitive \CI 492 GHz, $^{13}$CO(2--1) and $^{12}$CO(1--0) observations in the future. 

\item We derived the $X$(C~\textsc{i}) factors by dividing the 870 $\mu$m-based H$_{2}$ column densities by the \CI 492 GHz integrated intensities 
and found that $X$(C~\textsc{i}) (in units of cm$^{-2}$ (K km s$^{-1}$)$^{-1}$) 
ranges from 2.3 $\times$ 10$^{20}$ to 1.3 $\times$ 10$^{22}$ with a median of 1.7 $\times$ 10$^{21}$.  
Our median value is in good agreement with recent observations of the massive star-forming region RCW 38 (1.4 $\times$ 10$^{21}$), 
as well as predictions from numerical simulations of Galactic molecular clouds ($\sim$10$^{21}$).

\item In contrast to the tight correlation with $^{13}$CO(2--1), 
\CI 492 GHz shows a much more scattered relation with the 870 $\mu$m-traced molecular gas. 
We interpreted this as \CI 492 GHz and $^{13}$CO(2--1) probing warm molecular gas 
that surrounds denser and colder clumps traced by 870 $\mu$m
and cautioned not to use \CI 492 GHz as a tracer of total gas mass if dense and cold molecular gas dominates the cloud mass.

\item our LTE and non-LTE analyses showed that \CI emission arises from warm ($T_{\rm k}$ $\gtrsim$ 60 K), optically thin ($\tau$ $<$ 0.5), 
and highly pressurized regions ($P/k_{\rm B}$ $\sim$ a few (10$^{5}$--10$^{8}$) K cm$^{-3}$) in several of the Top100 sources. 


\end{enumerate}

Based on a large sample of sensitive \CI 492 GHz spectra, 
this paper investigated the physical properties of [C~\textsc{i}]-traced gas in a wide range of environments 
and shedded light on the origin of \CI 492 GHz emission by probing the relation with molecular gas tracers. 
One of our key results is the systematic variation in the ratio of $^{13}$CO(2--1) to \CI 492 GHz, 
which implies that the Top100 sources are in different stages of carbon phase transition (C$^{+}$-, C$^{0}$-, or CO-dominated).  
We will examine the exact stages of carbon phase transition and the underlying physical and energetic conditions in forthcoming papers 
where [C~\textsc{ii}], [C~\textsc{i}], [O~\textsc{i}], and CO observations from our multi-telescope campaign
are analyzed using state-of-the-art models of photodissociation regions and shocks (e.g., \citealt{LePetit06}; \citealt{Flower15}). 

\begin{acknowledgements} 
We would like to thank the anonymous referee for constructive comments that improved this work. 
We also thank A. Gusdorf and M. Gerin for providing the PRISMAS data and F. Le Petit for helpful comments.  
M.-Y.L. was partially supported through the sub-project of A6 of the Collaborative Research Council 956, 
funded by the Deutsche Forschungsgemeinschaft (DFG).
\end{acknowledgements}

\bibliographystyle{aa}
\bibliography{/media/mlee/Bibtex/myref}

\begin{thebibliography}{42}
\expandafter\ifx\csname natexlab\endcsname\relax\def\natexlab#1{#1}\fi

\bibitem[{{Braiding} {et~al.}(2018){Braiding}, {Wong}, {Maxted}, {Romano},
  {Burton}, {Blackwell}, {Filipovi{\'c}}, {Freeman}, {Indermuehle}, {Lau},
  {Rebolledo}, {Rowell}, {Snoswell}, {Tothill}, {Voisin}, \& {de
  Wilt}}]{Braiding18}
{Braiding}, C., {Wong}, G.~F., {Maxted}, N.~I., {et~al.} 2018, \pasa, 35, e029

\bibitem[{{Brown} {et~al.}(2014){Brown}, {Dickey}, {Dawson}, \&
  {McClure-Griffiths}}]{Brown14}
{Brown}, C., {Dickey}, J.~M., {Dawson}, J.~R., \& {McClure-Griffiths}, N.~M.
  2014, \apjs, 211, 29

\bibitem[{{Contreras} {et~al.}(2013){Contreras}, {Schuller}, {Urquhart},
  {Csengeri}, {Wyrowski}, {Beuther}, {Bontemps}, {Bronfman}, {Henning},
  {Menten}, {Schilke}, {Walmsley}, {Wienen}, {Tackenberg}, \&
  {Linz}}]{Contreras13}
{Contreras}, Y., {Schuller}, F., {Urquhart}, J.~S., {et~al.} 2013, \aap, 549,
  A45

\bibitem[{{Cox}(1981)}]{Cox81}
{Cox}, D.~P. 1981, \apj, 245, 534

\bibitem[{{Csengeri} {et~al.}(2016){Csengeri}, {Leurini}, {Wyrowski},
  {Urquhart}, {Menten}, {Walmsley}, {Bontemps}, {Wienen}, {Beuther}, {Motte},
  {Nguyen-Luong}, {Schilke}, {Schuller}, {Zavagno}, \& {Sanna}}]{Csengeri16}
{Csengeri}, T., {Leurini}, S., {Wyrowski}, F., {et~al.} 2016, \aap, 586, A149

\bibitem[{{Csengeri} {et~al.}(2014){Csengeri}, {Urquhart}, {Schuller}, {Motte},
  {Bontemps}, {Wyrowski}, {Menten}, {Bronfman}, {Beuther}, {Henning}, {Testi},
  {Zavagno}, \& {Walmsley}}]{Csengeri14}
{Csengeri}, T., {Urquhart}, J.~S., {Schuller}, F., {et~al.} 2014, \aap, 565,
  A75

\bibitem[{{Duarte-Cabral} {et~al.}(2021){Duarte-Cabral}, {Colombo}, {Urquhart},
  {Ginsburg}, {Russeil}, {Schuller}, {Anderson}, {Barnes}, {Beltr{\'a}n},
  {Beuther}, {Bontemps}, {Bronfman}, {Csengeri}, {Dobbs}, {Eden}, {Giannetti},
  {Kauffmann}, {Mattern}, {Medina}, {Menten}, {Lee}, {Pettitt}, {Riener},
  {Rigby}, {Traficante}, {Veena}, {Wienen}, {Wyrowski}, {Agurto}, {Azagra},
  {Cesaroni}, {Finger}, {Gonzalez}, {Henning}, {Hernandez}, {Kainulainen},
  {Leurini}, {Lopez}, {Mac-Auliffe}, {Mazumdar}, {Molinari}, {Motte}, {Muller},
  {Nguyen-Luong}, {Parra}, {Perez-Beaupuits}, {Montenegro-Montes}, {Moore},
  {Ragan}, {S{\'a}nchez-Monge}, {Sanna}, {Schilke}, {Schisano}, {Schneider},
  {Suri}, {Testi}, {Torstensson}, {Venegas}, {Wang}, \&
  {Zavagno}}]{Duarte-Cabral21}
{Duarte-Cabral}, A., {Colombo}, D., {Urquhart}, J.~S., {et~al.} 2021, \mnras,
  500, 3027

\bibitem[{{Flower} \& {Pineau des For{\^e}ts}(2015)}]{Flower15}
{Flower}, D.~R. \& {Pineau des For{\^e}ts}, G. 2015, \aap, 578, A63

\bibitem[{{Frerking} {et~al.}(1989){Frerking}, {Keene}, {Blake}, \&
  {Phillips}}]{Frerking89}
{Frerking}, M.~A., {Keene}, J., {Blake}, G.~A., \& {Phillips}, T.~G. 1989,
  \apj, 344, 311

\bibitem[{{Giannetti} {et~al.}(2014){Giannetti}, {Wyrowski}, {Brand},
  {Csengeri}, {Fontani}, {Walmsley}, {Nguyen Luong}, {Beuther}, {Schuller},
  {G{\"u}sten}, \& {Menten}}]{Giannetti14}
{Giannetti}, A., {Wyrowski}, F., {Brand}, J., {et~al.} 2014, \aap, 570, A65

\bibitem[{{Glover} {et~al.}(2015){Glover}, {Clark}, {Micic}, \&
  {Molina}}]{Glover15}
{Glover}, S.~C.~O., {Clark}, P.~C., {Micic}, M., \& {Molina}, F. 2015, \mnras,
  448, 1607

\bibitem[{{G{\"u}sten} {et~al.}(2008){G{\"u}sten}, {Baryshev}, {Bell},
  {Belloche}, {Graf}, {Hafok}, {Heyminck}, {Hochg{\"u}rtel}, {Honingh},
  {Jacobs}, {Kasemann}, {Klein}, {Klein}, {Korn}, {Kr{\"a}mer}, {Leinz},
  {Lundgren}, {Menten}, {Meyer}, {Muders}, {Pacek}, {Rabanus}, {Sch{\"a}fer},
  {Schilke}, {Schneider}, {Stutzki}, {Wieching}, {Wunsch}, \&
  {Wyrowski}}]{Gusten08}
{G{\"u}sten}, R., {Baryshev}, A., {Bell}, A., {et~al.} 2008, in Society of
  Photo-Optical Instrumentation Engineers (SPIE) Conference Series, Vol. 7020,
  Millimeter and Submillimeter Detectors and Instrumentation for Astronomy IV,
  ed. W.~D. {Duncan}, W.~S. {Holland}, S.~{Withington}, \& J.~{Zmuidzinas},
  702010

\bibitem[{{Henning} \& {Salama}(1998)}]{Henning98}
{Henning}, T. \& {Salama}, F. 1998, Science, 282, 2204

\bibitem[{{Hopkins} {et~al.}(2014){Hopkins}, {Kere{\v{s}}}, {O{\~n}orbe},
  {Faucher-Gigu{\`e}re}, {Quataert}, {Murray}, \& {Bullock}}]{Hopkins14}
{Hopkins}, P.~F., {Kere{\v{s}}}, D., {O{\~n}orbe}, J., {et~al.} 2014, \mnras,
  445, 581

\bibitem[{{Izumi} {et~al.}(2021){Izumi}, {Fukui}, {Tachihara}, {Fujita},
  {Torii}, {Kamazaki}, {Kaneko}, {Silva}, {Iono}, {Momose}, {Sugimoto},
  {Nakazato}, {Kosugi}, {Maekawa}, {Takahashi}, {Yoshino}, \&
  {Asayama}}]{Izumi21}
{Izumi}, N., {Fukui}, Y., {Tachihara}, K., {et~al.} 2021, \pasj, 73, 174

\bibitem[{{Keene} {et~al.}(1997){Keene}, {Lis}, {Phillips}, \&
  {Schilke}}]{Keene97}
{Keene}, J., {Lis}, D.~C., {Phillips}, T.~G., \& {Schilke}, P. 1997, IAU
  Symposium, 178, 129

\bibitem[{{Kennicutt} \& {Evans}(2012)}]{Kennicutt12}
{Kennicutt}, R.~C. \& {Evans}, N.~J. 2012, \araa, 50, 531

\bibitem[{{Kim} {et~al.}(2017){Kim}, {Wyrowski}, {Urquhart}, {Menten}, \&
  {Csengeri}}]{Kim17}
{Kim}, W.~J., {Wyrowski}, F., {Urquhart}, J.~S., {Menten}, K.~M., \&
  {Csengeri}, T. 2017, \aap, 602, A37

\bibitem[{Klein {et~al.}(2014)Klein, Ciechanowicz, Leinz, Heyminck, Güsten,
  Kasemann, Wunsch, Maier, \& Sekimoto}]{Klein14}
Klein, T., Ciechanowicz, M., Leinz, C., {et~al.} 2014, IEEE Transactions on
  Terahertz Science and Technology, 4, 588

\bibitem[{{K{\"o}nig} {et~al.}(2017){K{\"o}nig}, {Urquhart}, {Csengeri},
  {Leurini}, {Wyrowski}, {Giannetti}, {Wienen}, {Pillai}, {Kauffmann},
  {Menten}, \& {Schuller}}]{Koenig17}
{K{\"o}nig}, C., {Urquhart}, J.~S., {Csengeri}, T., {et~al.} 2017, \aap, 599,
  A139

\bibitem[{{Kramer} {et~al.}(2008){Kramer}, {Cubick}, {R{\"o}llig}, {Sun},
  {Yonekura}, {Aravena}, {Bensch}, {Bertoldi}, {Bronfman}, {Fujishita},
  {Fukui}, {Graf}, {Hitschfeld}, {Honingh}, {Ito}, {Jakob}, {Jacobs}, {Klein},
  {Koo}, {May}, {Miller}, {Miyamoto}, {Mizuno}, {Onishi}, {Park}, {Pineda},
  {Rabanus}, {Sasago}, {Schieder}, {Simon}, {Stutzki}, {Volgenau}, \&
  {Yamamoto}}]{Kramer08}
{Kramer}, C., {Cubick}, M., {R{\"o}llig}, M., {et~al.} 2008, \aap, 477, 547

\bibitem[{{Langer}(1976)}]{Langer76}
{Langer}, W. 1976, \apj, 206, 699

\bibitem[{{Le Petit} {et~al.}(2006){Le Petit}, {Nehm{\'e}}, {Le Bourlot}, \&
  {Roueff}}]{LePetit06}
{Le Petit}, F., {Nehm{\'e}}, C., {Le Bourlot}, J., \& {Roueff}, E. 2006, \apjs,
  164, 506

\bibitem[{{Lee} {et~al.}(2019){Lee}, {Madden}, {Le Petit}, {Gusdorf},
  {Lesaffre}, {Wu}, {Lebouteiller}, {Galliano}, \& {Chevance}}]{Lee19}
{Lee}, M.~Y., {Madden}, S.~C., {Le Petit}, F., {et~al.} 2019, \aap, 628, A113

\bibitem[{{Matteucci}(2021)}]{Matteucci21}
{Matteucci}, F. 2021, \aapr, 29, 5

\bibitem[{{McClure-Griffiths} {et~al.}(2005){McClure-Griffiths}, {Dickey},
  {Gaensler}, {Green}, {Haverkorn}, \& {Strasser}}]{McClure-Griffiths05}
{McClure-Griffiths}, N.~M., {Dickey}, J.~M., {Gaensler}, B.~M., {et~al.} 2005,
  \apjs, 158, 178

\bibitem[{{Motte} {et~al.}(2018){Motte}, {Bontemps}, \& {Louvet}}]{Motte18}
{Motte}, F., {Bontemps}, S., \& {Louvet}, F. 2018, \araa, 56, 41

\bibitem[{{Offner} {et~al.}(2014){Offner}, {Bisbas}, {Bell}, \&
  {Viti}}]{Offner14}
{Offner}, S.~S.~R., {Bisbas}, T.~G., {Bell}, T.~A., \& {Viti}, S. 2014, \mnras,
  440, L81

\bibitem[{{Oka} {et~al.}(2001){Oka}, {Yamamoto}, {Iwata}, {Maezawa}, {Ikeda},
  {Ito}, {Kamegai}, {Sakai}, {Sekimoto}, {Tatematsu}, {Arikawa}, {Aso},
  {Noguchi}, {Shi}, {Miyazawa}, {Saito}, {Ozeki}, {Fujiwara}, {Ohishi}, \&
  {Inatani}}]{Oka01}
{Oka}, T., {Yamamoto}, S., {Iwata}, M., {et~al.} 2001, \apj, 558, 176

\bibitem[{{Ostriker} {et~al.}(2010){Ostriker}, {McKee}, \&
  {Leroy}}]{Ostriker10}
{Ostriker}, E.~C., {McKee}, C.~F., \& {Leroy}, A.~K. 2010, \apj, 721, 975

\bibitem[{{Plume} {et~al.}(2000){Plume}, {Bensch}, {Howe}, {Ashby}, {Bergin},
  {Chin}, {Erickson}, {Goldsmith}, {Harwit}, {Kleiner}, {Koch}, {Neufeld},
  {Patten}, {Schieder}, {Snell}, {Stauffer}, {Tolls}, {Wang}, {Winnewisser},
  {Zhang}, {Reynolds}, {Joyce}, {Tavoletti}, {Jack}, {Rodkey}, \&
  {Melnick}}]{Plume00}
{Plume}, R., {Bensch}, F., {Howe}, J.~E., {et~al.} 2000, \apjl, 539, L133

\bibitem[{{Pon} {et~al.}(2014){Pon}, {Johnstone}, {Kaufman}, {Caselli}, \&
  {Plume}}]{Pon14}
{Pon}, A., {Johnstone}, D., {Kaufman}, M.~J., {Caselli}, P., \& {Plume}, R.
  2014, \mnras, 445, 1508

\bibitem[{{Schilke} {et~al.}(1995){Schilke}, {Keene}, {Le Bourlot}, {Pineau des
  Forets}, \& {Roueff}}]{Schilke95}
{Schilke}, P., {Keene}, J., {Le Bourlot}, J., {Pineau des Forets}, G., \&
  {Roueff}, E. 1995, \aap, 294, L17

\bibitem[{{Schuller} {et~al.}(2009){Schuller}, {Menten}, {Contreras},
  {Wyrowski}, {Schilke}, {Bronfman}, {Henning}, {Walmsley}, {Beuther},
  {Bontemps}, {Cesaroni}, {Deharveng}, {Garay}, {Herpin}, {Lefloch}, {Linz},
  {Mardones}, {Minier}, {Molinari}, {Motte}, {Nyman}, {Reveret}, {Risacher},
  {Russeil}, {Schneider}, {Testi}, {Troost}, {Vasyunina}, {Wienen}, {Zavagno},
  {Kovacs}, {Kreysa}, {Siringo}, \& {Wei{\ss}}}]{Schuller09}
{Schuller}, F., {Menten}, K.~M., {Contreras}, Y., {et~al.} 2009, \aap, 504, 415

\bibitem[{{Shimajiri} {et~al.}(2013){Shimajiri}, {Sakai}, {Tsukagoshi},
  {Kitamura}, {Momose}, {Saito}, {Oshima}, {Kohno}, \& {Kawabe}}]{Shimajiri13}
{Shimajiri}, Y., {Sakai}, T., {Tsukagoshi}, T., {et~al.} 2013, \apjl, 774, L20

\bibitem[{{Silk}(1997)}]{Silk97}
{Silk}, J. 1997, \apj, 481, 703

\bibitem[{{Tang} {et~al.}(2018){Tang}, {Henkel}, {Wyrowski}, {Giannetti},
  {Menten}, {Csengeri}, {Leurini}, {Urquhart}, {K{\"o}nig}, {G{\"u}sten},
  {Lin}, {Zheng}, {Esimbek}, \& {Zhou}}]{Tang18}
{Tang}, X.~D., {Henkel}, C., {Wyrowski}, F., {et~al.} 2018, \aap, 611, A6

\bibitem[{{Umemoto} {et~al.}(2017){Umemoto}, {Minamidani}, {Kuno}, {Fujita},
  {Matsuo}, {Nishimura}, {Torii}, {Tosaki}, {Kohno}, {Kuriki}, {Tsuda},
  {Hirota}, {Ohashi}, {Yamagishi}, {Handa}, {Nakanishi}, {Omodaka}, {Koide},
  {Matsumoto}, {Onishi}, {Tokuda}, {Seta}, {Kobayashi}, {Tachihara}, {Sano},
  {Hattori}, {Onodera}, {Oasa}, {Kamegai}, {Tsuboi}, {Sofue}, {Higuchi},
  {Chibueze}, {Mizuno}, {Honma}, {Muller}, {Inoue}, {Morokuma-Matsui},
  {Shinnaga}, {Ozawa}, {Takahashi}, {Yoshiike}, {Costes}, \&
  {Kuwahara}}]{Umemoto17}
{Umemoto}, T., {Minamidani}, T., {Kuno}, N., {et~al.} 2017, \pasj, 69, 78

\bibitem[{{Urquhart} {et~al.}(2014){Urquhart}, {Csengeri}, {Wyrowski},
  {Schuller}, {Bontemps}, {Bronfman}, {Menten}, {Walmsley}, {Contreras},
  {Beuther}, {Wienen}, \& {Linz}}]{Urquhart14}
{Urquhart}, J.~S., {Csengeri}, T., {Wyrowski}, F., {et~al.} 2014, \aap, 568,
  A41

\bibitem[{{Urquhart} {et~al.}(2021){Urquhart}, {Figura}, {Cross}, {Wells},
  {Moore}, {Eden}, {Ragan}, {Pettitt}, {Duarte-Cabral}, {Colombo}, {Schuller},
  {Csengeri}, {Mattern}, {Beuther}, {Menten}, {Wyrowski}, {Anderson}, {Barnes},
  {Beltr{\'a}n}, {Billington}, {Bronfman}, {Giannetti}, {Kainulainen},
  {Kauffmann}, {Lee}, {Leurini}, {Medina}, {Montenegro-Montes}, {Riener},
  {Rigby}, {S{\'a}nchez-Monge}, {Schilke}, {Schisano}, {Traficante}, \&
  {Wienen}}]{Urquhart21}
{Urquhart}, J.~S., {Figura}, C., {Cross}, J.~R., {et~al.} 2021, \mnras, 500,
  3050

\bibitem[{{van der Tak} {et~al.}(2007){van der Tak}, {Black}, {Sch{\"o}ier},
  {Jansen}, \& {van Dishoeck}}]{vanderTak07}
{van der Tak}, F.~F.~S., {Black}, J.~H., {Sch{\"o}ier}, F.~L., {Jansen}, D.~J.,
  \& {van Dishoeck}, E.~F. 2007, \aap, 468, 627

\bibitem[{{Zmuidzinas} {et~al.}(1988){Zmuidzinas}, {Betz}, {Boreiko}, \&
  {Goldhaber}}]{Zmuidzinas88}
{Zmuidzinas}, J., {Betz}, A.~L., {Boreiko}, R.~T., \& {Goldhaber}, D.~M. 1988,
  \apj, 335, 774

\end{thebibliography}

\begin{appendix} 

\section{Tables} 
\label{s:appendix_tables} 

\onecolumn
\begin{ThreePartTable}
\begin{TableNotes}
\item \textbf{Notes.} The physical characteristics are from \cite{Koenig17}, and the columns are as follows. 
Name: ATLASGAL source name; $l$: Galactic longtidue; $b$: Galactic latitude; $d$: distance; $\varv_{\rm src}$: source velocity;  
log$_{10}$($N$(H$_{2}$)): dust-based $N$(H$_{2}$) in log; Class: source classification; \CI 809 GHz?: presence of the \CI 809 GHz mapping data 
\end{TableNotes}  

\begin{center}
\begin{longtable}{c c c c c c c c} 
\caption{Characteristics of the Top100 sources.}  
\label{t:appendix_table1} \\ 
\toprule \toprule 
Name & $l$ & $b$ & $d$ & $\varv_{\rm src}$ & log$_{10}$($N$(H$_{2}$)) & Class. & \CI 809 GHz?\\ 
       & (deg) & (deg) & (kpc) & (km s$^{-1}$) & (cm$^{-2}$) & & \\ \midrule 
\endfirsthead

\caption* {Table \ref{t:appendix_table1}. (continued)} \\
\toprule \toprule
Name & $l$ & $b$ & $d$ & $\varv_{\rm src}$ & log$_{10}$($N$(H$_{2}$)) & Class. & \CI 809 GHz?\\ 
       & (deg) & (deg) & (kpc) & (km s$^{-1}$) & (cm$^{-2}$) & & \\ \midrule 
\endhead

\addlinespace
\insertTableNotes 
\endlastfoot 

G08.68$-$0.37 &   8.68 &  $-$0.37 &  4.8 &   38.0 & 22.9 & IRw & \\ 
G08.71$-$0.41 &   8.71 &  $-$0.41 &  4.8 &   39.4 & 22.8 & IRw & \\
G10.45$-$0.02 &  10.44 &  $-$0.02 &  8.6 &   75.9 & 22.7 & IRw & \\
G10.47$+$0.03 &  10.47 &   0.03 &  8.6 &   67.6 & 23.7 & HII & \\
G10.62$-$0.38 &  10.62 &  $-$0.38 &  5.0 &   $-$2.9 & 23.6 & HII & \\
G12.81$-$0.20 &  12.81 &  $-$0.20 &  2.4 &   36.2 & 23.6 & HII & \\
G13.18$+$0.06 &  13.18 &   0.06 &  2.4 &   50.4 & 22.8 & 70w & \\
G13.66$-$0.60 &  13.66 &  $-$0.60 &  4.5 &   48.4 & 22.8 & IRb & \\
G14.11$-$0.57 &  14.11 &  $-$0.57 &  2.6 &   20.9 & 22.9 & IRw & \\
G14.19$-$0.19 &  14.19 &  $-$0.19 &  3.9 &   39.2 & 23.0 & IRw & \\
G14.49$-$0.14 &  14.49 &  $-$0.14 &  3.9 &   39.5 & 23.1 & 70w & \\
G14.63$-$0.58 &  14.63 &  $-$0.58 &  1.8 &   18.5 & 23.0 & IRw & \\
G15.03$-$0.67 &  15.03 &  $-$0.67 &  2.0 &   19.6 & 23.3 & IRb & \\
G18.61$-$0.07 &  18.61 &  $-$0.07 &  4.3 &   46.6 & 22.8 & IRw & \\
G18.73$-$0.23 &  18.73 &  $-$0.23 & 12.5 &   42.6 & 22.9 & IRw &\\
G18.89$-$0.47 &  18.89 &  $-$0.47 &  4.7 &   66.1 & 23.2 & IRw & \\
G19.88$-$0.54 &  19.88 &  $-$0.53 &  3.7 &   45.0 & 23.1 & IRb & \\
G22.37$+$0.45 &  22.38 &   0.45 &  4.0 &   54.0 & 22.9 & IRw & \\
G23.21$-$0.38 &  23.21 &  $-$0.38 &  4.6 &   78.2 & 23.2 & IRw & \\
G24.63$+$0.17 &  24.63 &   0.17 &  7.7 &  116.0 & 22.6 & IRw & \cmark \\
G28.56$-$0.24 &  28.56 &  $-$0.24 &  5.5 &   87.2 & 23.1 & IRw & \cmark \\
G30.82$-$0.06 &  30.82 &  $-$0.06 &  4.9 &   97.8 & 23.7 & IRb & \\
G30.85$-$0.08 &  30.85 &  $-$0.08 &  4.9 &   94.3 & 22.7 & 70w & \\
G30.89$+$0.14 &  30.89 &   0.14 &  4.9 &   97.3 & 23.0 & 70w & \\
G31.41$+$0.31 &  31.41 &   0.31 &  4.9 &   98.2 & 23.6 & HII & \\
G34.26$+$0.15 &  34.26 &   0.15 &  1.6 &   58.5 & 23.8 & HII & \\
G34.40$+$0.23 &  34.40 &   0.23 &  1.6 &   57.4 & 23.1 & HII & \\
G34.41$+$0.23 &  34.41 &   0.23 &  1.6 &   58.0 & 23.2 & IRb & \\
G34.82$+$0.35 &  34.82 &   0.35 &  1.6 &   58.3 & 22.7 & IRb & \\
G35.20$-$0.74 &  35.20 &  $-$0.74 &  2.2 &   34.7 & 23.2 & IRb &\\
G37.55$+$0.20 &  37.55 &   0.20 &  6.7 &   86.1 & 22.7 & IRb & \\
G43.17$+$0.01 &  43.17 &   0.01 & 11.1 &    5.3 & 23.8 & HII & \\
G49.49$-$0.39 &  49.49 &  $-$0.39 &  5.4 &   57.2 & 24.0 & HII & \\
G53.14$+$0.07 &  53.14 &   0.07 &  1.6 &   22.3 & 22.9 & IRb & \\
G301.14$-$0.23 & 301.14 &  $-$0.23 &  4.4 &  $-$39.1 & 23.4 & HII & \\
G305.19$-$0.01 & 305.19 &  $-$0.01 &  3.8 &  $-$34.2 & 22.7 & IRw & \\
G305.21$+$0.21 & 305.21 &   0.21 &  3.8 &  $-$42.5 & 23.3 & IRb & \\
G305.56$+$0.01 & 305.56 &   0.01 &  3.8 &  $-$39.8 & 22.7 & HII & \\
G305.80$-$0.10 & 305.79 &  $-$0.10 &  3.8 &  $-$40.8 & 22.7 & 70w & \\
G309.38$-$0.13 & 309.38 &  $-$0.13 &  5.3 &  $-$51.3 & 22.8 & IRb & \\
G310.01$+$0.39 & 310.01 &   0.39 &  3.6 &  $-$41.3 & 22.7 & IRb & \\
G317.87$-$0.15 & 317.87 &  $-$0.15 &  3.0 &  $-$40.2 & 23.0 & IRw & \\
G318.78$-$0.14 & 318.78 &  $-$0.14 &  2.8 &  $-$38.1 & 22.6 & IRw & \\
G320.88$-$0.40 & 320.88 &  $-$0.40 & 10.0 &  $-$45.3 & 22.7 & 70w & \\
G326.66$+$0.52 & 326.66 &   0.52 &  1.8 &  $-$39.8 & 22.7 & IRb & \\
G326.99$-$0.03 & 326.99 &  $-$0.03 &  4.0 &  $-$58.6 & 22.9 & IRw & \\
G327.12$+$0.51 & 327.12 &   0.51 &  5.5 &  $-$83.7 & 22.6 & IRb & \\
G327.29$-$0.60 & 327.29 &  $-$0.58 &  3.1 &  $-$44.7 & 23.9 & IRb & \\
G327.39$+$0.20 & 327.39 &   0.20 &  5.9 &  $-$89.2 & 22.8 & IRb & \\
G328.81$+$0.63 & 328.81 &   0.63 &  3.0 &  $-$41.9 & 23.4 & HII & \\
G329.03$-$0.20 & 329.03 &  $-$0.20 & 11.5 &  $-$43.5 & 23.3 & IRw & \\
G329.07$-$0.31 & 329.07 &  $-$0.31 & 11.6 &  $-$41.9 & 22.8 & IRb & \\
G330.88$-$0.37 & 330.88 &  $-$0.37 &  4.2 &  $-$62.5 & 23.3 & HII & \\
G330.95$-$0.18 & 330.95 &  $-$0.18 &  9.3 &  $-$91.2 & 23.8 & HII & \\
G331.71$+$0.60 & 331.71 &   0.58 & 10.5 &  $-$67.8 & 22.9 & IRw & \\
G332.09$-$0.42 & 332.09 &  $-$0.42 &  3.6 &  $-$56.9 & 22.9 & IRb & \\
G332.83$-$0.55 & 332.83 &  $-$0.55 &  3.6 &  $-$57.1 & 23.5 & HII & \\
G333.13$-$0.43 & 333.13 &  $-$0.43 &  3.6 &  $-$53.1 & 23.4 & HII & \\
G333.28$-$0.39 & 333.28 &  $-$0.39 &  3.6 &  $-$52.4 & 23.2 & HII & \\
G333.31$+$0.11 & 333.31 &   0.11 &  3.6 &  $-$46.5 & 22.8 & IRb & \\
G333.60$-$0.21 & 333.60 &  $-$0.21 &  3.6 &  $-$48.4 & 23.5 & HII & \\
G333.66$+$0.06 & 333.66 &   0.06 &  5.3 &  $-$85.0 & 22.7 & 70w & \\
G335.79$+$0.17 & 335.79 &   0.17 &  3.7 &  $-$50.1 & 23.2 & IRw & \\
G336.96$-$0.23 & 336.96 &  $-$0.22 & 10.9 &  $-$71.3 & 22.8 & IRw & \\
G337.17$-$0.03 & 337.18 &  $-$0.03 & 11.0 &  $-$67.8 & 22.7 & IRw & \cmark \\
G337.26$-$0.10 & 337.26 &  $-$0.10 & 11.0 &  $-$68.3 & 22.7 & IRw & \\
G337.28$+$0.01 & 337.29 &   0.01 &  9.4 & $-$106.6 & 22.9 & 70w & \\
G337.40$-$0.40 & 337.41 &  $-$0.40 &  3.3 &  $-$41.0 & 23.4 & HII & \\
G337.70$-$0.02 & 337.70 &  $-$0.05 & 12.3 &  $-$47.4 & 23.4 & HII & \\
G337.92$-$0.48 & 337.92 &  $-$0.48 &  3.2 &  $-$39.5 & 23.4 & IRb & \\
G338.07$+$0.01 & 338.07 &   0.04 &  4.7 &  $-$69.2 & 22.5 & 70w & \cmark \\
G338.78$+$0.48 & 338.79 &   0.48 &  4.5 &  $-$63.8 & 22.9 & 70w & \cmark \\
G338.92$+$0.56 & 338.92 &   0.55 &  4.4 &  $-$61.6 & 23.5 & IRw & \\
G339.62$-$0.12 & 339.62 &  $-$0.12 &  3.0 &  $-$34.6 & 22.7 & IRb & \\
G340.37$-$0.39 & 340.37 &  $-$0.39 &  3.6 &  $-$43.5 & 22.9 & IRw & \\
G340.75$-$1.00 & 340.75 &  $-$1.00 &  2.8 &  $-$29.4 & 22.6 & IRb & \\
G340.78$-$0.10 & 340.78 &  $-$0.10 & 10.0 & $-$101.5 & 22.8 & IRw & \\
G341.22$-$0.21 & 341.22 &  $-$0.21 &  3.7 &  $-$43.5 & 22.9 & IRb & \\
G342.48$+$0.18 & 342.48 &   0.18 & 12.6 &  $-$41.5 & 22.8 & IRw & \\
G343.13$-$0.06 & 343.13 &  $-$0.06 &  3.0 &  $-$30.6 & 23.4 & HII & \\
G343.75$-$0.16 & 343.76 &  $-$0.16 &  2.9 &  $-$27.8 & 23.3 & IRw & \\
G344.23$-$0.57 & 344.23 &  $-$0.57 &  2.5 &  $-$22.3 & 23.6 & IRw & \cmark \\
G345.00$-$0.23 & 345.00 &  $-$0.22 &  3.0 &  $-$27.6 & 23.4 & HII & \cmark \\
G345.49$+$0.32 & 345.49 &   0.31 &  2.2 &  $-$17.6 & 23.3 & HII & \cmark \\
G345.51$+$0.35 & 345.50 &   0.35 &  2.2 &  $-$17.9 & 23.1 & IRb & \cmark \\
G345.72$+$0.82 & 345.72 &   0.82 &  1.6 &  $-$11.2 & 22.8 & IRb & \\
G351.13$+$0.77 & 351.13 &   0.77 &  1.8 &   $-$5.3 & 22.5 & 70w & \\
G351.16$+$0.70 & 351.16 &   0.70 &  1.8 &   $-$6.7 & 23.7 & IRb & \\
G351.25$+$0.67 & 351.24 &   0.67 &  1.8 &   $-$3.3 & 23.3 & IRb & \\
G351.42$+$0.65 & 351.42 &   0.65 &  1.3 &   $-$7.4 & 23.8 & HII & \\
G351.45$+$0.66 & 351.44 &   0.66 &  1.3 &   $-$4.3 & 23.9 & IRw & \\
G351.57$+$0.77 & 351.57 &   0.76 &  1.3 &   $-$3.1 & 22.7 & 70w & \\
G351.58$-$0.35 & 351.58 &  $-$0.35 &  6.8 &  $-$95.9 & 23.6 & IRb & \\
G351.77$-$0.54 & 351.77 &  $-$0.54 &  1.0 &   $-$2.8 & 23.8 & IRb & \\
G353.07$+$0.45 & 353.07 &   0.45 &  0.9 &    1.7 & 22.6 & IRw & \\
G353.41$-$0.36 & 353.41 &  $-$0.36 &  3.4 &  $-$16.1 & 23.5 & IRb & \\
G353.42$-$0.08 & 353.42 &  $-$0.08 &  6.1 &  $-$54.4 & 22.4 & 70w & \\
G354.95$-$0.54 & 354.94 &  $-$0.54 &  1.9 &   $-$5.4 & 22.6 & 70w & \\ \bottomrule 
\end{longtable}
\end{center}
\end{ThreePartTable}
\twocolumn

\onecolumn
\begin{ThreePartTable}
\begin{TableNotes}
\item \textbf{Notes.} 
These are the data that were used for the analyses in Sect. \ref{s:CI_CO}. 
Among all fitted components, only those with clean reference positions were selected. 
When one of the two lines is not detected, 5$\sigma$ of the undetected transition is integrated over the velocity range 
where the other transition is seen, providing a upper limit on the integrated intensity (indicated as the downward arrows). 
The columns are as follows. 
Name: ATLASGAL source name; $\varv_{\rm c}$: representative velocity of the component (determined based on Gaussian fitting); 
$\varv_{\rm l}$: lower velocity for emission integration; $\varv_{\rm u}$: upper velocity for emission integration; 
$I$: integrated intensity; $\sigma(I)$: 1$\sigma$ on the integrated intensity.
\end{TableNotes} 

\begin{landscape} 
\begin{center}
\begin{longtable}{c c c c c c c c c c c c} 
\caption{Derived line parameters for \CI 492 GHz and $^{13}$CO(2--1).} 
\label{t:appendix_table2} \\
\toprule \toprule 
Name & \multicolumn{5}{c}{[C~\textsc{i}] 492 GHz} & & \multicolumn{5}{c}{$^{13}$CO(2--1)} \\ \cline{2-6} \cline{8-12}
       & $\varv_{\rm c}$ & $\varv_{\rm l}$ & $\varv_{\rm u}$ & $I$ & $\sigma(I)$ & & $\varv_{\rm c}$ & $\varv_{\rm l}$ & $\varv_{\rm u}$ & $I$ & $\sigma(I)$ \\ 
       & (km s$^{-1}$) & (km s$^{-1}$) & (km s$^{-1}$) & (K km s$^{-1}$) & (K km s$^{-1}$) & & (km s$^{-1}$) & (km s$^{-1}$) & (km s$^{-1}$) & (K km s$^{-1}$) & (K km s$^{-1}$) \\ \midrule
\endfirsthead

\caption* {Table \ref{t:appendix_table2}. (continued)} \\
\toprule \toprule 
Name & \multicolumn{5}{c}{[C~\textsc{i}] 492 GHz} & & \multicolumn{5}{c}{$^{13}$CO(2--1)} \\ \cline{2-6} \cline{8-12}
       & $\varv_{\rm c}$ & $\varv_{\rm l}$ & $\varv_{\rm u}$ & $I$ & $\sigma(I)$ & & $\varv_{\rm c}$ & $\varv_{\rm l}$ & $\varv_{\rm u}$ & $I$ & $\sigma(I)$ \\ 
       & (km s$^{-1}$) & (km s$^{-1}$) & (km s$^{-1}$) & (K km s$^{-1}$) & (K km s$^{-1}$) & & (km s$^{-1}$) & (km s$^{-1}$) & (km s$^{-1}$) & (K km s$^{-1}$) & (K km s$^{-1}$) \\ \midrule
\endhead

\addlinespace
\insertTableNotes 
\endlastfoot

G08.68$-$0.37 &   36.55 &  30.0 &  45.0 & 31.03 &  0.47 & &   36.35 &  30.0 &  50.0 & 60.90 &  0.13 \\ 
G08.71$-$0.41 &   38.49 &  30.0 &  50.0 & 16.44 &  0.44 & &   38.57 &  30.0 &  47.0 & 22.83 &  0.42 \\
G10.45$-$0.02 &   72.72 &  60.0 &  83.0 & 31.34 &  0.75 & &   72.52 &  60.0 &  85.0 & 40.92 &  0.19 \\
G10.45$-$0.02 &   34.42 &  30.0 &  40.0 &  5.33 &  0.48 & &   33.90 &  30.0 &  40.0 &  1.23 &  0.12 \\
G10.45$-$0.02 & -- &  $-$7.0 &   4.0 & 10.54$\downarrow$ & -- & &   $-$8.03 &  $-$7.0 &   4.0 &  1.70 &  0.11 \\
G10.45$-$0.02 & -- & $-$12.0 &   0.0 & 13.70$\downarrow$ & -- & &  146.83 & $-$12.0 &   0.0 &  1.93 &  0.13 \\
G10.47$+$0.03 &   66.19 &  50.0 &  85.0 & 56.98 &  1.13 & &   66.67 &  50.0 &  85.0 &118.70 &  0.28 \\
G10.62$-$0.38 &   $-$2.72 & $-$20.0 &  10.0 & 97.99 &  0.99 & &   $-$3.19 & $-$15.0 &  10.0 &248.43 &  0.19 \\
G12.81$-$0.20 &   34.68 &  20.0 &  50.0 & 83.29 &  1.74 & &   35.53 &  20.0 &  53.0 &221.74 &  0.30 \\
G12.81$-$0.20 & -- &   5.0 &  14.0 & 20.56$\downarrow$ & -- & &    9.09 &   5.0 &  14.0 &  2.48 &  0.16 \\
G13.18$+$0.06 &   35.75 &  33.0 &  38.0 &  2.61 &  0.32 & &   35.90 &  34.0 &  38.0 &  6.42 &  0.06 \\
G13.18$+$0.06 & -- &   5.0 &  22.0 & 18.86$\downarrow$ & -- & &   13.25 &   5.0 &  22.0 &  8.01 &  0.13 \\
G13.18$+$0.06 & -- &  22.0 &  28.0 &  6.63$\downarrow$ & -- & &   25.53 &  22.0 &  28.0 &  3.53 &  0.08 \\
G14.11$-$0.57 &   19.40 &  13.0 &  25.0 & 26.50 &  0.44 & &   19.32 &  15.0 &  25.0 & 34.81 &  0.27 \\
G14.49$-$0.14 &   38.91 &  30.0 &  50.0 & 24.06 &  0.51 & &   40.23 &  30.0 &  50.0 & 36.41 &  0.24 \\
G14.49$-$0.14 & -- &  21.0 &  30.0 &  7.32$\downarrow$ & -- & &   25.55 &  21.0 &  30.0 &  3.35 &  0.17 \\
G15.03$-$0.67 &   20.69 &  13.0 &  28.0 &129.81 &  0.82 & &   19.91 &   5.0 &  27.0 &241.84 &  0.16 \\
G18.61$-$0.07 &   45.57 &  40.0 &  50.0 & 11.81 &  0.60 & &   45.10 &  40.0 &  60.0 & 33.89 &  0.25 \\
G18.73$-$0.23 &   40.88 &  30.0 &  60.0 & 16.42 &  0.87 & &   41.36 &  30.0 &  55.0 & 34.65 &  0.20 \\
G18.89$-$0.47 &   65.45 &  57.0 &  72.0 & 44.74 &  0.48 & &   65.43 &  55.0 &  80.0 & 96.07 &  0.24 \\
G18.89$-$0.47 &   52.76 &  50.0 &  56.0 &  4.64 &  0.29 & &   52.54 &  51.0 &  55.0 &  4.25 &  0.10 \\
G18.89$-$0.47 & -- &  33.0 &  36.0 &  2.53$\downarrow$ & -- & &   34.41 &  33.0 &  36.0 &  1.28 &  0.09 \\
G19.88$-$0.54 &   44.09 &  35.0 &  54.0 & 37.50 &  1.22 & &   44.08 &  35.0 &  55.0 & 62.21 &  0.16 \\
G19.88$-$0.54 &   24.42 &  22.0 &  26.0 &  3.09 &  0.54 & &   24.60 &  22.0 &  26.0 &  3.82 &  0.07 \\
G22.37$+$0.45 &   52.68 &  45.0 &  70.0 & 19.78 &  1.03 & &   53.00 &  50.0 &  70.0 & 27.36 &  0.20 \\
G22.37$+$0.45 & -- &  80.0 &  93.0 & 19.86$\downarrow$ & -- & &   83.83 &  80.0 &  93.0 &  6.14 &  0.16 \\
G22.37$+$0.45 & -- & 110.0 & 121.0 & 16.31$\downarrow$ & -- & &  113.44 & 110.0 & 121.0 &  2.26 &  0.15 \\
G24.63$+$0.17 &   38.43 &  35.0 &  43.0 &  3.82 &  0.48 & &   39.76 &  35.0 &  45.0 &  4.68 &  0.14 \\
G24.63$+$0.17 &   53.24 &  50.0 &  57.0 &  3.73 &  0.45 & &   53.13 &  49.0 &  56.0 &  5.84 &  0.12 \\
G24.63$+$0.17 & -- &  76.0 &  81.0 &  5.76$\downarrow$ & -- & &   78.57 &  76.0 &  81.0 &  0.88 &  0.10 \\
G28.56$-$0.24 & -- &  $-$6.0 &   0.0 & 19.00$\downarrow$ & -- & &   $-$3.41 &  $-$6.0 &   0.0 &  1.68 &  0.09 \\
G28.56$-$0.24 & -- &  61.5 &  66.0 & 13.15$\downarrow$ & -- & &   63.42 &  61.5 &  66.0 &  4.32 &  0.08 \\
G30.82$-$0.06 & -- & 110.0 & 120.0 & 20.75$\downarrow$ & -- & &  114.77 & 110.0 & 120.0 &  5.57 &  0.14 \\
G30.89$+$0.14 &  107.14 & 101.0 & 110.0 & 22.62 &  0.70 & &  106.92 & 101.5 & 111.5 & 30.83 &  0.13 \\
G31.41$+$0.31 &   98.26 &  90.0 & 105.0 & 38.83 &  0.52 & &   97.38 &  90.0 & 110.0 & 49.66 &  0.22 \\
G34.26$+$0.15 &   56.85 &  50.0 &  68.0 & 67.75 &  0.69 & &   56.09 &  50.0 &  80.0 &130.86 &  0.36 \\
G34.26$+$0.15 & -- &  40.0 &  50.0 & 12.14$\downarrow$ & -- & &   47.52 &  40.0 &  50.0 &  3.53 &  0.21 \\
G34.40$+$0.23 &   56.78 &  50.0 &  65.0 & 41.09 &  0.50 & &   55.64 &  50.0 &  70.0 & 56.48 &  0.30 \\
G34.40$+$0.23 &  104.98 & 103.0 & 107.0 &  2.50 &  0.24 & &  104.82 & 102.5 & 107.0 &  1.96 &  0.13 \\
G34.40$+$0.23 & -- &  97.0 & 102.5 &  4.75$\downarrow$ & -- & &   99.97 &  97.0 & 102.5 &  1.17 &  0.16 \\
G34.41$+$0.23 &   56.72 &  50.0 &  65.0 & 24.53 &  0.53 & &   57.87 &  50.0 &  66.0 & 38.75 &  0.24 \\
G34.41$+$0.23 &   68.13 &  65.0 &  73.0 &  2.82 &  0.38 & &   68.15 &  66.0 &  72.0 &  2.71 &  0.15 \\
G34.41$+$0.23 & -- & 102.0 & 107.0 &  4.64$\downarrow$ & -- & &  104.33 & 102.0 & 107.0 &  1.19 &  0.13 \\
G34.82$+$0.35 & -- &  85.0 &  91.0 &  8.12$\downarrow$ & -- & &   88.08 &  85.0 &  91.0 &  1.48 &  0.08 \\
G37.55$+$0.20 &   86.20 &  78.0 &  95.0 & 18.30 &  0.29 & &   85.90 &  70.0 & 100.0 & 33.21 &  0.18 \\
G37.55$+$0.20 &   65.97 &  64.0 &  68.0 &  0.90 &  0.14 & &   65.47 &  62.0 &  67.5 &  0.70 &  0.08 \\
G37.55$+$0.20 & -- &   2.5 &   5.0 &  1.20$\downarrow$ & -- & &    3.52 &   2.5 &   5.0 &  0.33 &  0.05 \\
G43.17$+$0.01 &    7.45 & $-$10.0 &  20.0 &125.88 &  1.51 & &    6.30 & $-$20.0 &  30.0 &402.53 &  0.21 \\
G49.49$-$0.39 &   55.86 &  47.0 &  73.0 &140.48 &  1.34 & &   56.31 &  47.5 &  78.0 &355.20 &  0.38 \\
G53.14$+$0.07 &   21.58 &  15.0 &  33.0 & 23.81 &  0.31 & &   21.31 &  10.0 &  33.0 & 42.06 &  0.15 \\
G53.14$+$0.07 & -- &  51.0 &  55.0 &  2.24$\downarrow$ & -- & &   53.31 &  51.0 &  55.0 &  0.56 &  0.06 \\
G301.14$-$0.23 &  $-$38.67 & $-$50.0 & $-$30.0 & 67.07 &  0.86 & &  $-$39.47 & $-$65.0 & $-$10.0 &183.12 &  0.27 \\
G305.19$-$0.01 &  $-$32.33 & $-$47.0 & $-$20.0 & 30.47 &  1.10 & &  $-$33.44 & $-$41.5 & $-$27.0 & 55.34 &  0.12 \\
G305.19$-$0.01 &   31.67 &  27.0 &  35.0 &  3.66 &  0.60 & &   31.65 &  26.0 &  37.0 &  6.67 &  0.10 \\
G305.21$+$0.21 &  $-$42.83 & $-$50.0 & $-$35.0 & 33.64 &  1.50 & &  $-$42.08 & $-$60.0 & $-$25.0 &141.77 &  0.18 \\
G305.56$+$0.01 &  $-$39.60 & $-$45.0 & $-$34.0 & 39.94 &  0.67 & &  $-$39.57 & $-$55.0 & $-$20.0 &105.42 &  0.18 \\
G305.80$-$0.10 &  $-$40.79 & $-$45.0 & $-$35.0 & 16.97 &  1.04 & &  $-$41.00 & $-$47.0 & $-$37.0 & 18.60 &  0.09 \\
G310.01$+$0.39 &  $-$41.32 & $-$52.0 & $-$36.0 & 21.63 &  0.91 & &  $-$41.80 & $-$53.0 & $-$27.0 & 54.41 &  0.19 \\
G317.87$-$0.15 & -- &  20.0 &  26.0 &  7.58$\downarrow$ & -- & &   22.73 &  20.0 &  26.0 &  1.85 &  0.11 \\
G318.78$-$0.14 &  $-$37.83 & $-$44.0 & $-$33.0 & 11.38 &  0.82 & &  $-$37.63 & $-$44.5 & $-$30.0 & 26.20 &  0.15 \\
G318.78$-$0.14 & -- & $-$51.0 & $-$44.5 & 11.72$\downarrow$ & -- & &  $-$48.14 & $-$51.0 & $-$44.5 &  3.15 &  0.10 \\
G320.88$-$0.40 &  $-$45.29 & $-$49.0 & $-$42.0 & 16.68 &  0.61 & &  $-$45.37 & $-$50.0 & $-$38.0 & 43.07 &  0.15 \\
G326.66$+$0.52 & -- & $-$24.0 & $-$18.0 & 11.25$\downarrow$ & -- & &  $-$21.92 & $-$24.0 & $-$18.0 &  2.30 &  0.16 \\
G326.66$+$0.52 & -- &  $-$4.0 &   0.0 &  7.79$\downarrow$ & -- & &   $-$2.04 &  $-$4.0 &   0.0 &  1.04 &  0.13 \\
G326.99$-$0.03 &  $-$57.59 & $-$62.0 & $-$53.0 &  6.73 &  0.68 & &  $-$58.73 & $-$65.0 & $-$52.5 & 23.26 &  0.12 \\
G327.12$+$0.51 &  $-$83.93 & $-$90.0 & $-$75.0 & 13.61 &  1.11 & &  $-$84.01 & $-$96.0 & $-$73.0 & 25.17 &  0.20 \\
G327.29$-$0.60 & $-$46.06 & $-$55.0 & $-$35.0 & 89.03 &  1.58 & &  $-$45.37 & $-$58.0 & $-$28.0 &159.91 &  0.59 \\
G327.39$+$0.20 &  $-$90.05 & $-$95.0 & $-$85.0 &  9.23 &  0.57 & &  $-$89.27 & $-$94.0 & $-$81.0 & 42.75 &  0.14 \\
G328.81$+$0.63 &  $-$42.79 & $-$49.0 & $-$36.0 & 27.20 &  1.07 & &  $-$42.99 & $-$60.0 & $-$30.0 &127.16 &  0.25 \\
G329.07$-$0.31 &  $-$42.10 & $-$47.0 & $-$36.0 &  8.92 &  0.75 & &  $-$42.81 & $-$50.0 & $-$33.0 & 29.10 &  0.19 \\
G330.88$-$0.37 &  $-$62.41 & $-$72.0 & $-$52.0 & 60.48 &  0.97 & &  $-$62.98 & $-$80.0 & $-$52.0 &166.41 &  0.22 \\
G330.95$-$0.18 &  $-$91.56 &$-$105.0 & $-$80.0 &104.73 &  0.67 & &  $-$91.74 &$-$110.0 & $-$70.0 &215.67 &  0.31 \\
G331.71$+$0.60 &  $-$67.37 & $-$71.0 & $-$58.0 & 15.57 &  0.52 & &  $-$67.37 & $-$80.0 & $-$55.0 & 38.15 &  0.22 \\
G332.09$-$0.42 &  $-$57.12 & $-$63.0 & $-$52.0 & 39.75 &  0.55 & &  $-$56.36 & $-$75.0 & $-$45.0 & 73.00 &  0.25 \\
G332.09$-$0.42 & -- & $-$38.0 & $-$33.0 &  6.33$\downarrow$ & -- & &  $-$36.42 & $-$38.0 & $-$33.0 &  1.84 &  0.10 \\
G332.83$-$0.55 &  $-$58.18 & $-$70.0 & $-$50.0 & 80.75 &  1.26 & &  $-$56.93 & $-$72.0 & $-$44.0 &152.45 &  0.23 \\
G333.13$-$0.43 &  $-$56.58 & $-$75.0 & $-$40.0 & 57.42 &  1.06 & &  $-$53.85 & $-$80.0 & $-$40.0 &302.85 &  0.32 \\
G333.28$-$0.39 &  $-$52.08 & $-$60.0 & $-$45.0 &101.65 &  0.53 & &  $-$52.05 & $-$62.0 & $-$46.0 &150.48 &  0.21 \\
G333.31$+$0.11 & -- & $-$95.0 & $-$85.0 &  6.97$\downarrow$ & -- & &  $-$93.26 & $-$95.0 & $-$85.0 &  2.38 &  0.12 \\
G333.31$+$0.11 &  $-$80.05 & $-$85.0 & $-$75.0 &  6.10 &  0.29 & &  $-$80.21 & $-$85.0 & $-$73.0 &  5.08 &  0.13 \\
G333.31$+$0.11 &  $-$13.40 & $-$14.0 & $-$12.5 &  0.76 &  0.11 & &  $-$13.37 & $-$15.0 & $-$12.0 &  1.24 &  0.07 \\
G333.60$-$0.21 &  $-$44.78 & $-$70.0 & $-$25.0 & 90.46 &  0.93 & &  $-$47.32 & $-$70.0 & $-$20.0 &252.83 &  0.27 \\
G333.60$-$0.21 &  $-$93.82 & $-$96.0 & $-$91.0 &  2.58 &  0.31 & &  $-$93.73 & $-$96.0 & $-$92.0 &  2.48 &  0.08 \\
G333.60$-$0.21 &  $-$89.43 & $-$91.0 & $-$88.0 &  1.95 &  0.25 & &  $-$89.54 & $-$92.0 & $-$87.5 &  5.47 &  0.08 \\
G333.66$+$0.06 &  $-$84.92 &$-$100.0 & $-$80.0 & 40.81 &  0.38 & &  $-$85.13 & $-$97.0 & $-$75.0 & 49.68 &  0.19 \\
G333.66$+$0.06 &  $-$70.29 & $-$73.0 & $-$67.0 &  4.50 &  0.21 & &  $-$70.34 & $-$74.0 & $-$67.0 &  2.38 &  0.10 \\
G333.66$+$0.06 &  $-$50.72 & $-$57.0 & $-$46.0 &  8.14 &  0.29 & &  $-$49.51 & $-$56.0 & $-$46.0 & 15.06 &  0.13 \\
G333.66$+$0.06 &  $-$42.44 & $-$44.0 & $-$40.0 &  1.28 &  0.17 & &  $-$42.72 & $-$46.0 & $-$35.0 &  5.00 &  0.13 \\
G335.79$+$0.17 &  $-$50.19 & $-$58.0 & $-$46.0 & 38.67 &  0.35 & &  $-$50.13 & $-$63.0 & $-$30.0 & 84.60 &  0.31 \\
G336.96$-$0.23 &  $-$71.63 & $-$75.0 & $-$67.0 & 20.58 &  0.46 & &  $-$71.44 & $-$78.0 & $-$67.0 & 22.64 &  0.22 \\
G336.96$-$0.23 & $-$128.68 &$-$131.0 &$-$120.0 &  5.52 &  0.55 & & $-$126.33 &$-$130.0 &$-$114.0 &  4.57 &  0.26 \\
G336.96$-$0.23 & $-$100.82 &$-$103.0 & $-$99.0 &  2.37 &  0.34 & & -- &$-$103.0 & $-$99.0 &  1.57$\downarrow$ & -- \\
G336.96$-$0.23 &  $-$82.80 & $-$85.0 & $-$80.0 &  5.03 &  0.37 & &  $-$82.78 & $-$84.5 & $-$80.5 &  3.89 &  0.13 \\
G337.17$-$0.03 &  $-$65.94 & $-$73.0 & $-$60.0 & 23.56 &  0.94 & &  $-$67.83 & $-$73.0 & $-$58.0 & 48.71 &  0.23 \\
G337.17$-$0.03 & -- &$-$123.0 &$-$116.5 & 11.54$\downarrow$ & -- & & $-$119.81 &$-$123.0 &$-$116.5 &  2.84 &  0.16 \\
G337.17$-$0.03 & -- &$-$116.5 &$-$107.0 & 17.75$\downarrow$ & -- & & $-$110.57 &$-$116.5 &$-$107.0 & 10.31 &  0.18 \\
G337.17$-$0.03 & -- & $-$81.0 & $-$74.0 & 13.31$\downarrow$ & -- & &  $-$78.90 & $-$81.0 & $-$74.0 &  5.33 &  0.16 \\
G337.26$-$0.10 & $-$119.57 &$-$122.0 &$-$116.0 &  4.13 &  0.48 & & $-$119.50 &$-$122.0 &$-$115.0 &  8.02 &  0.14 \\
G337.26$-$0.10 & -- & $-$55.0 & $-$47.0 & 11.39$\downarrow$ & -- & &  $-$50.00 & $-$55.0 & $-$47.0 &  3.29 &  0.16 \\
G337.26$-$0.10 & -- & $-$37.0 & $-$28.0 & 12.73$\downarrow$ & -- & &  $-$33.73 & $-$37.0 & $-$28.0 &  2.17 &  0.16 \\
G337.28$+$0.01 & $-$105.11 &$-$117.5 &$-$102.0 & 10.81 &  0.53 & & $-$108.82 &$-$118.0 &$-$100.0 & 14.33 &  0.24 \\
G337.28$+$0.01 & -- &$-$131.0 &$-$125.0 &  5.95$\downarrow$ & -- & & $-$127.62 &$-$131.0 &$-$125.0 &  1.55 &  0.14 \\
G337.40$-$0.40 &  $-$41.37 & $-$47.0 & $-$36.5 & 49.71 &  0.62 & &  $-$42.07 & $-$50.0 & $-$30.0 & 95.38 &  0.29 \\
G337.70$-$0.02 &  $-$47.66 & $-$60.0 & $-$40.0 & 52.18 &  0.69 & &  $-$48.22 & $-$69.0 & $-$33.0 & 84.18 &  0.37 \\
G337.92$-$0.48 & -- & $-$35.0 & $-$24.0 & 10.17$\downarrow$ & -- & &  $-$32.35 & $-$35.0 & $-$24.0 &  5.65 &  0.28 \\
G337.92$-$0.48 &  $-$18.95 & $-$21.0 & $-$17.0 &  2.30 &  0.25 & & -- & $-$21.0 & $-$17.0 &  2.08$\downarrow$ & -- \\
G338.07$+$0.01 &  $-$65.43 & $-$72.0 & $-$50.0 & 33.38 &  0.64 & &  $-$68.51 & $-$75.0 & $-$47.0 & 35.67 &  0.22 \\
G338.07$+$0.01 & $-$121.18 &$-$122.5 &$-$115.5 &  5.54 &  0.36 & & $-$121.36 &$-$125.0 &$-$117.0 &  3.35 &  0.12 \\
G338.07$+$0.01 & -- & $-$88.0 & $-$82.0 &  6.03$\downarrow$ & -- & &  $-$85.49 & $-$88.0 & $-$82.0 &  1.26 &  0.10 \\
G338.07$+$0.01 &  $-$40.39 & $-$50.0 & $-$30.0 & 35.07 &  0.61 & &  $-$39.89 & $-$47.0 & $-$30.0 & 35.78 &  0.17 \\
G338.78$+$0.48 & -- & $-$40.0 & $-$36.0 &  2.85$\downarrow$ & -- & &  $-$37.66 & $-$40.0 & $-$36.0 &  0.54 &  0.10 \\
G338.78$+$0.48 & -- & $-$36.0 & $-$26.0 &  6.64$\downarrow$ & -- & &  $-$30.12 & $-$36.0 & $-$26.0 &  1.61 &  0.16 \\
G338.92$+$0.56 &  $-$60.35 & $-$70.0 & $-$56.0 & 48.54 &  0.41 & &  $-$63.89 & $-$80.0 & $-$50.0 &122.80 &  0.44 \\
G338.92$+$0.56 & -- & $-$25.0 & $-$20.0 &  4.16$\downarrow$ & -- & &  $-$22.27 & $-$25.0 & $-$20.0 &  1.88 &  0.19 \\
G339.62$-$0.12 &  $-$93.58 & $-$95.0 & $-$92.5 &  1.58 &  0.12 & &  $-$93.64 & $-$98.0 & $-$92.0 &  1.44 &  0.12 \\
G339.62$-$0.12 & -- & $-$74.0 & $-$68.5 &  3.35$\downarrow$ & -- & &  $-$70.65 & $-$74.0 & $-$68.5 &  1.44 &  0.12 \\
G340.75$-$1.00 &  $-$29.30 & $-$35.0 & $-$25.0 & 27.79 &  0.45 & &  $-$29.32 & $-$36.0 & $-$20.0 & 61.91 &  0.13 \\
G340.78$-$0.10 & $-$101.53 &$-$107.0 & $-$98.0 & 13.09 &  0.55 & & $-$101.81 &$-$110.0 & $-$96.0 & 28.35 &  0.20 \\
G340.78$-$0.10 & -- & $-$80.0 & $-$76.0 &  5.01$\downarrow$ & -- & &  $-$78.04 & $-$80.0 & $-$76.0 &  1.01 &  0.10 \\
G342.48$+$0.18 &  $-$41.29 & $-$50.0 & $-$36.0 & 27.74 &  0.41 & &  $-$41.50 & $-$50.0 & $-$34.0 & 41.96 &  0.19 \\
G343.13$-$0.06 &  $-$30.76 & $-$55.0 & $-$20.0 & 62.55 &  0.84 & &  $-$31.21 & $-$50.0 & $-$10.0 &115.64 &  0.65 \\
G343.75$-$0.16 &  $-$26.80 & $-$33.0 & $-$20.5 & 18.72 &  0.60 & &  $-$28.11 & $-$35.0 & $-$19.0 & 35.62 &  0.13 \\
G344.23$-$0.57 &  $-$22.96 & $-$27.5 & $-$15.0 & 22.44 &  0.70 & &  $-$21.96 & $-$40.0 & $-$10.0 & 44.18 &  0.33 \\
G344.23$-$0.57 & -- &  $-$4.5 &  $-$1.0 &  4.73$\downarrow$ & -- & &   $-$2.93 &  $-$4.5 &  $-$1.0 &  1.13 &  0.11 \\
G345.00$-$0.23 &  $-$29.03 & $-$35.0 & $-$26.5 &  9.59 &  0.55 & &  $-$30.39 & $-$36.5 & $-$12.0 & 34.01 &  0.43 \\
G345.49$+$0.32 &  $-$17.36 & $-$24.0 & $-$10.0 & 88.32 &  0.54 & &  $-$17.69 & $-$27.0 & $-$10.2 &142.40 &  0.18 \\
G345.51$+$0.35 &  $-$18.89 & $-$25.0 &   5.0 & 68.51 &  0.65 & &  $-$17.21 & $-$33.0 &   5.0 &128.31 &  0.33 \\
G345.51$+$0.35 & -- & $-$41.5 & $-$33.0 &  7.31$\downarrow$ & -- & &  $-$38.09 & $-$41.5 & $-$33.0 &  2.29 &  0.16 \\
G345.72$+$0.82 &  $-$11.39 & $-$16.0 &  $-$8.0 & 19.50 &  0.42 & &  $-$10.90 & $-$21.0 &  $-$5.0 & 27.03 &  0.17 \\
G351.13$+$0.77 &   $-$5.09 & $-$10.0 &   0.0 & 21.22 &  0.79 & &   $-$5.39 & $-$12.0 &   1.0 & 44.39 &  0.18 \\
G351.16$+$0.70 &   $-$7.36 & $-$22.0 &   6.0 & 66.27 &  0.51 & &   $-$6.51 & $-$30.0 &   8.0 &190.41 &  0.37 \\
G351.25$+$0.67 &   $-$3.47 & $-$10.0 &   5.0 & 63.40 &  0.38 & &   $-$3.03 & $-$13.0 &  10.0 &175.29 &  0.25 \\
G351.42$+$0.65 &   $-$6.96 & $-$17.0 &   3.0 &108.45 &  1.81 & &   $-$7.12 & $-$17.0 &   7.0 &218.95 &  0.35 \\
G351.42$+$0.65 & -- & $-$25.0 & $-$17.0 & 24.83$\downarrow$ & -- & &  $-$20.71 & $-$25.0 & $-$17.0 &  7.32 &  0.20 \\
G351.42$+$0.65 & -- &  30.0 &  40.0 & 28.97$\downarrow$ & -- & &   33.44 &  30.0 &  40.0 &  3.80 &  0.23 \\
G351.45$+$0.66 &   $-$4.53 & $-$13.0 &   3.0 & 61.06 &  0.82 & &   $-$4.75 & $-$15.0 &   7.0 &141.01 &  0.22 \\
G351.57$+$0.77 &   $-$3.72 & $-$10.0 &   2.0 & 39.80 &  0.41 & &   $-$3.62 &  $-$8.0 &   1.0 & 37.28 &  0.17 \\
G351.57$+$0.77 &  $-$15.71 & $-$19.0 & $-$13.0 &  3.77 &  0.29 & & -- & $-$19.0 & $-$13.0 &  2.07$\downarrow$ & -- \\
G351.58$-$0.35 &  $-$97.74 &$-$106.0 & $-$77.0 & 38.39 &  1.06 & &  $-$98.00 &$-$110.0 & $-$84.0 & 50.25 &  0.27 \\
G351.77$-$0.54 &   $-$2.16 & $-$15.0 &  10.0 & 43.40 &  0.82 & &   $-$2.77 & $-$28.0 &  19.0 &191.80 &  0.56 \\
G353.41$-$0.36 &  $-$15.20 & $-$21.0 &  $-$9.0 & 56.45 &  0.64 & &  $-$15.50 & $-$25.0 &  $-$3.0 &110.96 &  0.22 \\
G354.95$-$0.54 &   $-$5.46 & $-$12.0 &   1.0 & 22.43 &  0.36 & &   $-$5.43 & $-$10.0 &   0.0 & 31.24 &  0.18 \\ \bottomrule 
\end{longtable}
\end{center}
\end{landscape}
\end{ThreePartTable}
\twocolumn

\section{Spectra} 
\label{s:appendix_spectra}

\begin{figure*}
\centering
\includegraphics[scale=0.46]{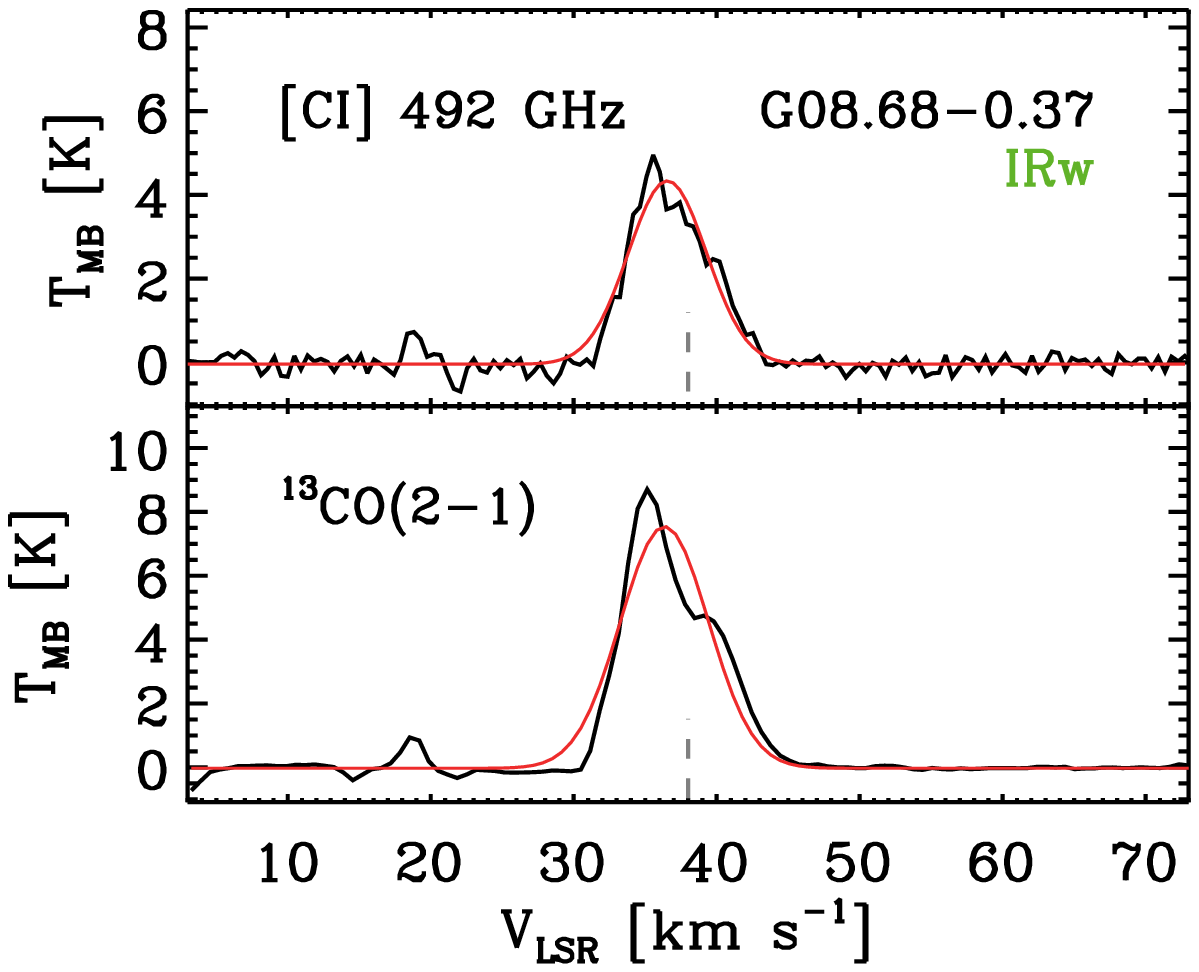} 
\includegraphics[scale=0.46]{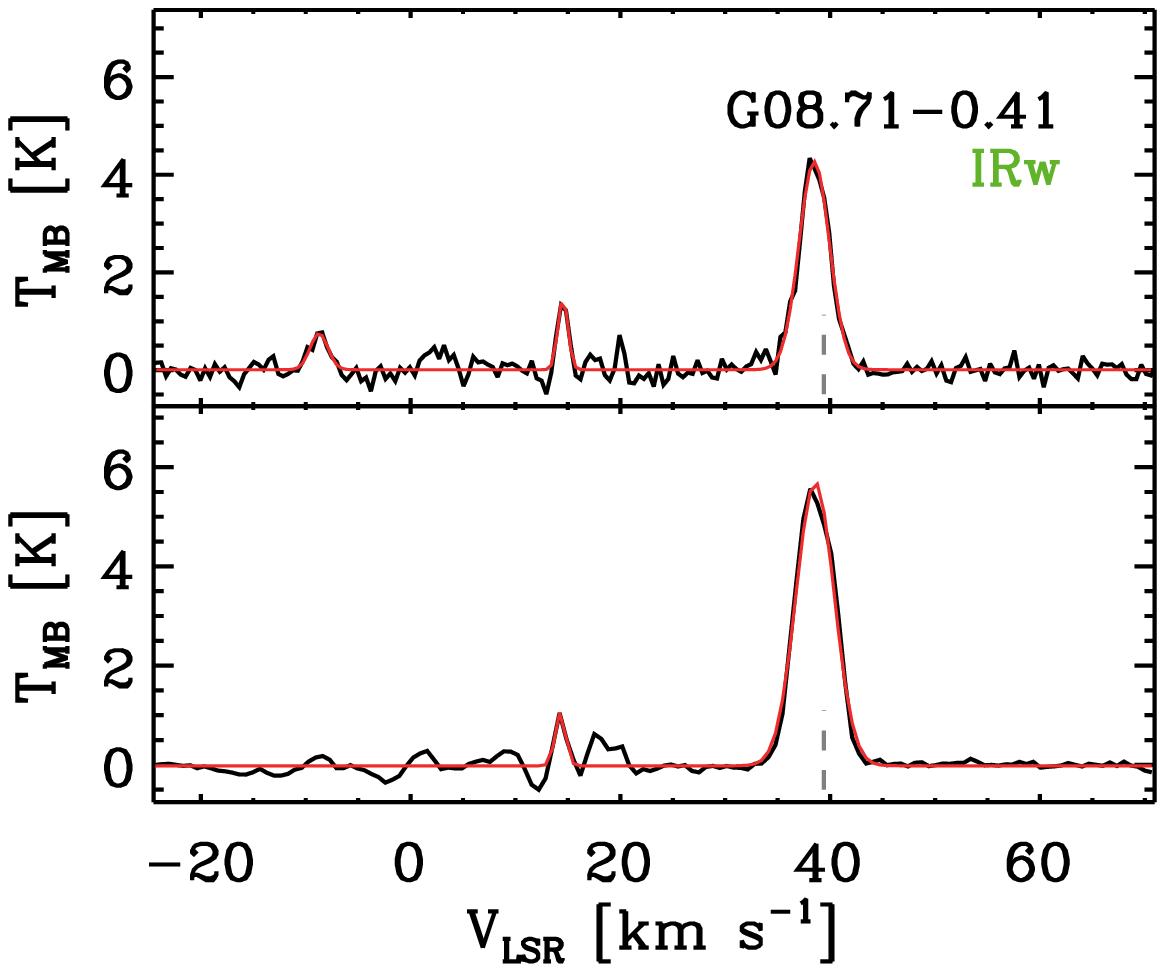} 
\includegraphics[scale=0.46]{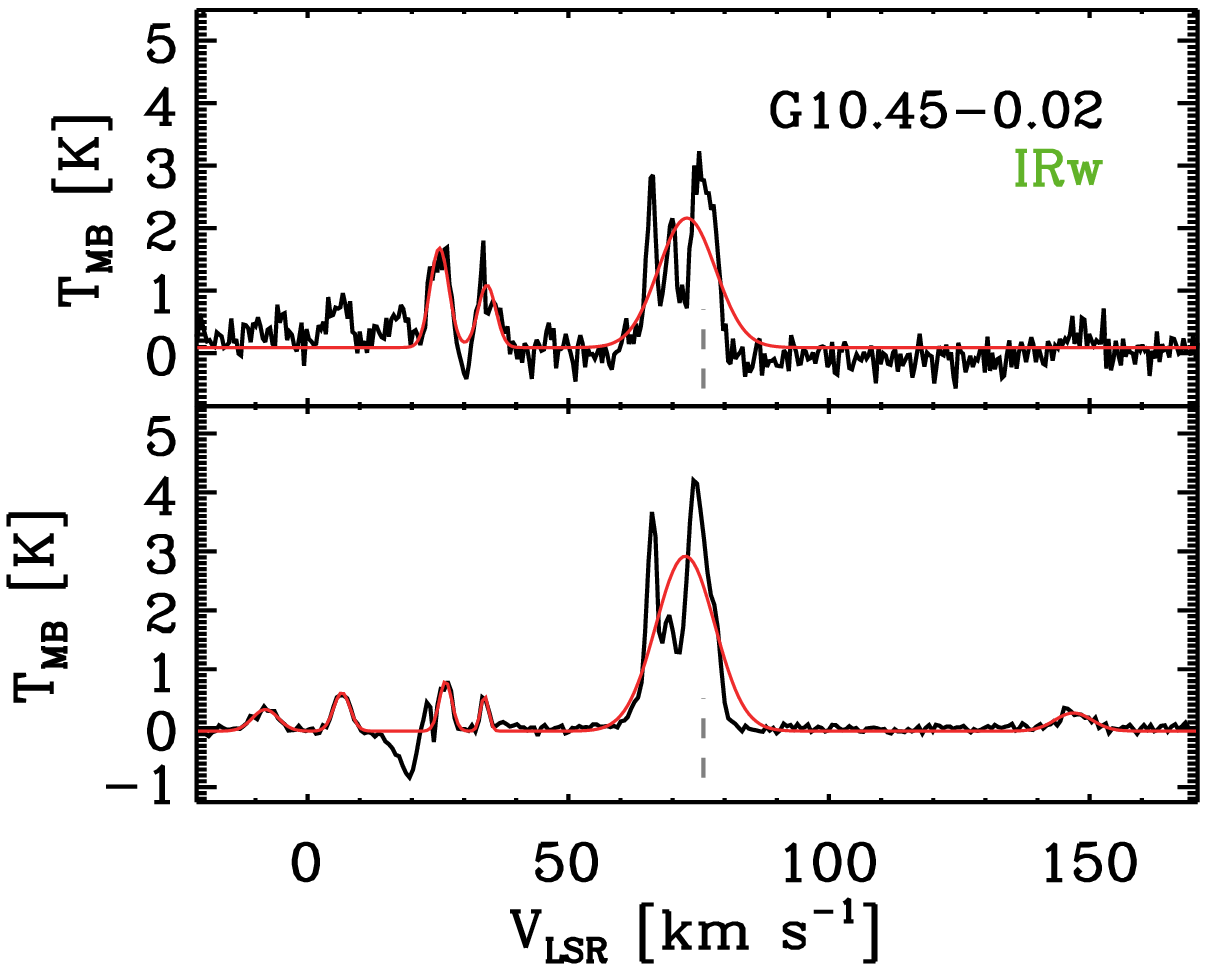} 
\includegraphics[scale=0.46]{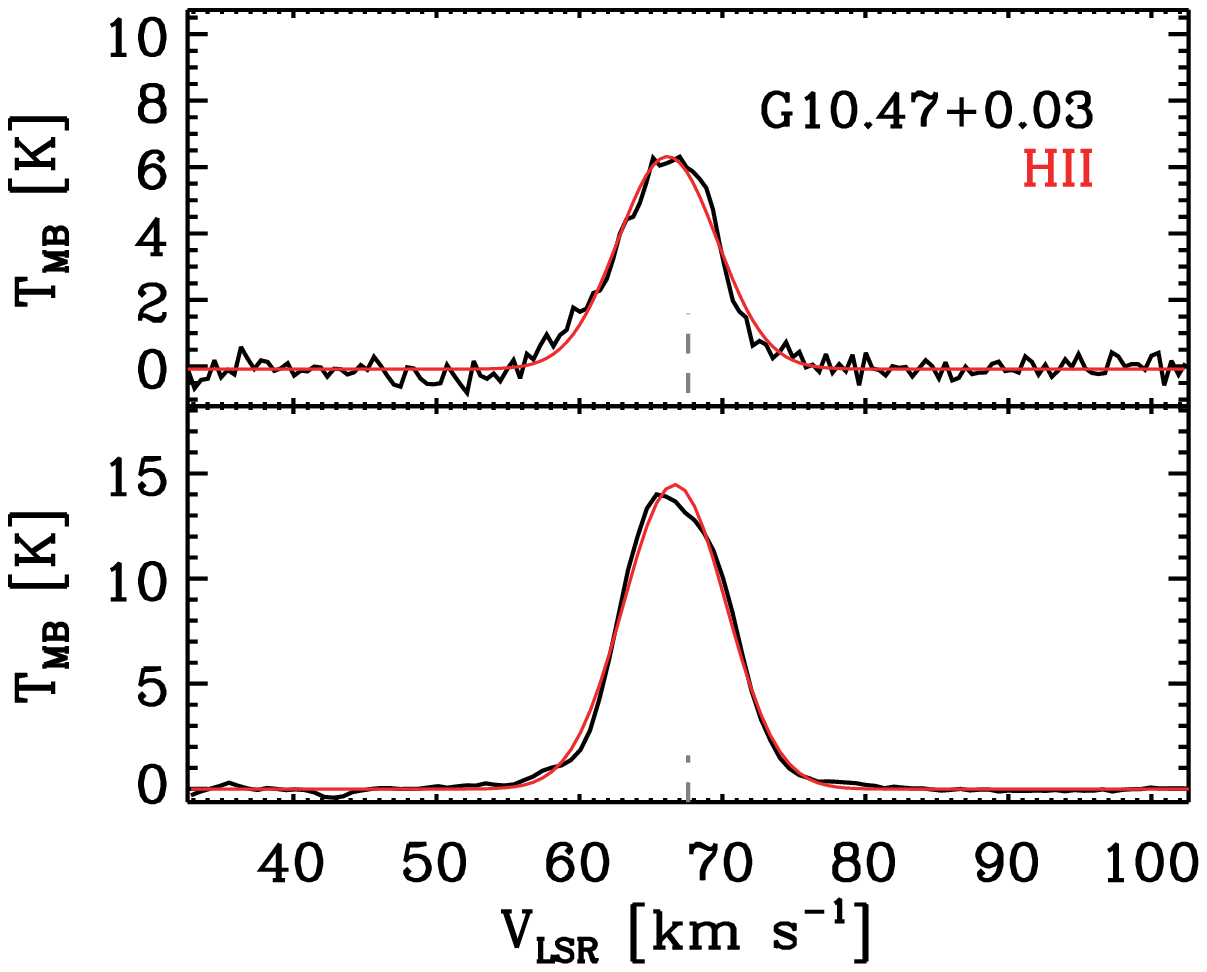}
\includegraphics[scale=0.46]{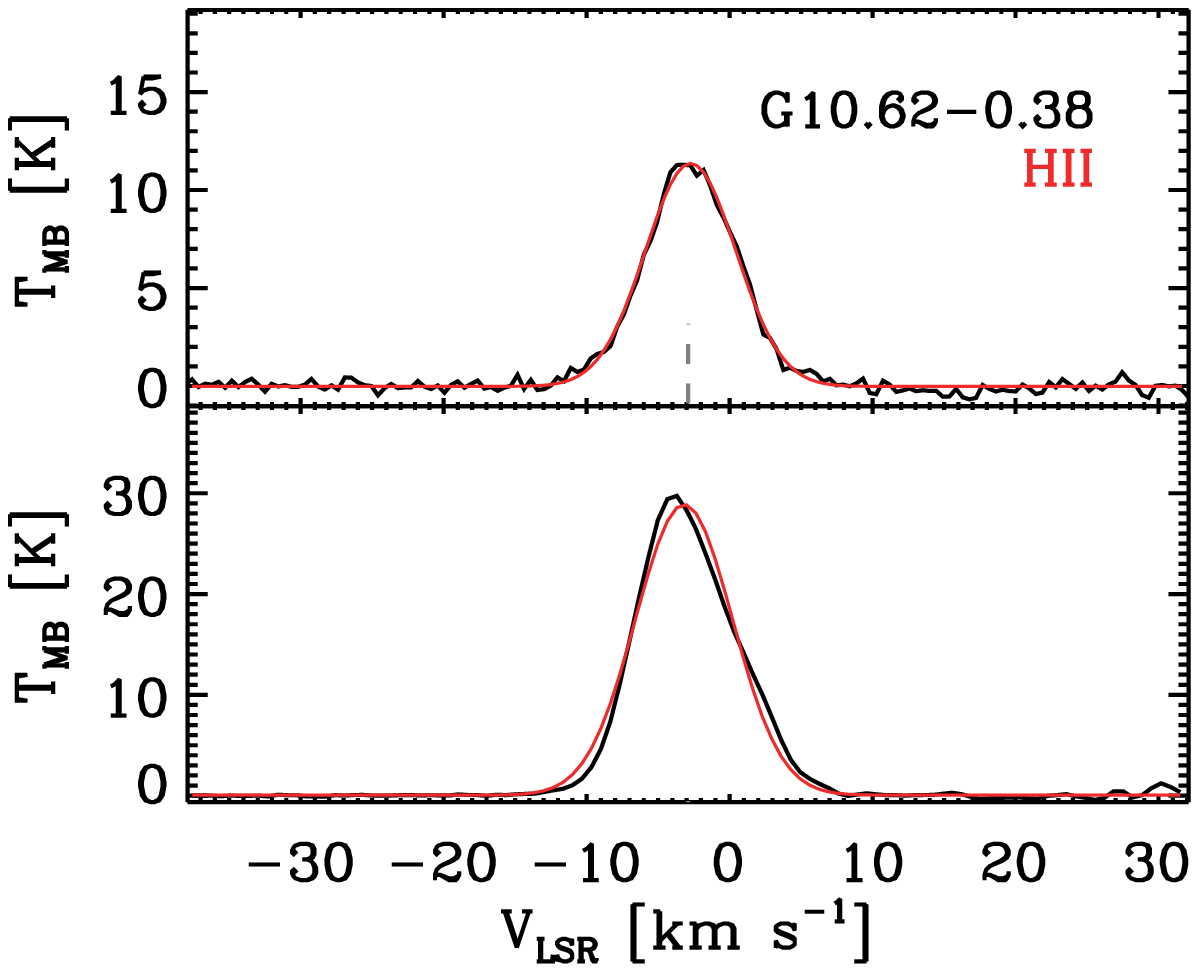}
\includegraphics[scale=0.46]{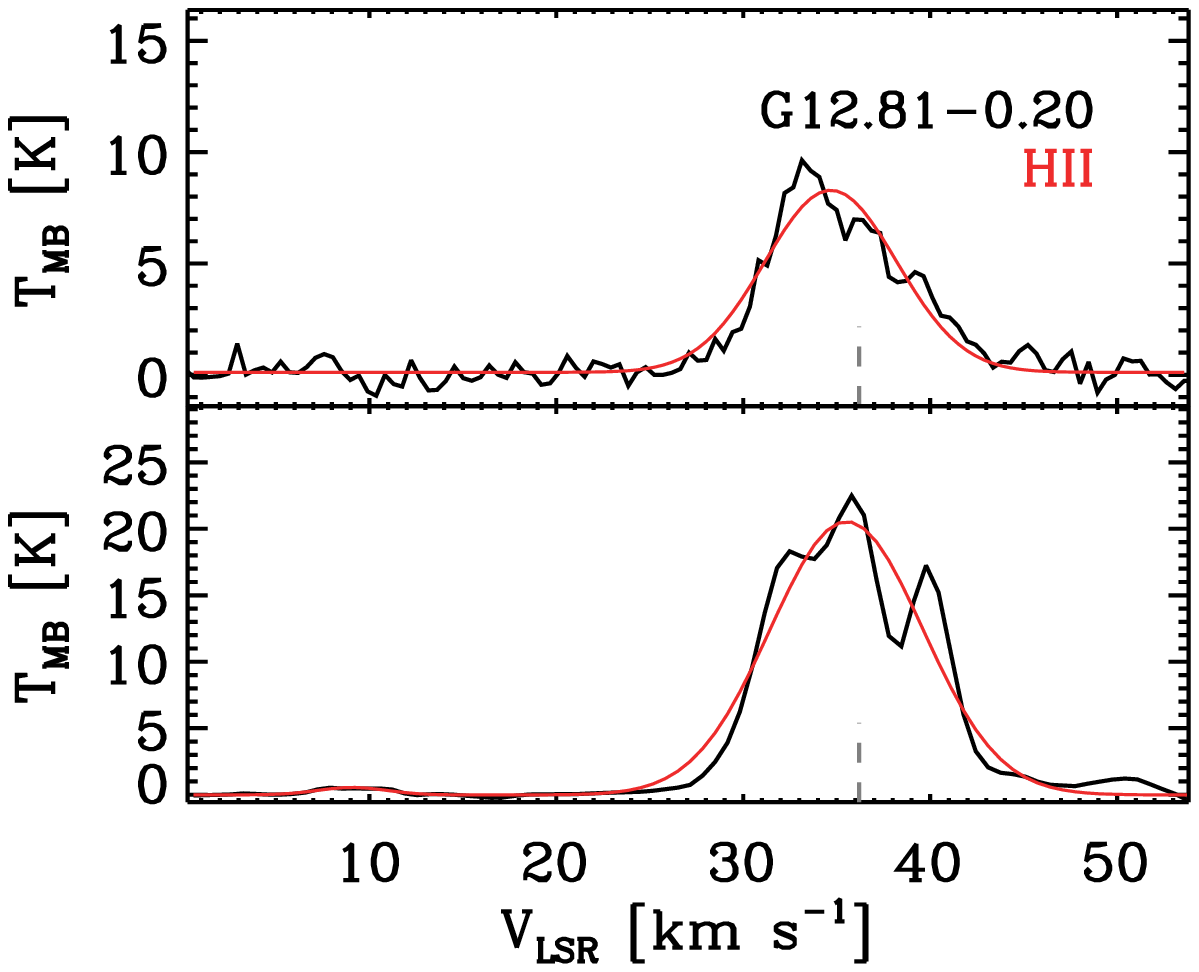}
\includegraphics[scale=0.46]{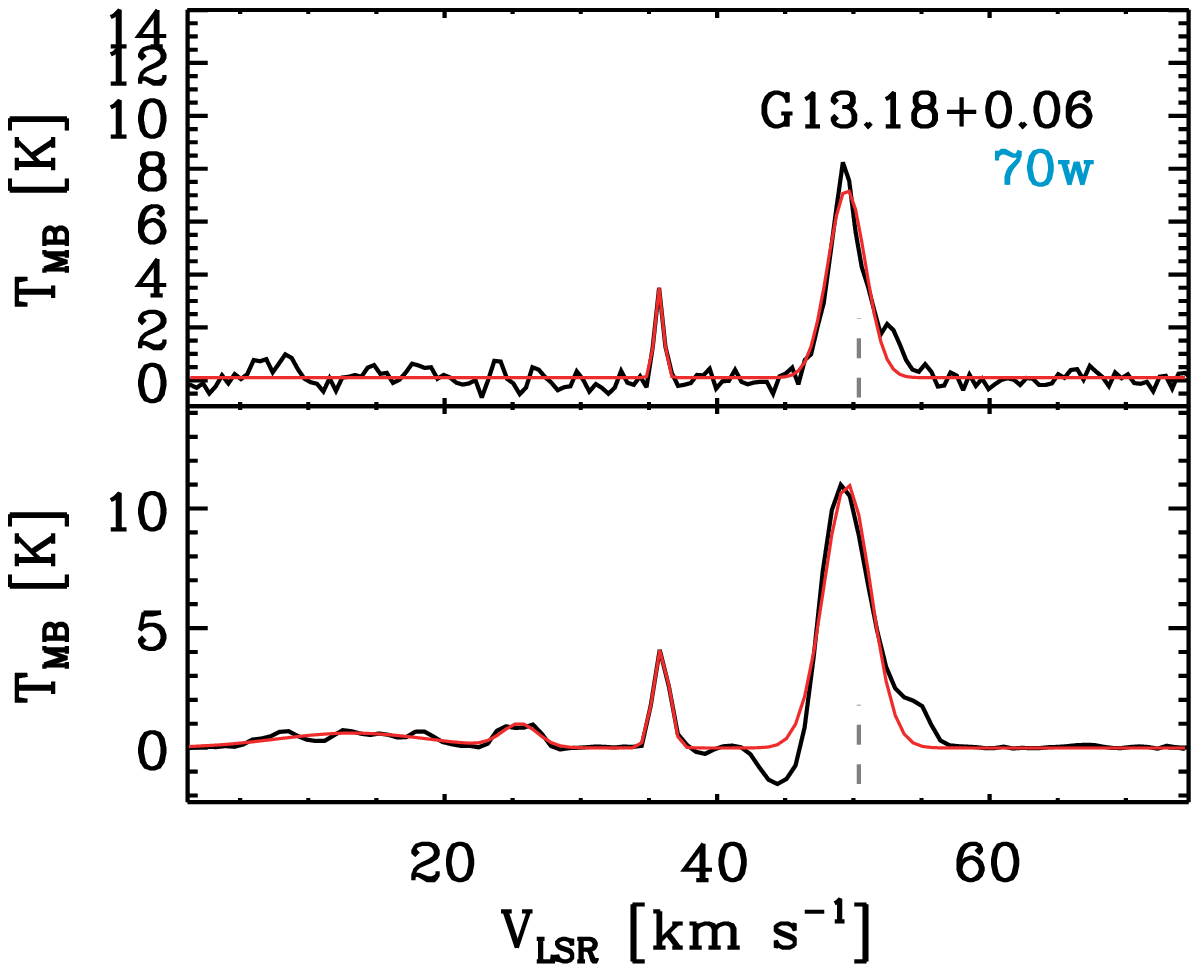}
\includegraphics[scale=0.46]{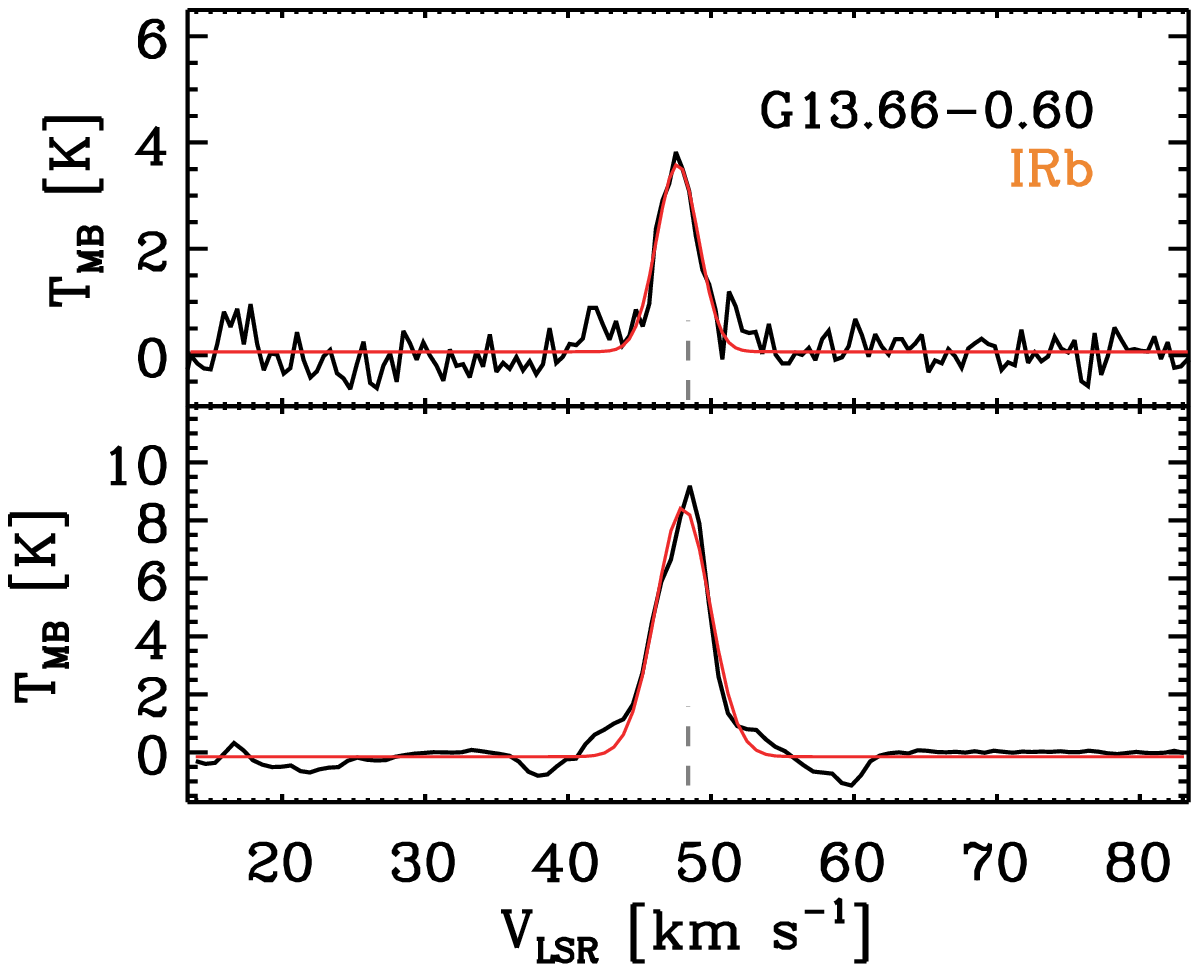}
\includegraphics[scale=0.46]{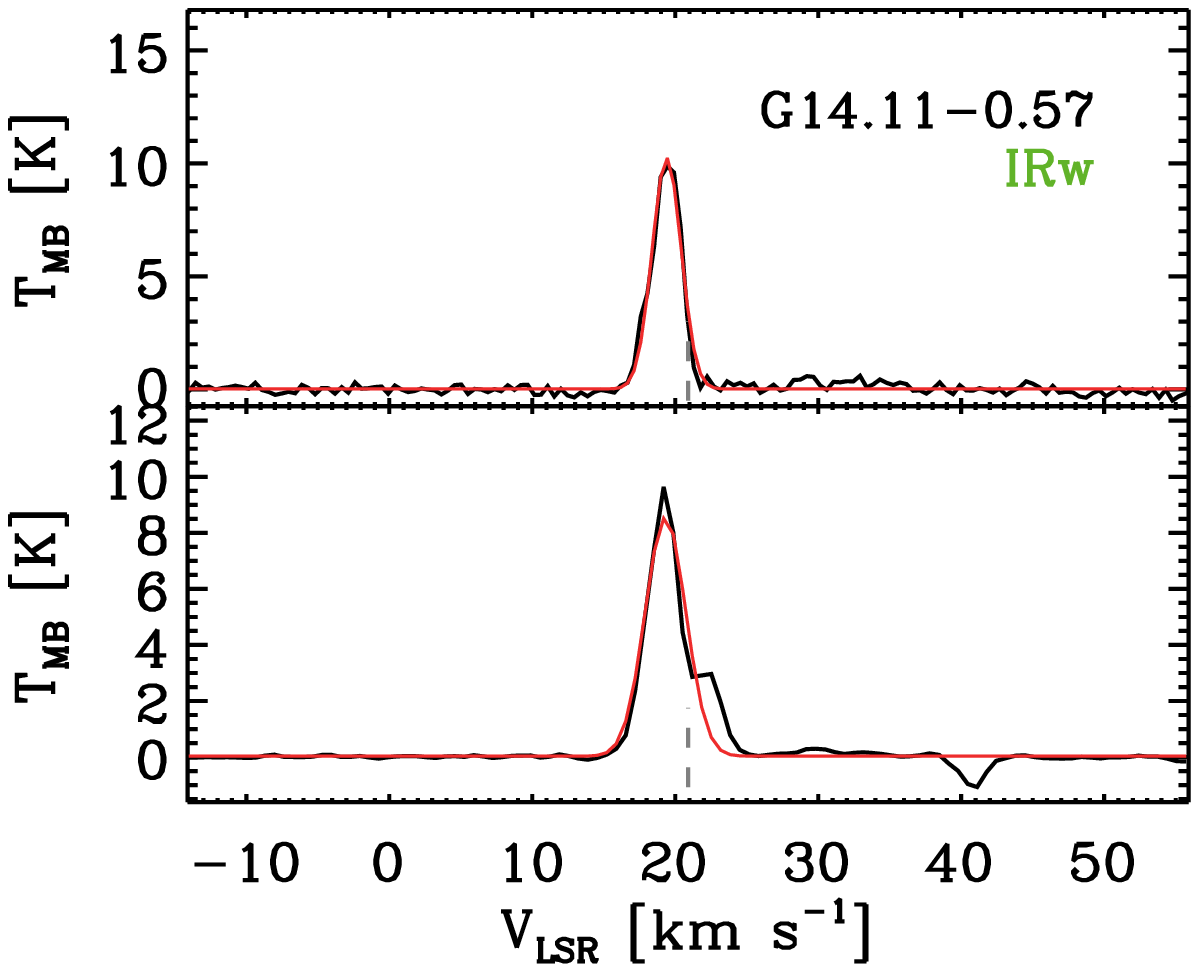}
\includegraphics[scale=0.46]{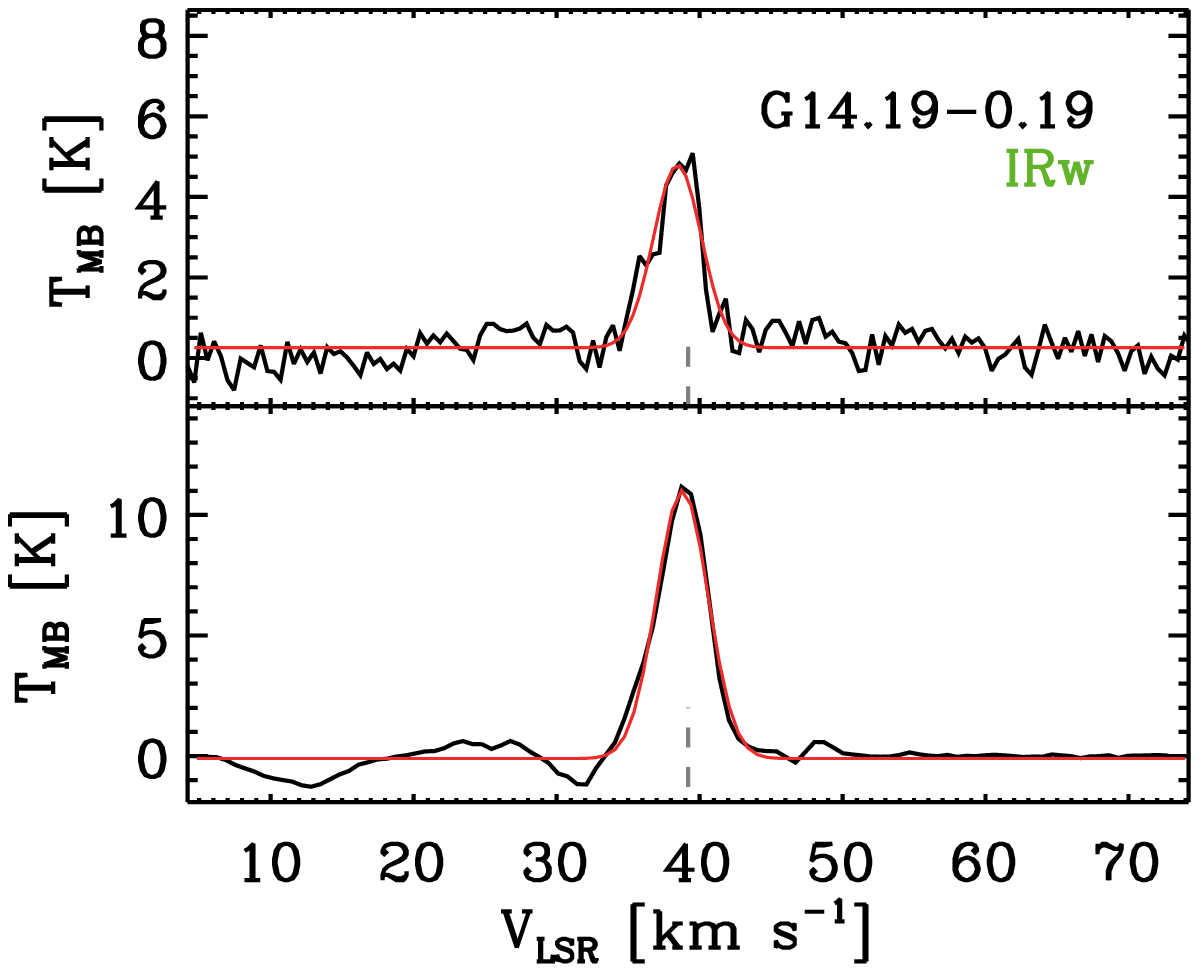}
\includegraphics[scale=0.46]{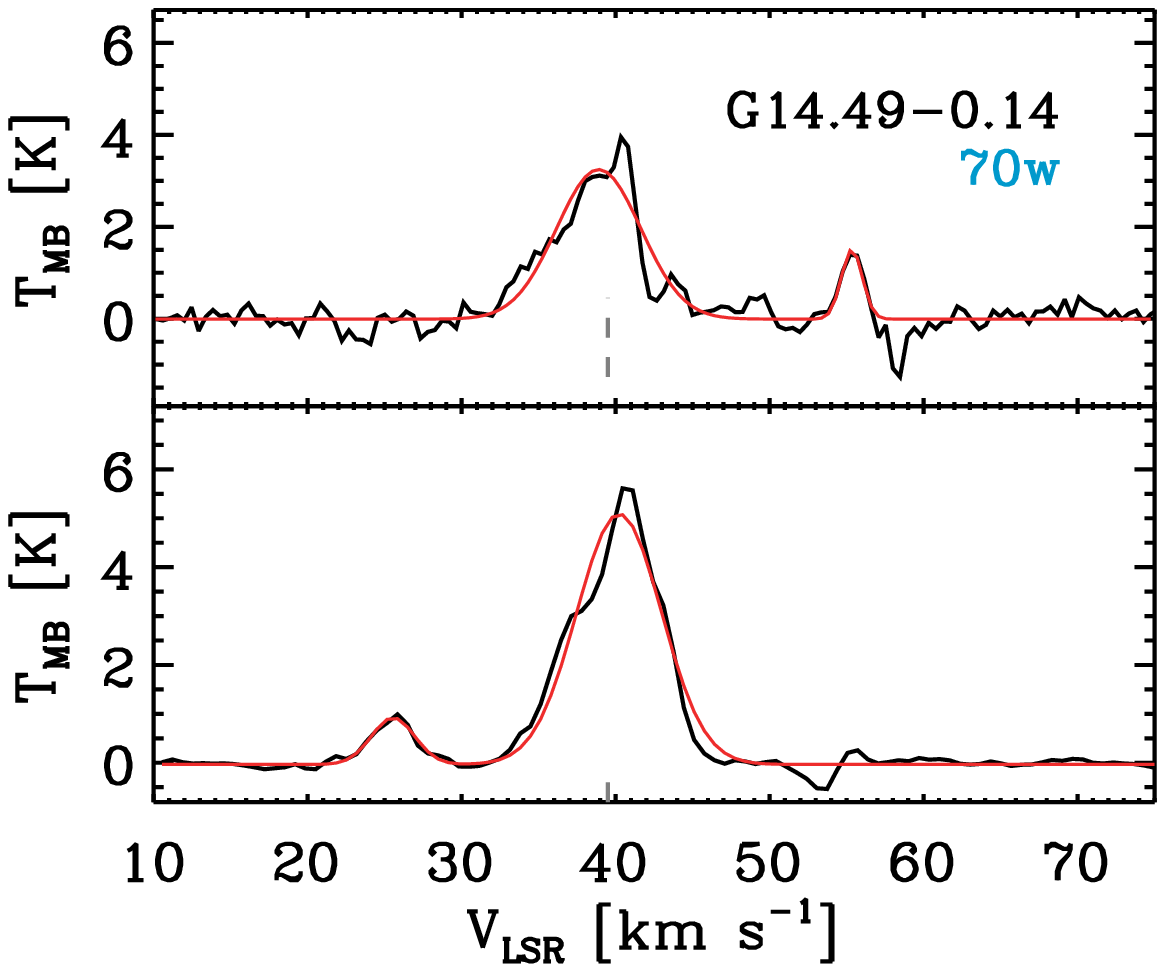}
\includegraphics[scale=0.46]{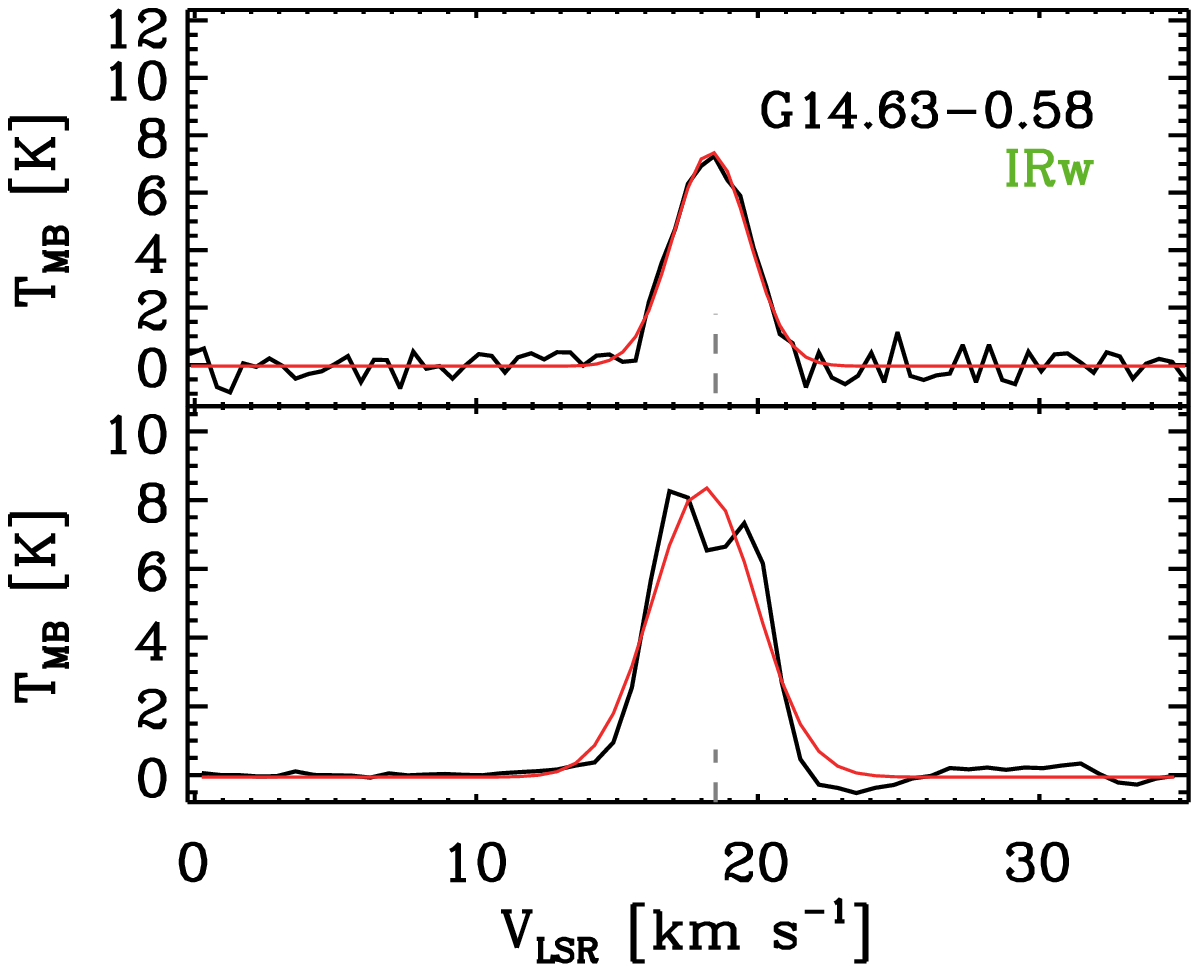}
\includegraphics[scale=0.46]{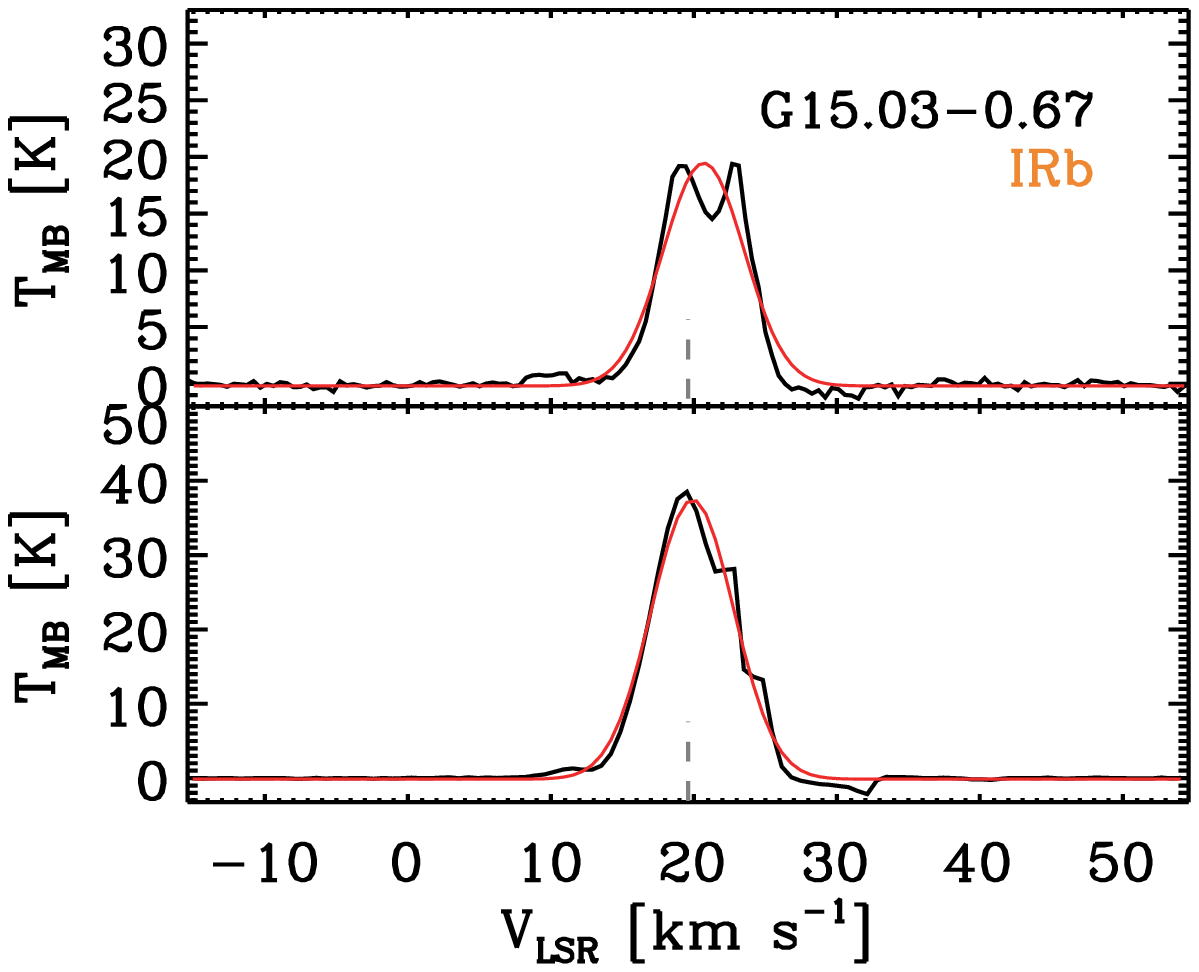}
\includegraphics[scale=0.46]{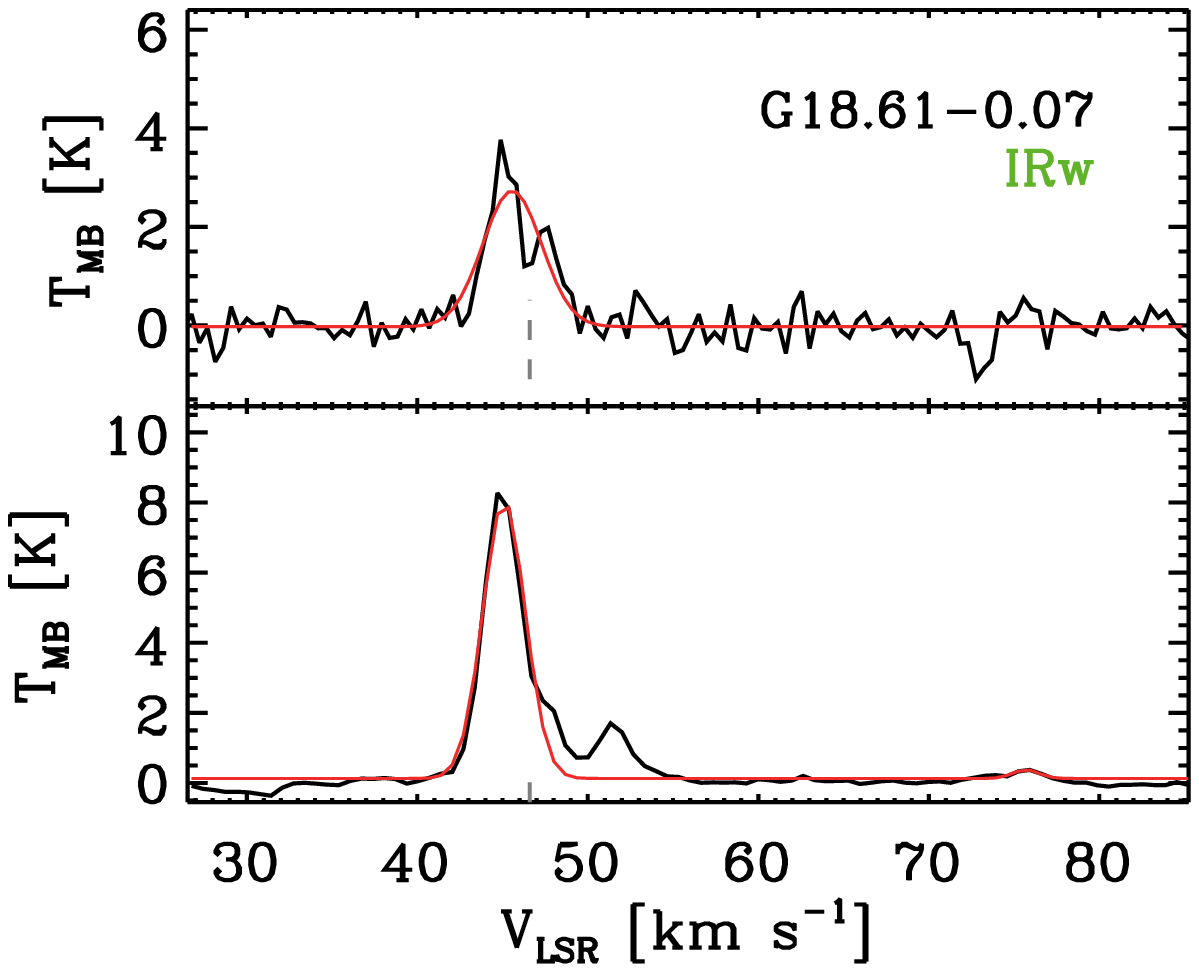}
\includegraphics[scale=0.46]{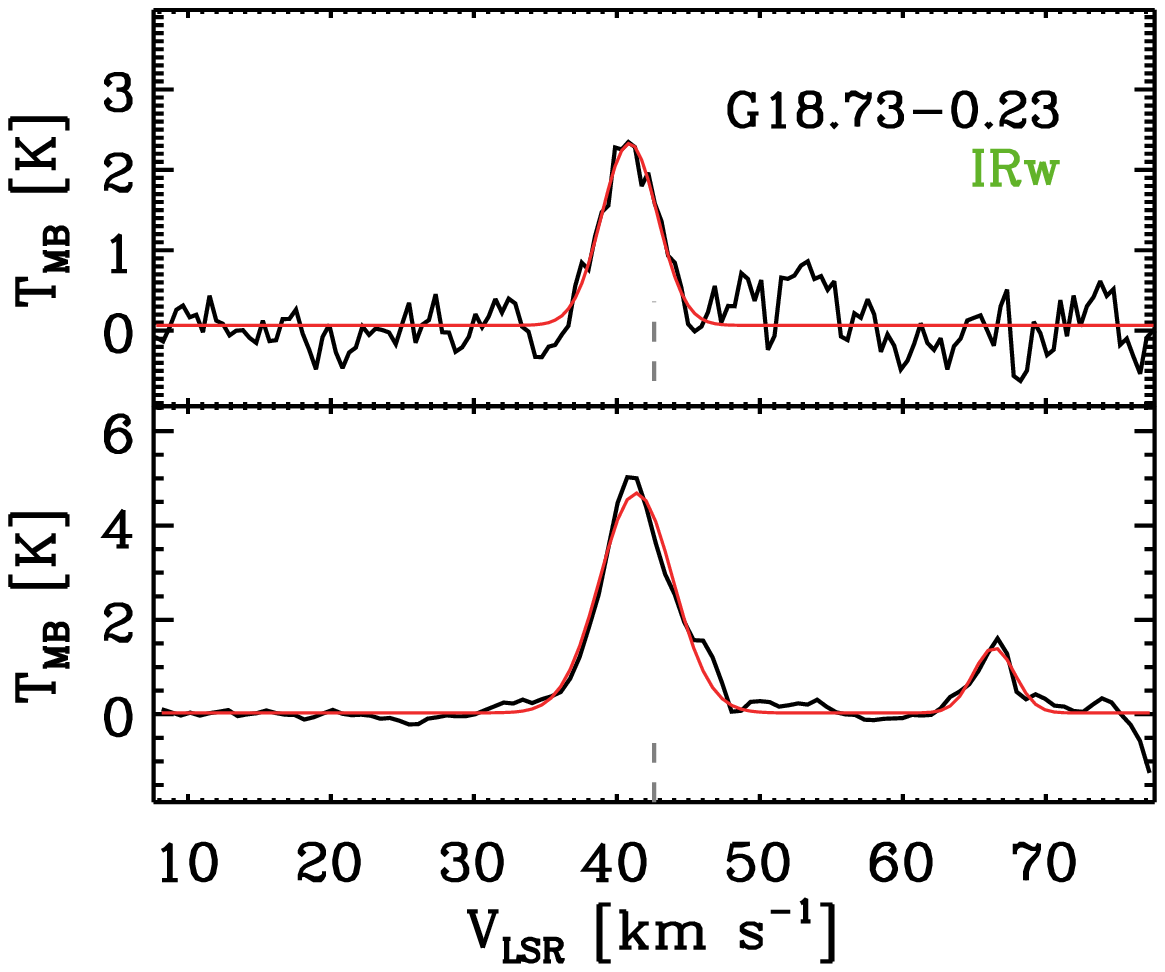}
\caption{\label{f:CI_CO_plot_comp}
APEX \CI 492 GHz and $^{13}$CO(2--1) spectra (\textit{top} and \textit{bottom} for each panel). 
The fitted Gaussians are overlaid in red, and the source velocities are indicated as the gray dashed lines. 
Several spectra (in particular $^{13}$CO(2--1)) show absorption features, which most likely result from contaminated reference positions. 
On the other hand, [C~\textsc{i}]-absorbing components toward G43.17$+$0.01 and G330.95$-$0.18 are labelled with the word ``ABS''.}
\end{figure*}
 
\begin{figure*}
\centering
\ContinuedFloat
\includegraphics[scale=0.46]{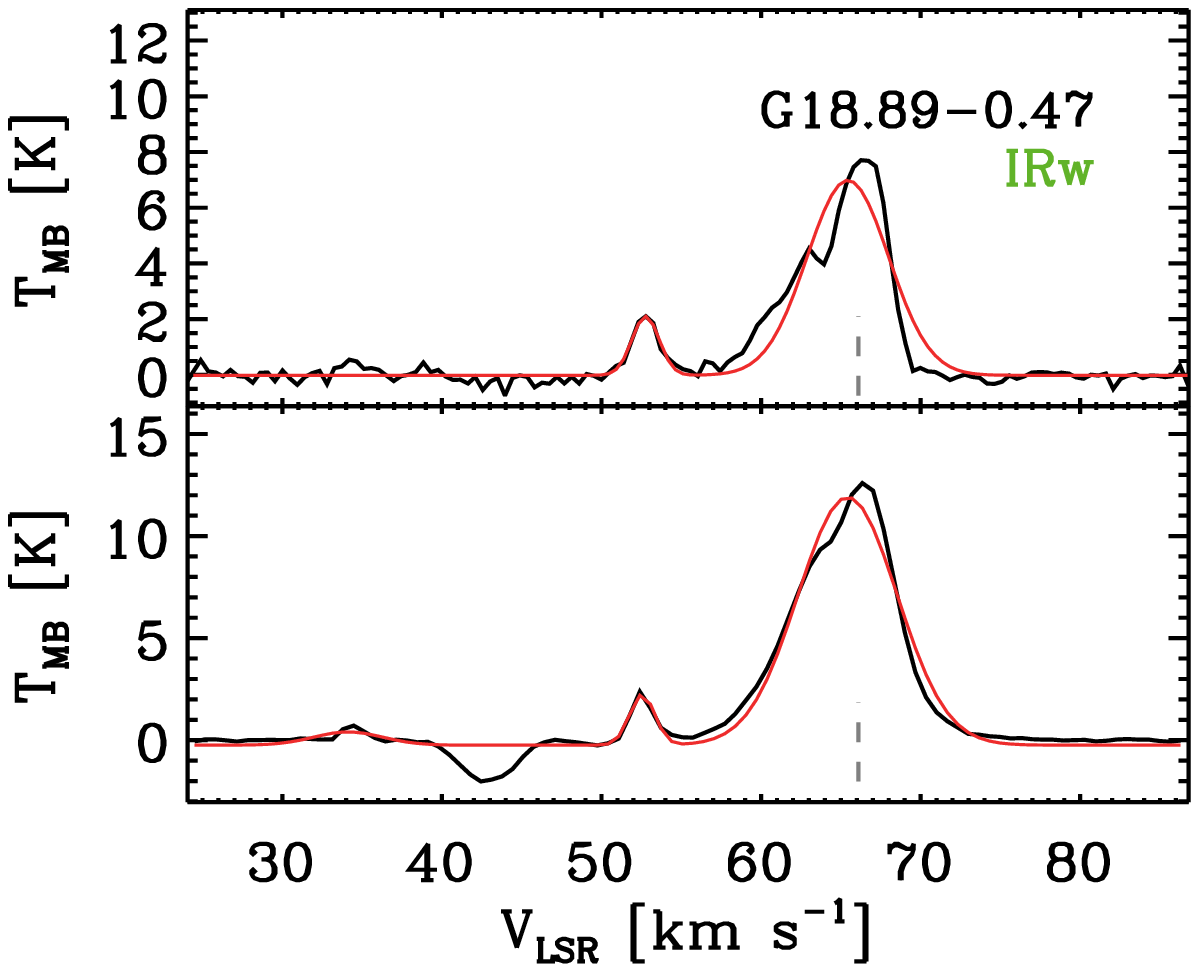}
\includegraphics[scale=0.46]{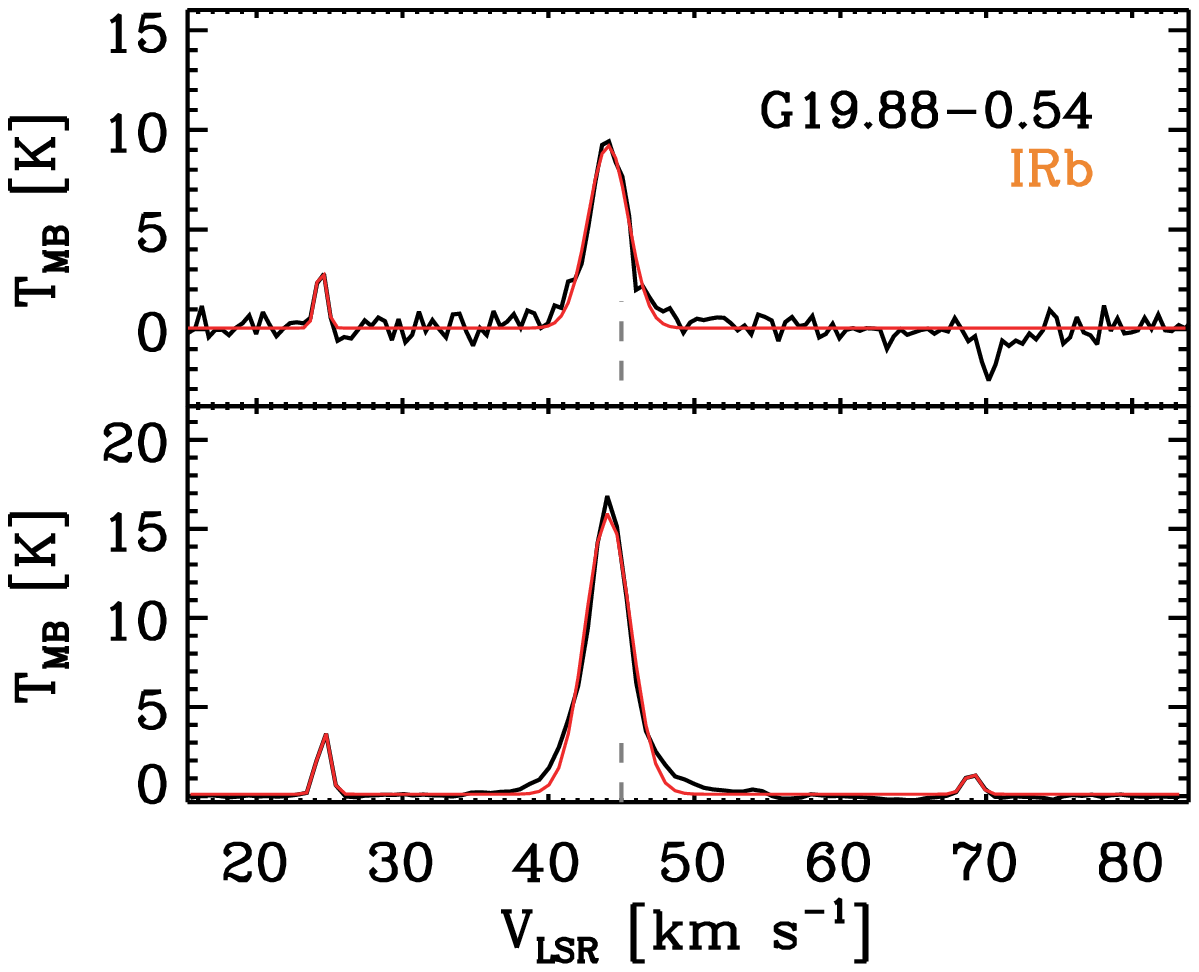}
\includegraphics[scale=0.46]{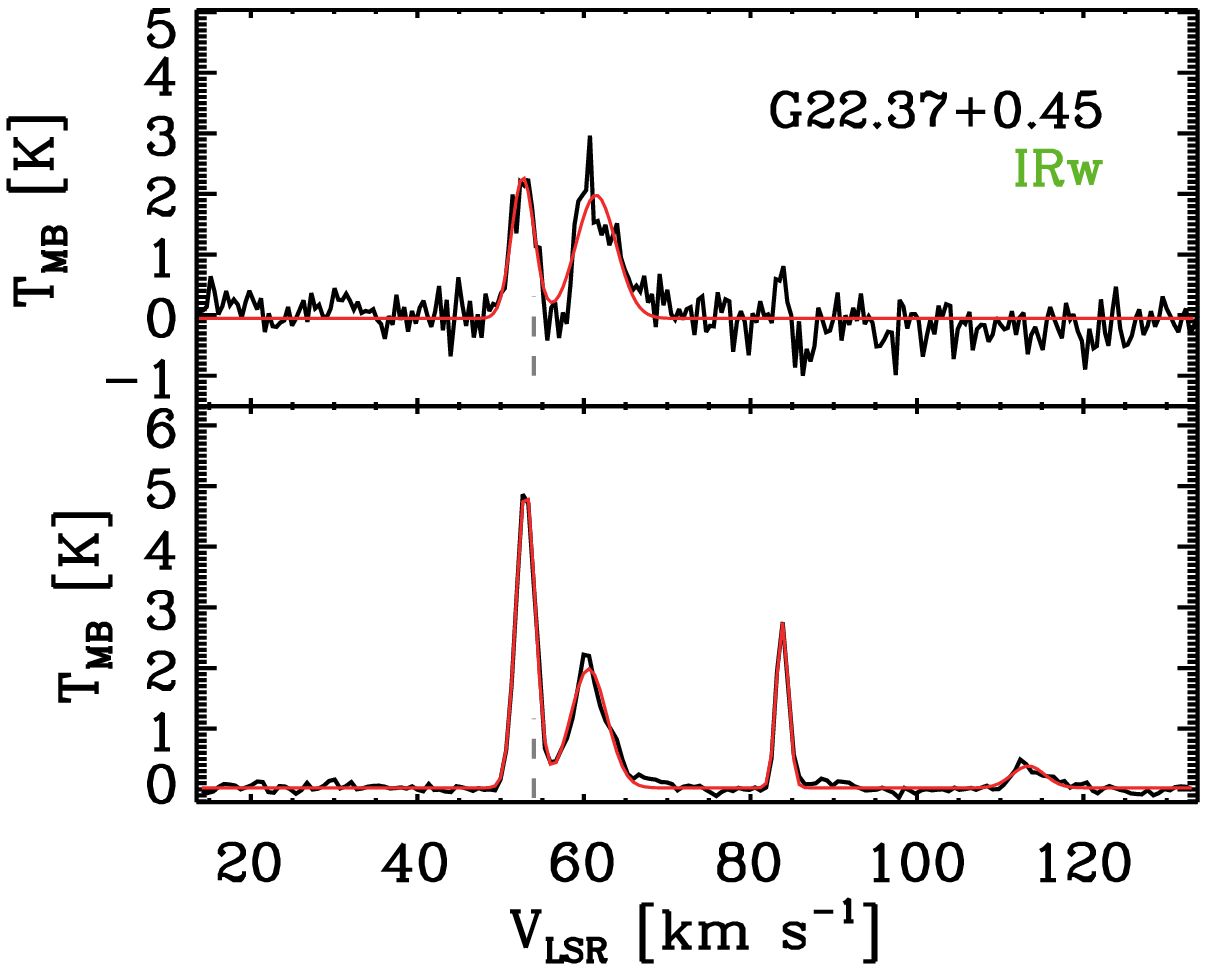}
\includegraphics[scale=0.46]{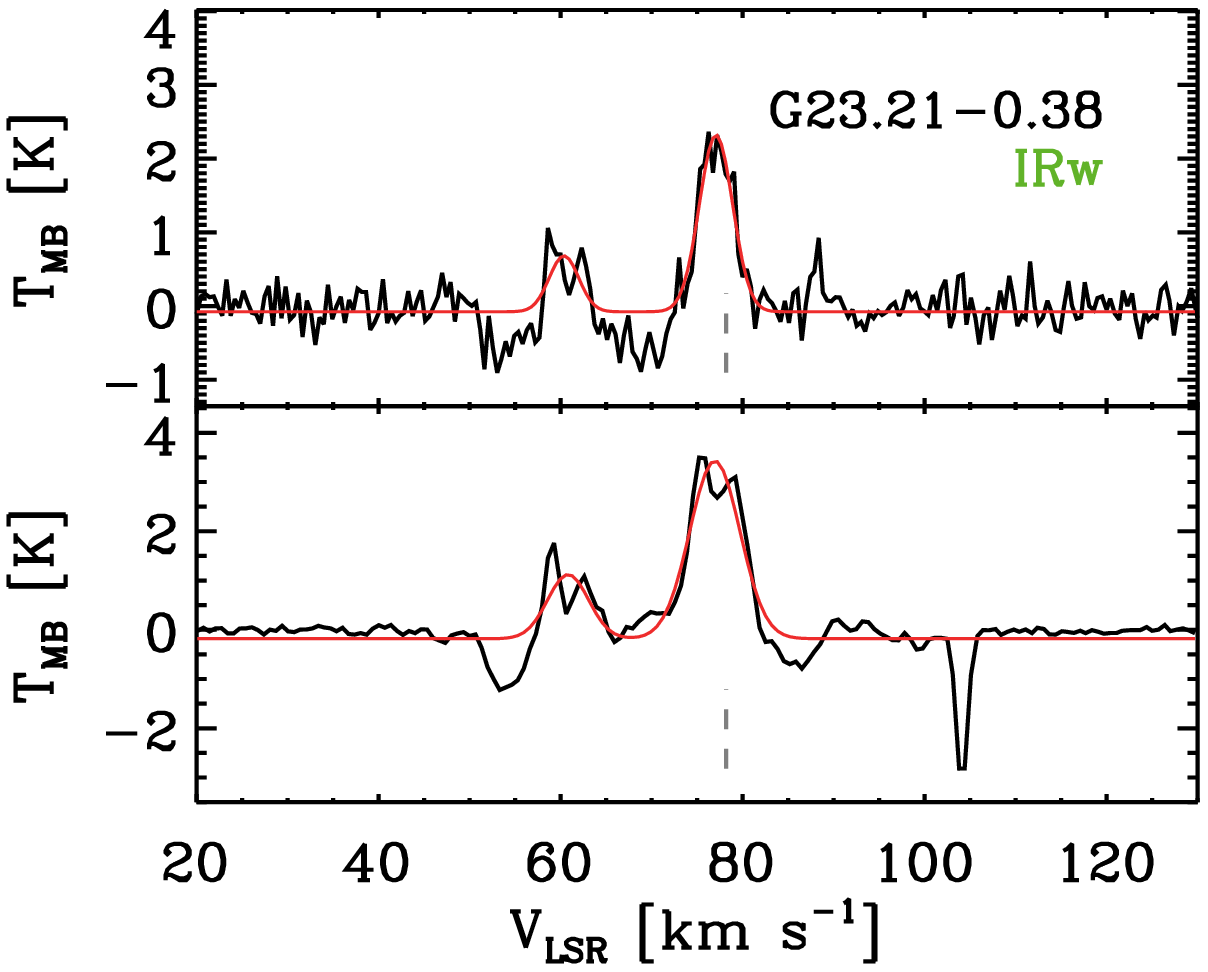}
\includegraphics[scale=0.46]{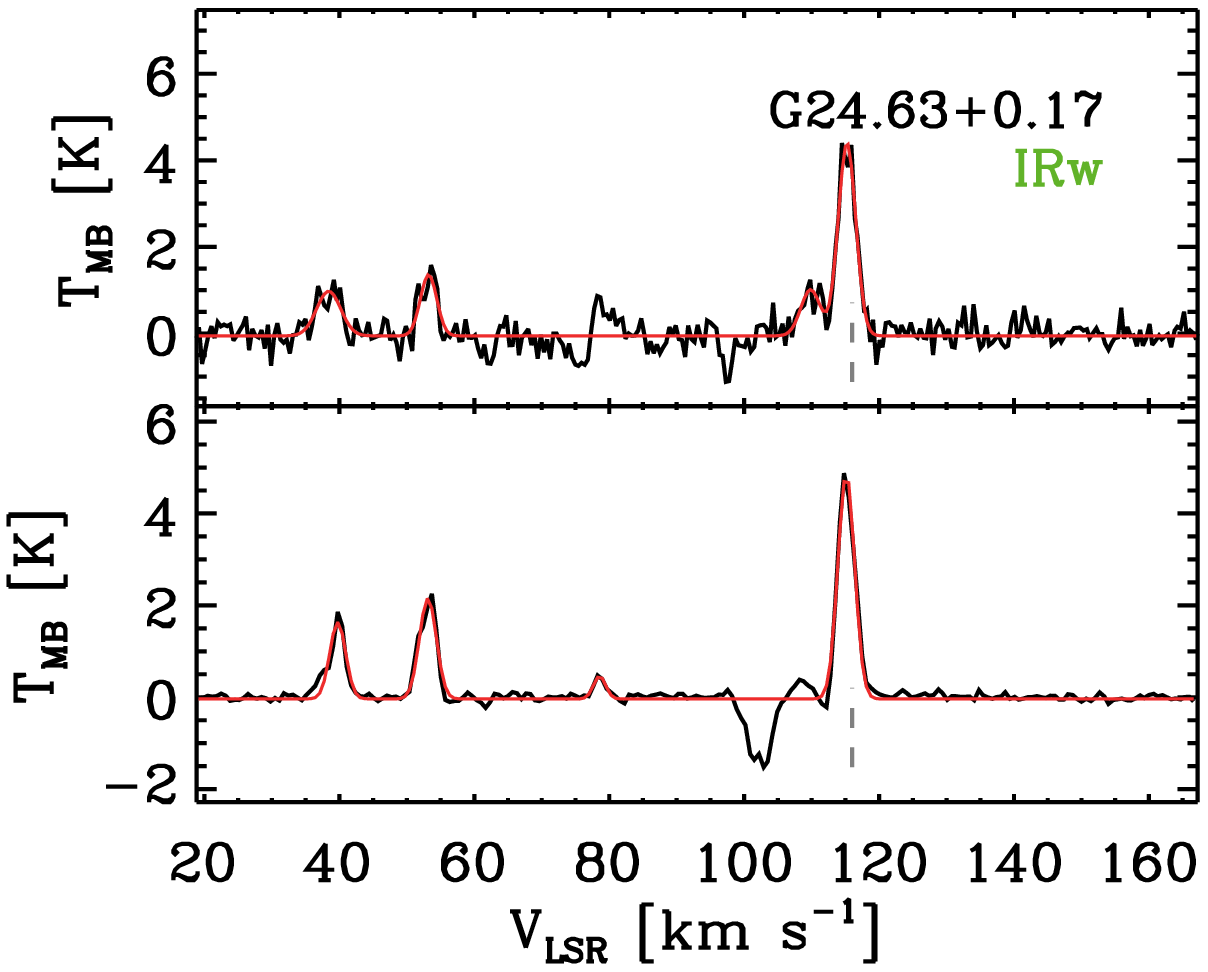}
\includegraphics[scale=0.46]{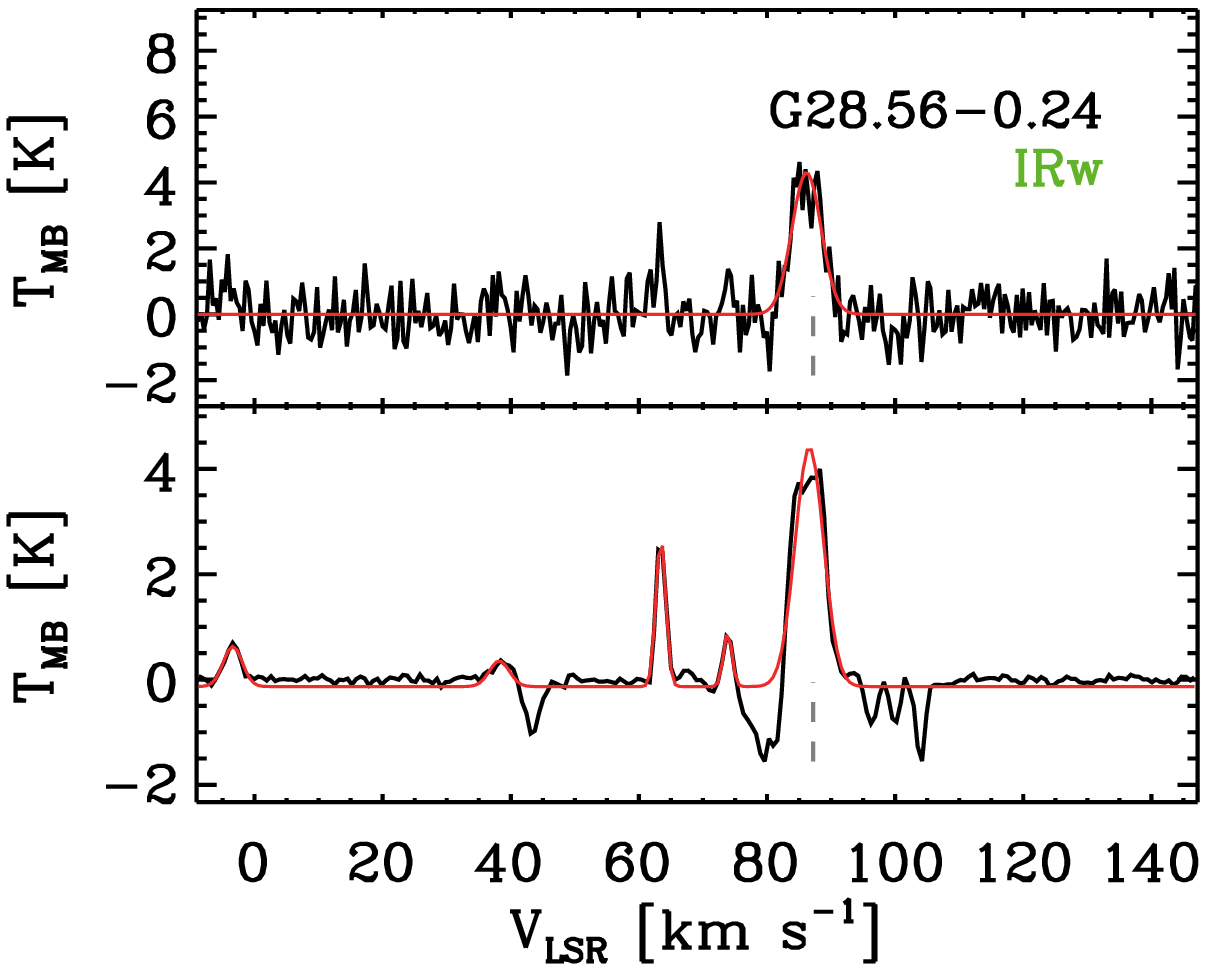}
\includegraphics[scale=0.46]{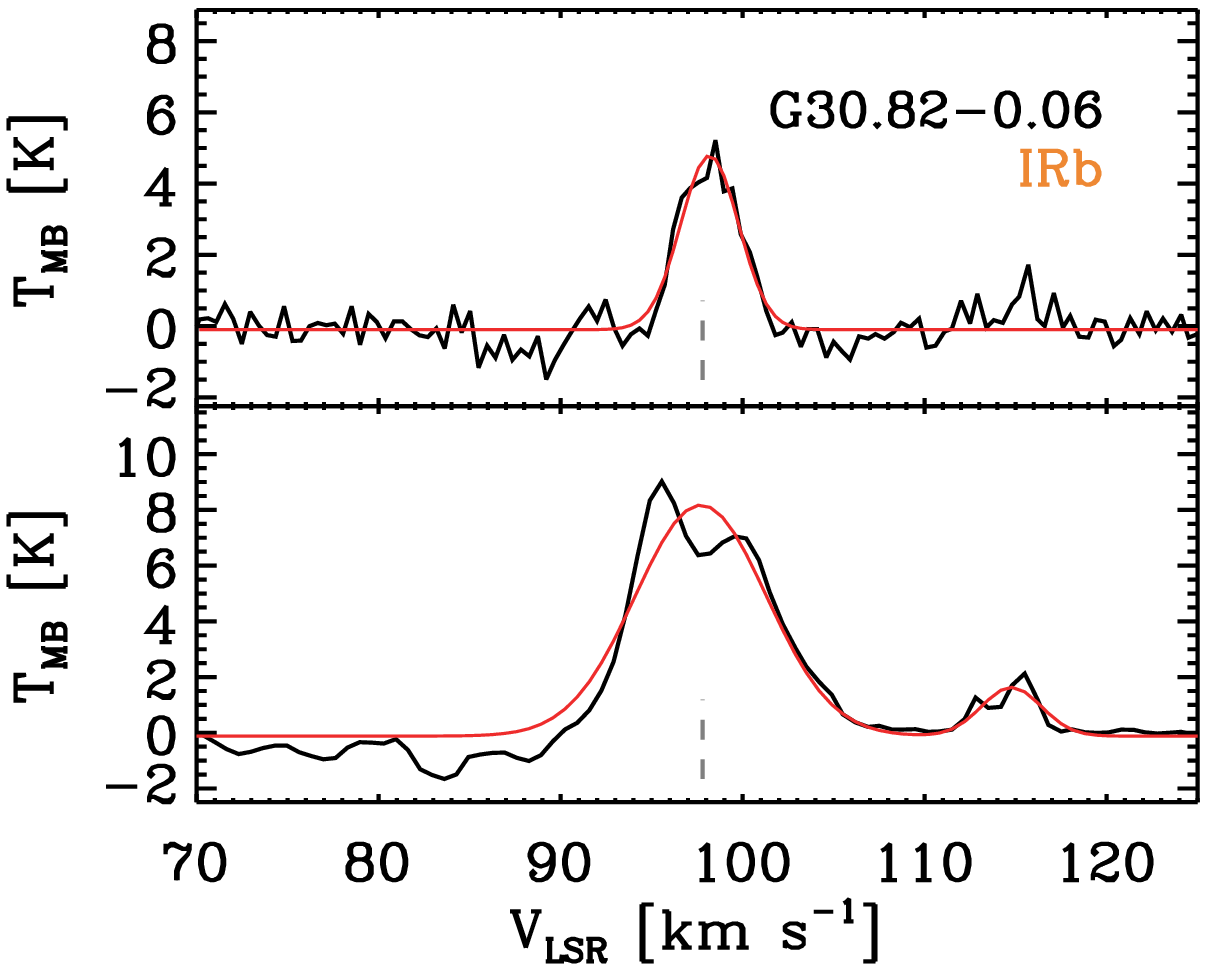}
\includegraphics[scale=0.46]{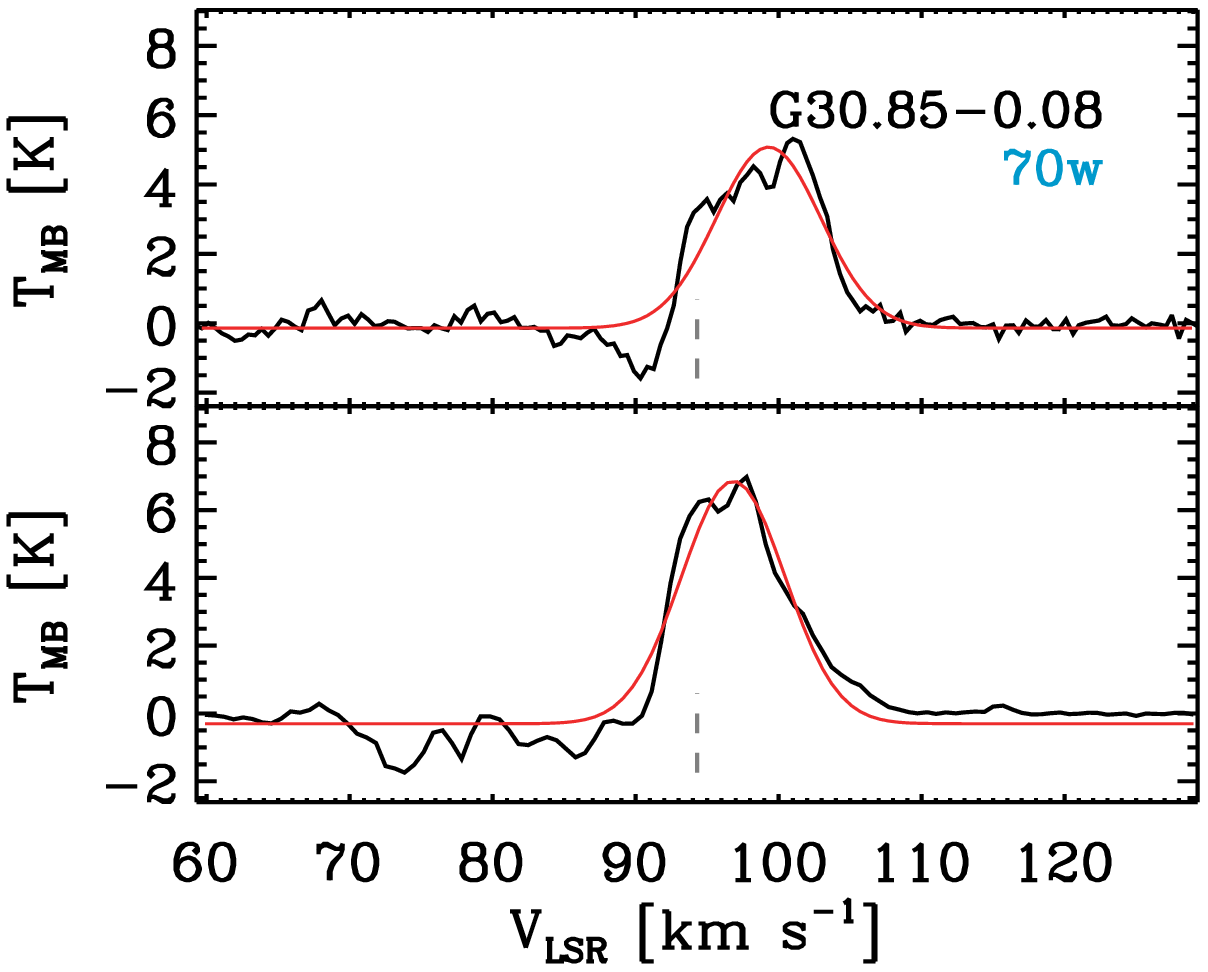}
\includegraphics[scale=0.46]{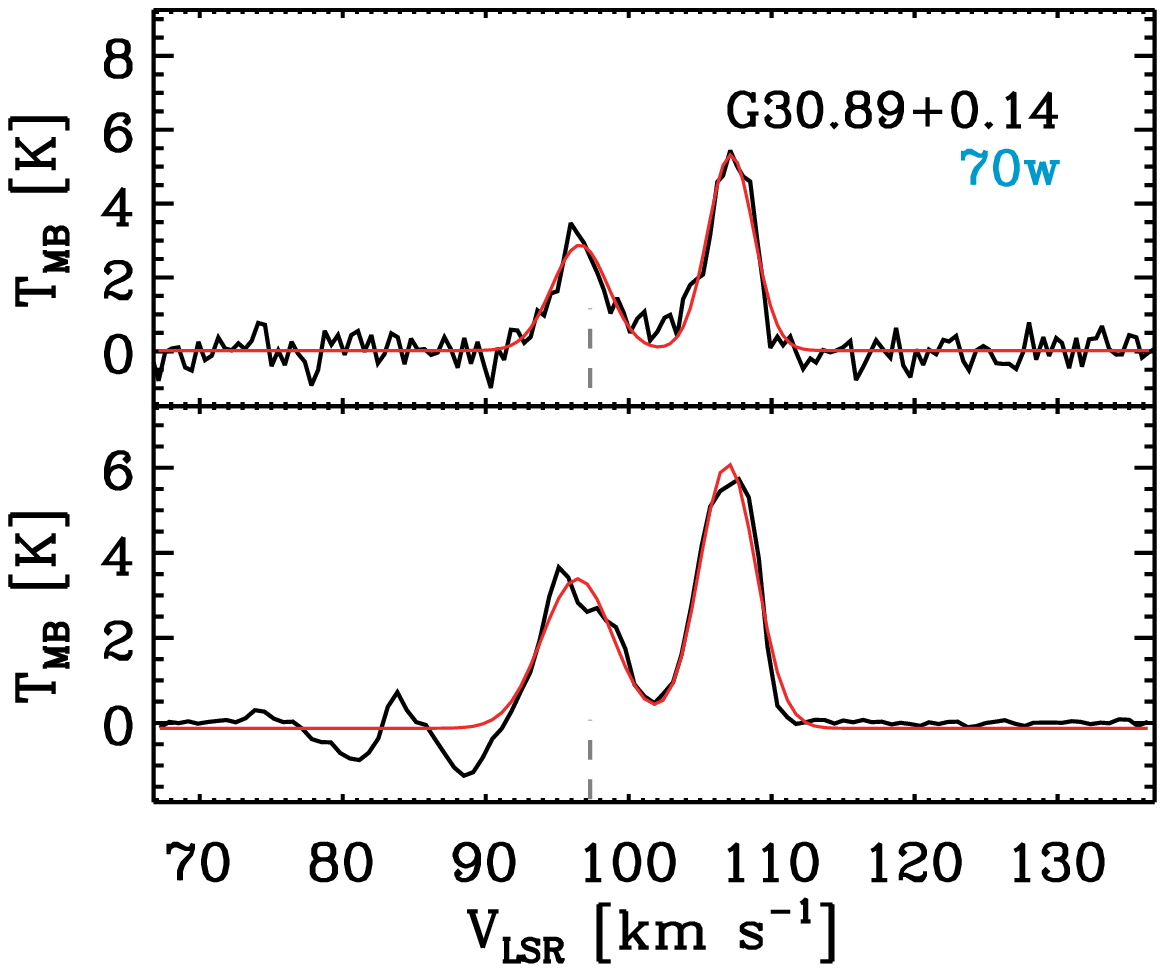}
\includegraphics[scale=0.46]{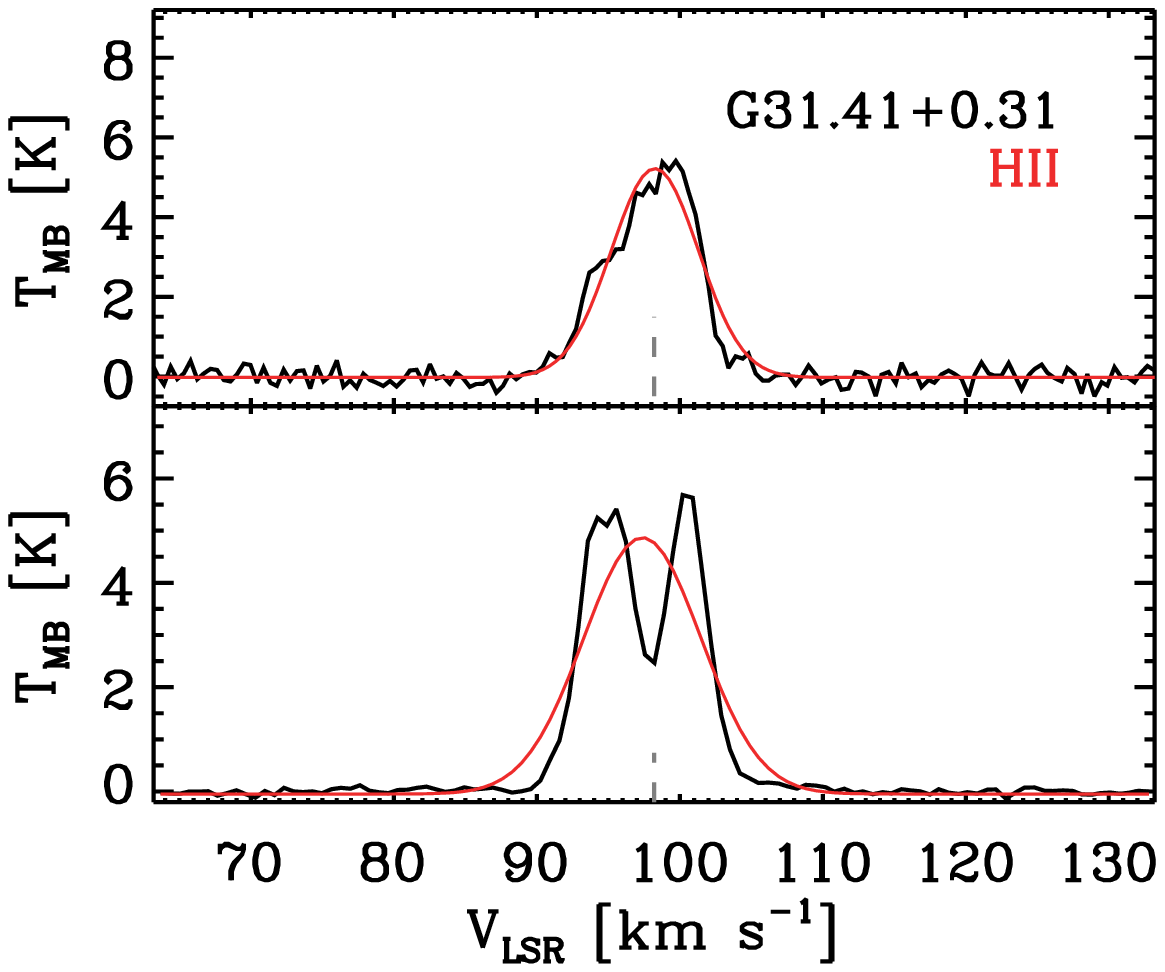}
\includegraphics[scale=0.46]{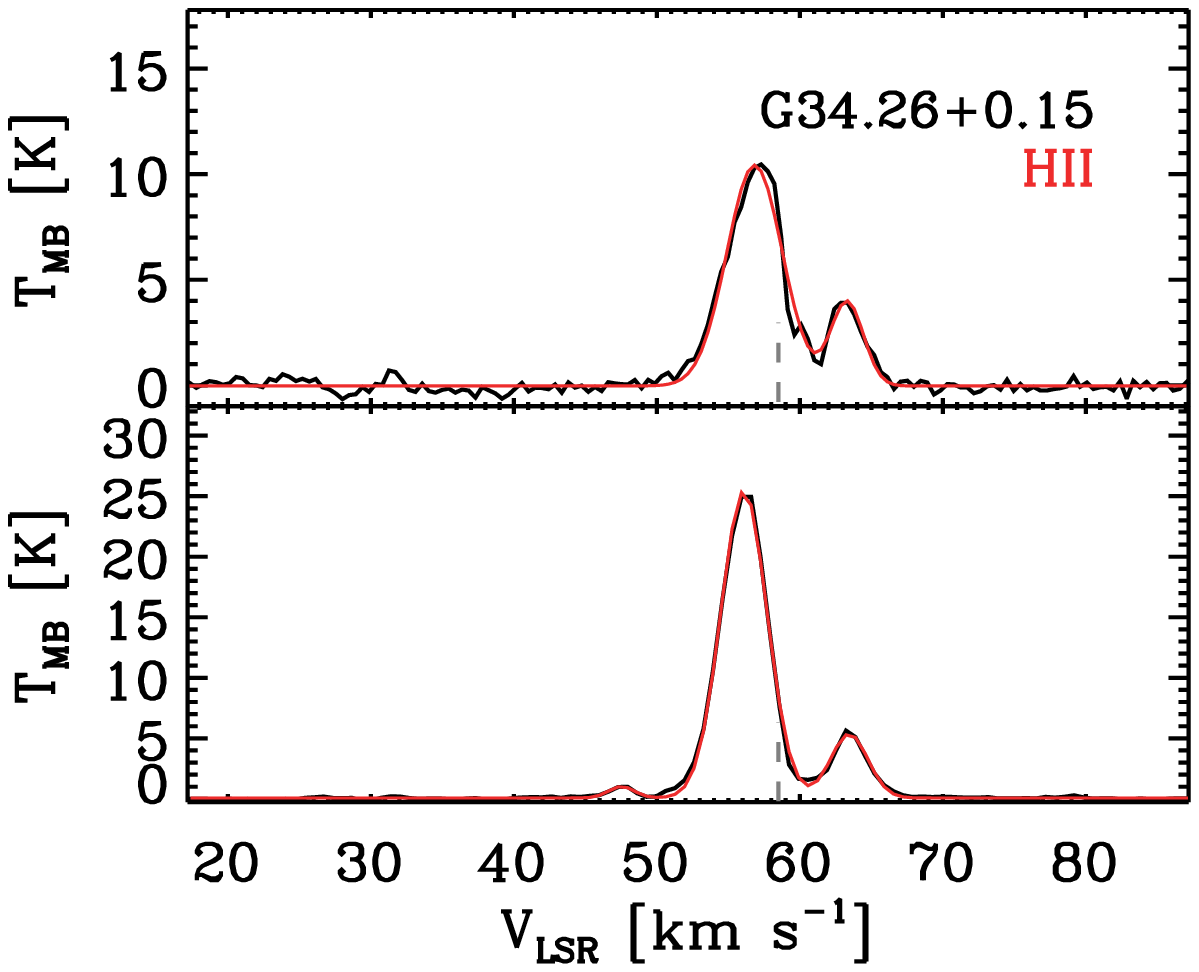}
\includegraphics[scale=0.46]{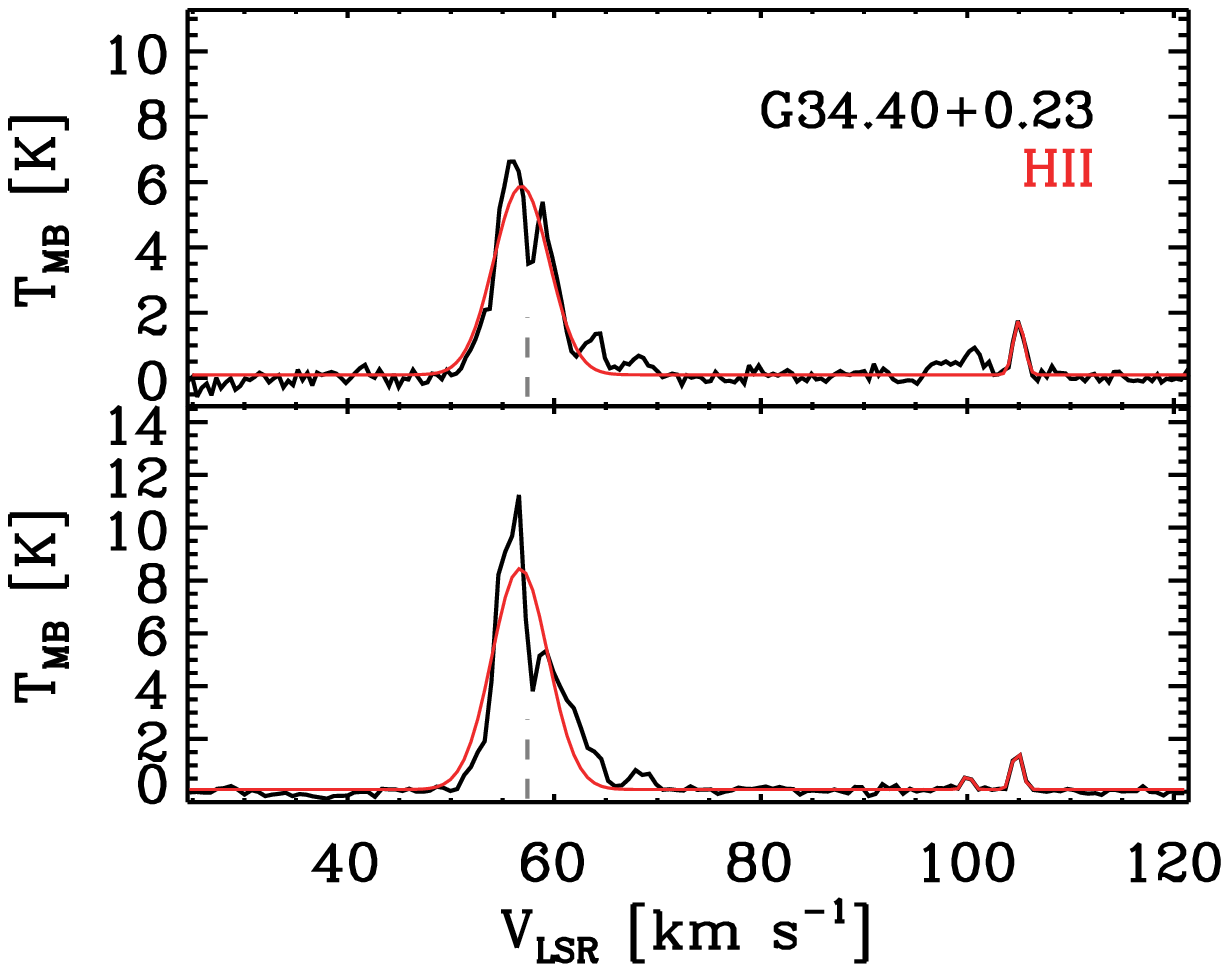}
\includegraphics[scale=0.46]{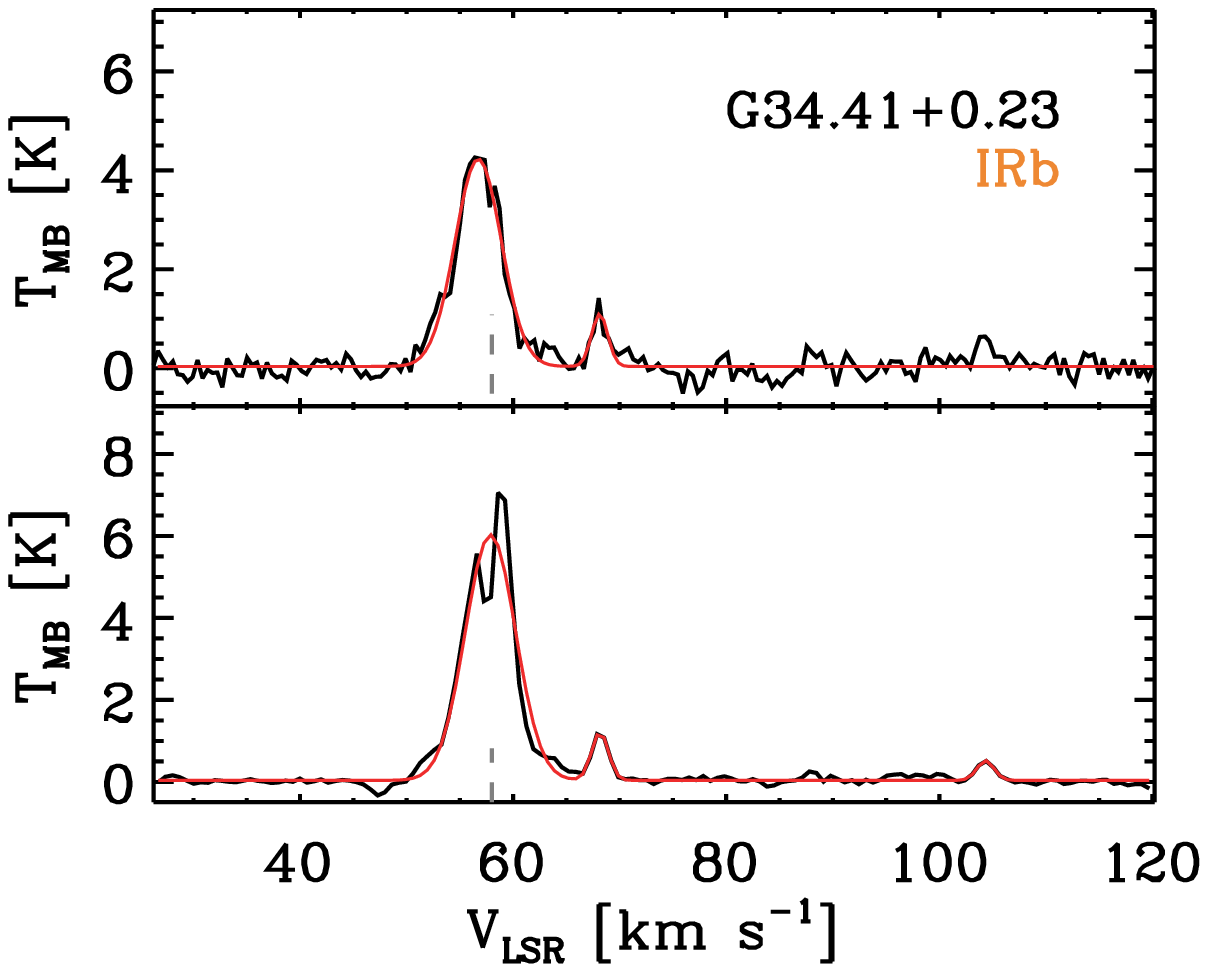}
\includegraphics[scale=0.46]{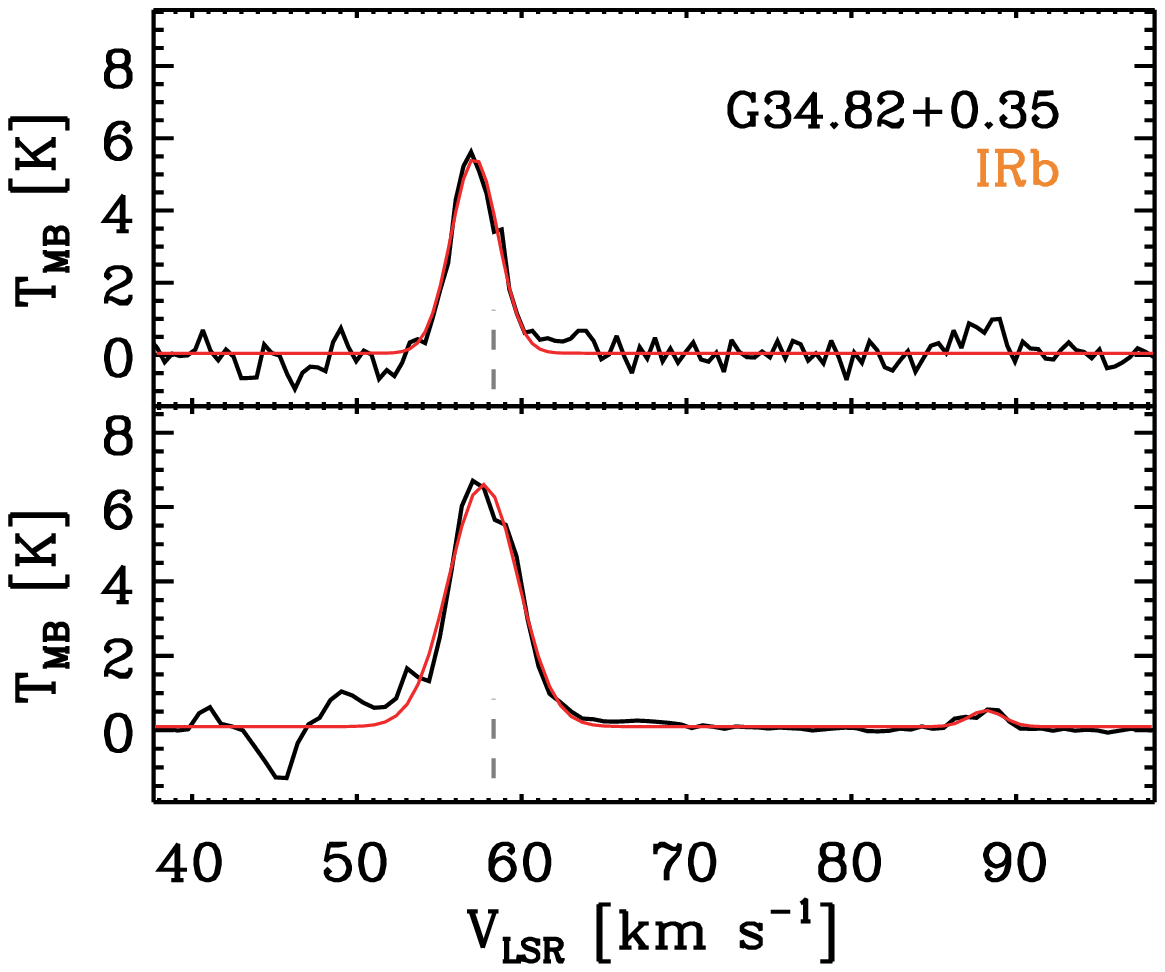}
\includegraphics[scale=0.46]{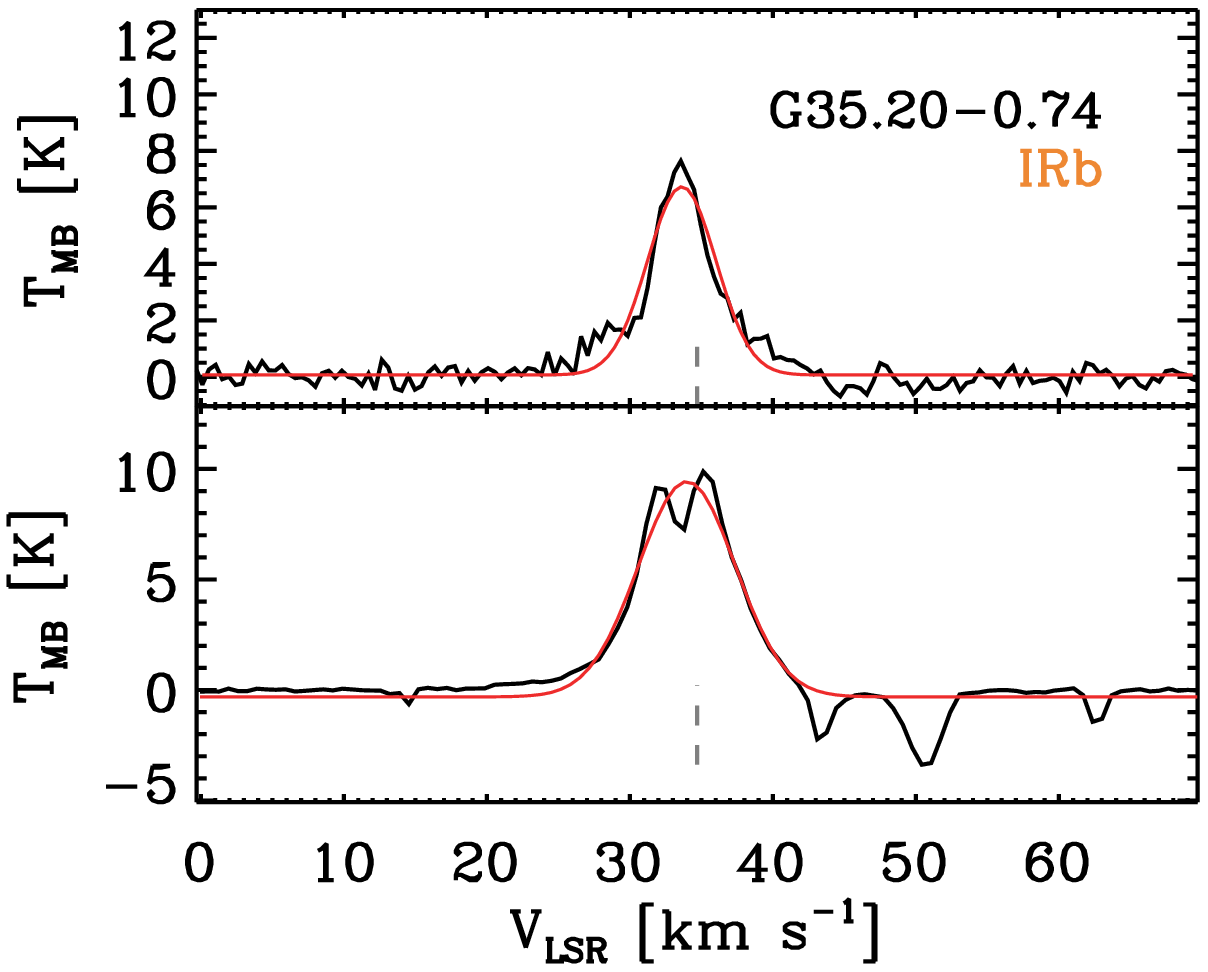}
\caption[]{(continued)}
\end{figure*}

\begin{figure*}
\centering
\ContinuedFloat
\includegraphics[scale=0.46]{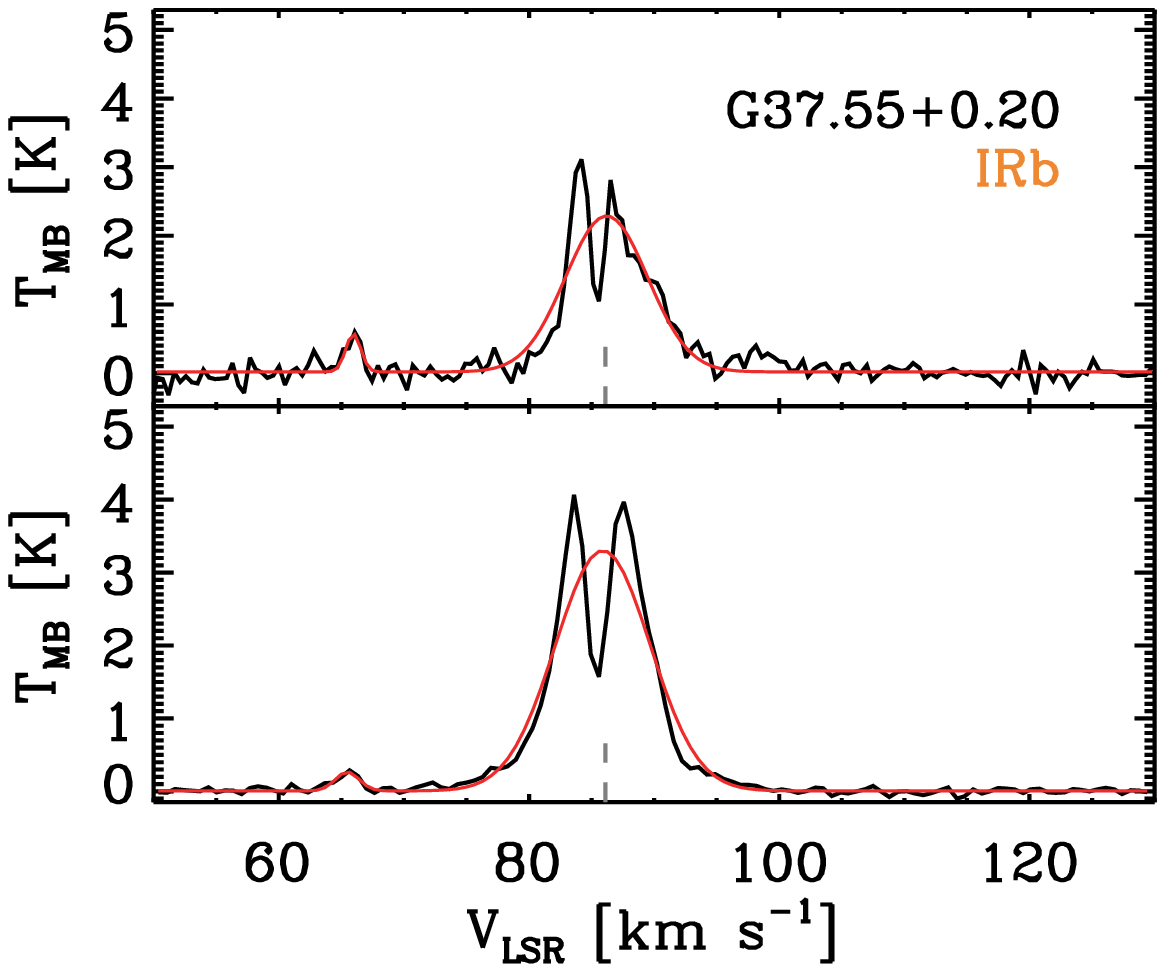}
\includegraphics[scale=0.46]{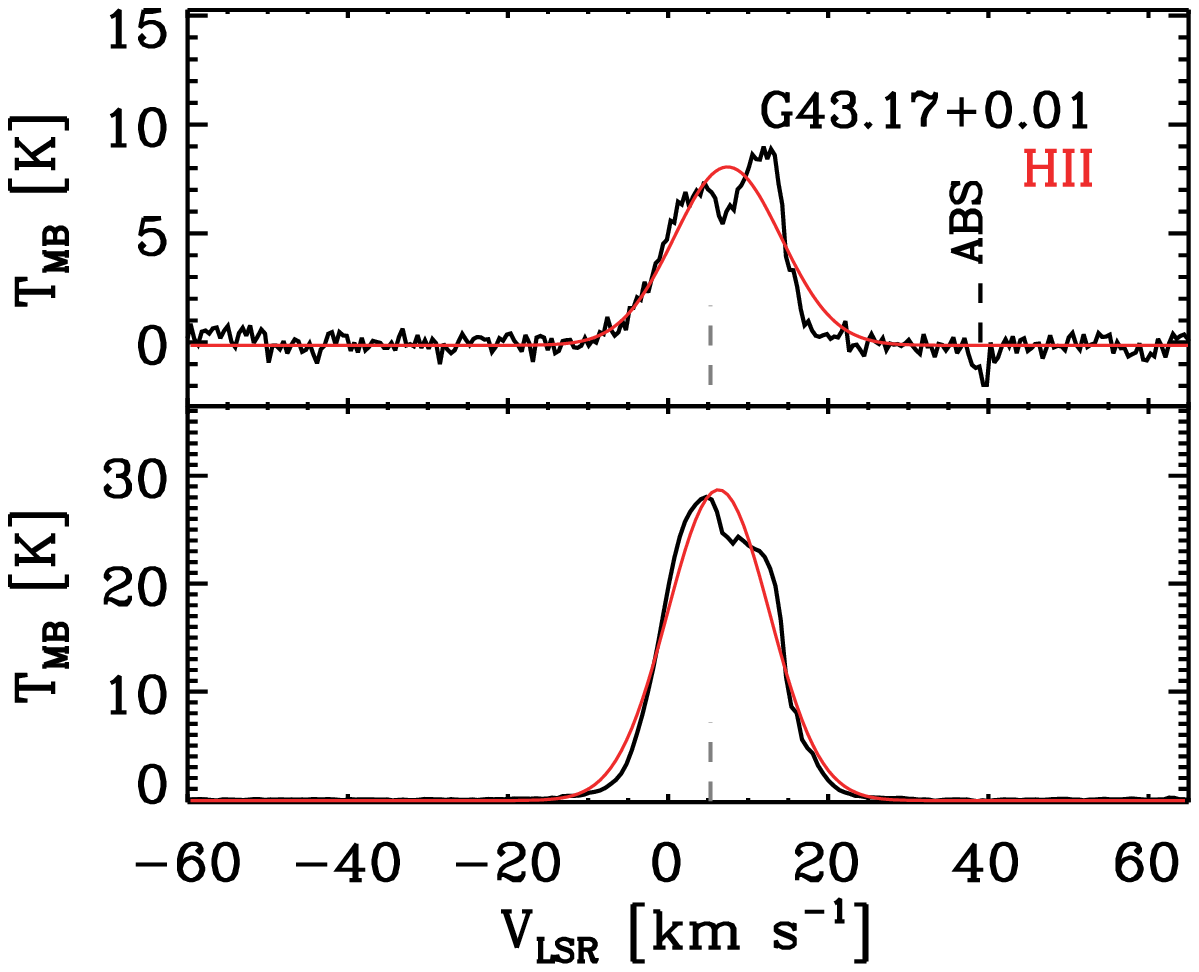}
\includegraphics[scale=0.46]{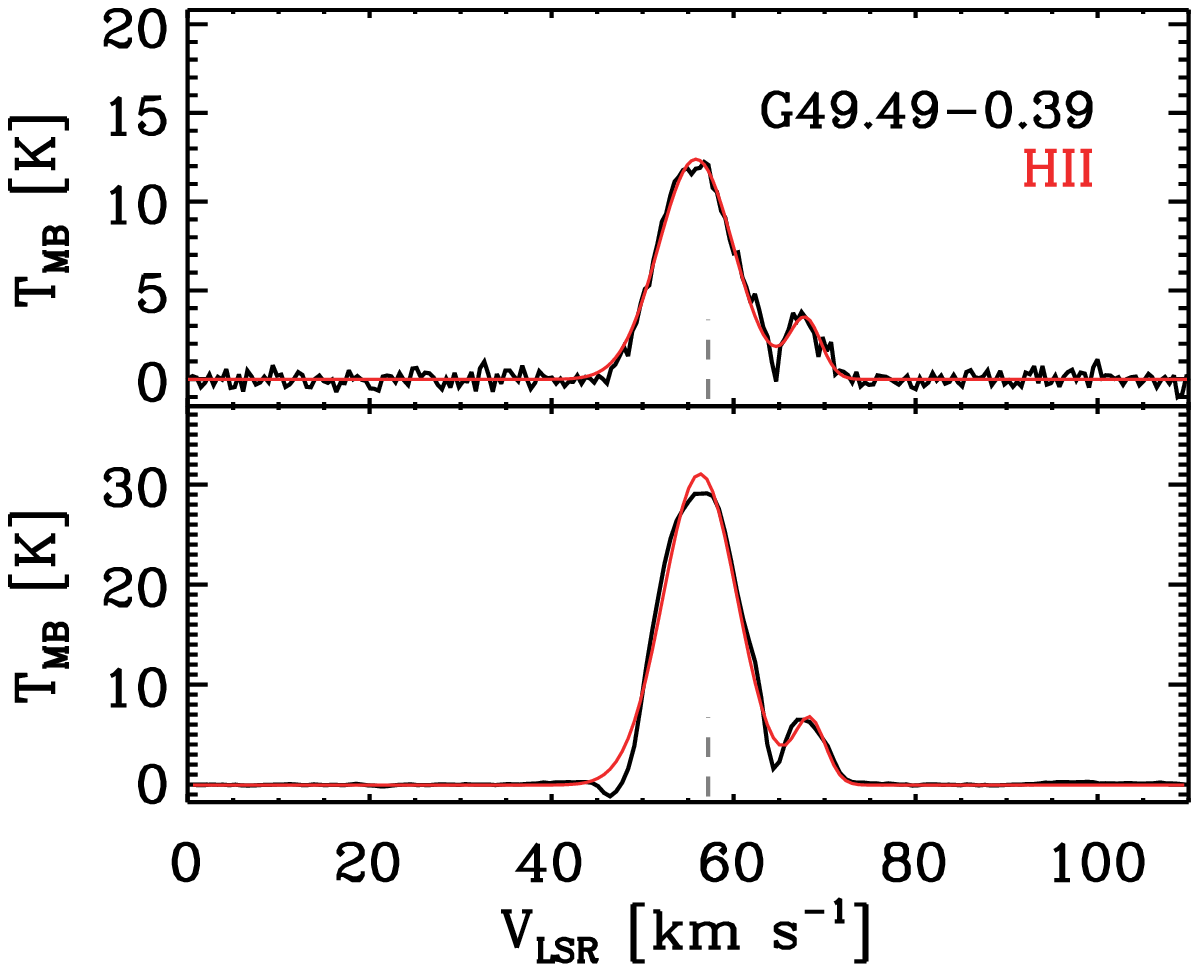}
\includegraphics[scale=0.46]{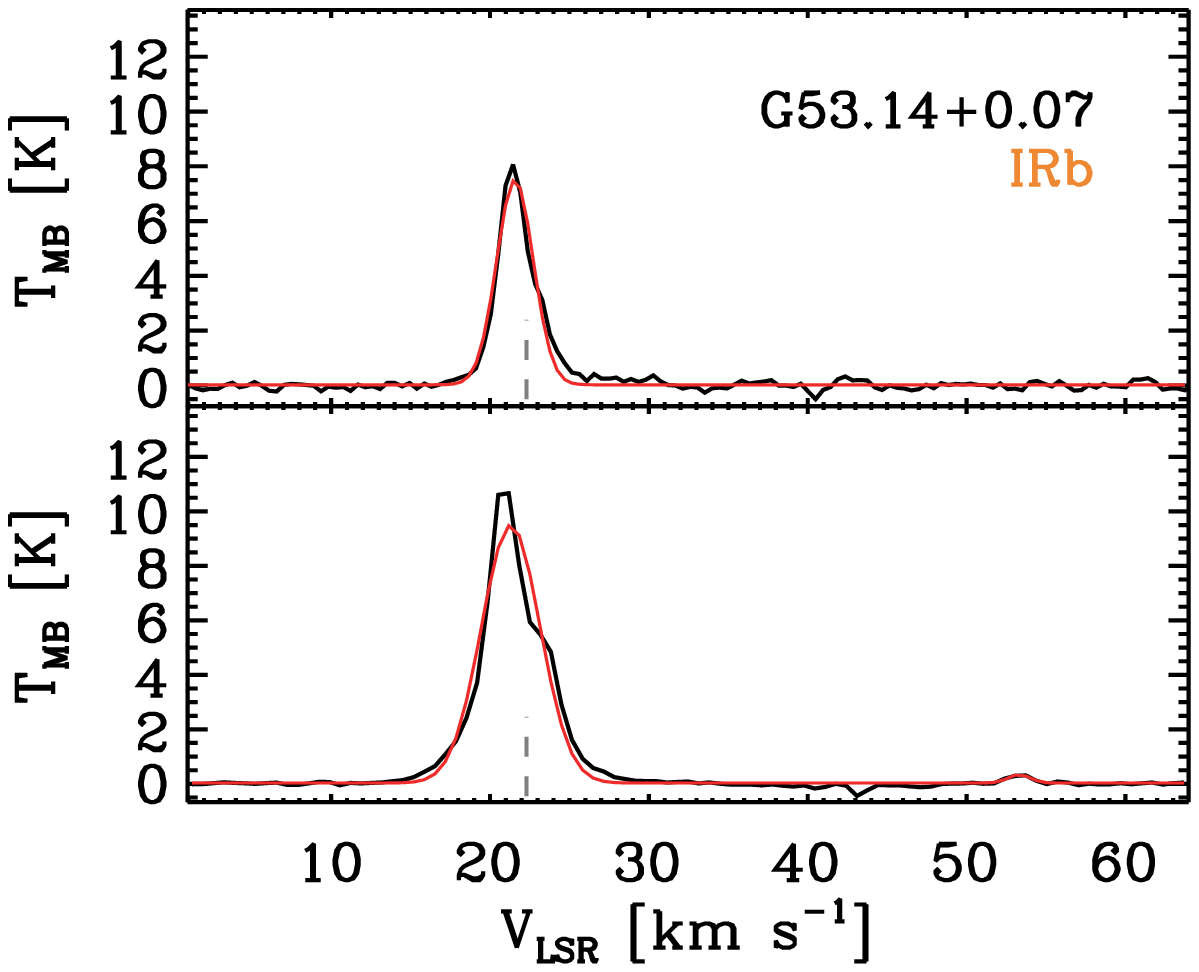}
\includegraphics[scale=0.46]{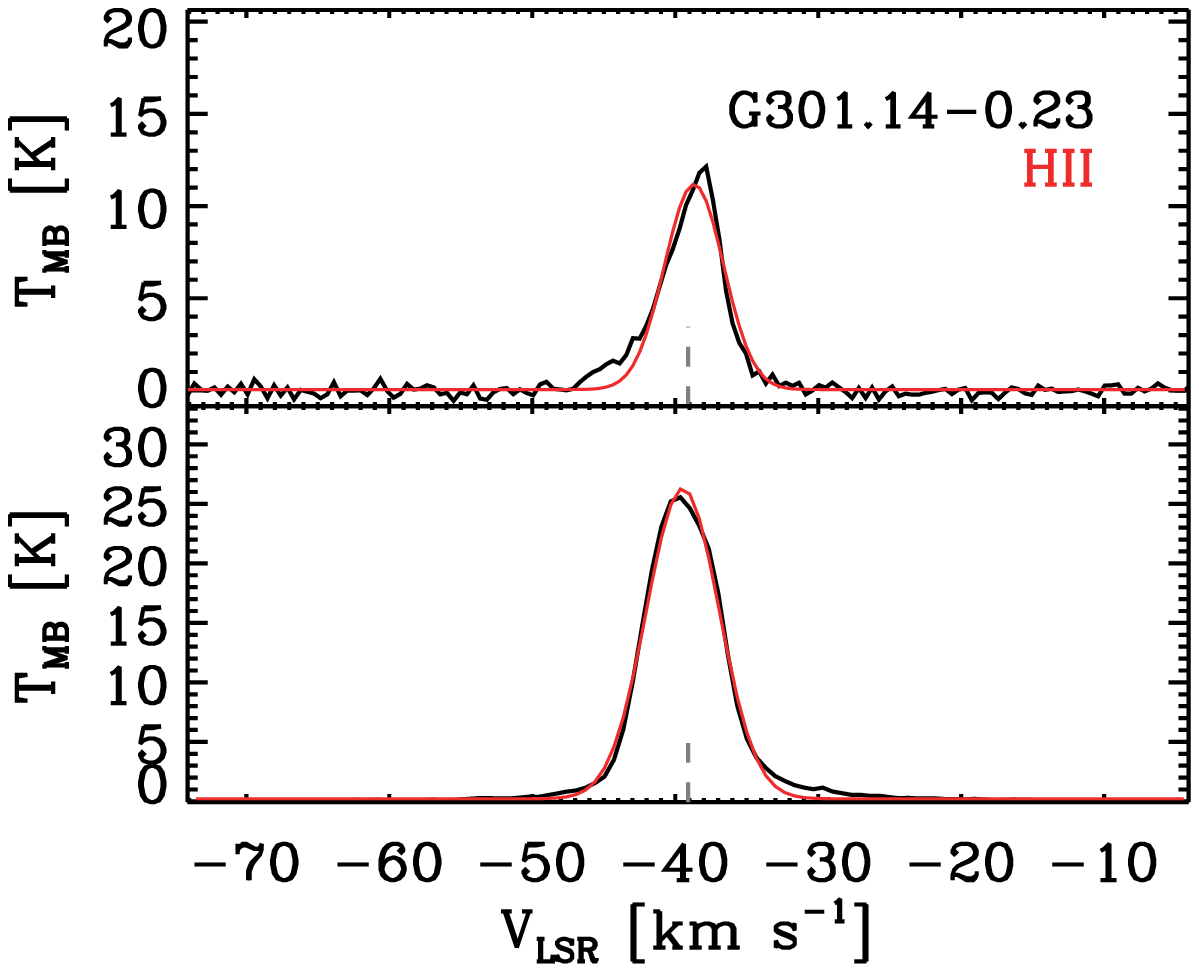}
\includegraphics[scale=0.46]{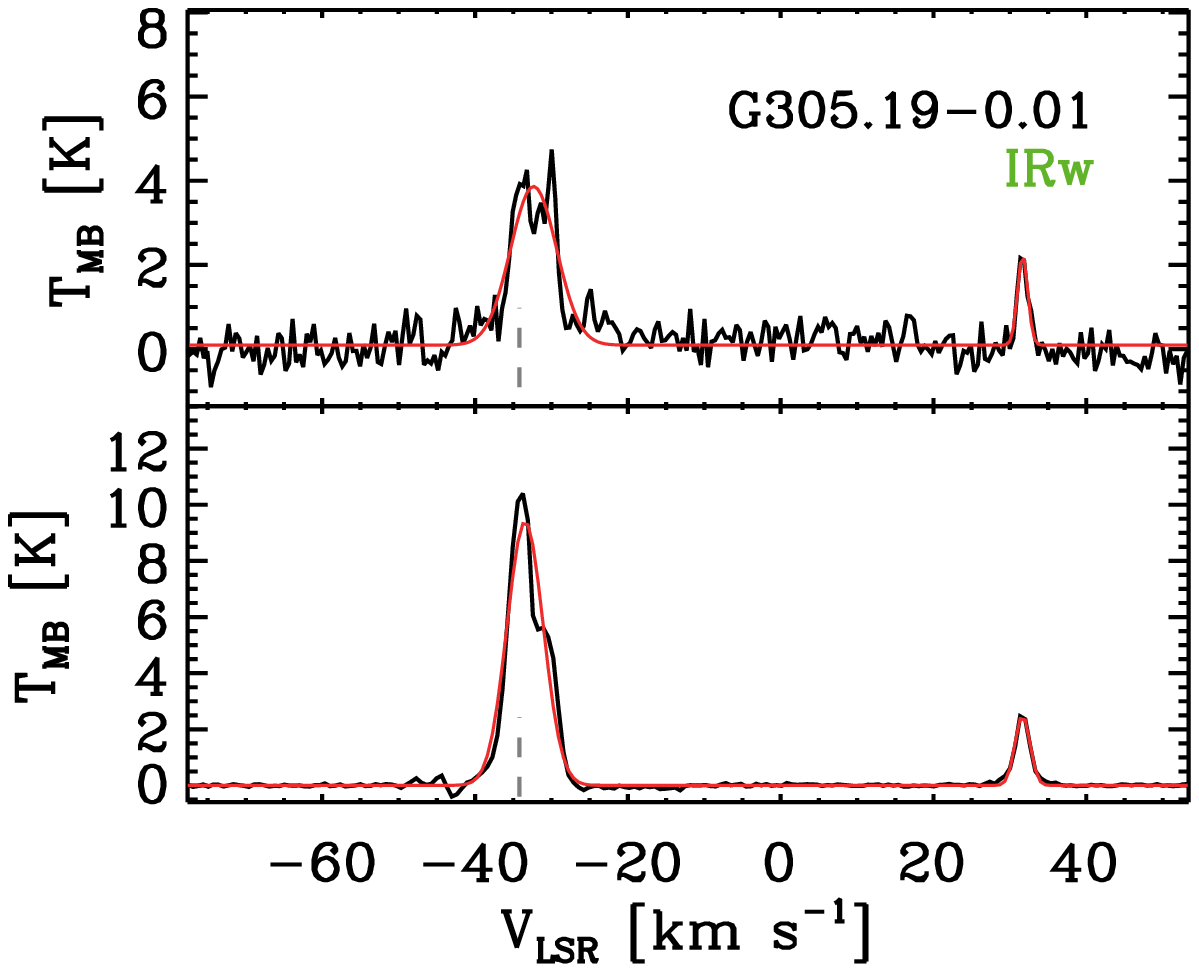}
\includegraphics[scale=0.46]{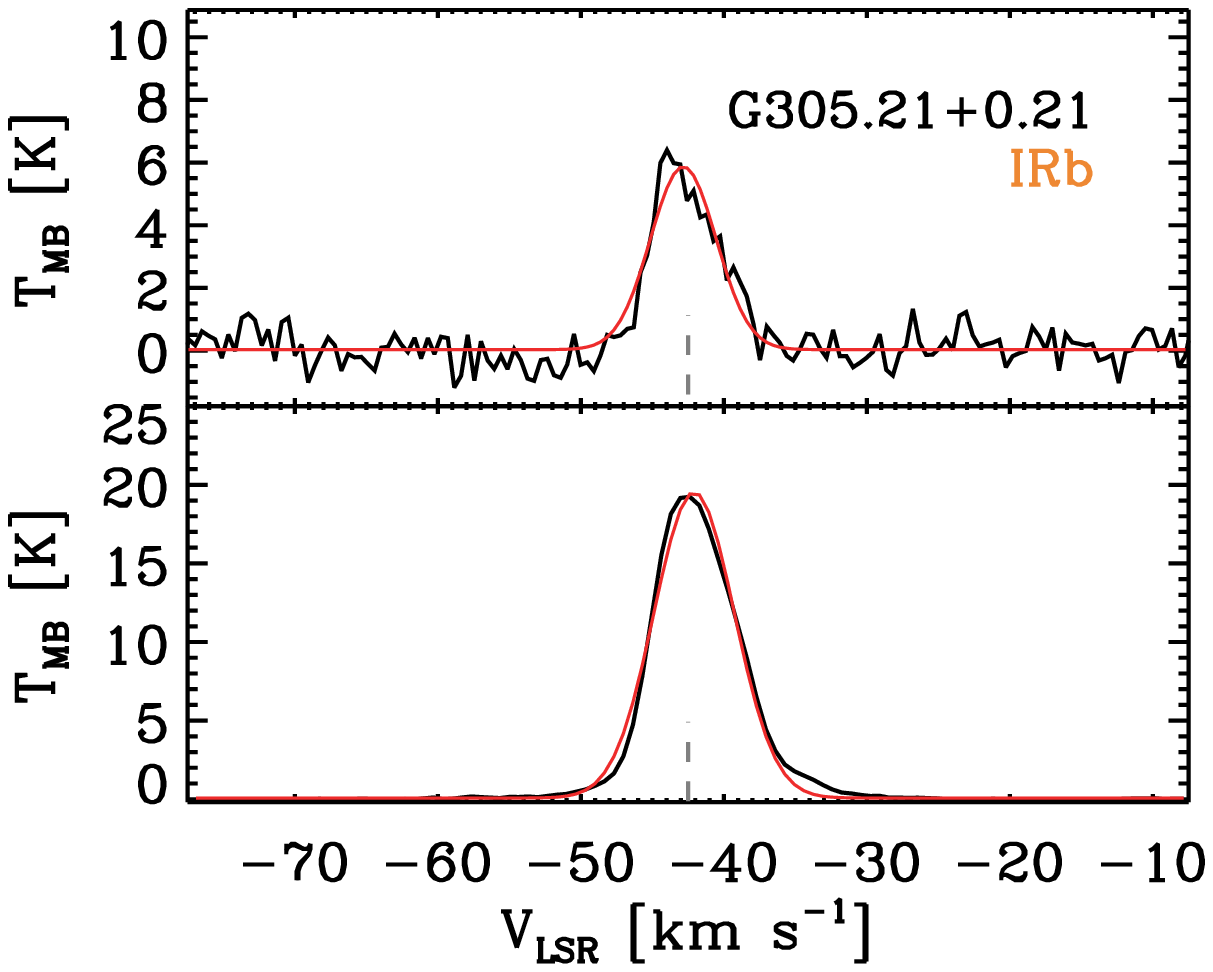}
\includegraphics[scale=0.46]{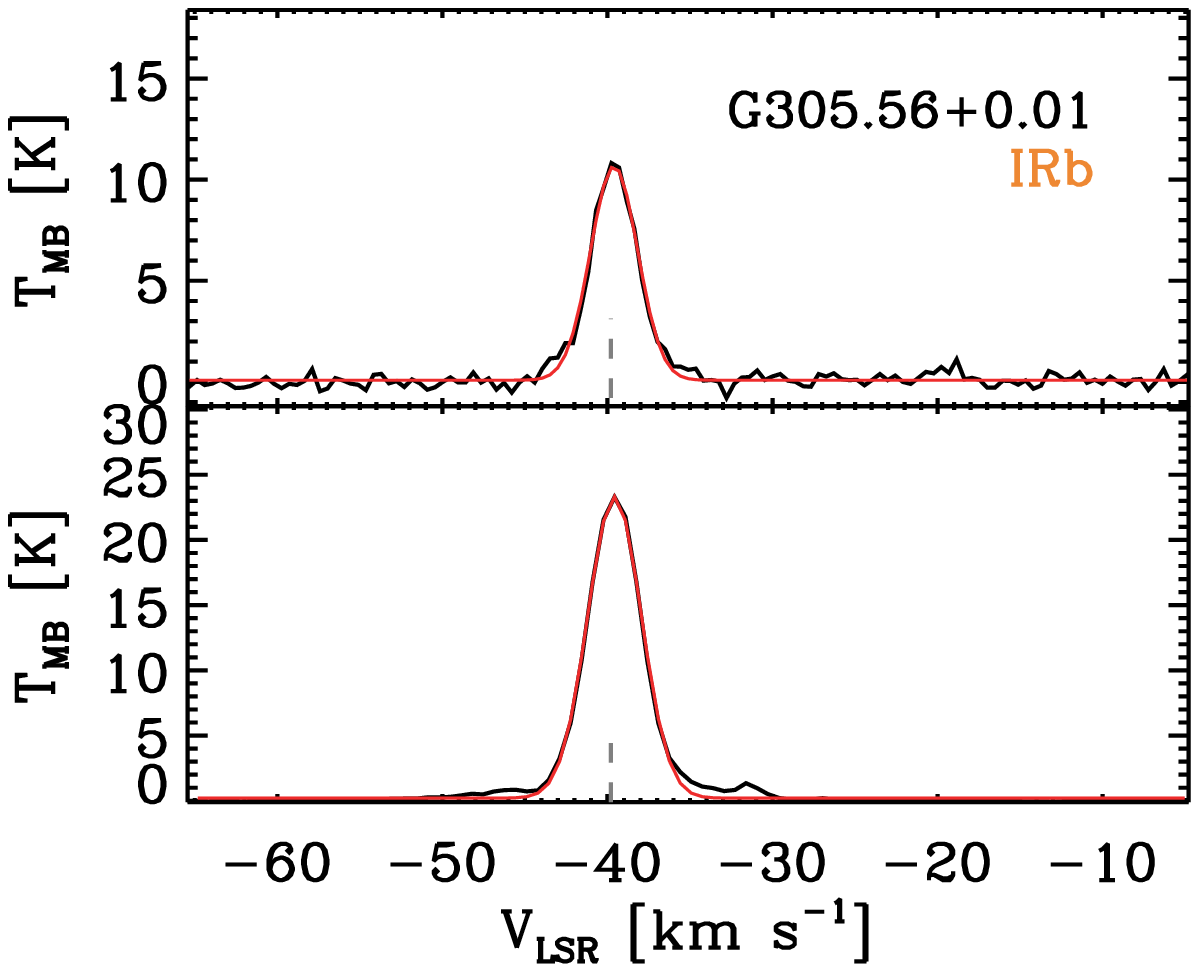}
\includegraphics[scale=0.46]{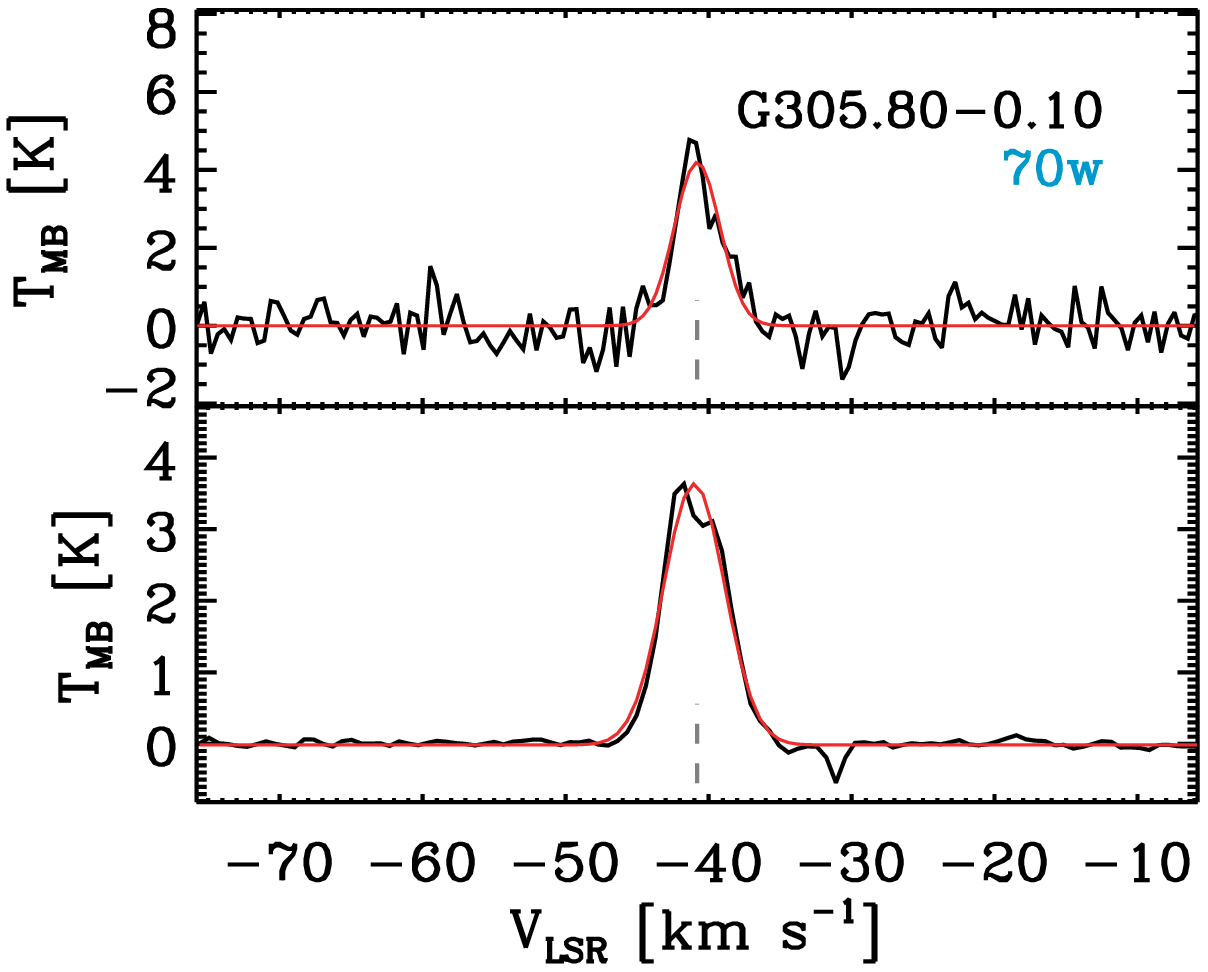}
\includegraphics[scale=0.46]{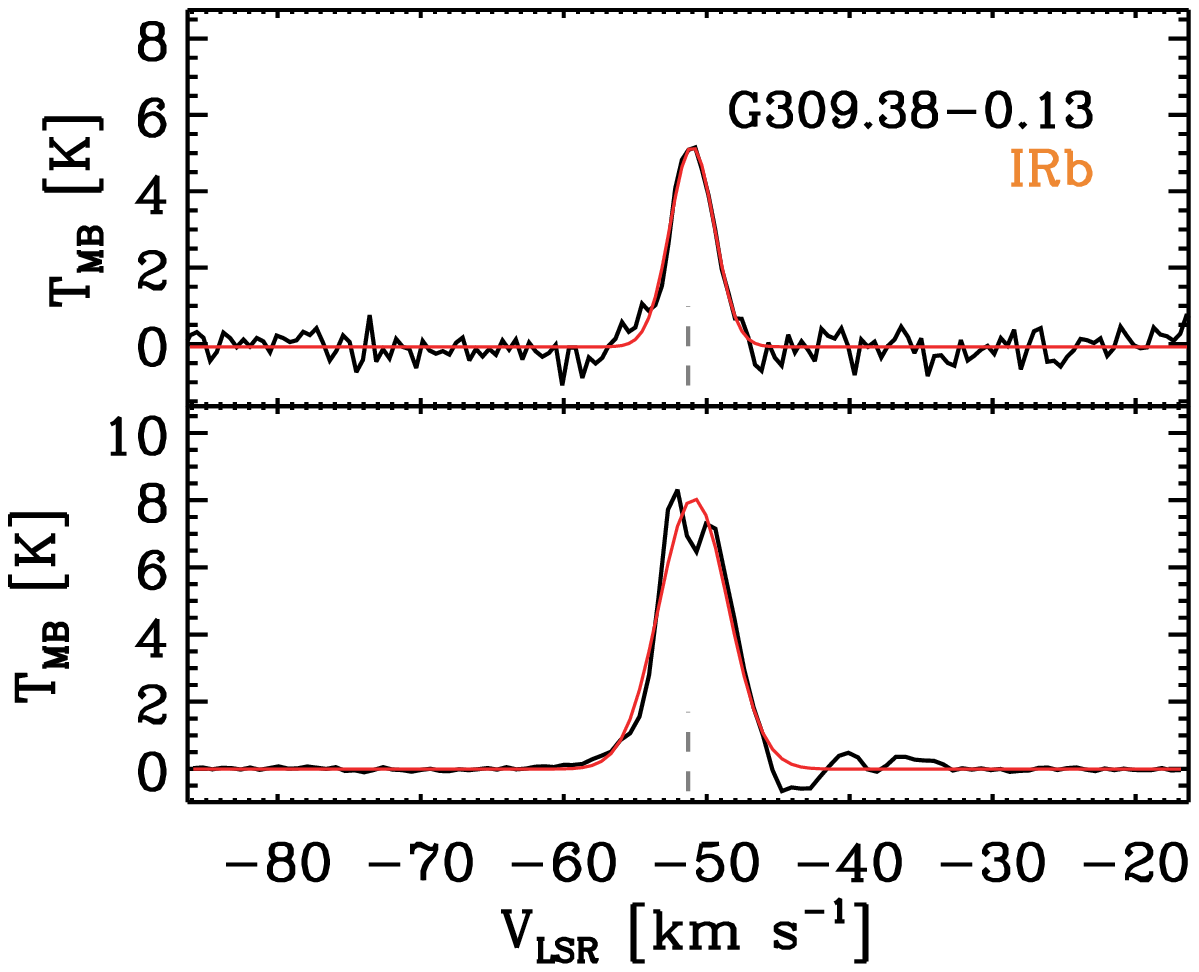}
\includegraphics[scale=0.46]{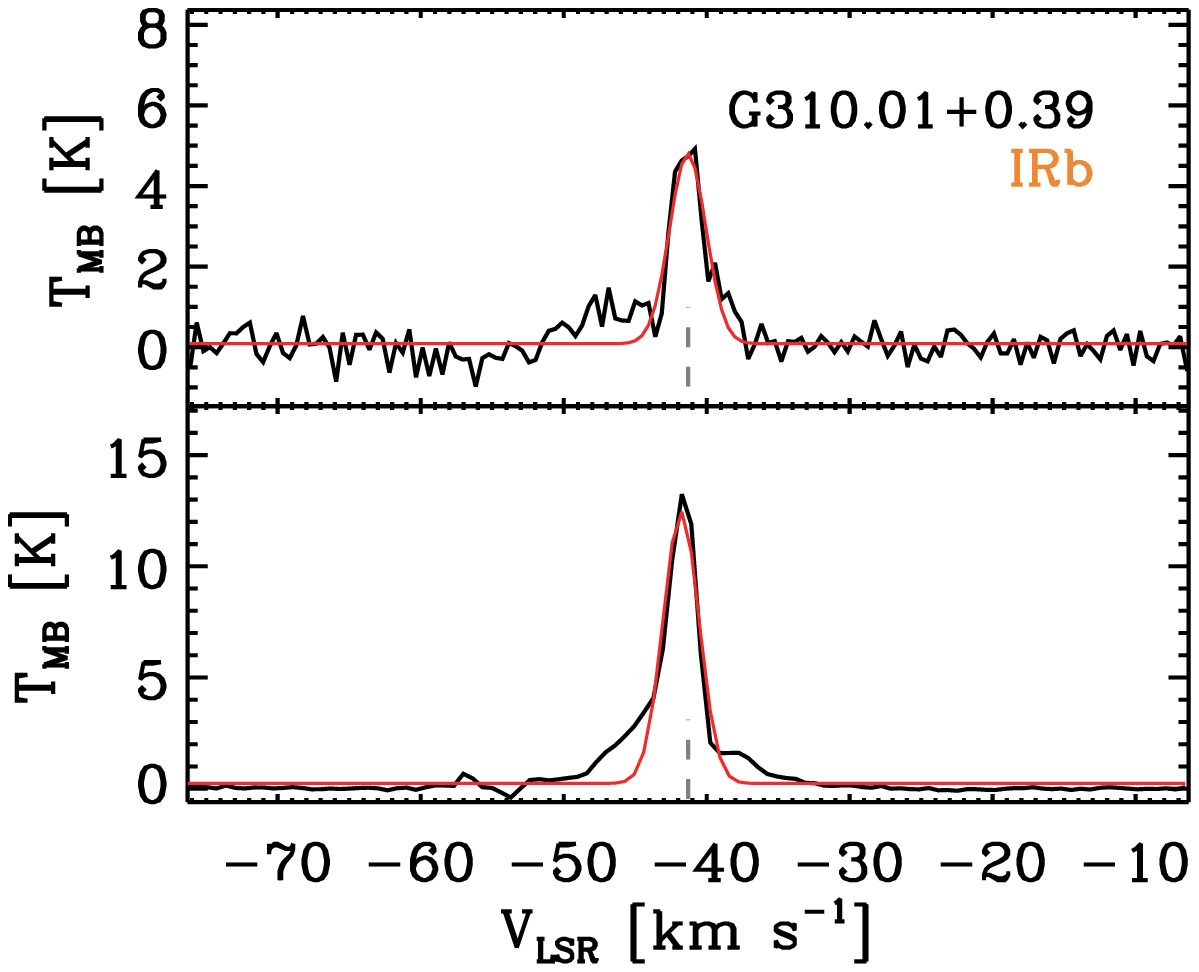}
\includegraphics[scale=0.46]{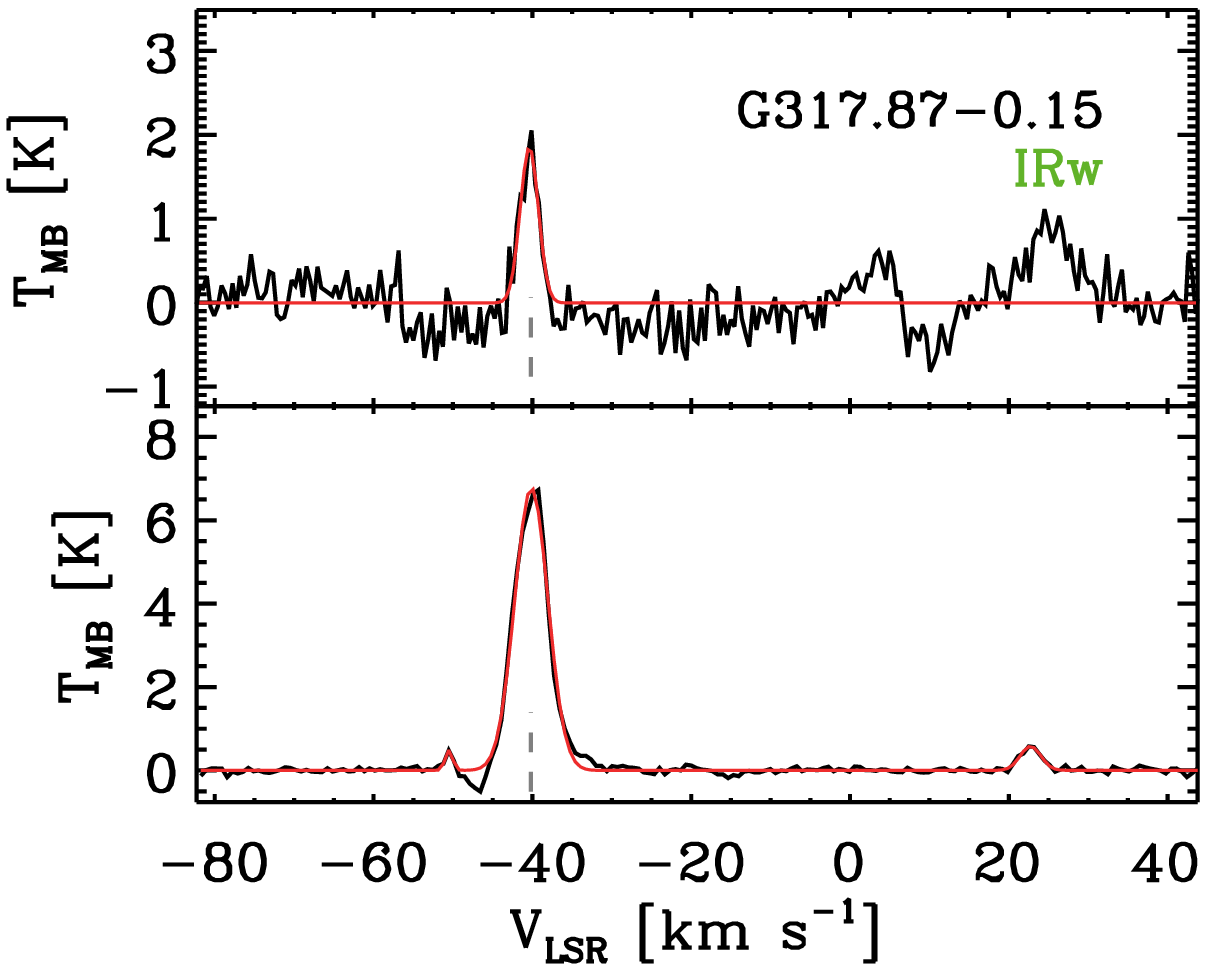}
\includegraphics[scale=0.46]{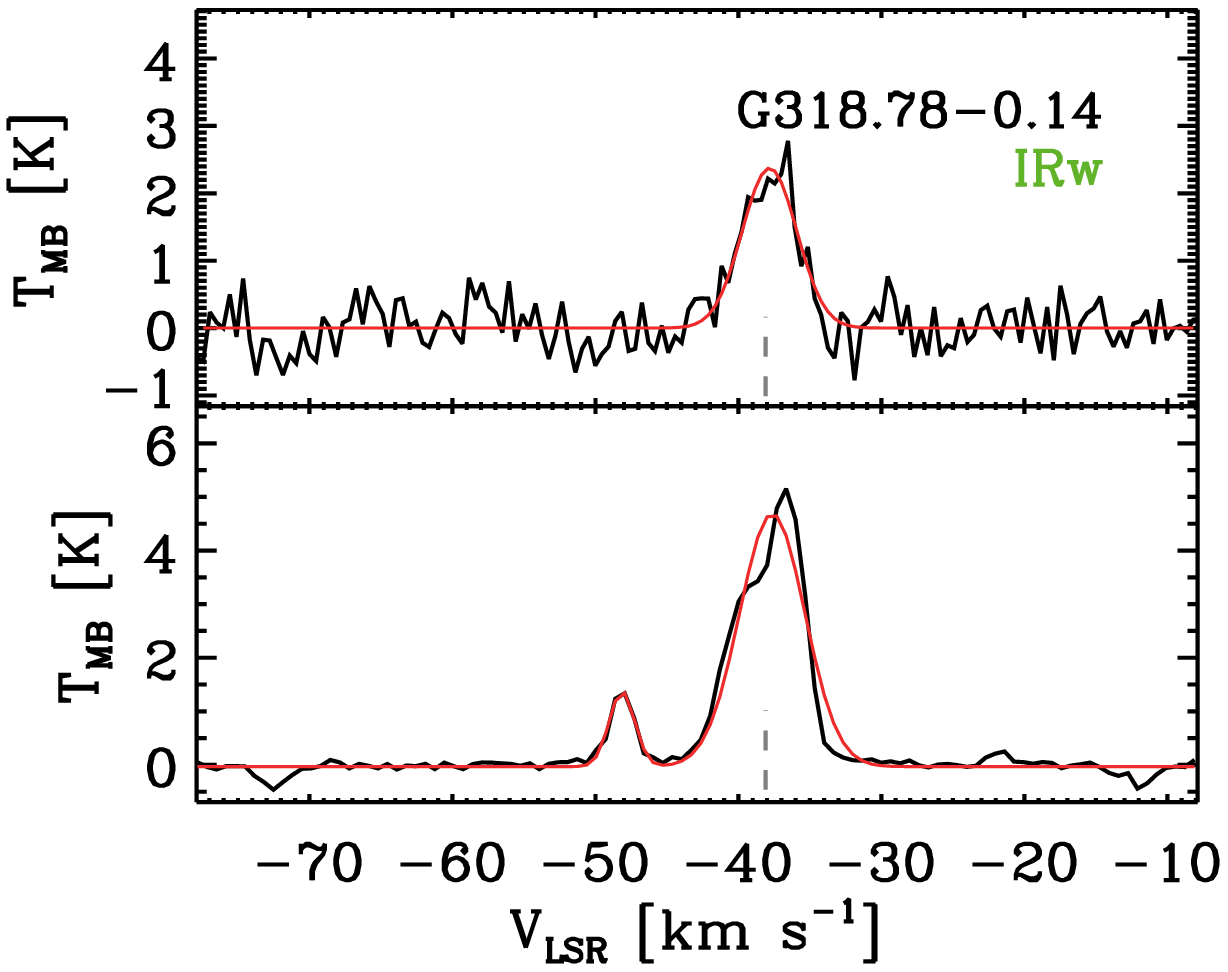}
\includegraphics[scale=0.46]{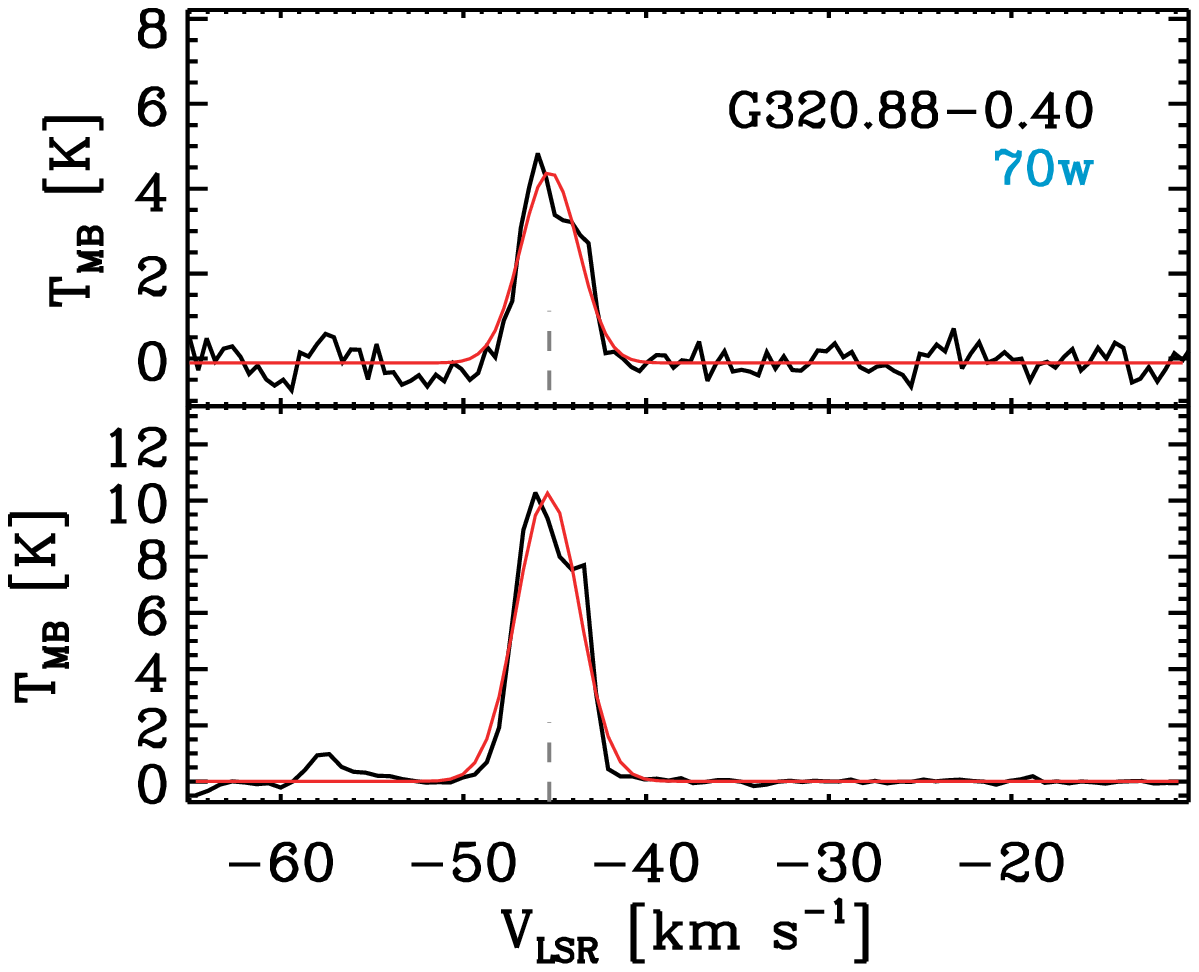}
\includegraphics[scale=0.46]{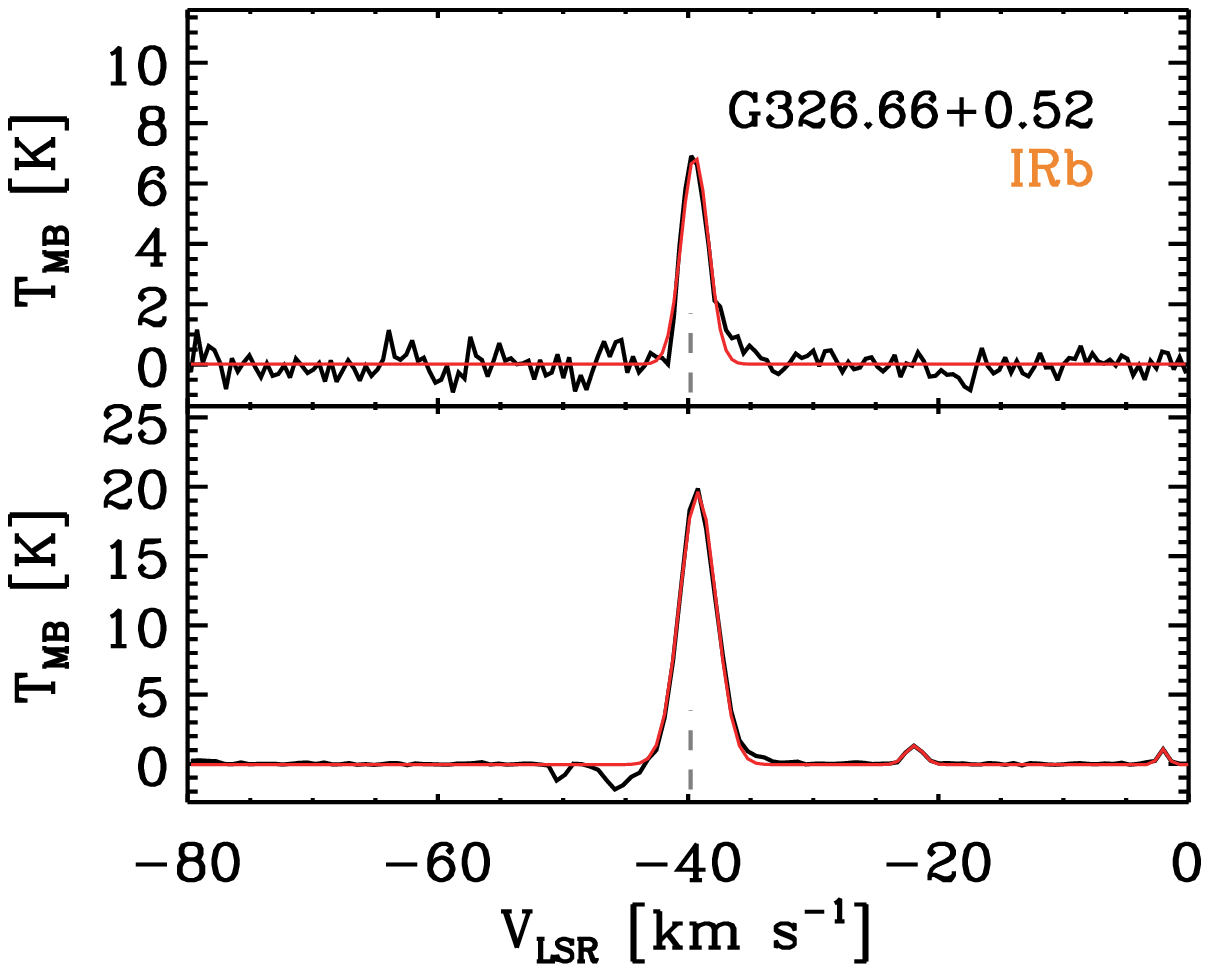}
\caption[]{(continued)}
\end{figure*}

\begin{figure*}
\centering
\ContinuedFloat
\includegraphics[scale=0.46]{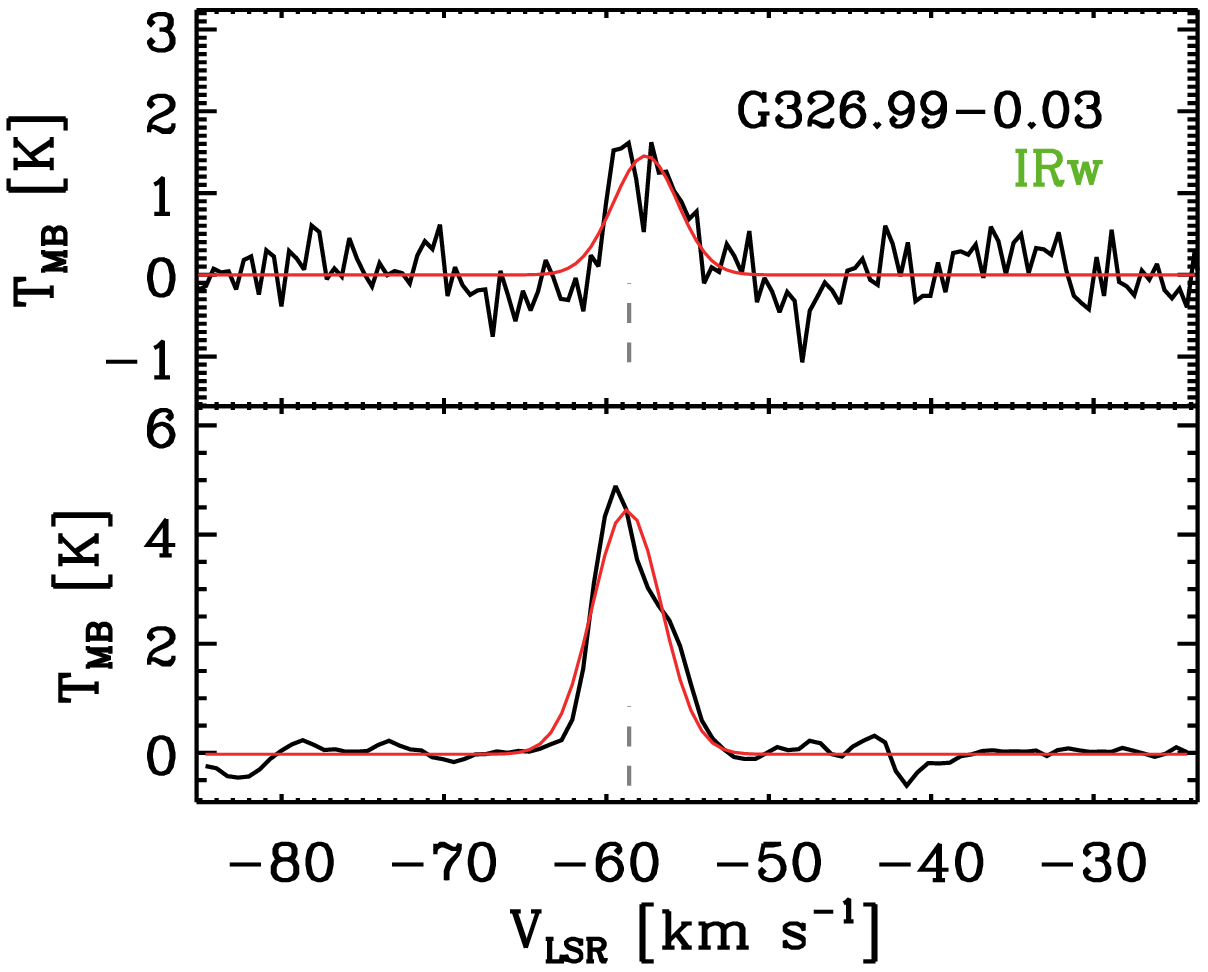}
\includegraphics[scale=0.46]{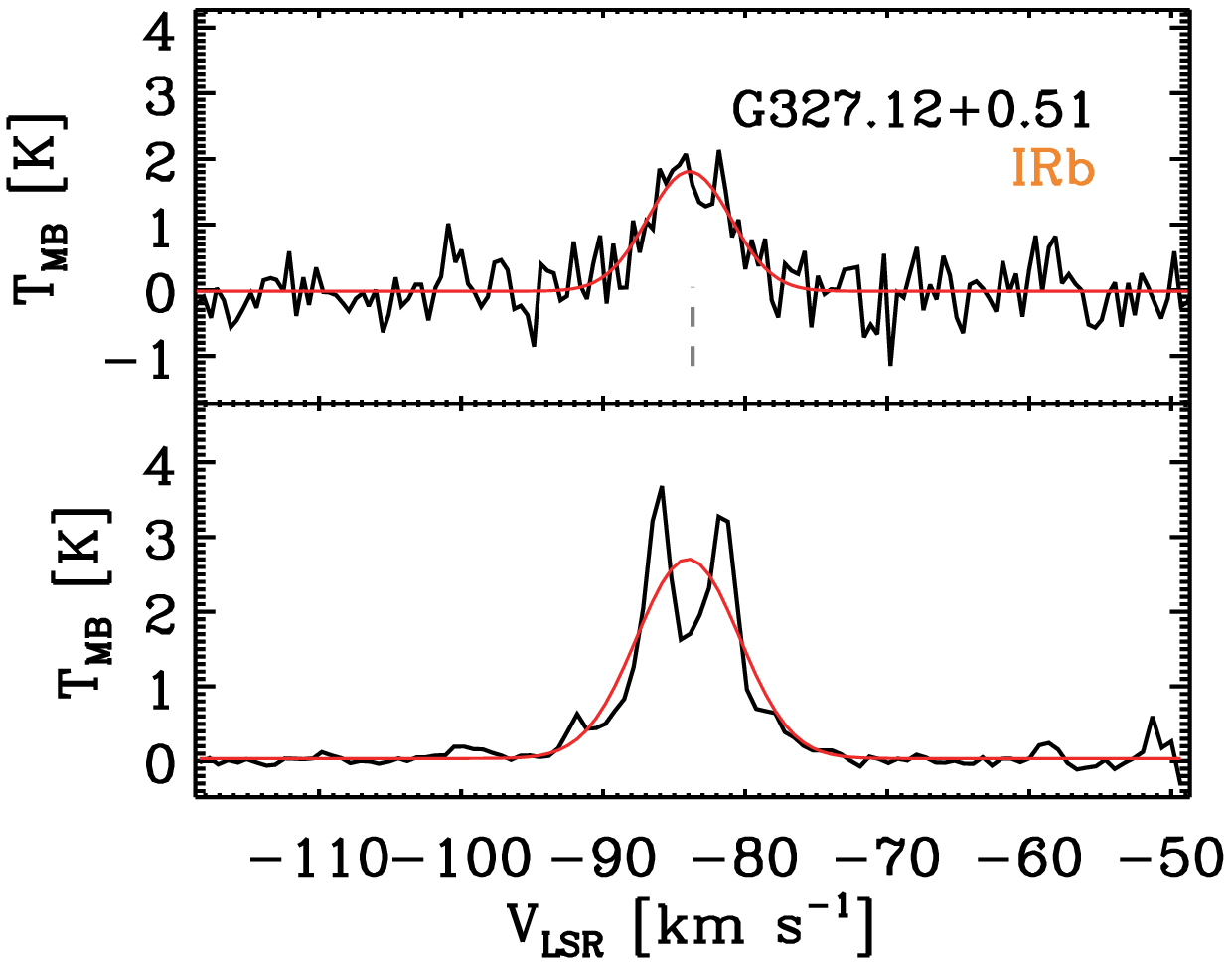}
\includegraphics[scale=0.46]{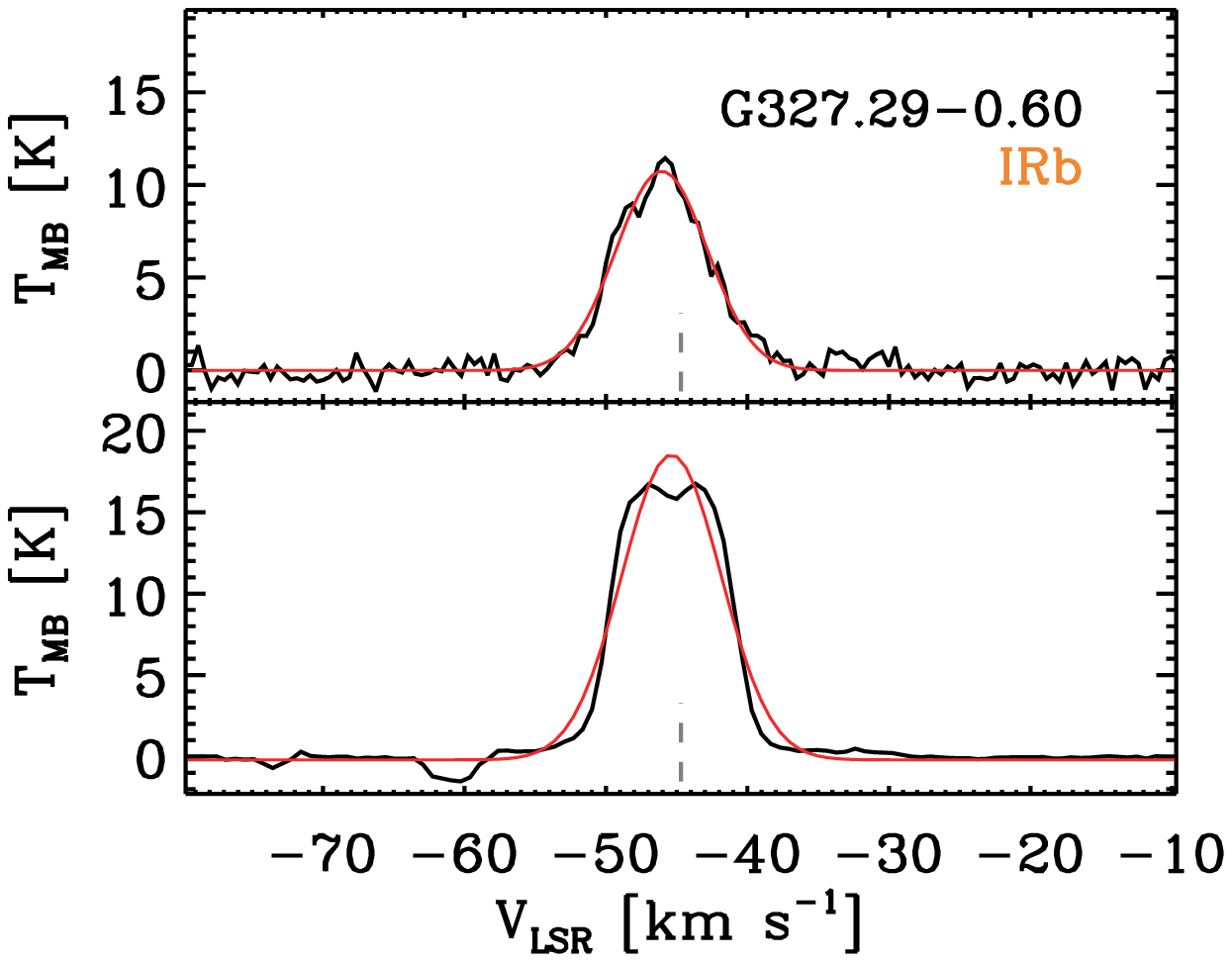}
\includegraphics[scale=0.46]{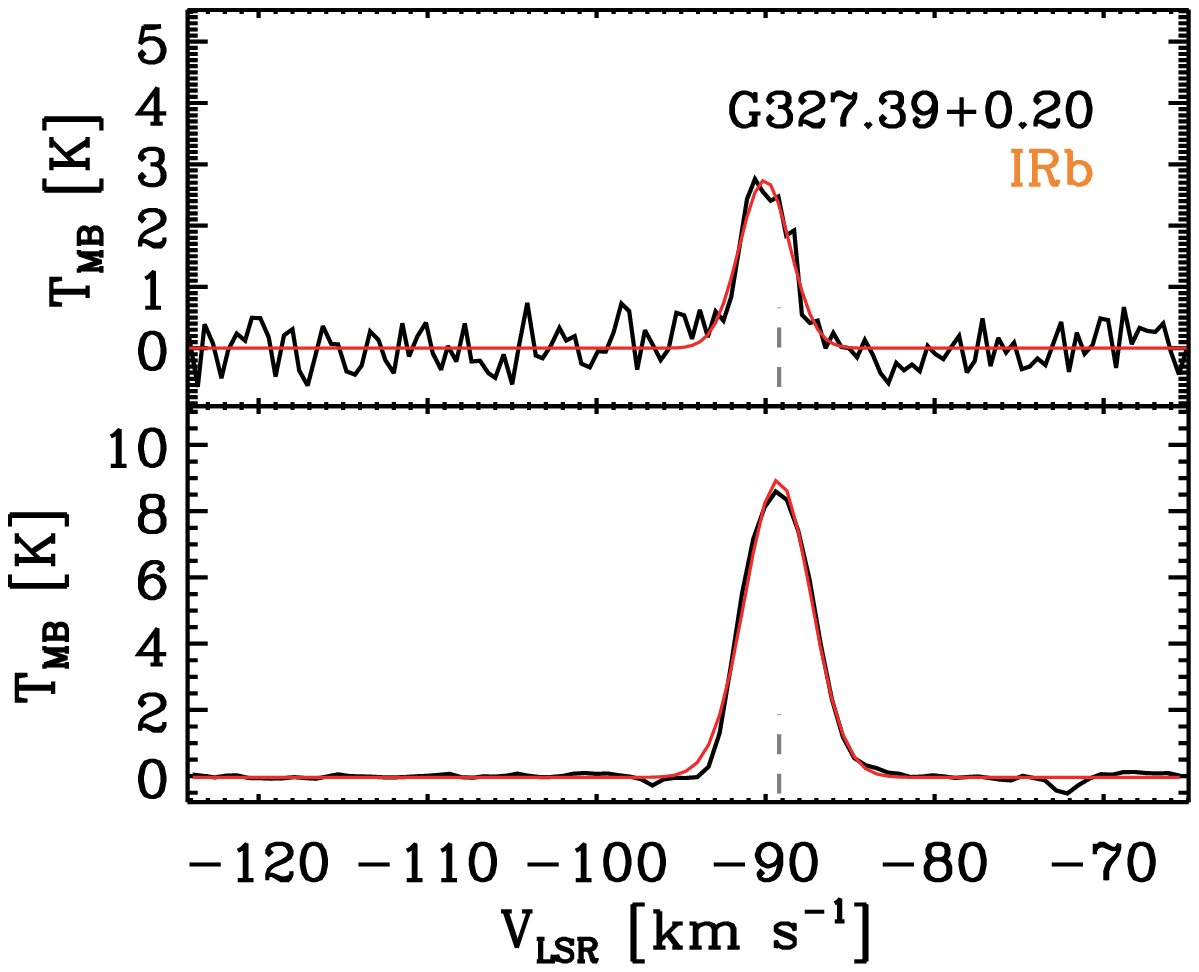}
\includegraphics[scale=0.46]{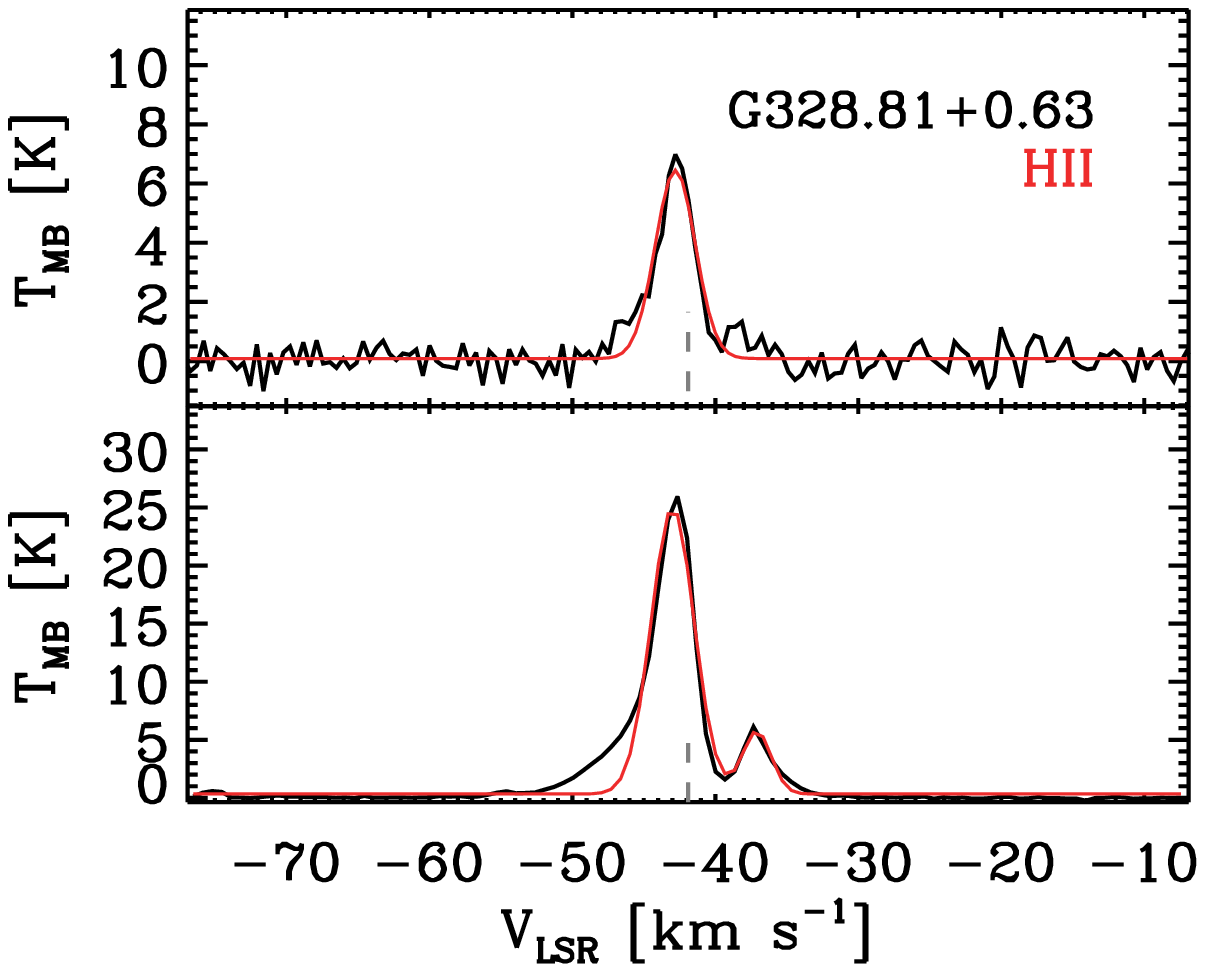}
\includegraphics[scale=0.46]{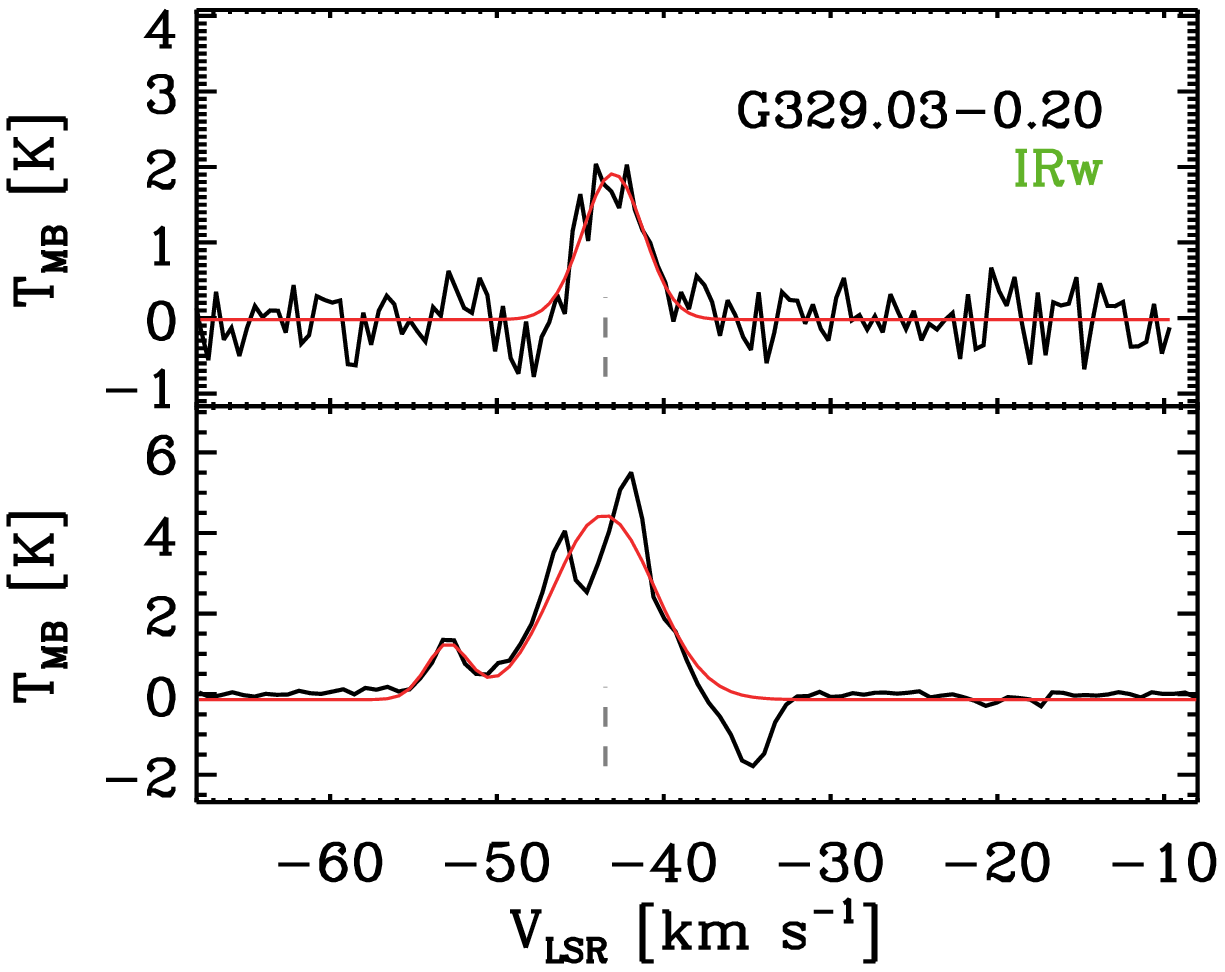}
\includegraphics[scale=0.46]{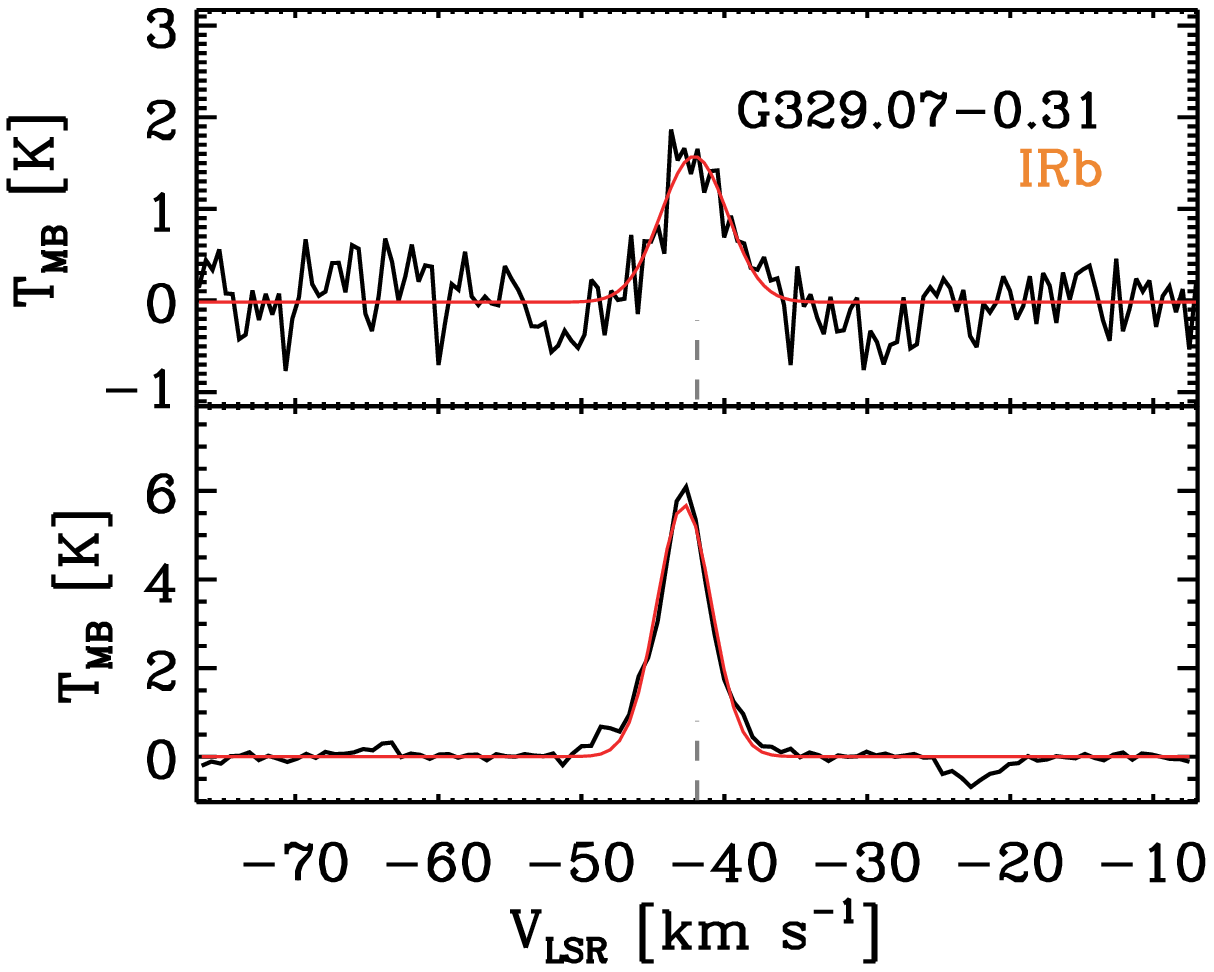}
\includegraphics[scale=0.46]{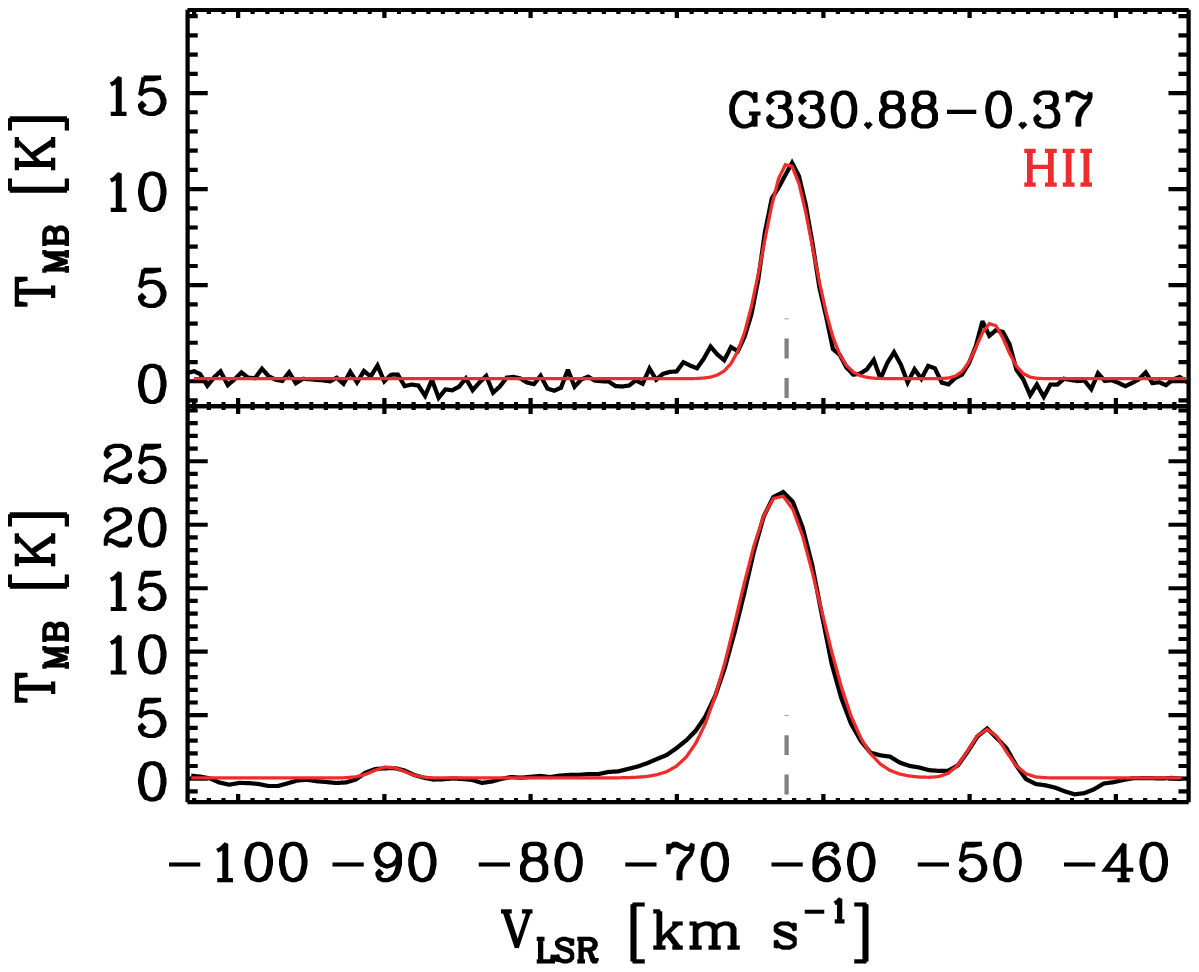}
\includegraphics[scale=0.46]{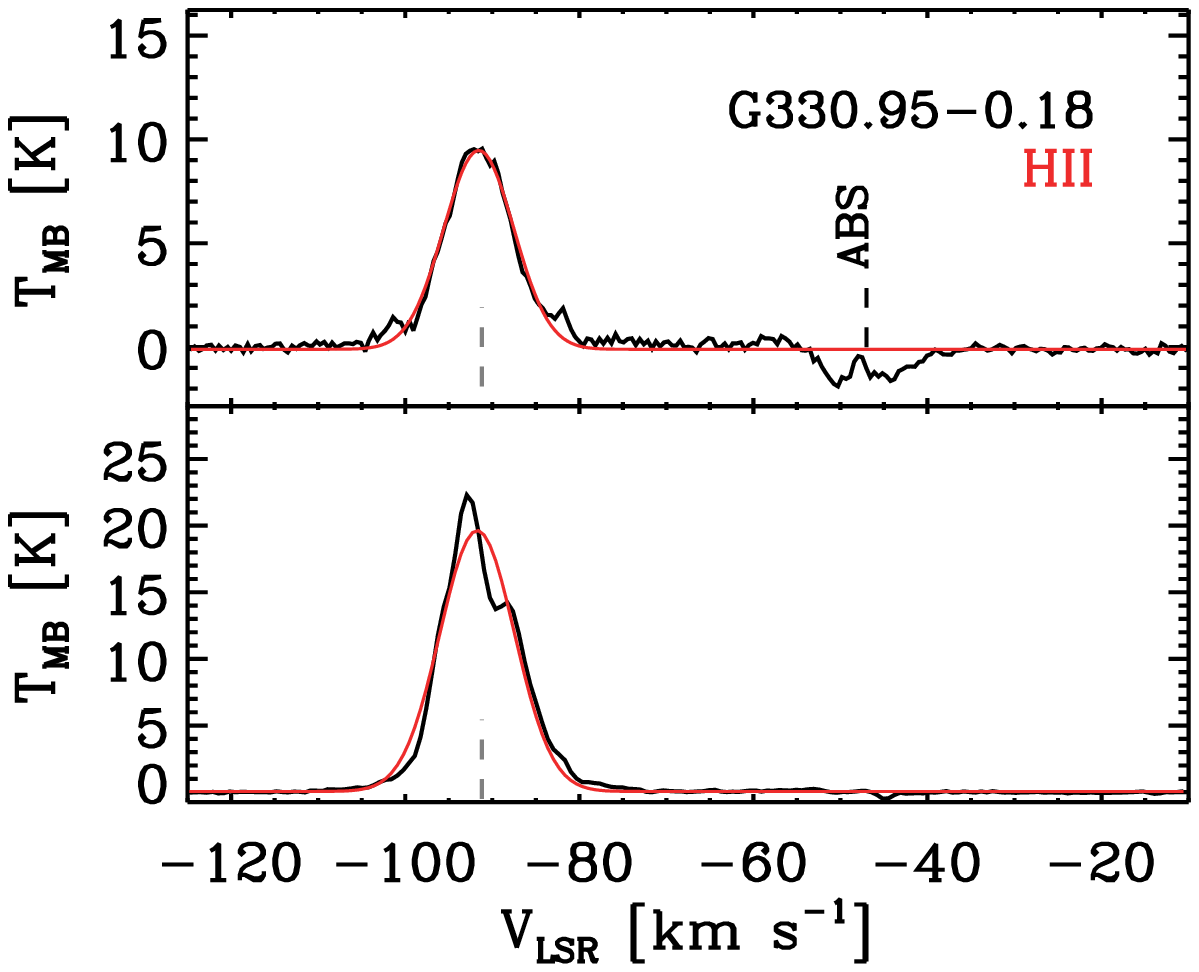}
\includegraphics[scale=0.46]{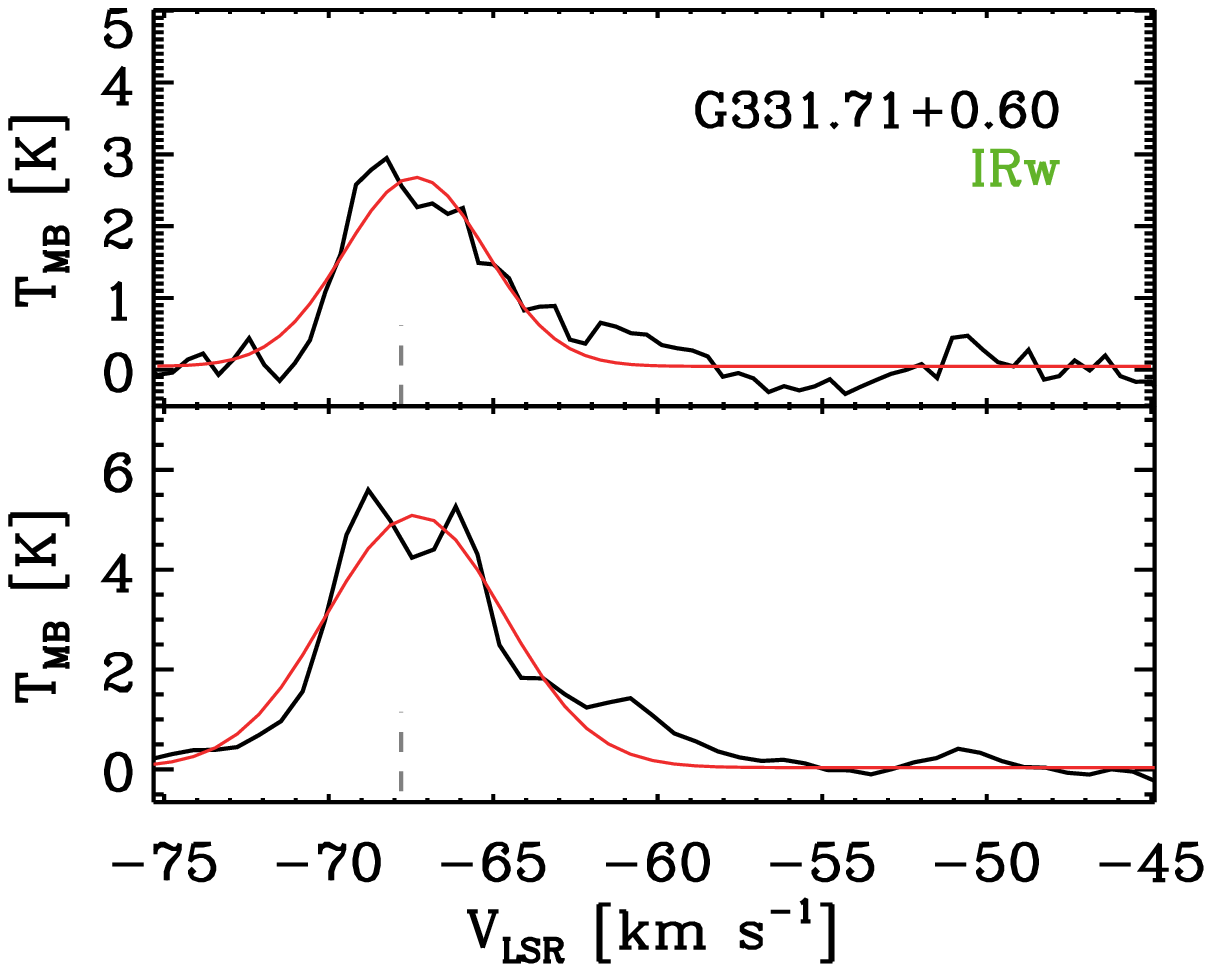}
\includegraphics[scale=0.46]{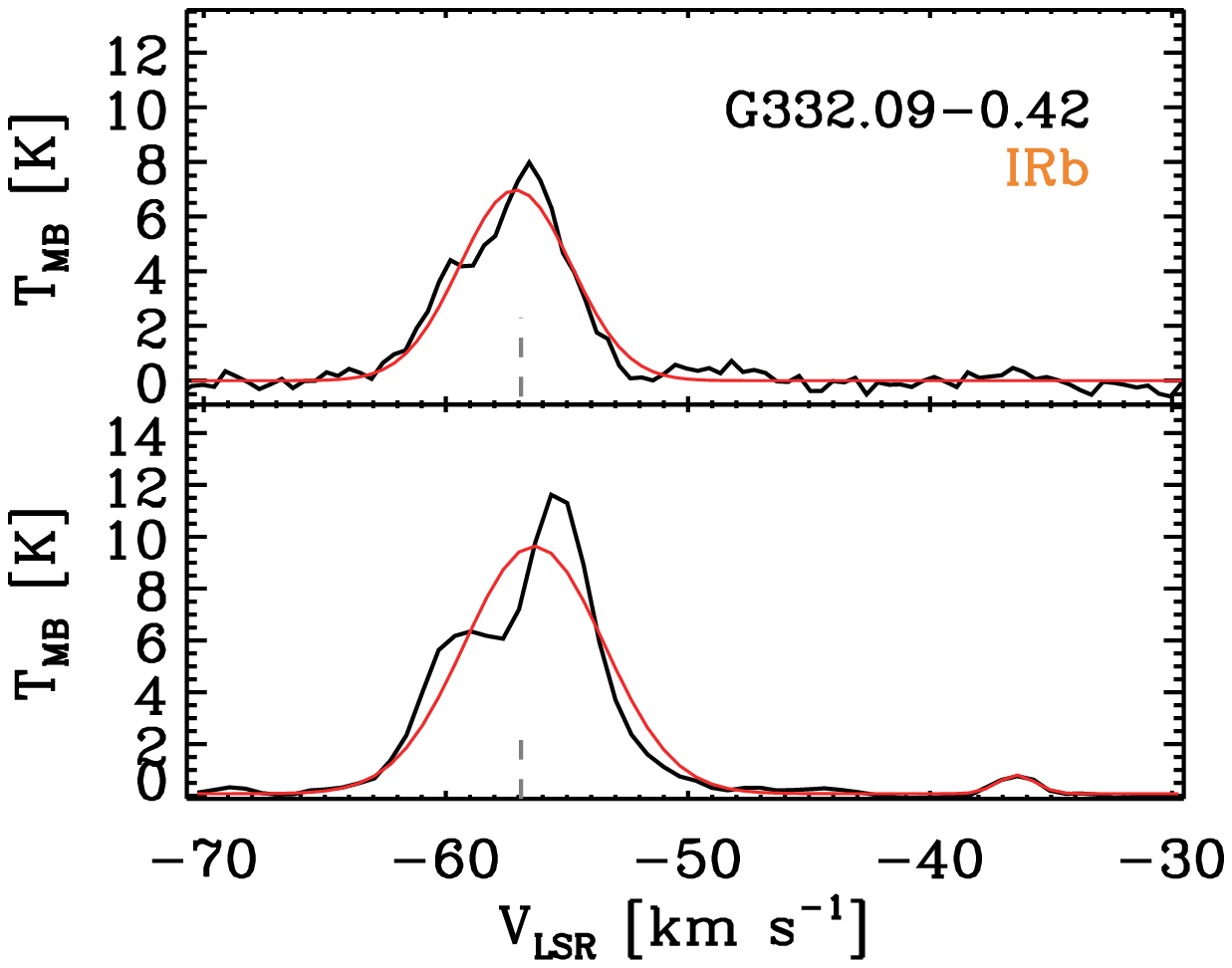}
\includegraphics[scale=0.46]{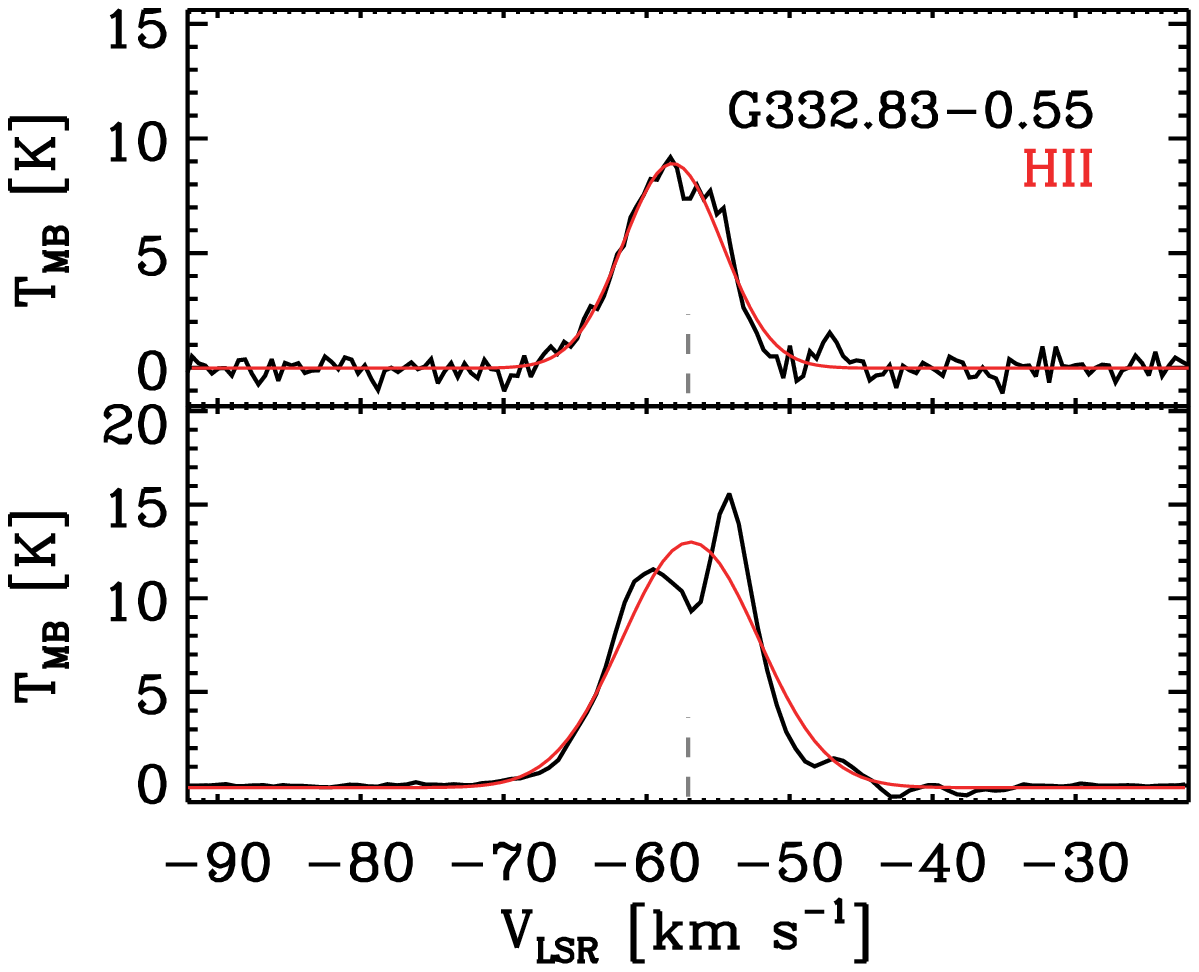}
\includegraphics[scale=0.46]{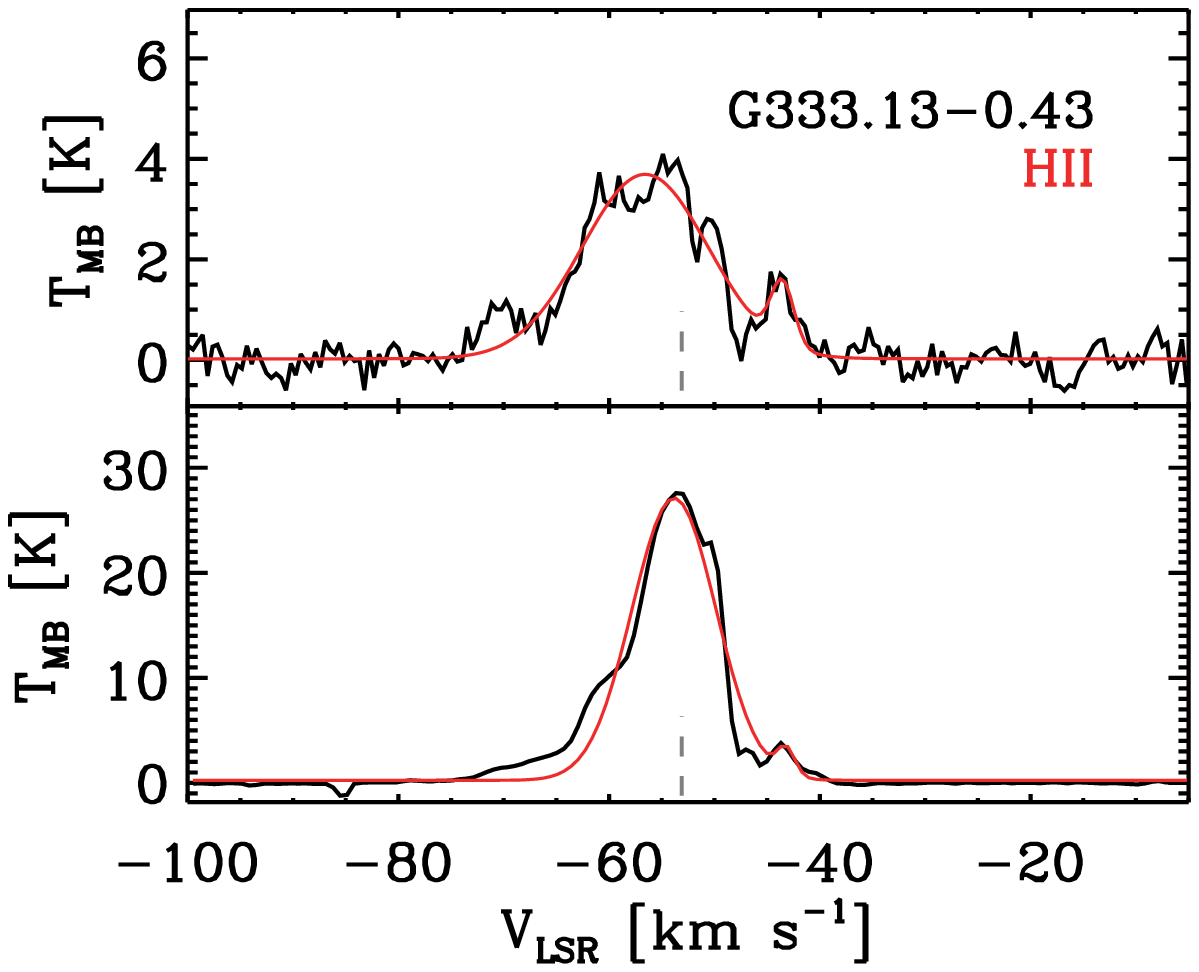}
\includegraphics[scale=0.46]{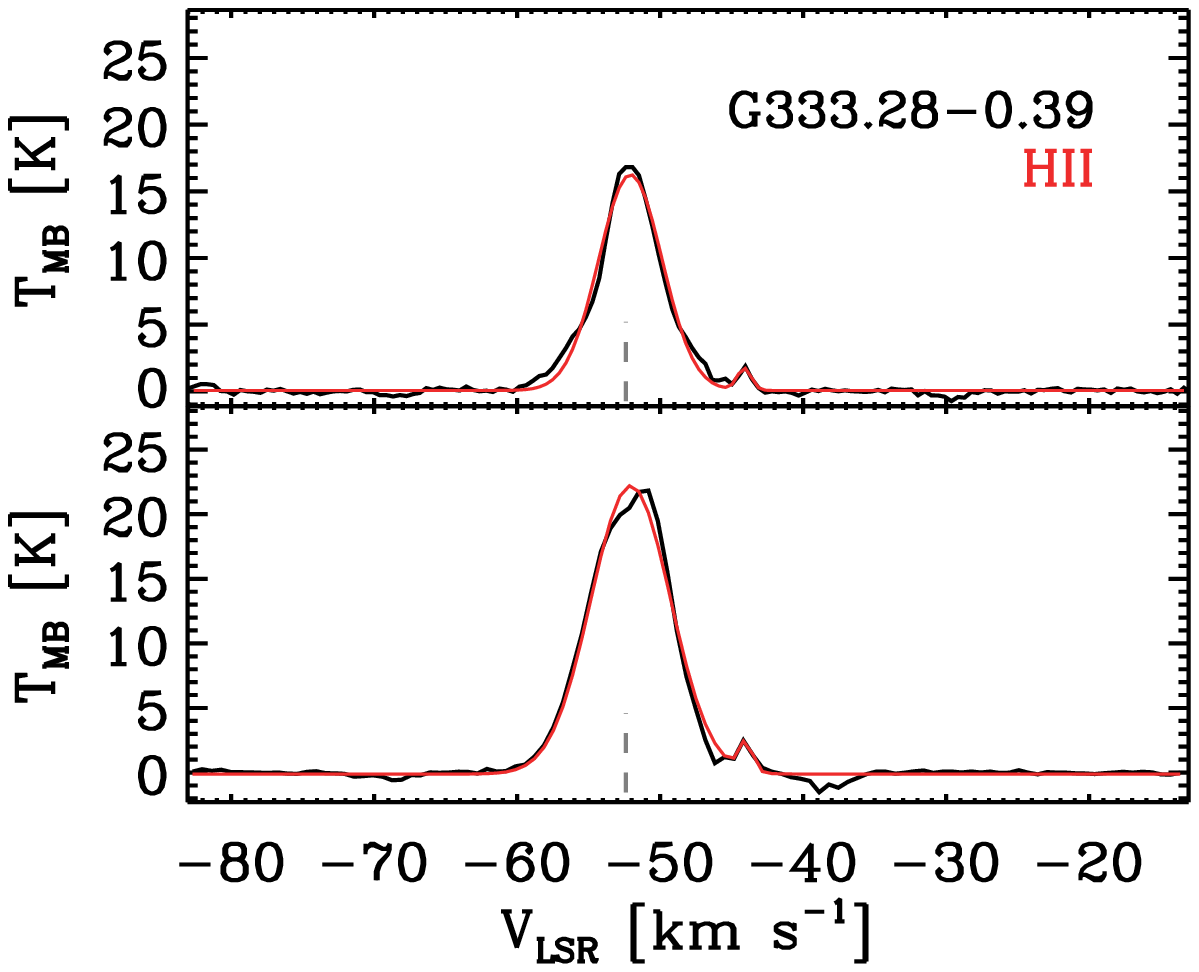}
\includegraphics[scale=0.46]{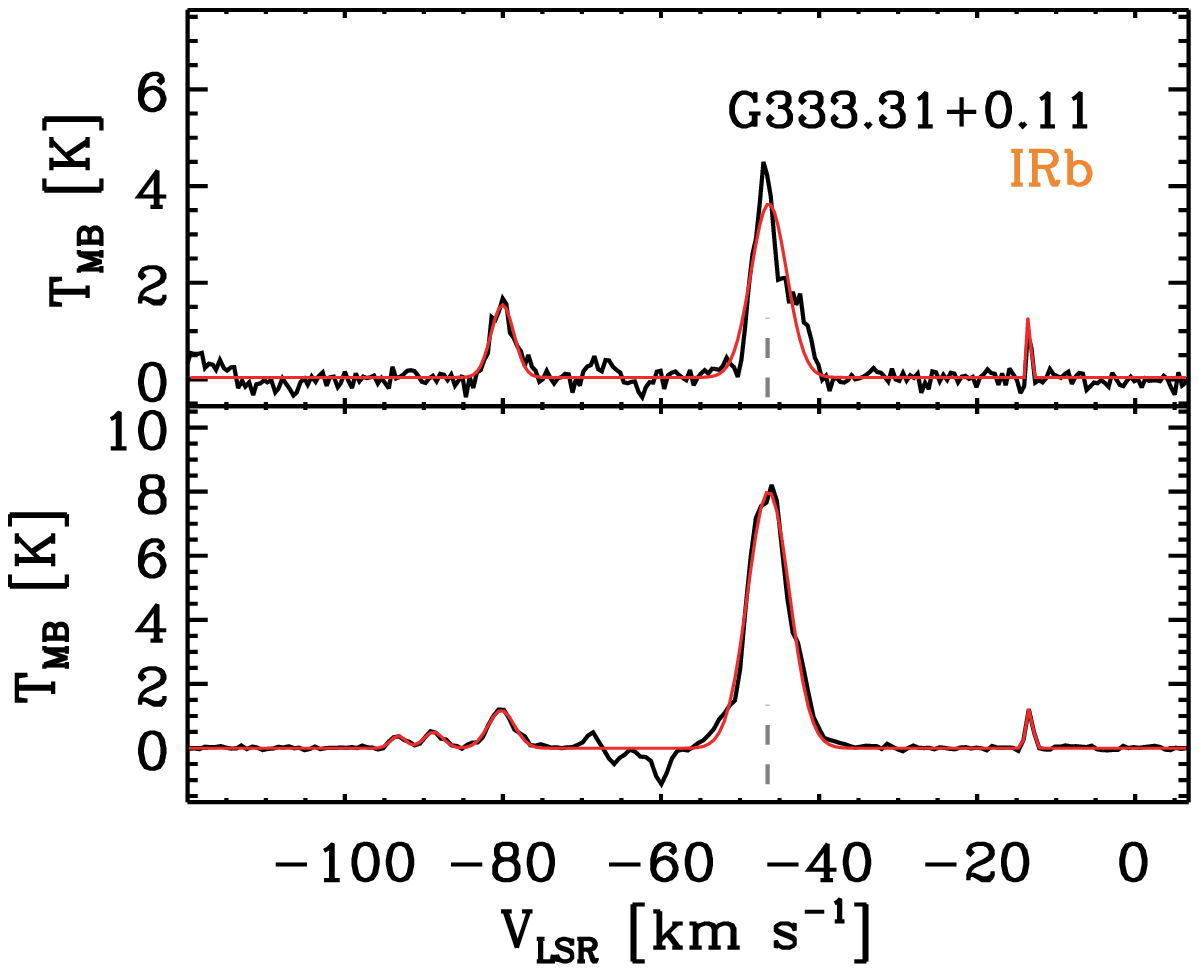}
\caption[]{(continued)}
\end{figure*}

\begin{figure*}
\centering
\ContinuedFloat
\includegraphics[scale=0.46]{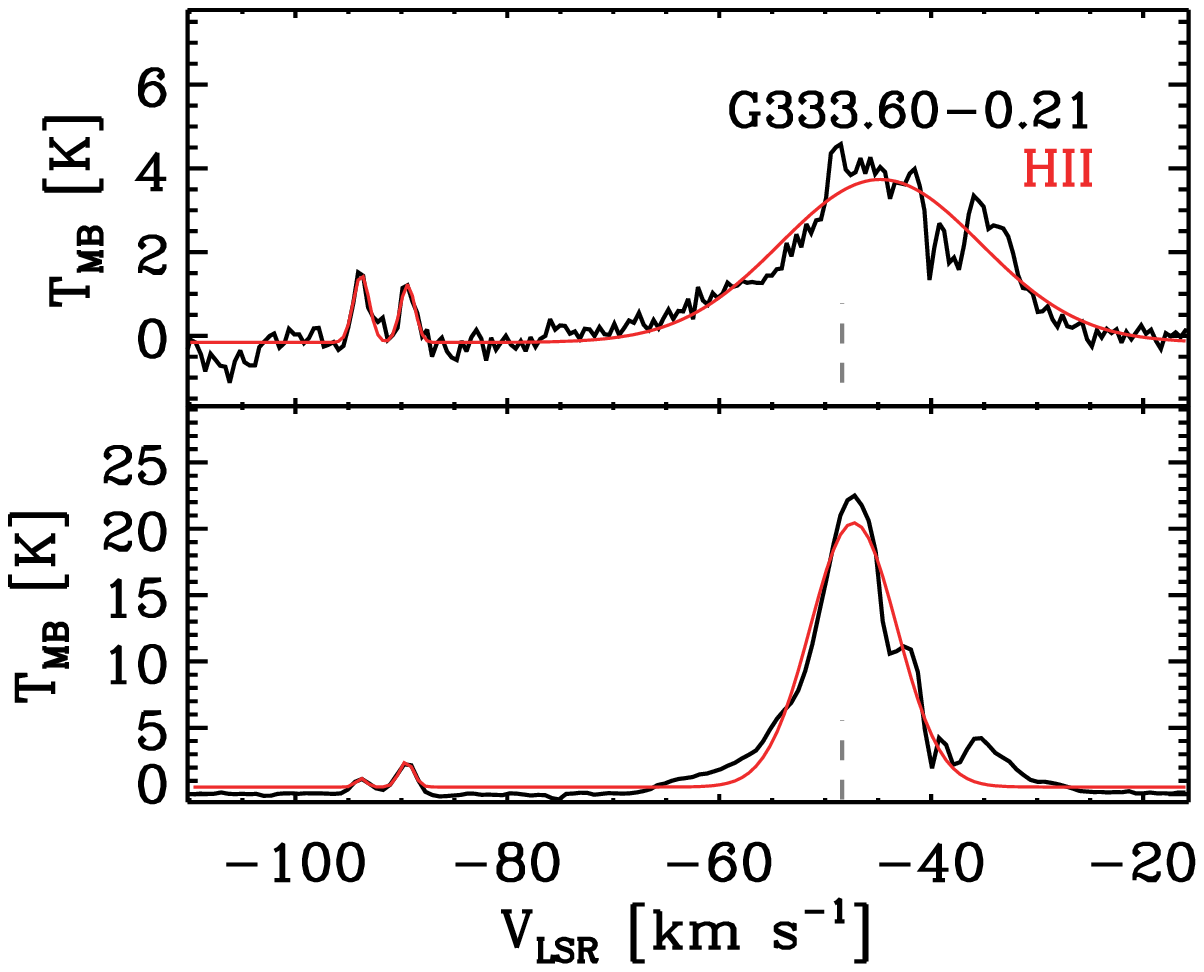}
\includegraphics[scale=0.46]{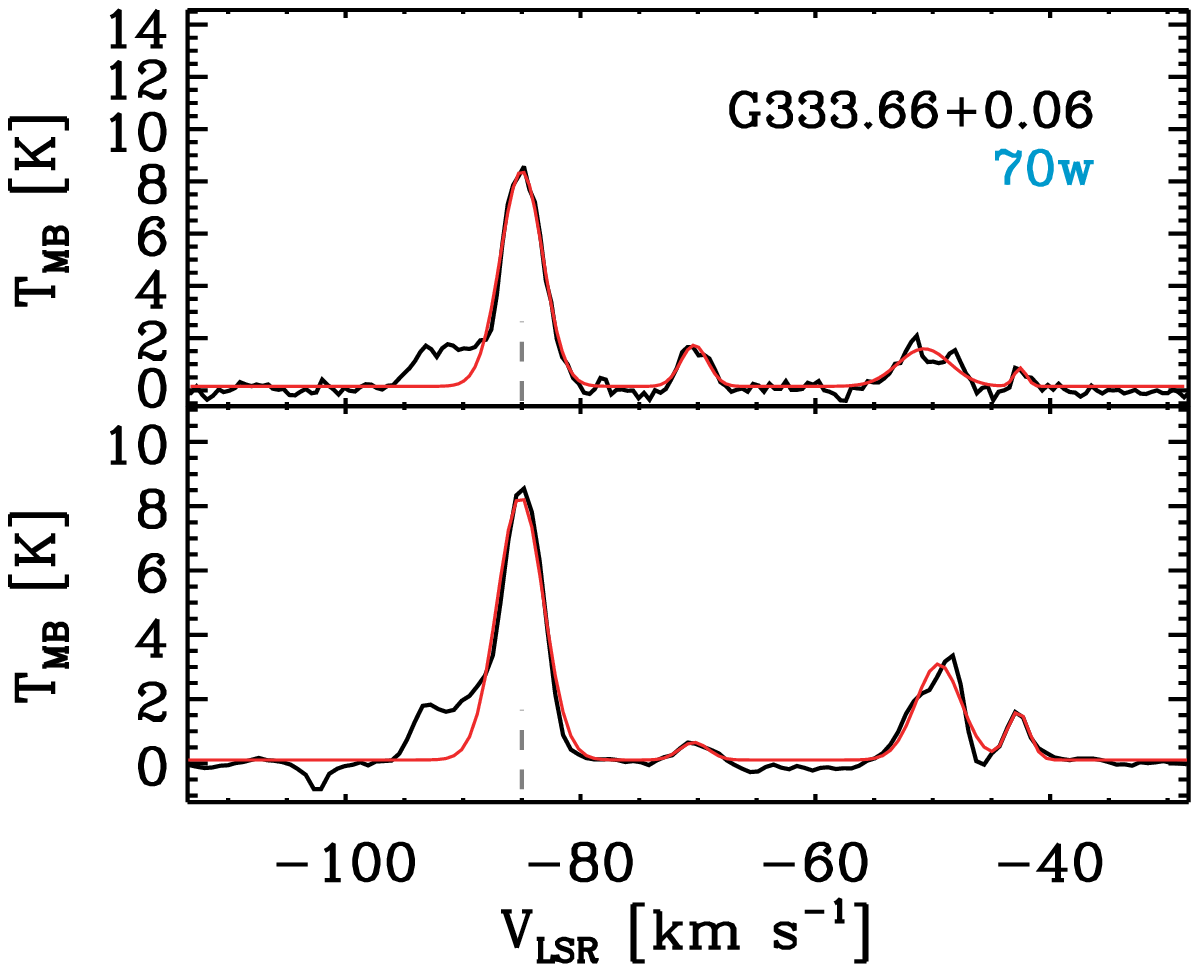}
\includegraphics[scale=0.46]{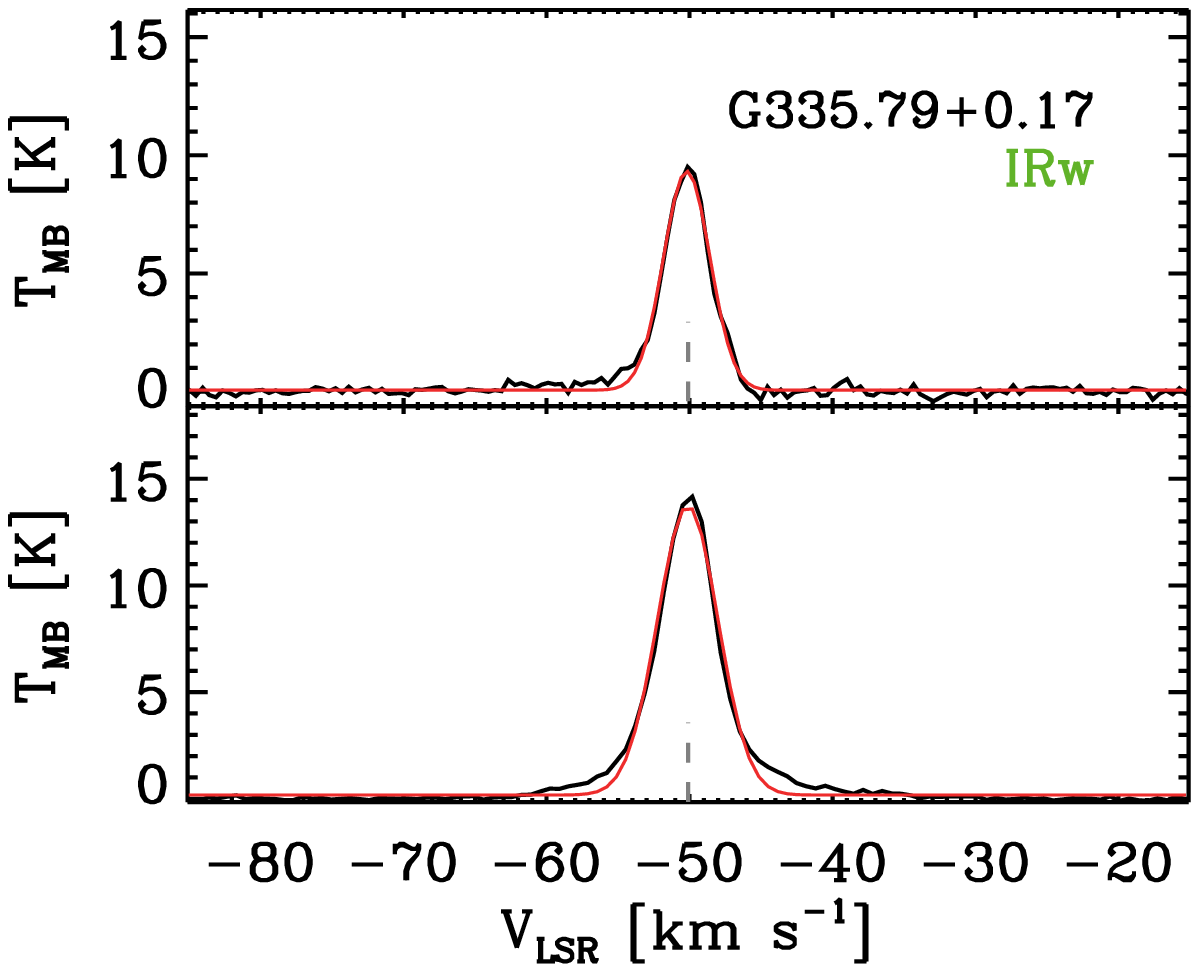}
\includegraphics[scale=0.46]{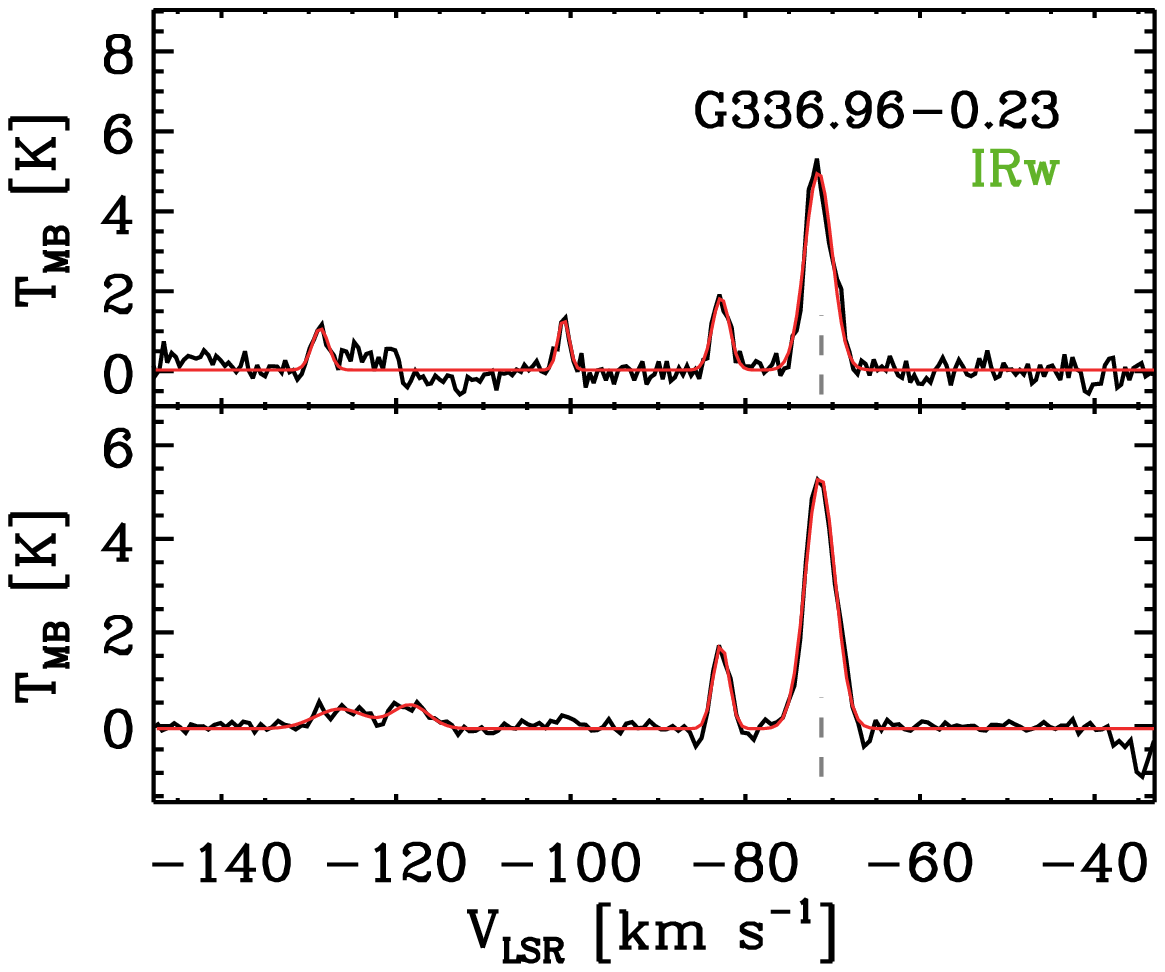}
\includegraphics[scale=0.46]{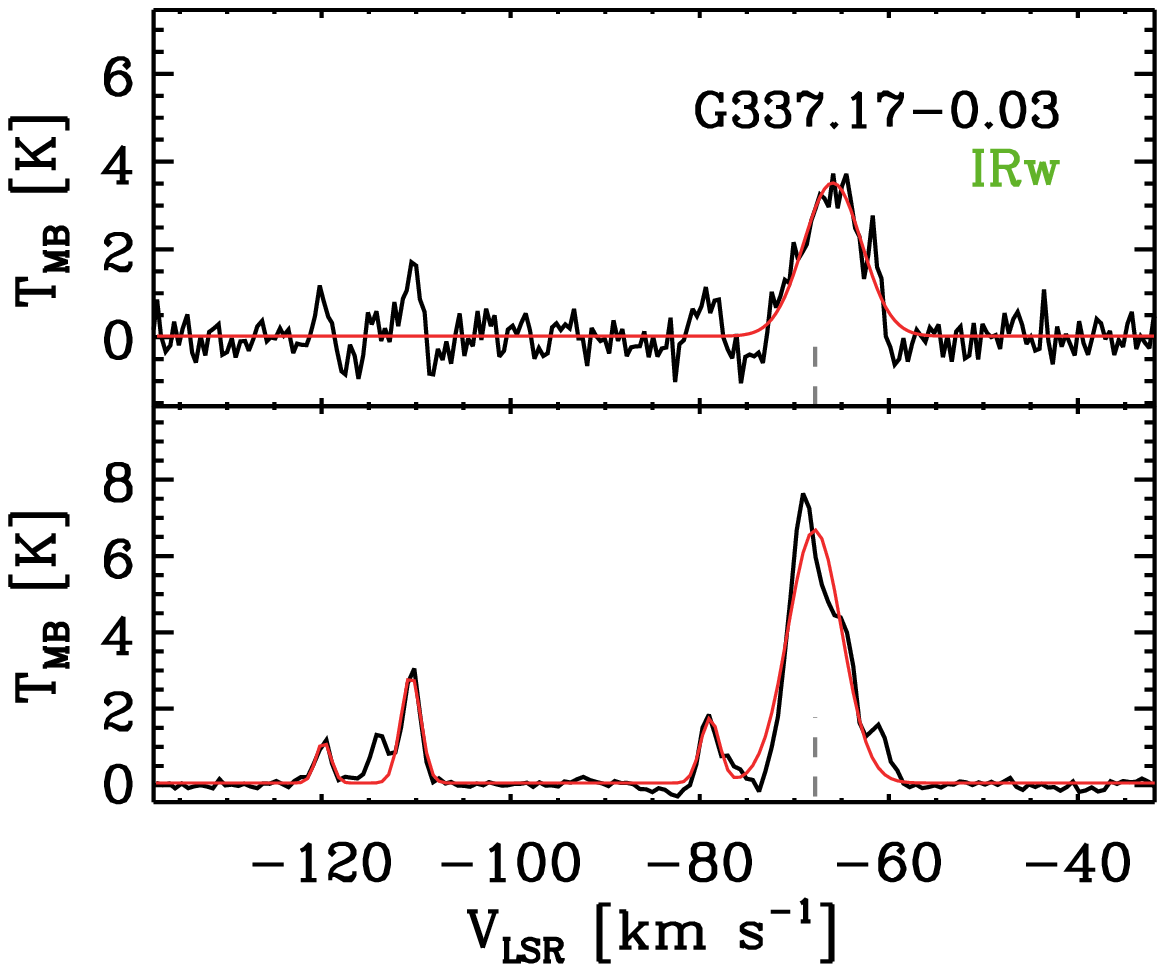}
\includegraphics[scale=0.46]{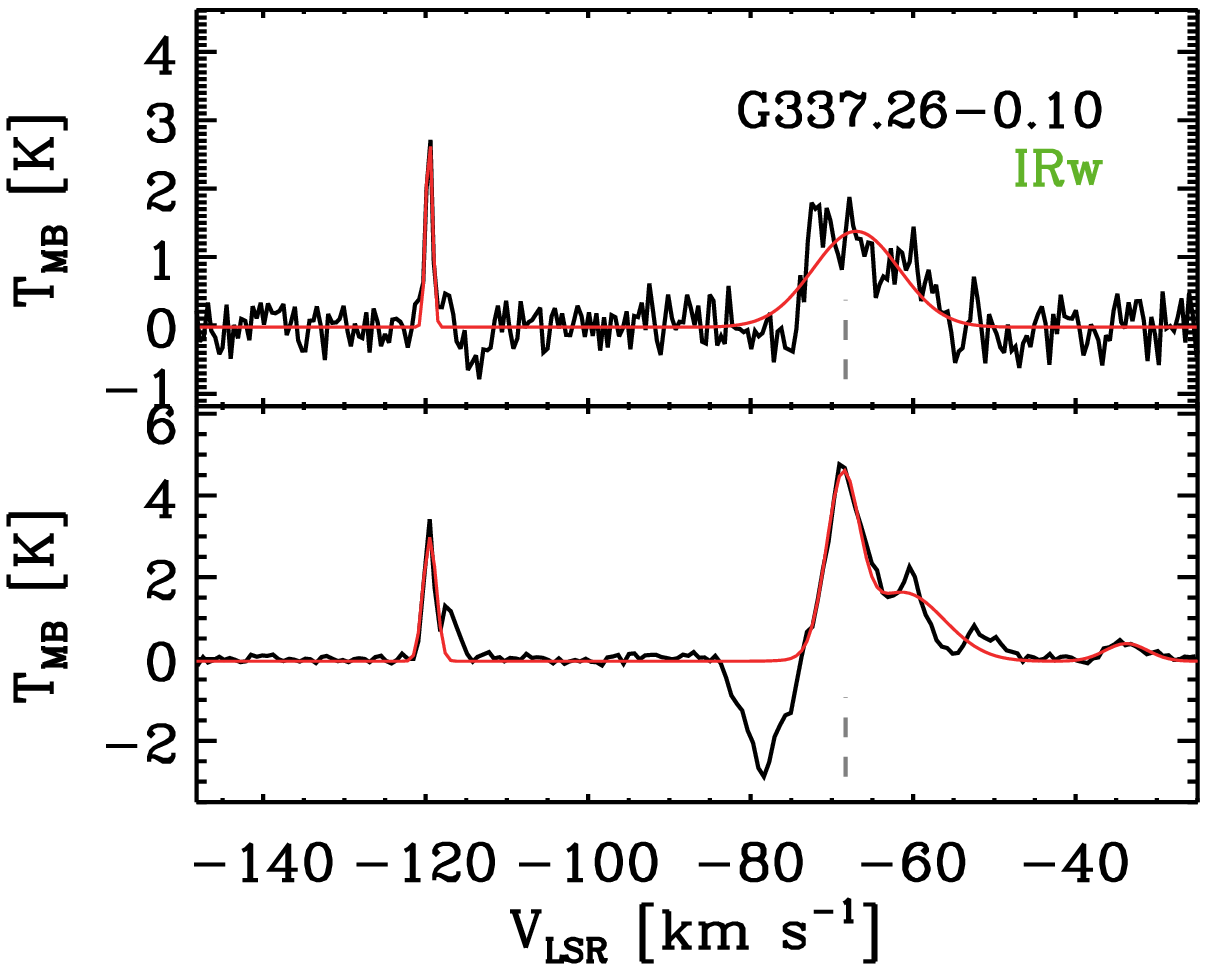}
\includegraphics[scale=0.46]{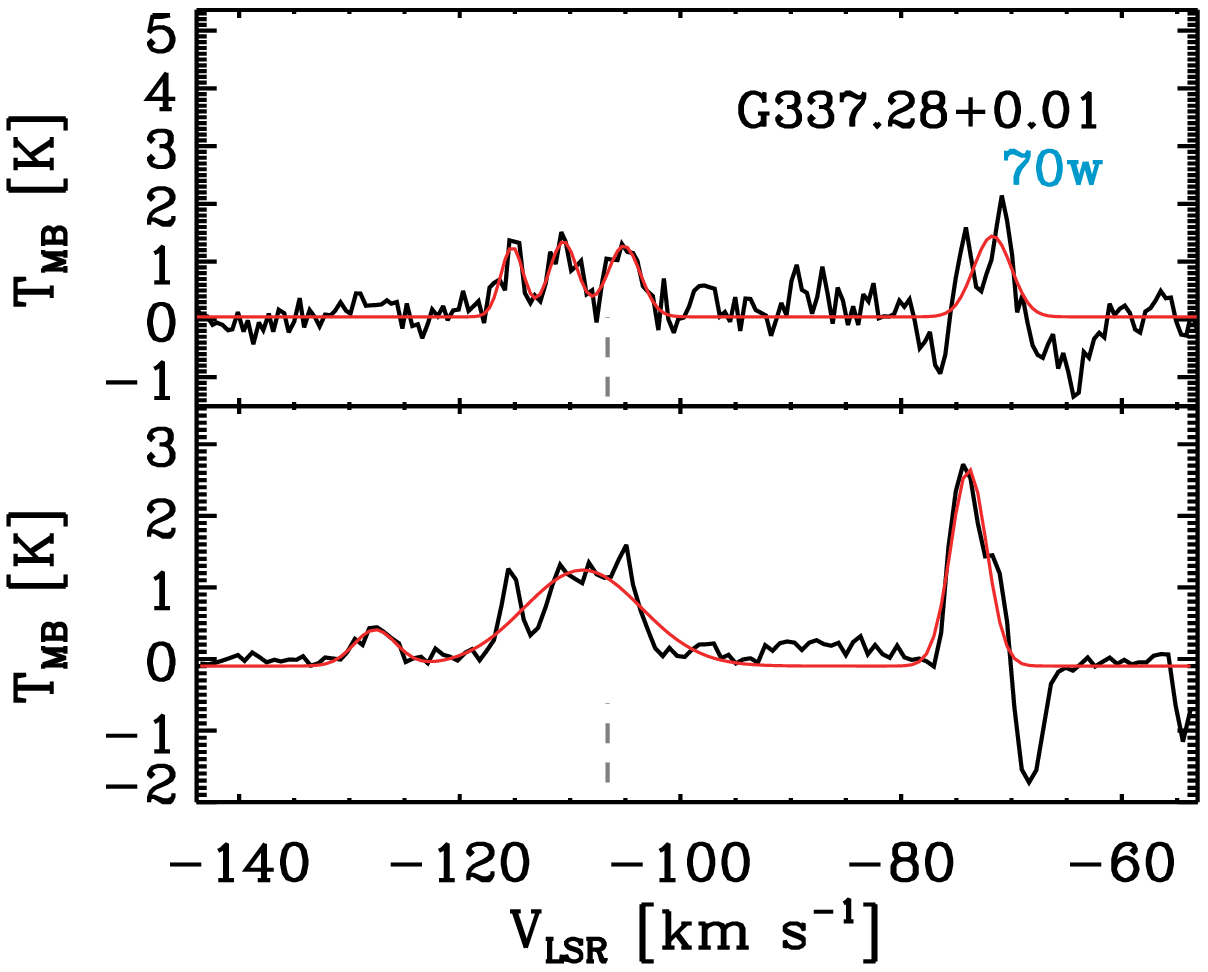}
\includegraphics[scale=0.46]{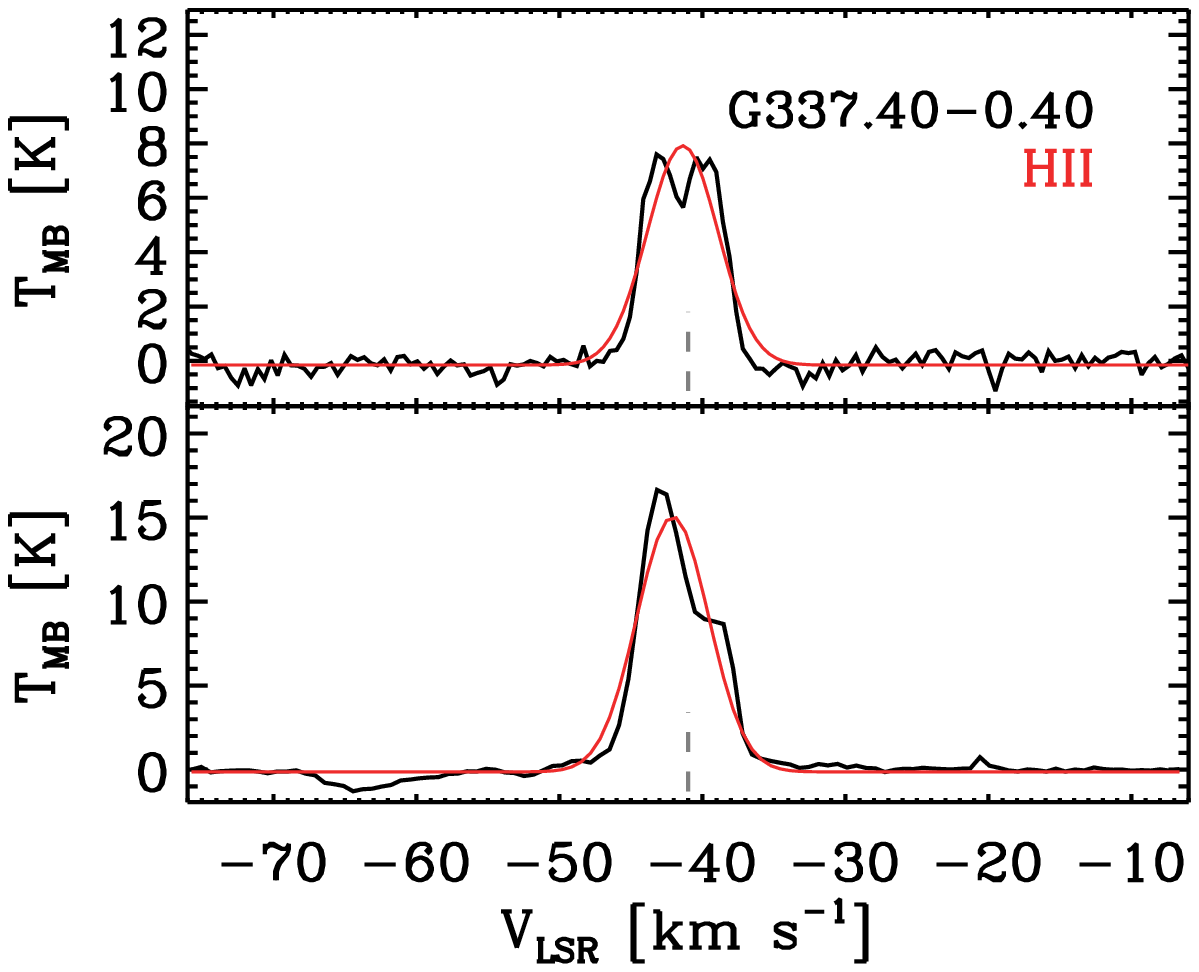}
\includegraphics[scale=0.46]{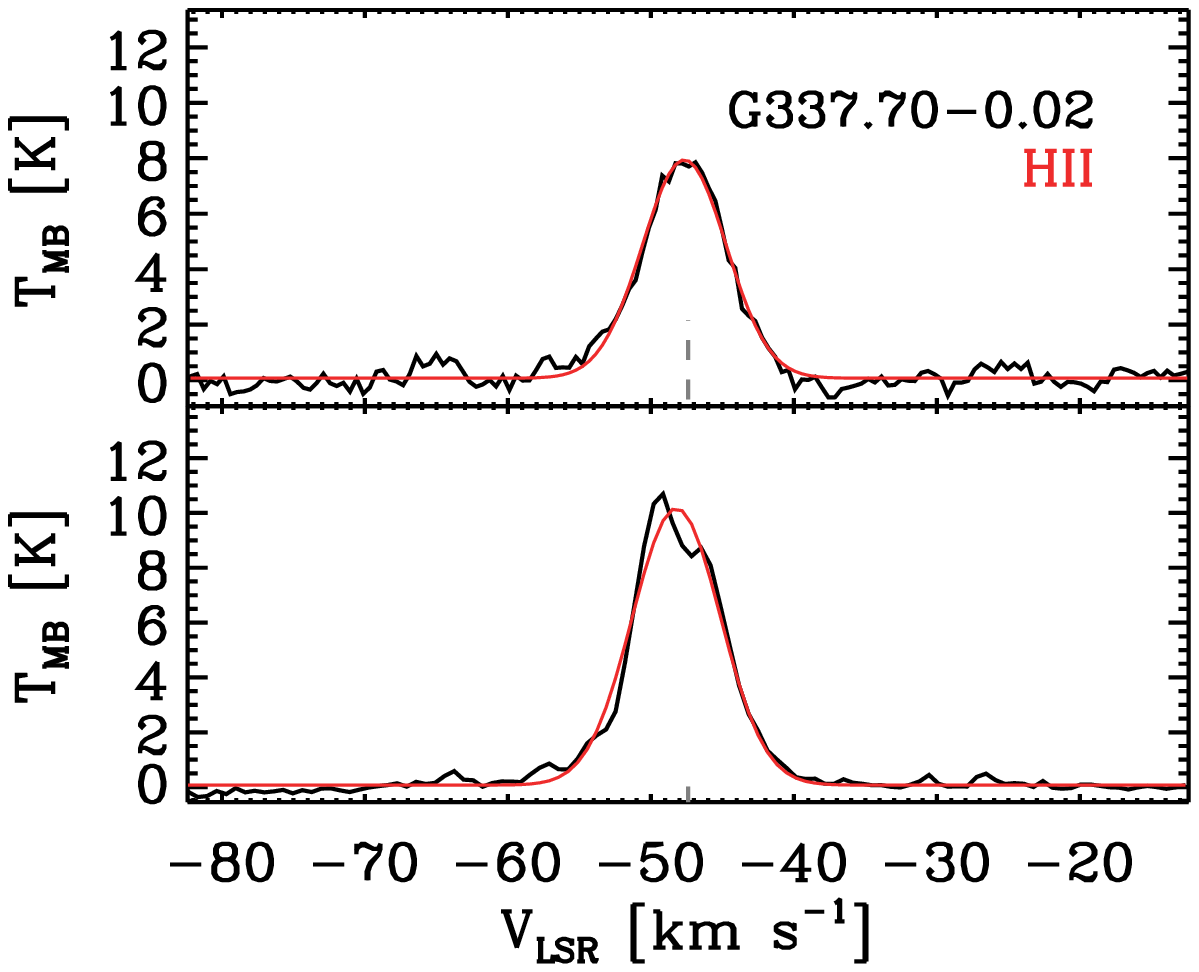}
\includegraphics[scale=0.46]{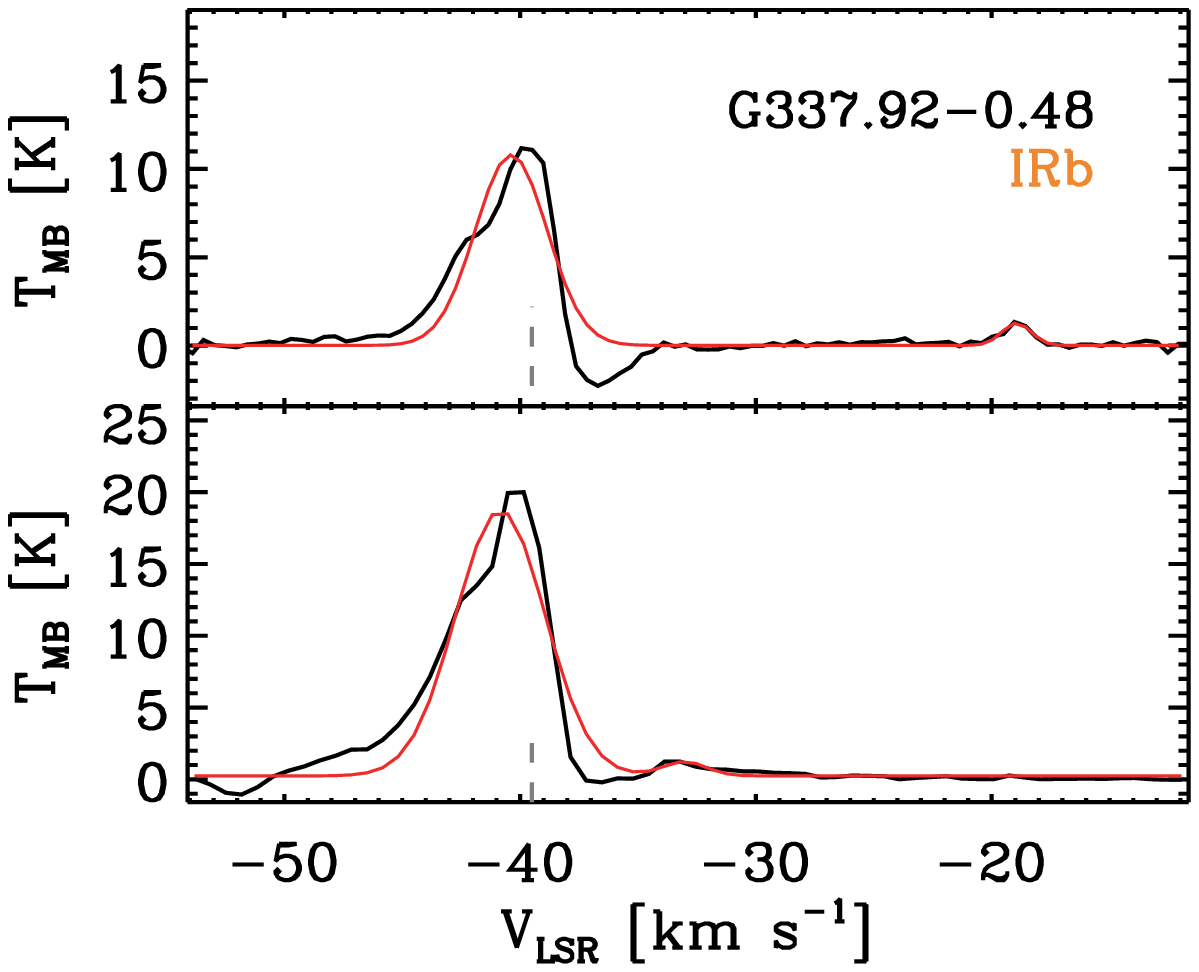}
\includegraphics[scale=0.46]{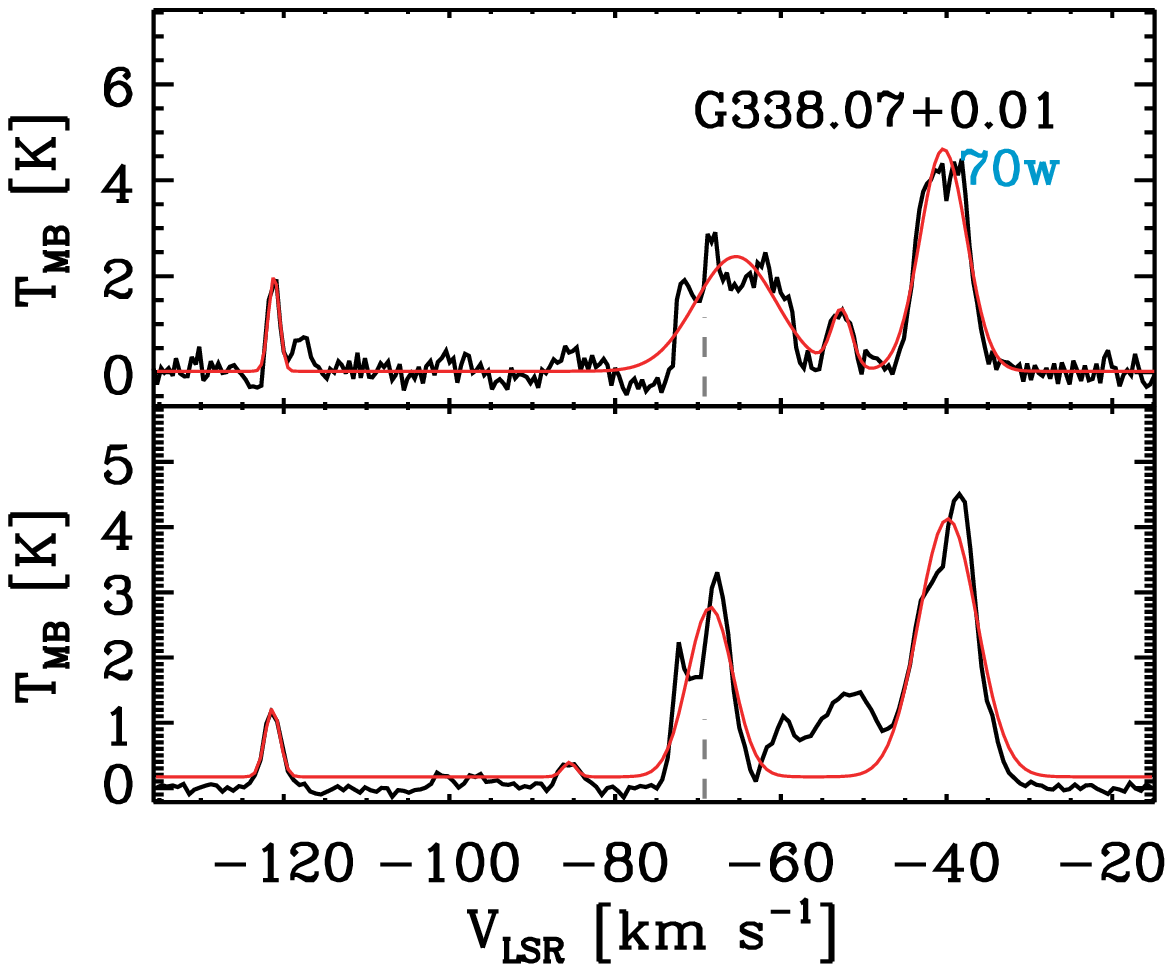}
\includegraphics[scale=0.46]{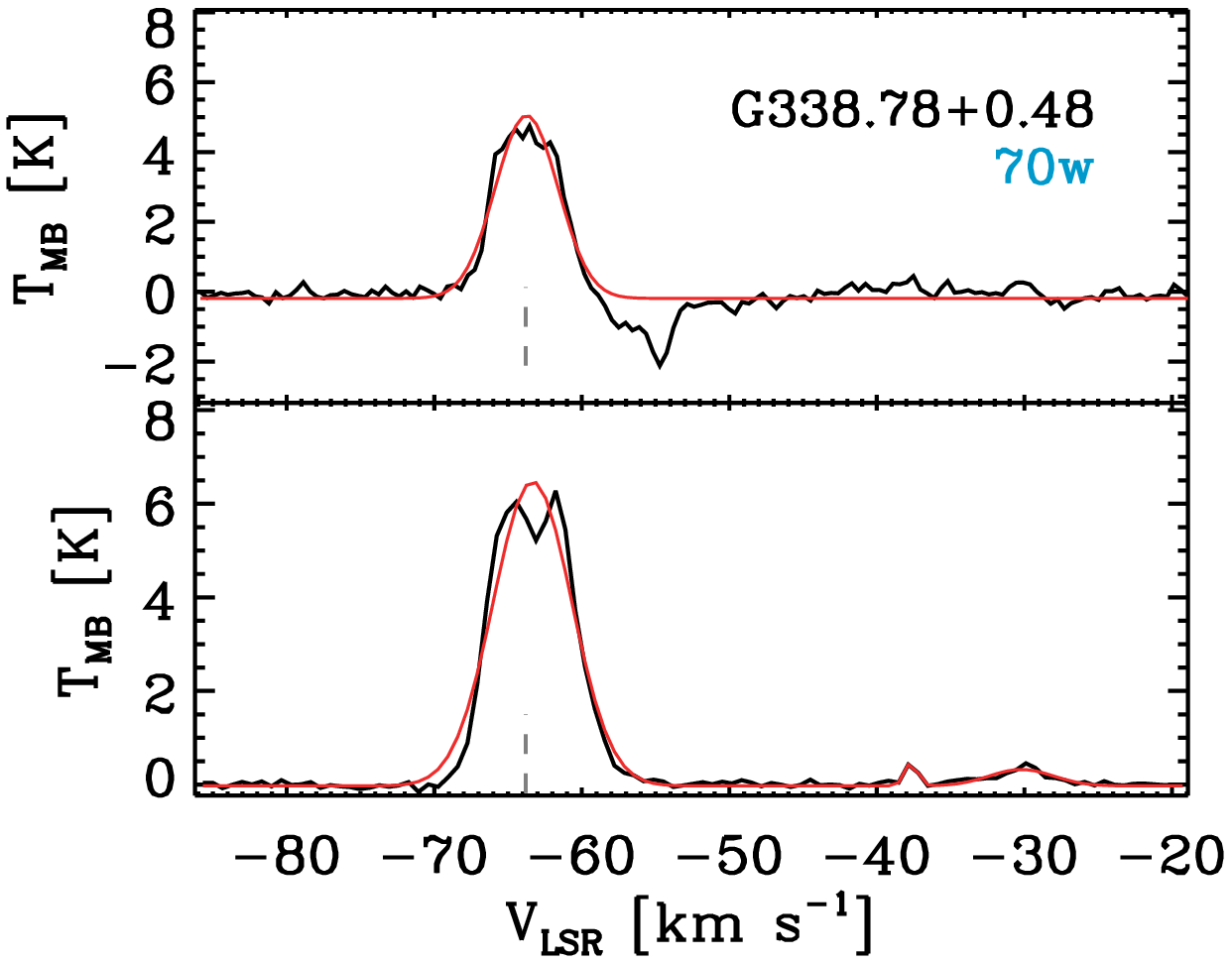}
\includegraphics[scale=0.46]{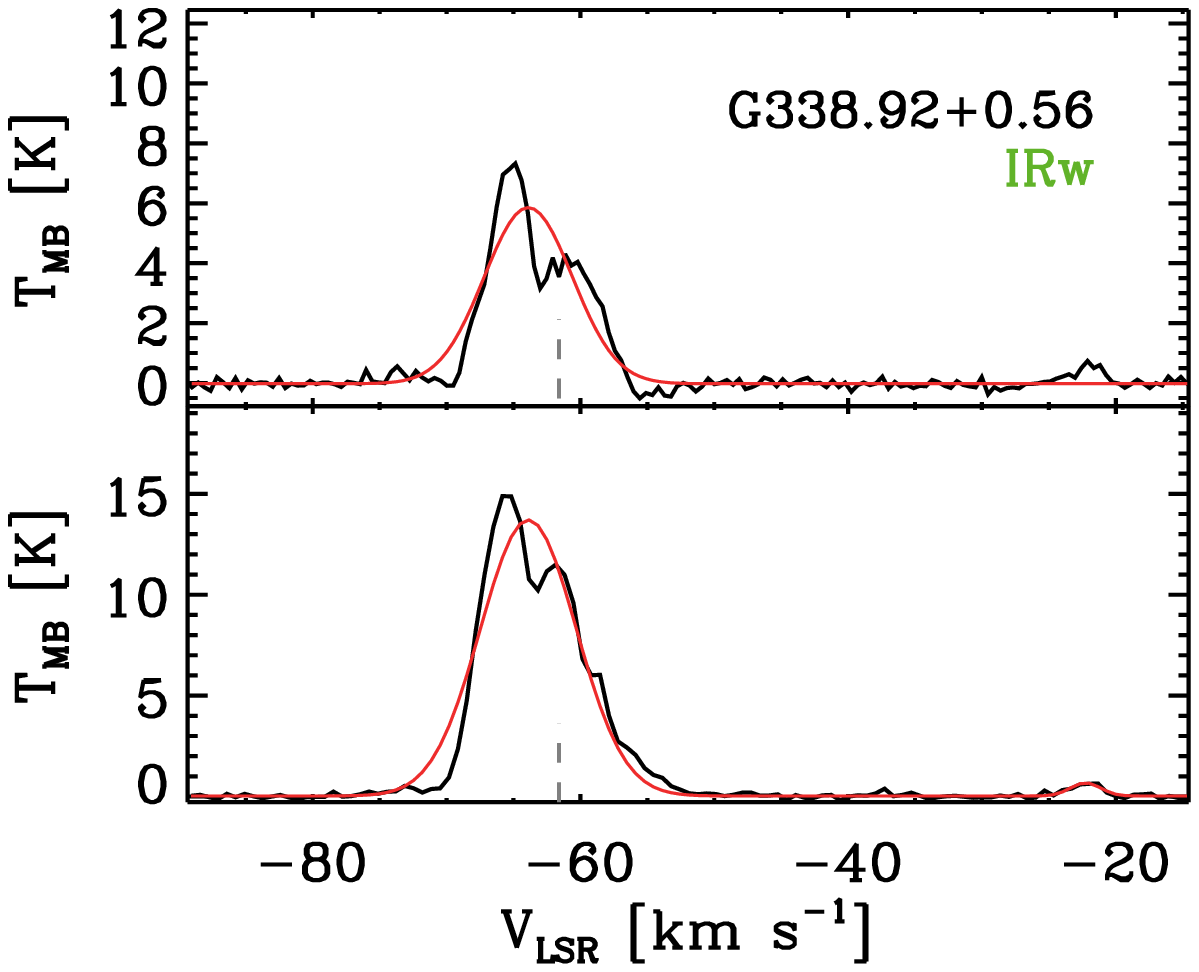}
\includegraphics[scale=0.46]{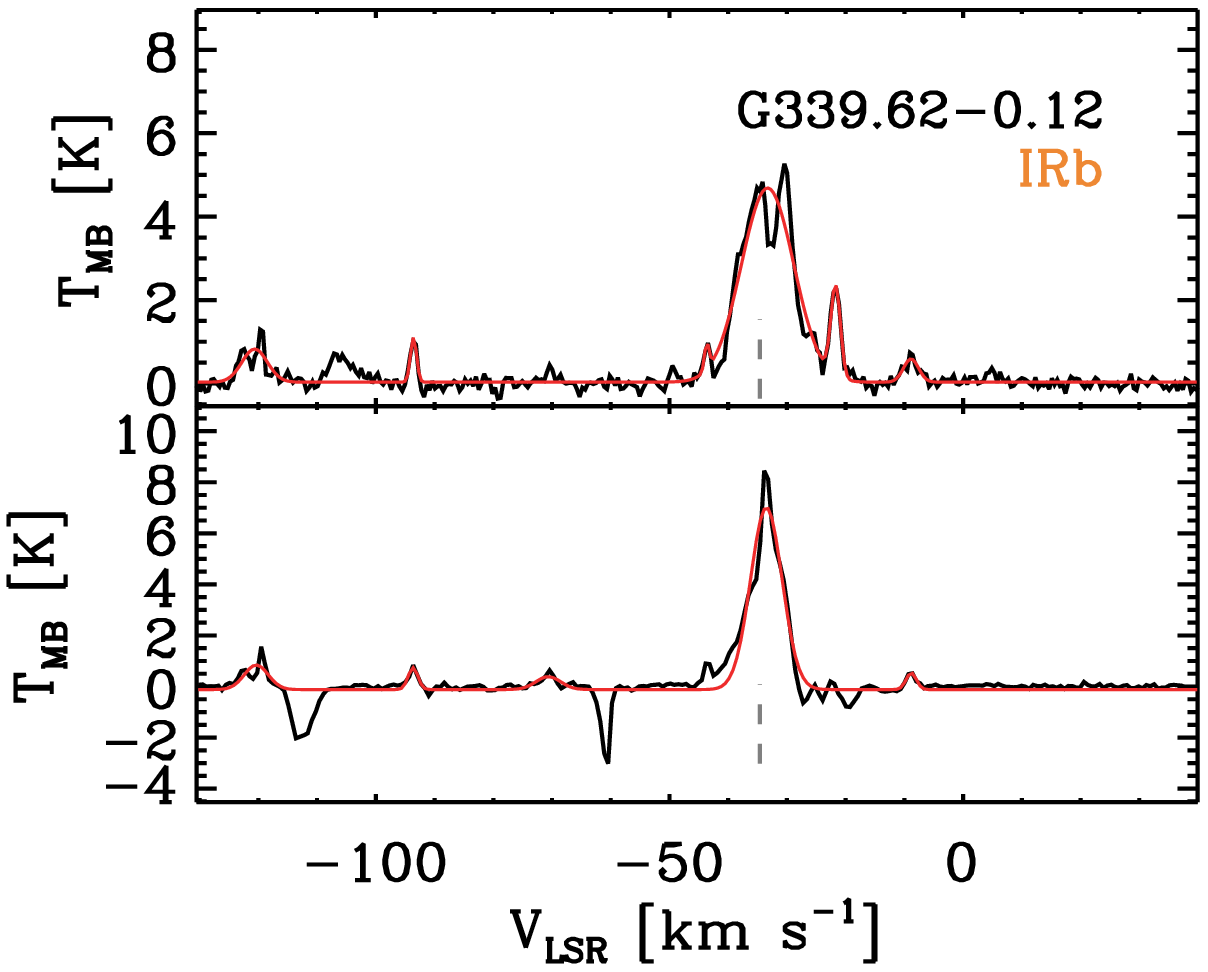}
\includegraphics[scale=0.46]{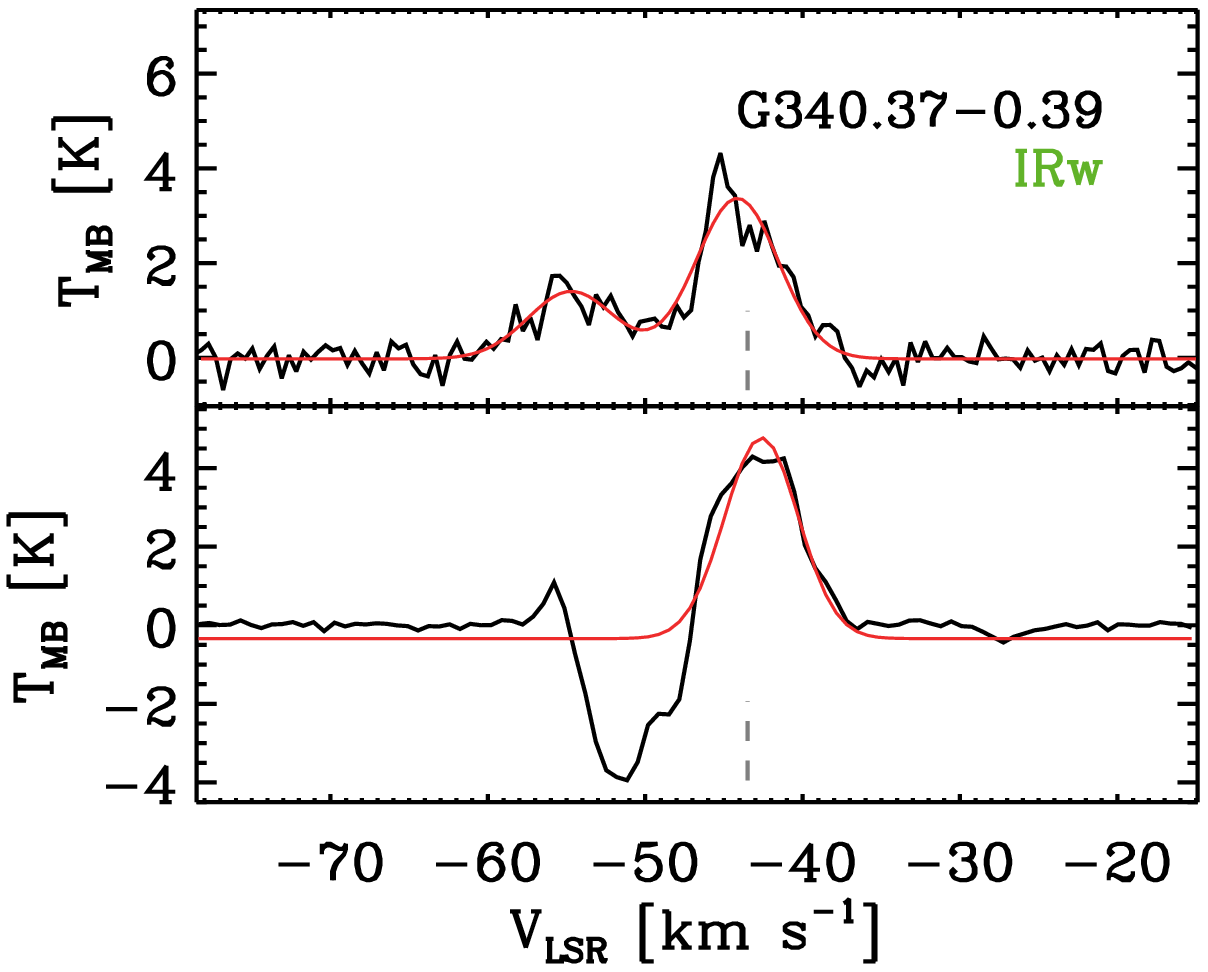}
\caption[]{(continued)}
\end{figure*}

\begin{figure*}
\centering
\ContinuedFloat
\includegraphics[scale=0.46]{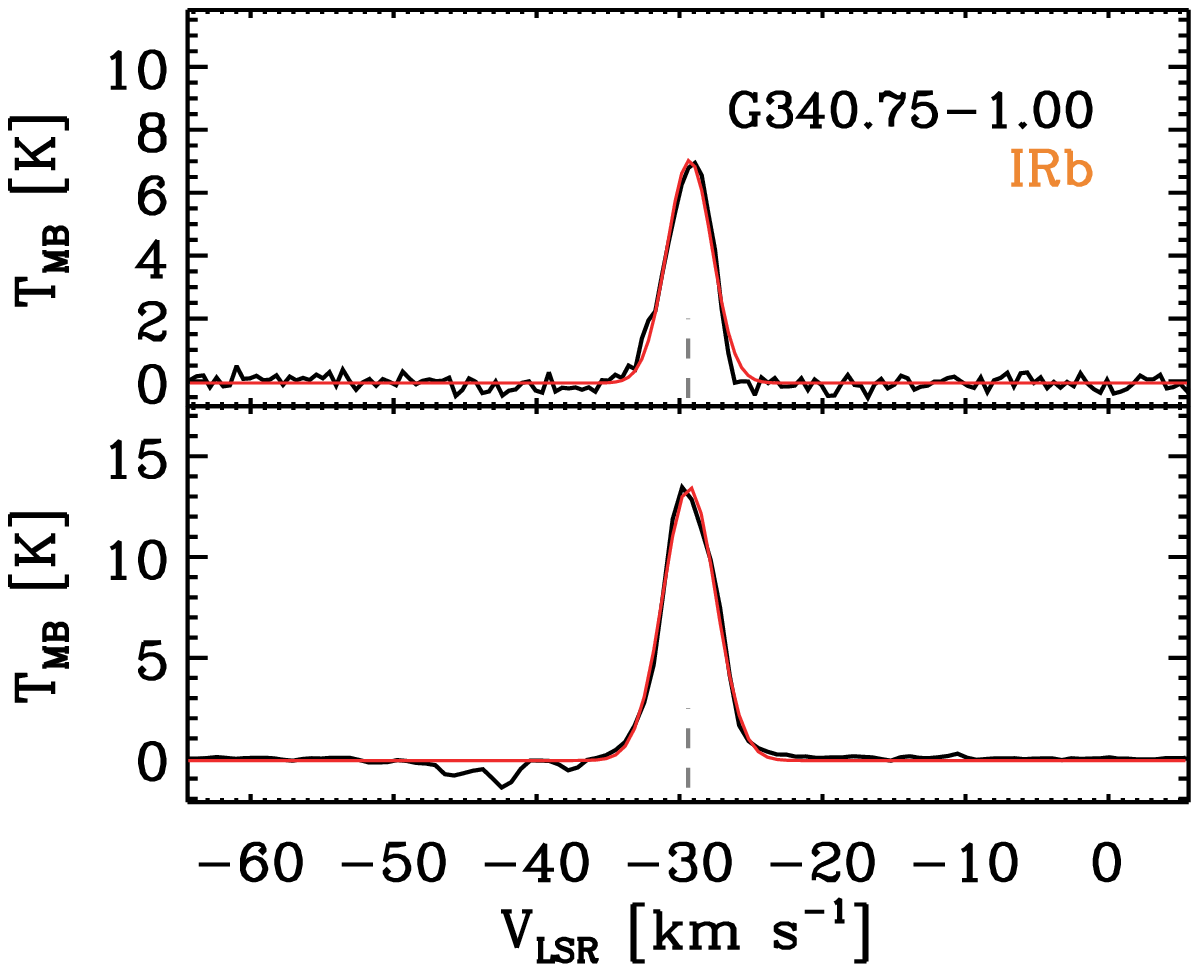}
\includegraphics[scale=0.46]{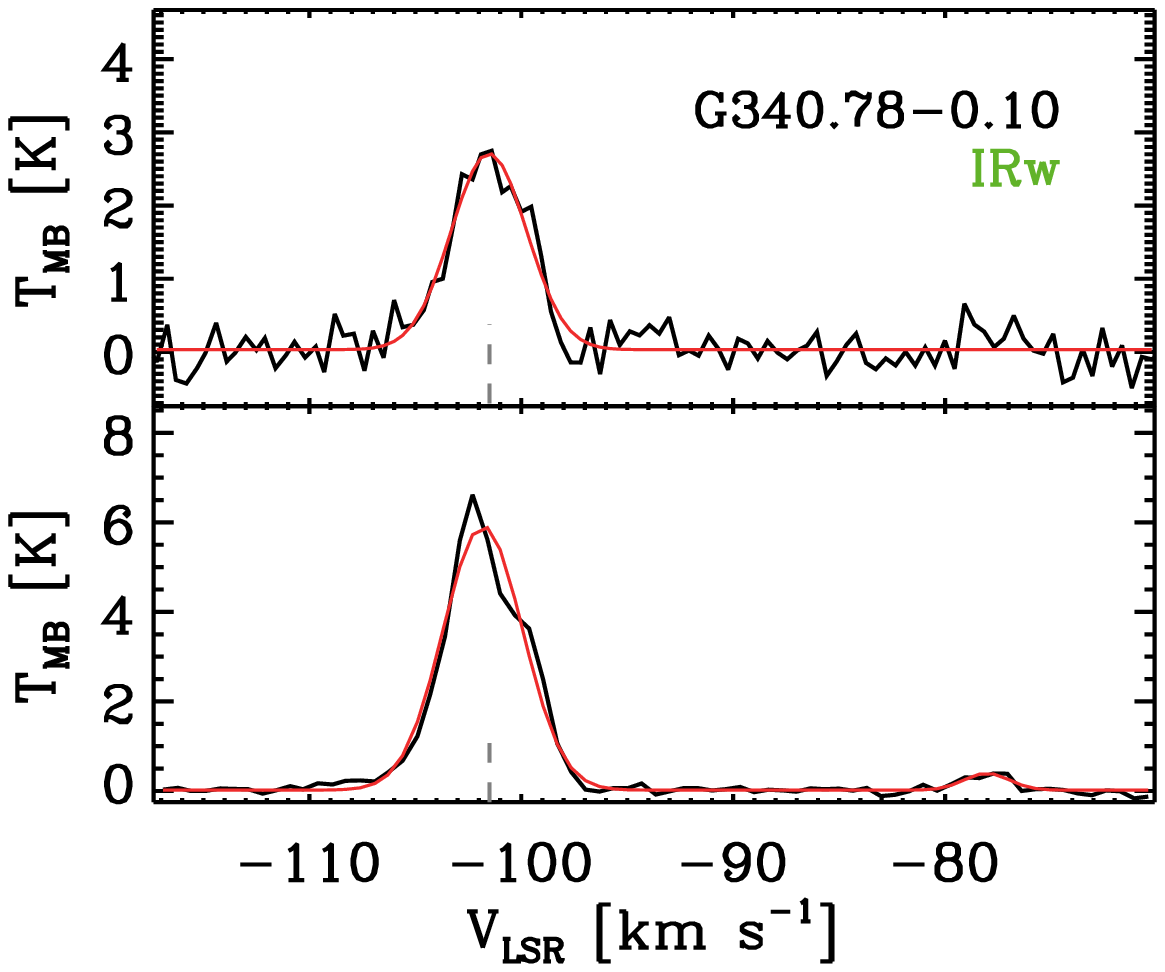}
\includegraphics[scale=0.46]{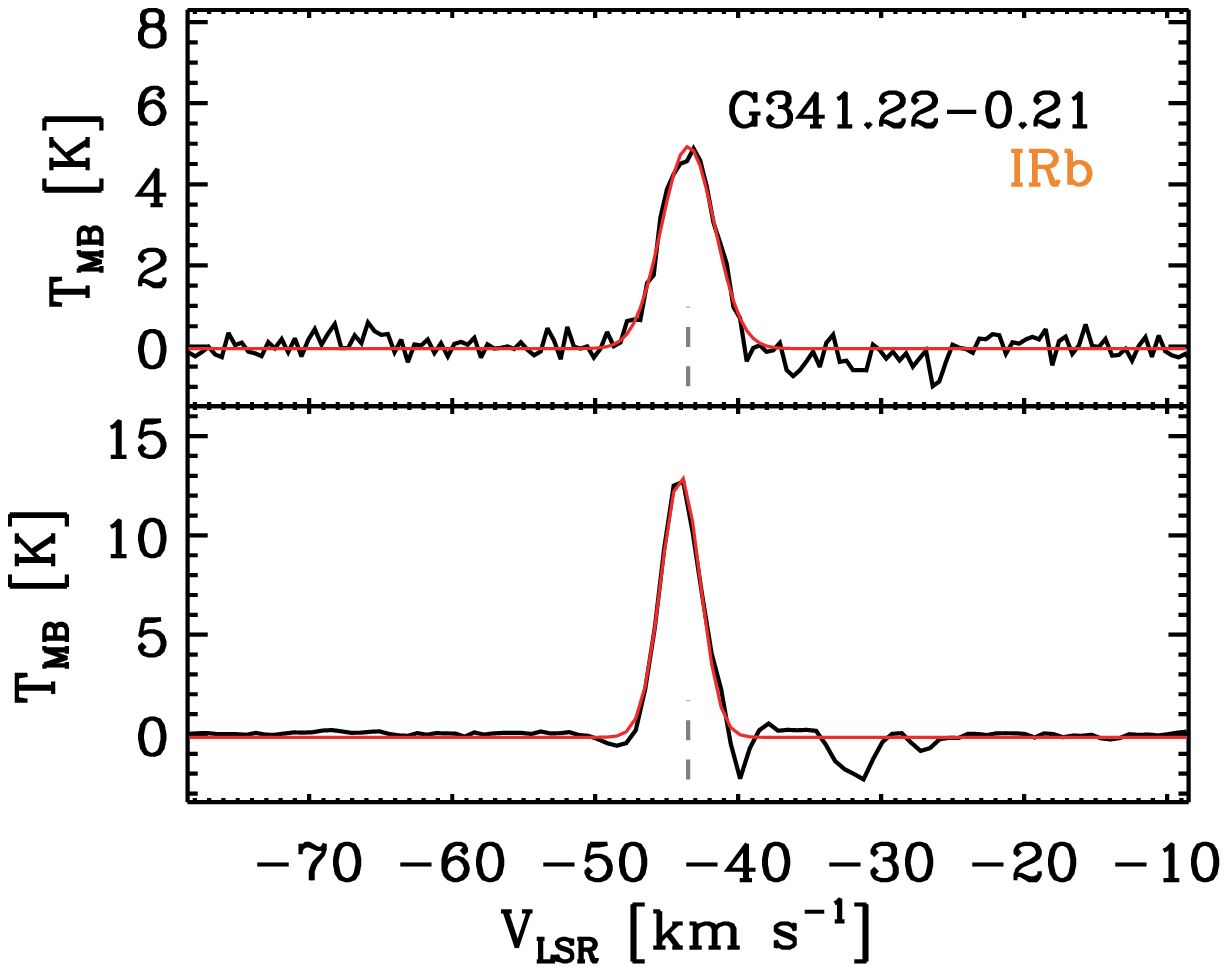}
\includegraphics[scale=0.46]{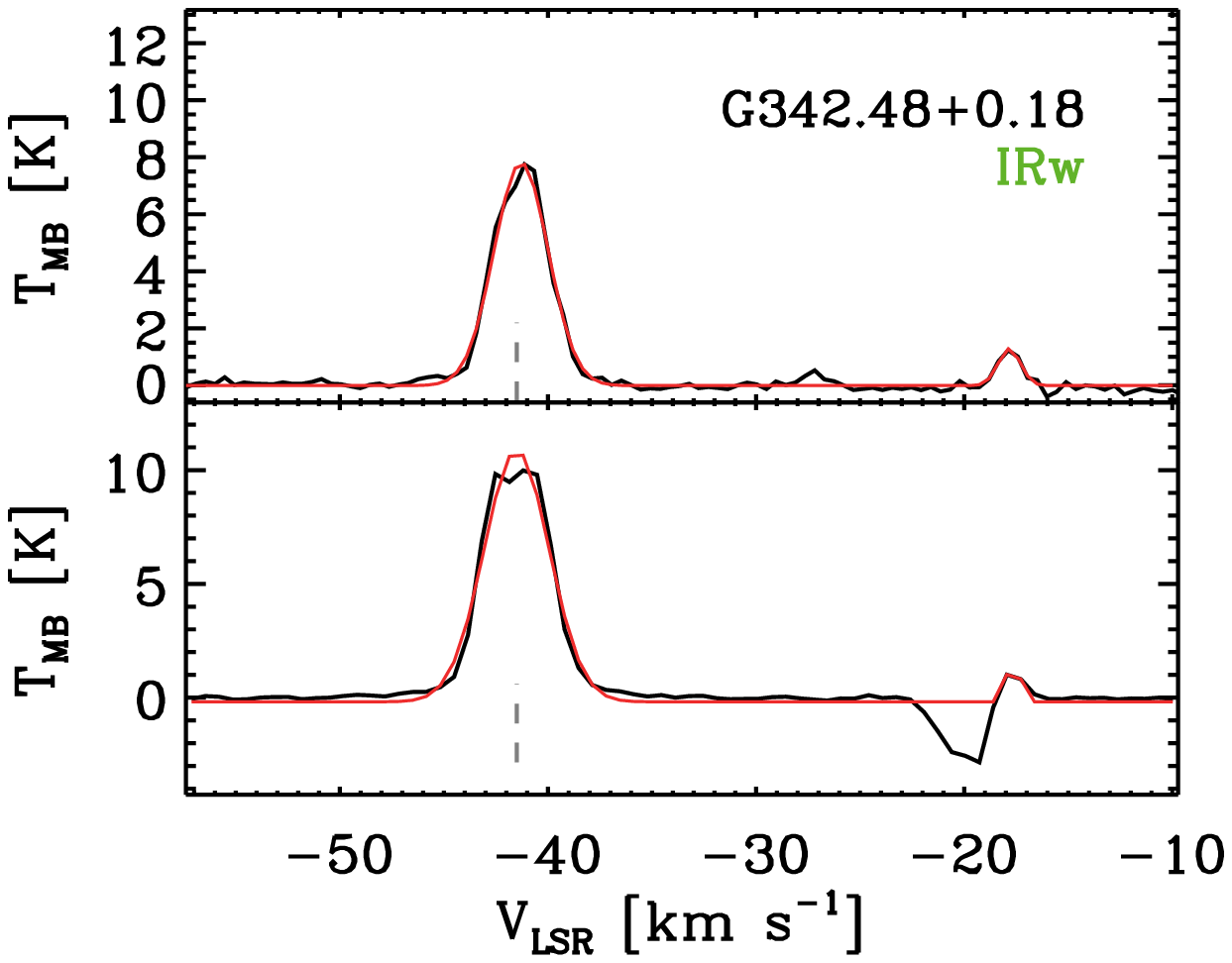}
\includegraphics[scale=0.46]{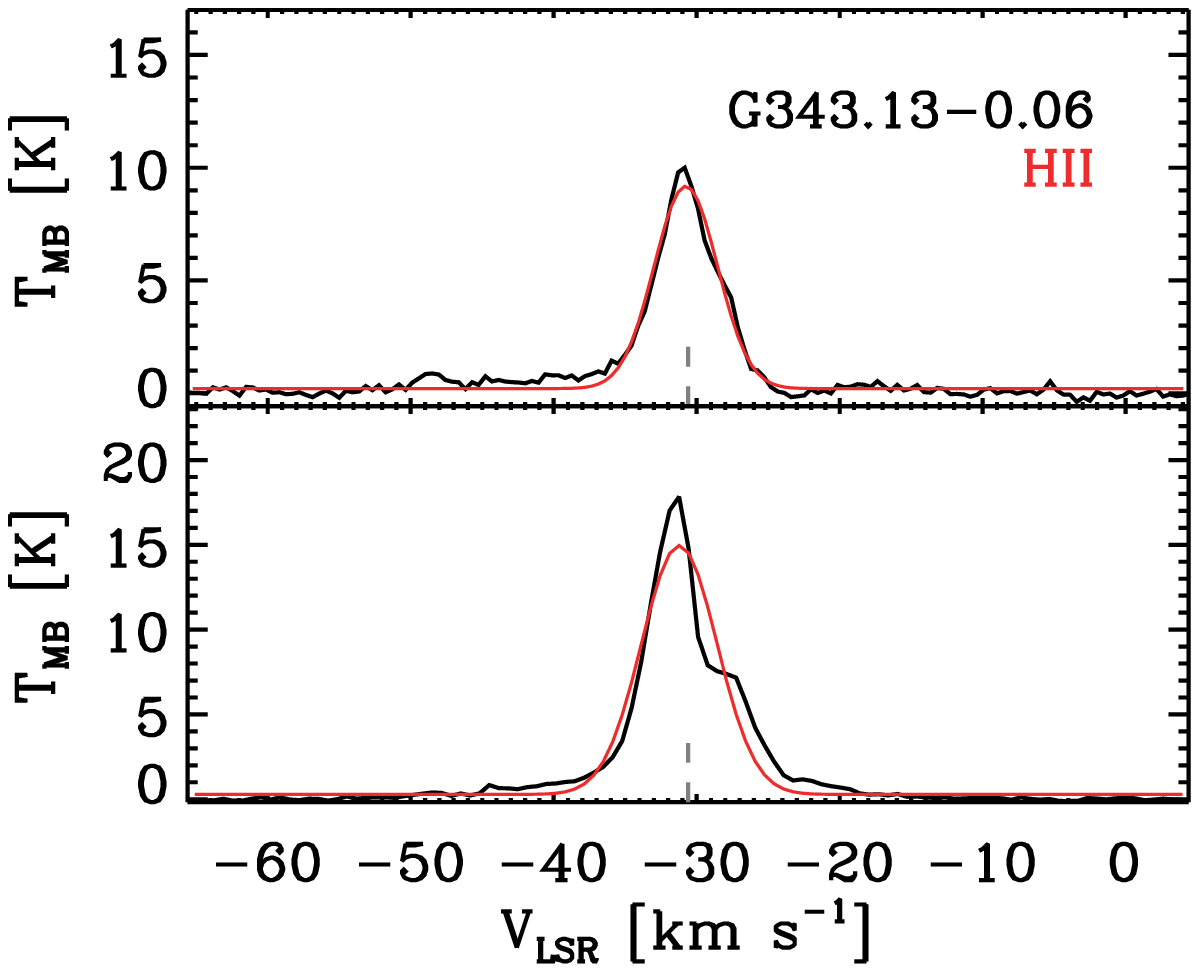}
\includegraphics[scale=0.46]{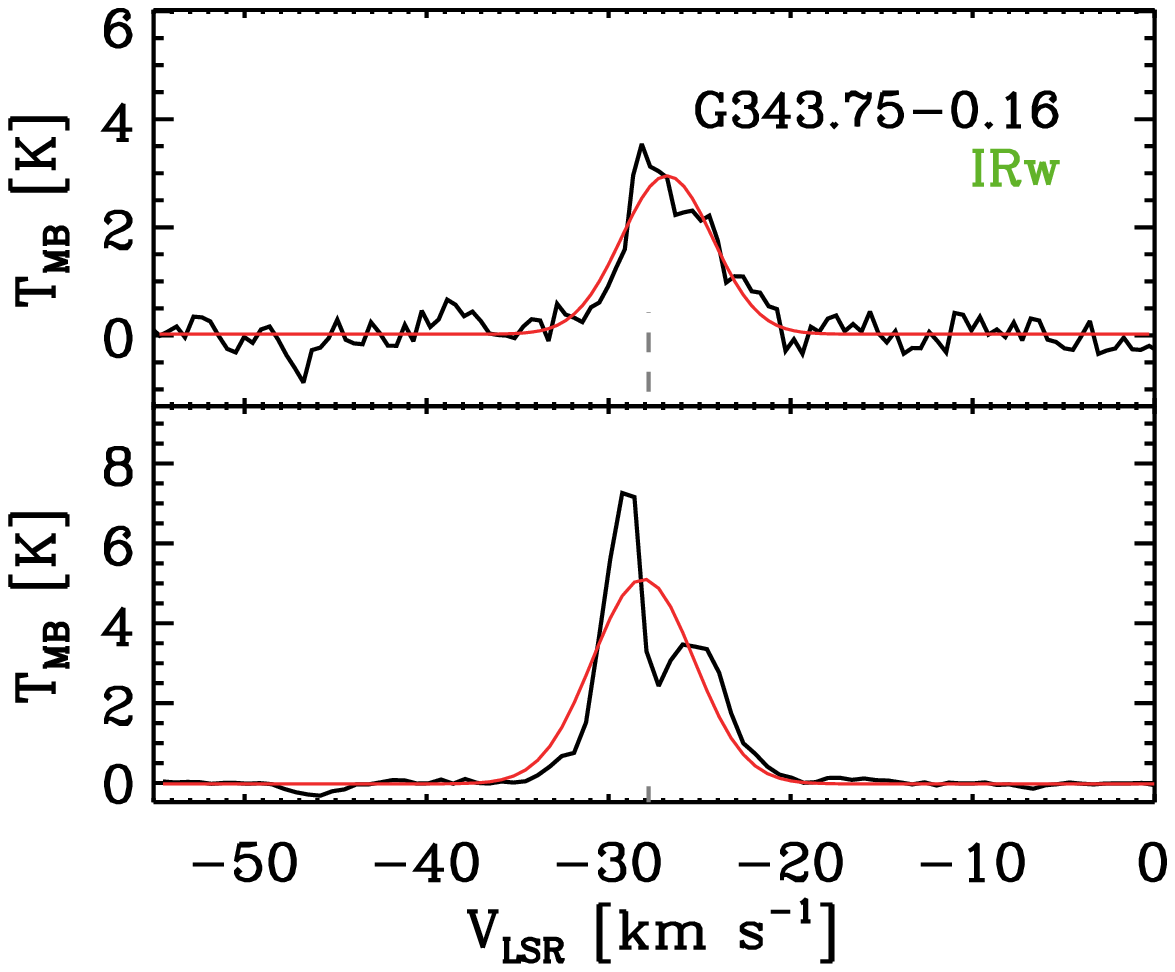}
\includegraphics[scale=0.46]{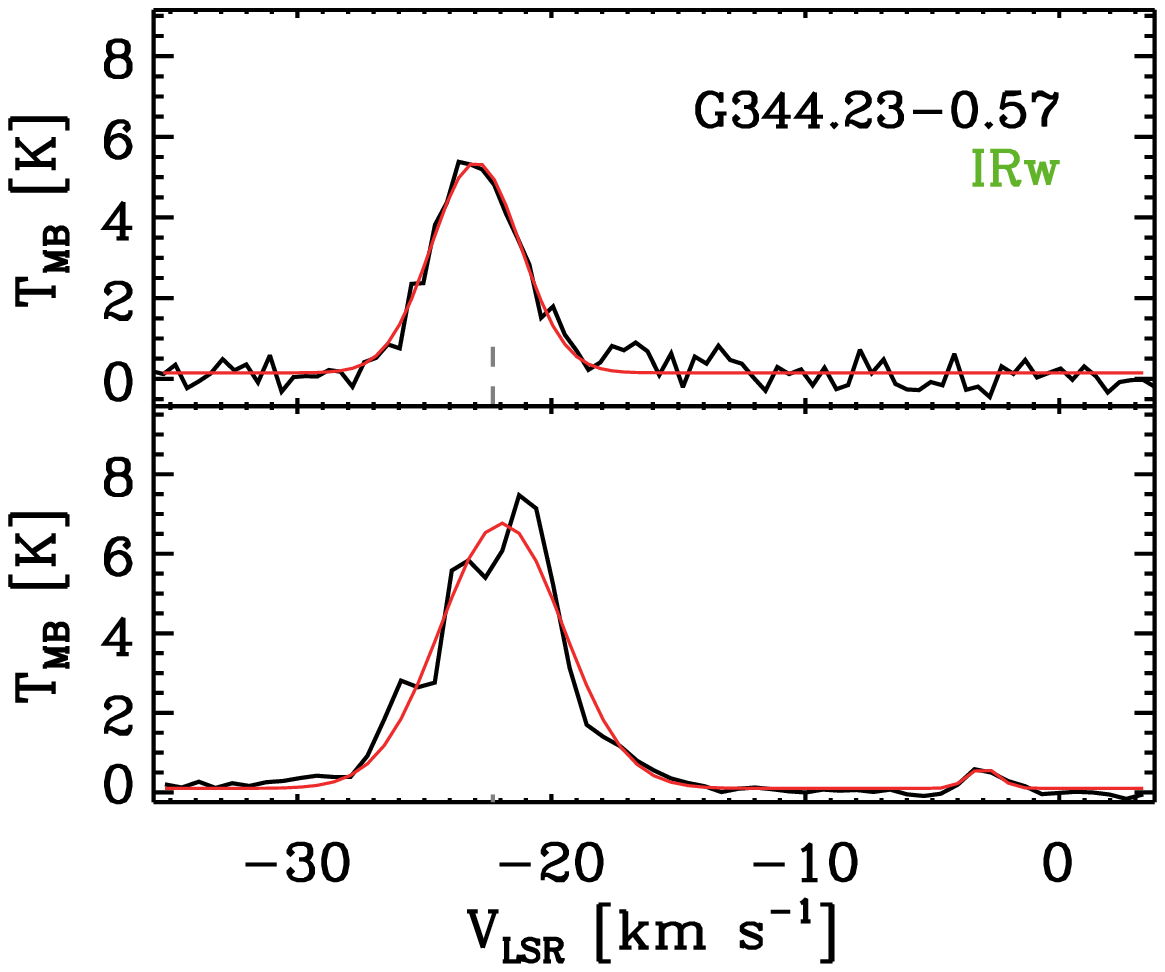}
\includegraphics[scale=0.46]{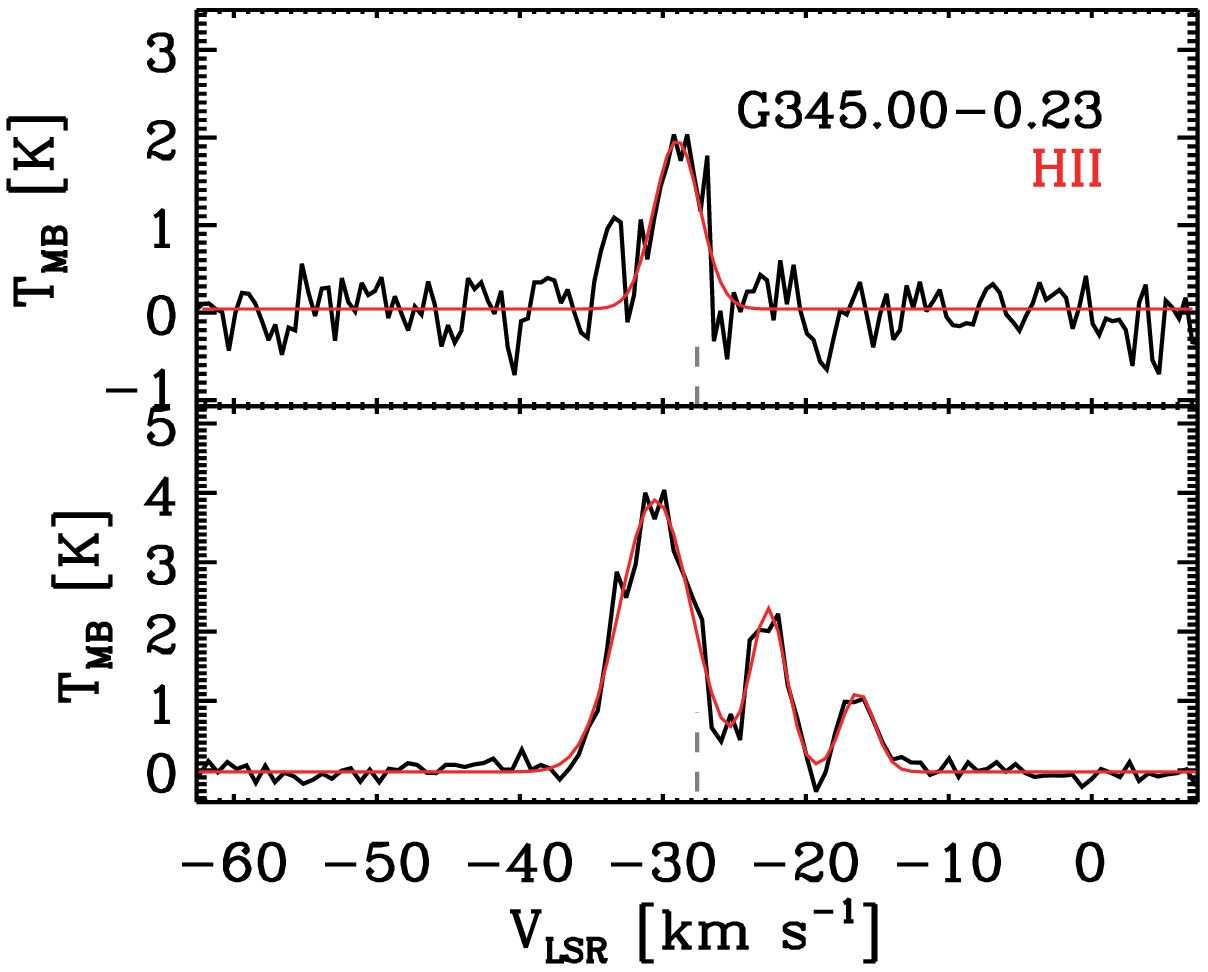}
\includegraphics[scale=0.46]{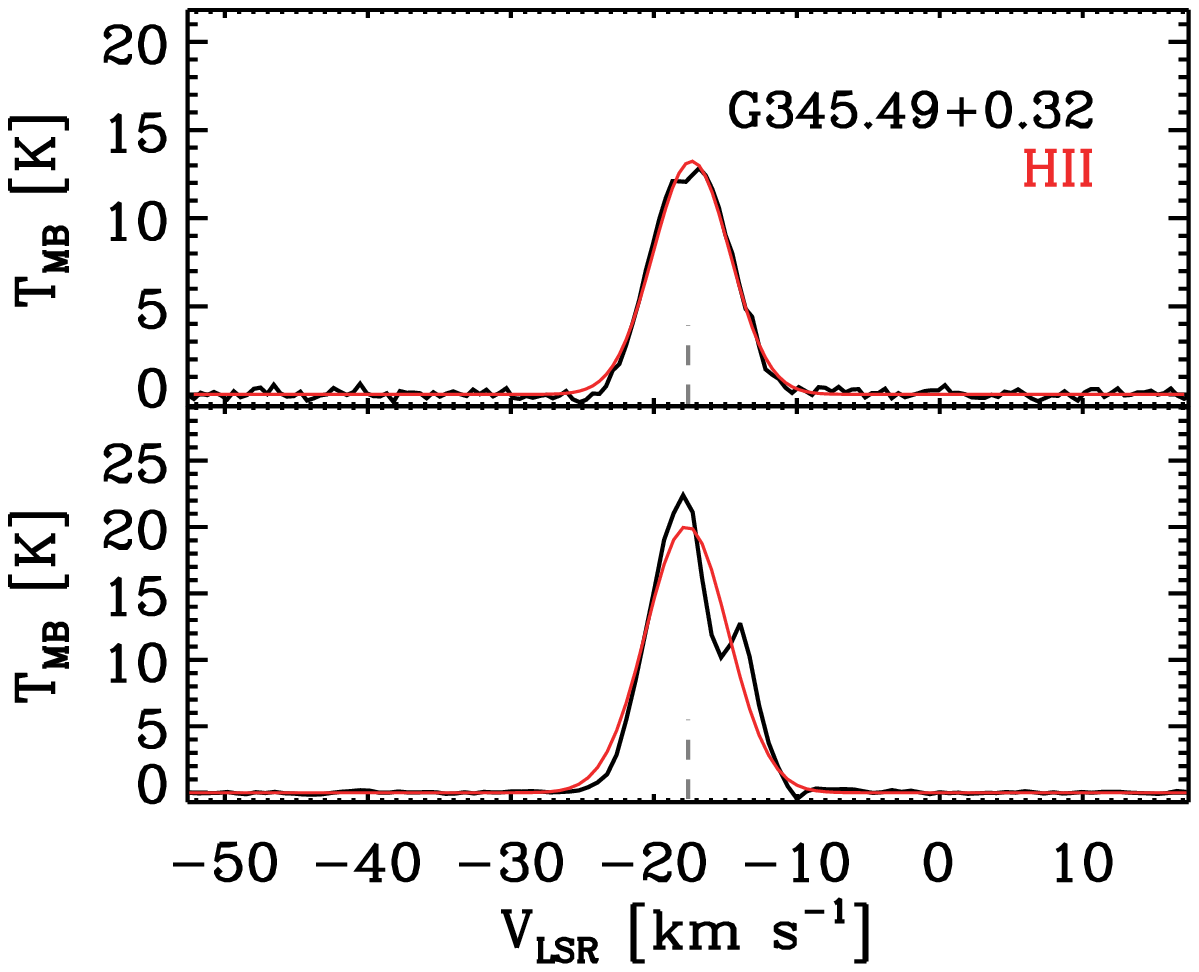}
\includegraphics[scale=0.46]{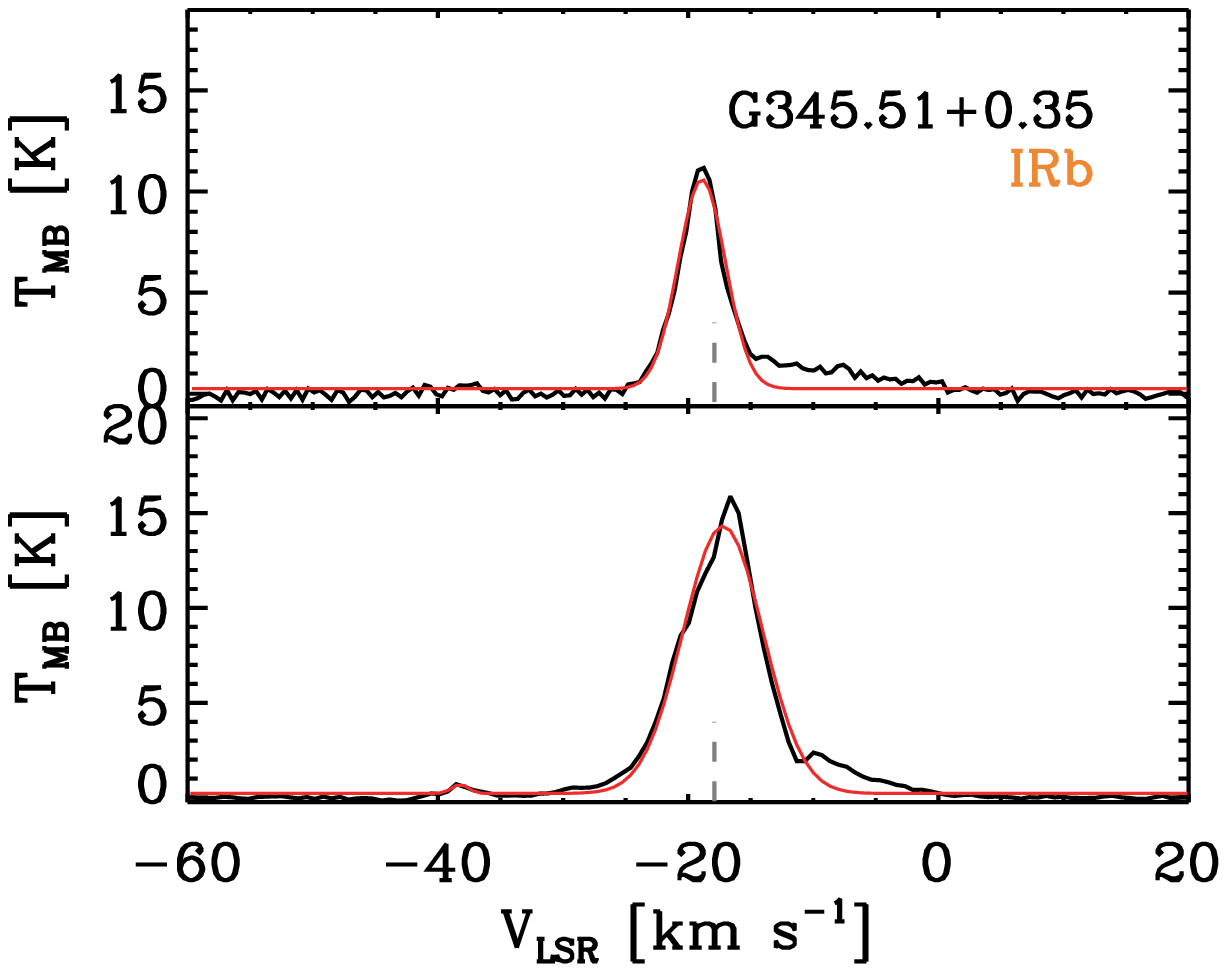}
\includegraphics[scale=0.46]{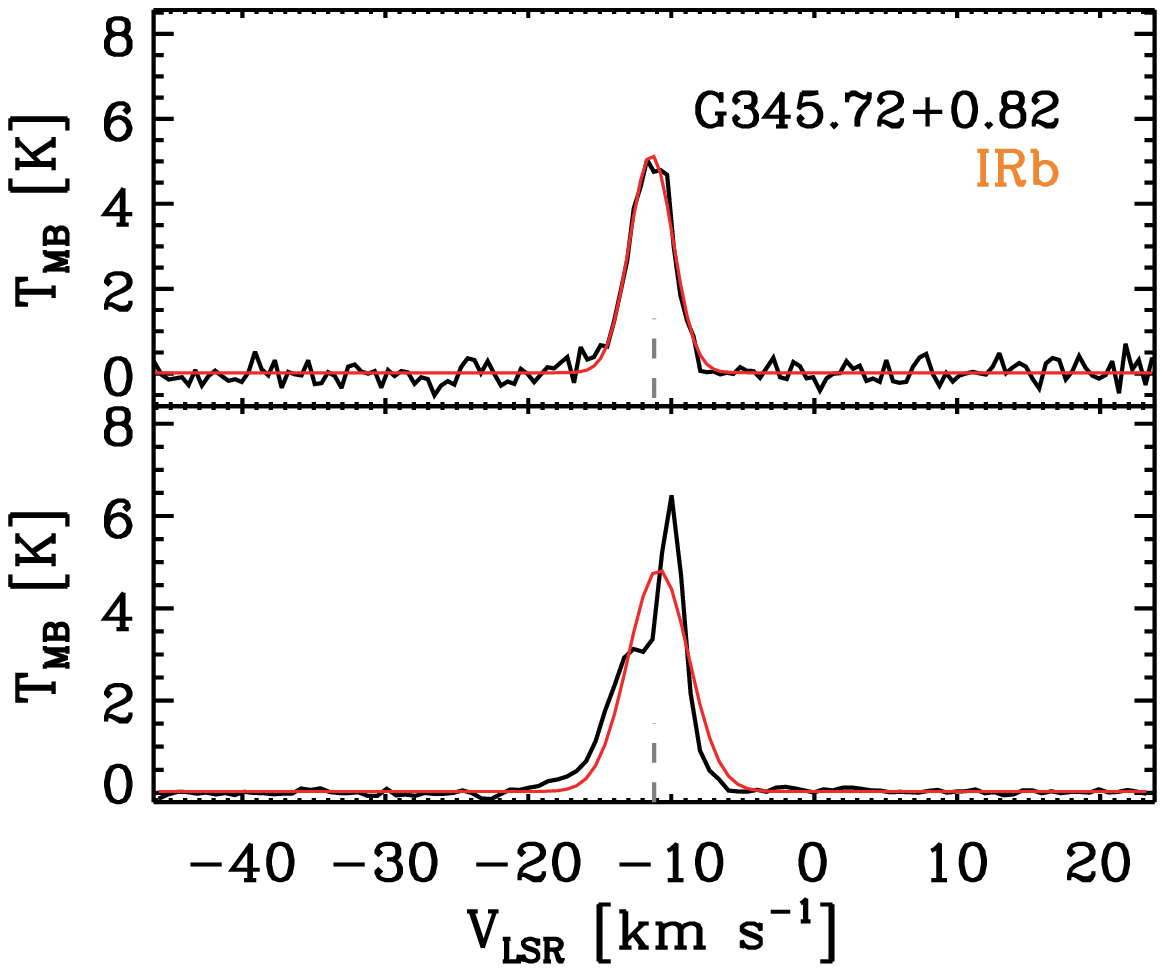}
\includegraphics[scale=0.46]{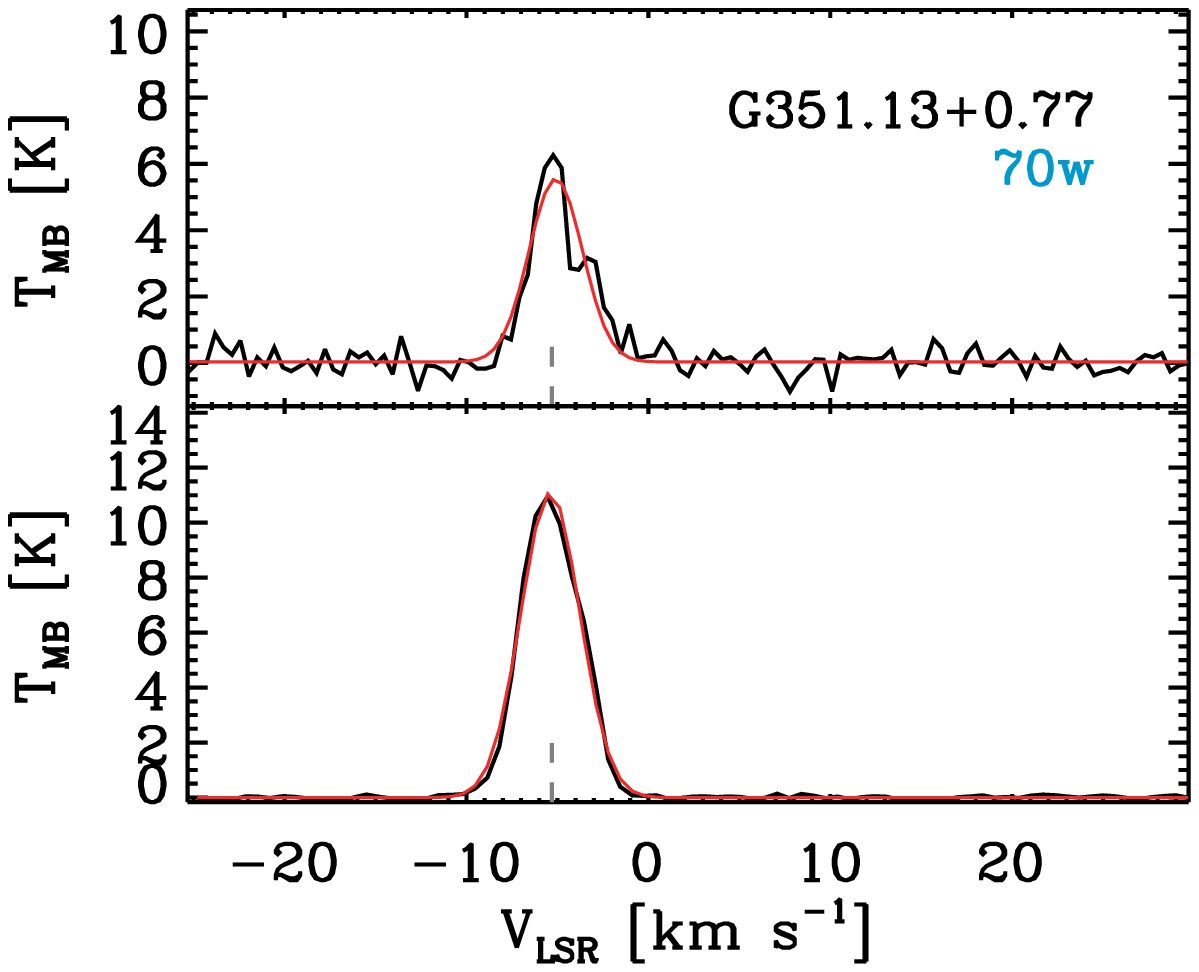}
\includegraphics[scale=0.46]{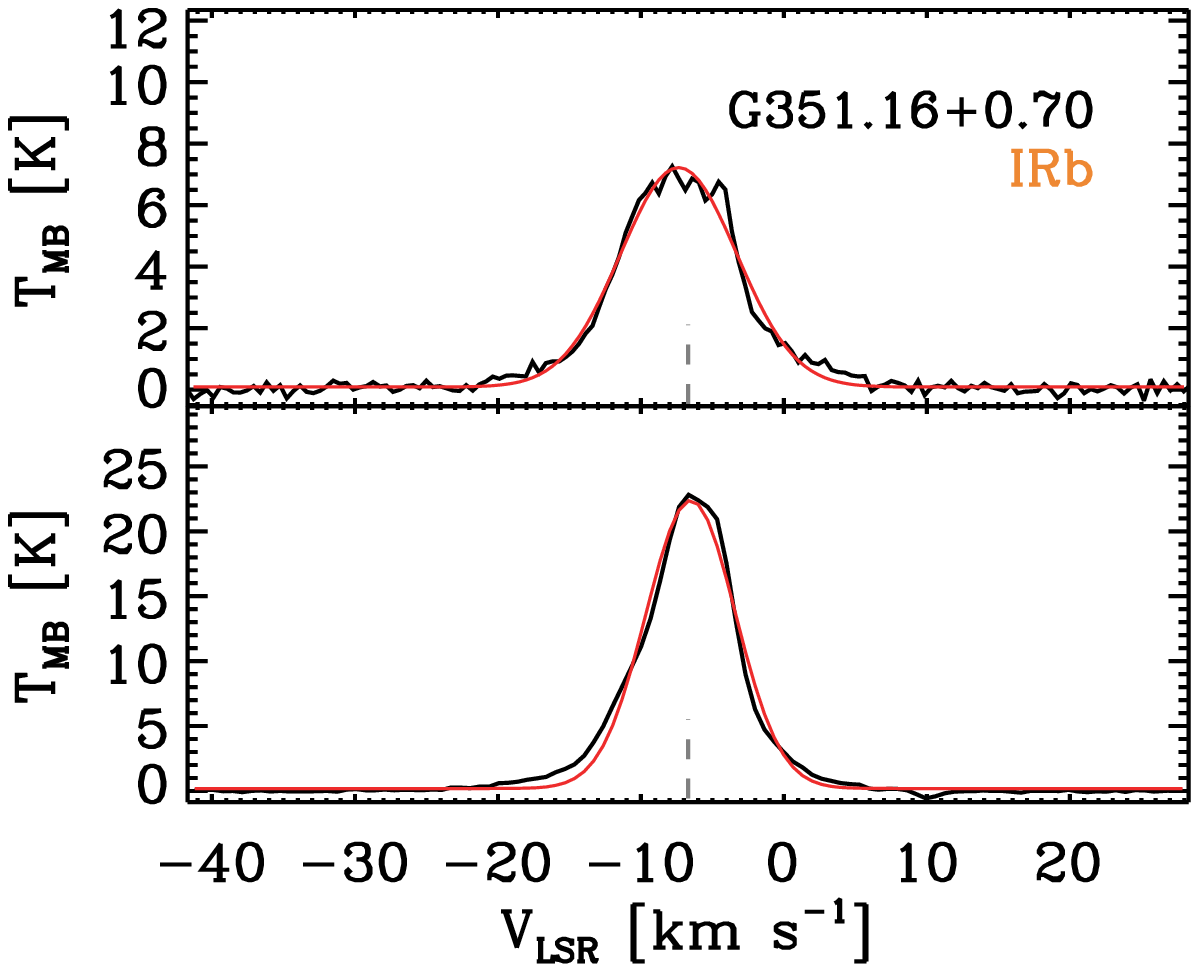}
\includegraphics[scale=0.46]{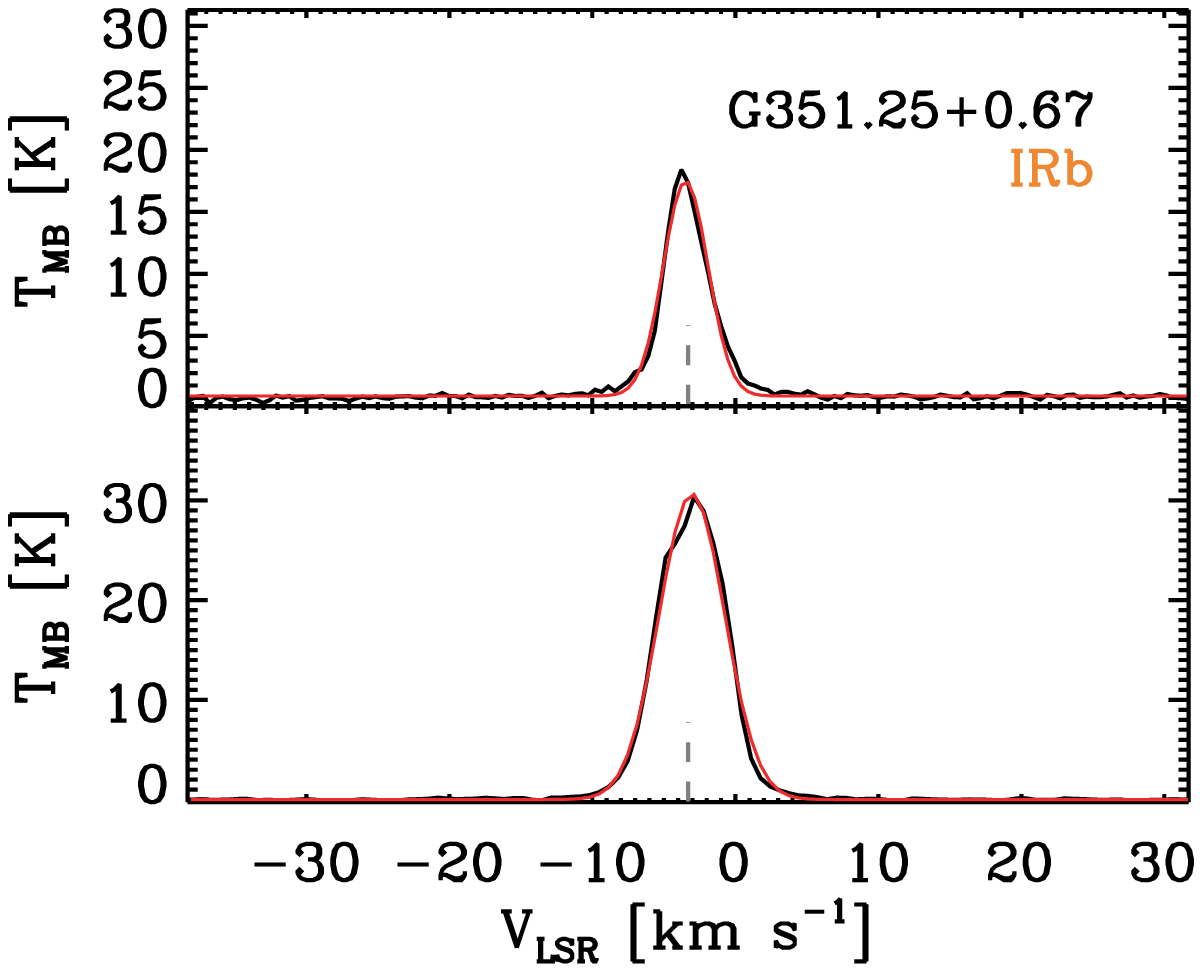}
\includegraphics[scale=0.46]{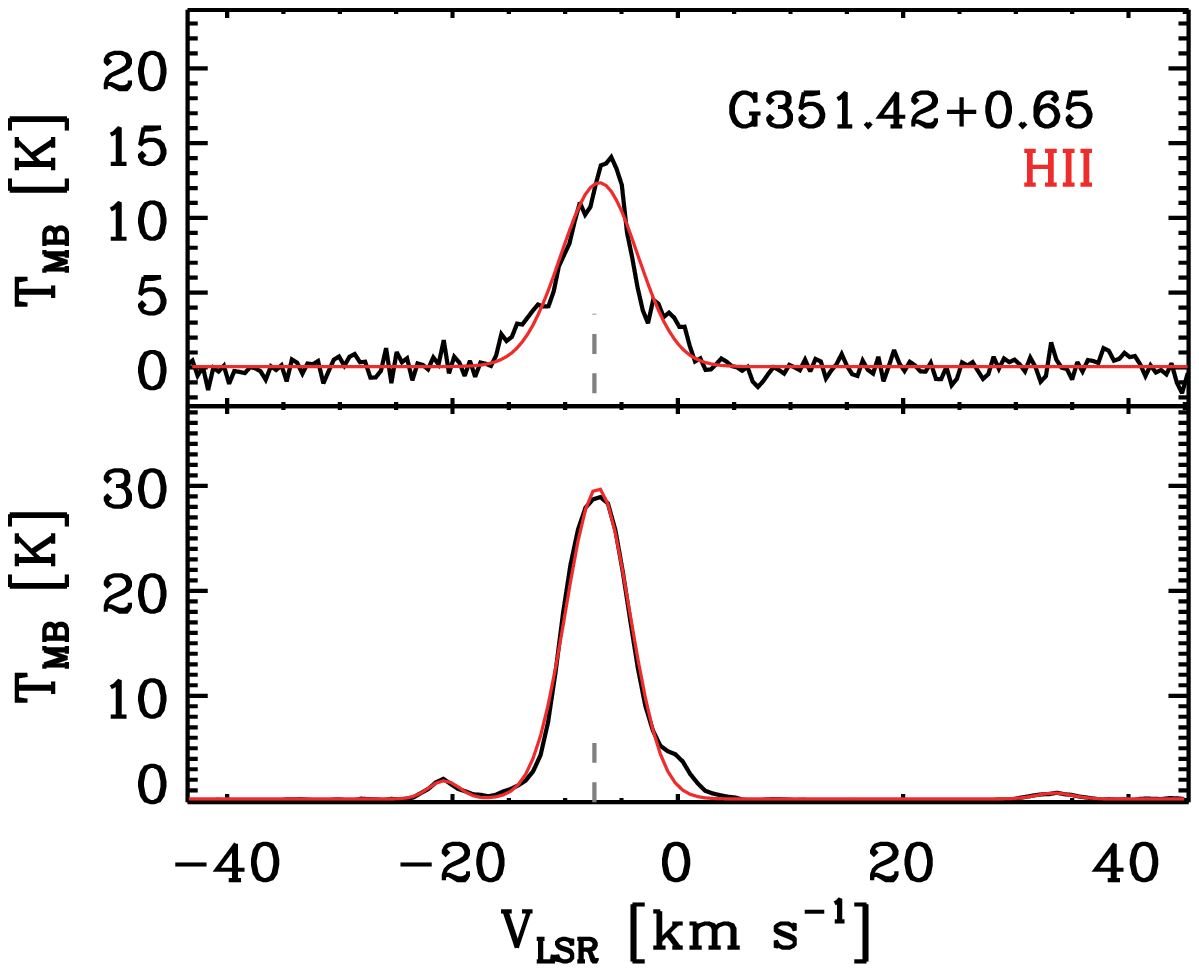}
\caption[]{(continued)}
\end{figure*}

\begin{figure*}
\centering
\ContinuedFloat
\includegraphics[scale=0.46]{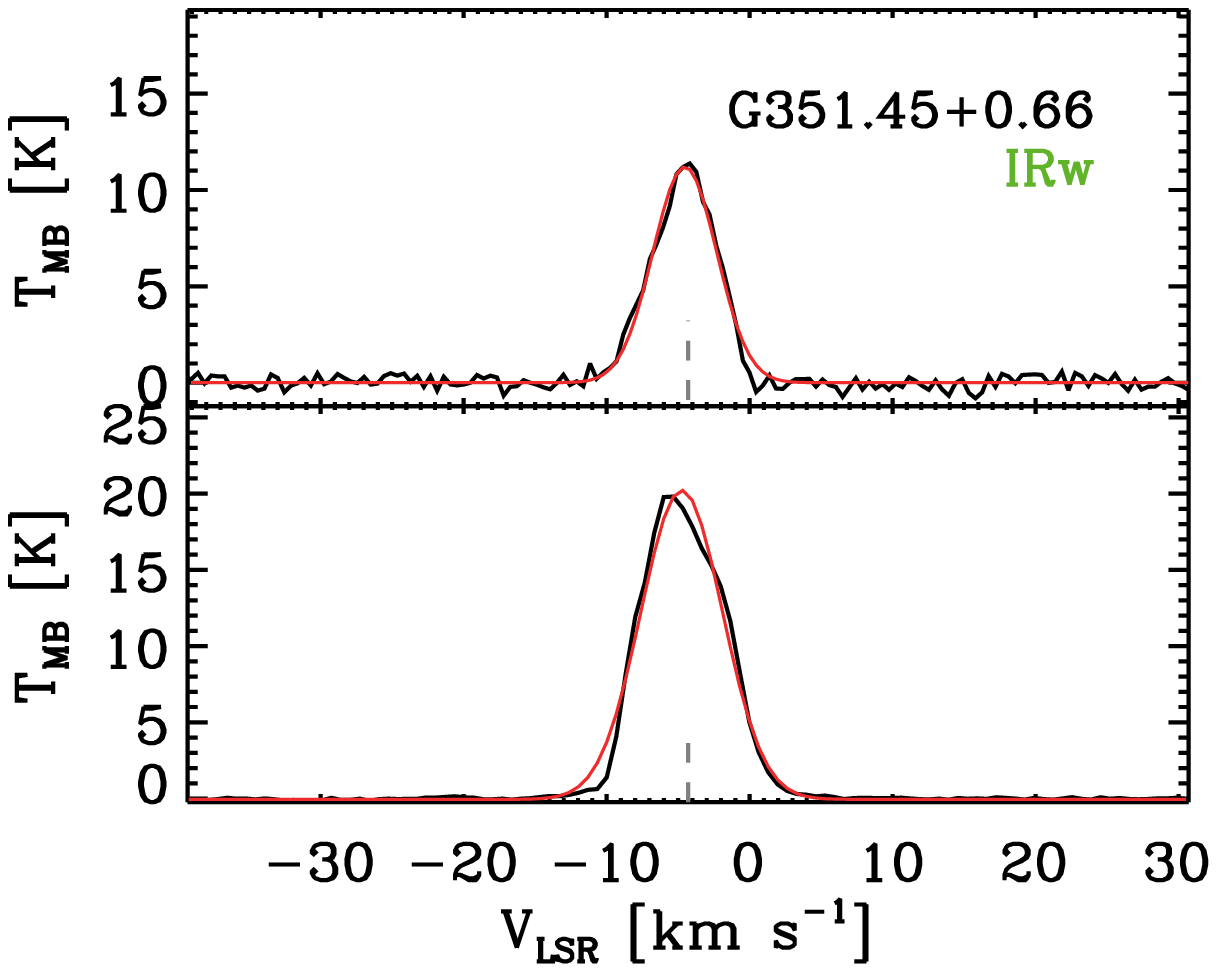}
\includegraphics[scale=0.46]{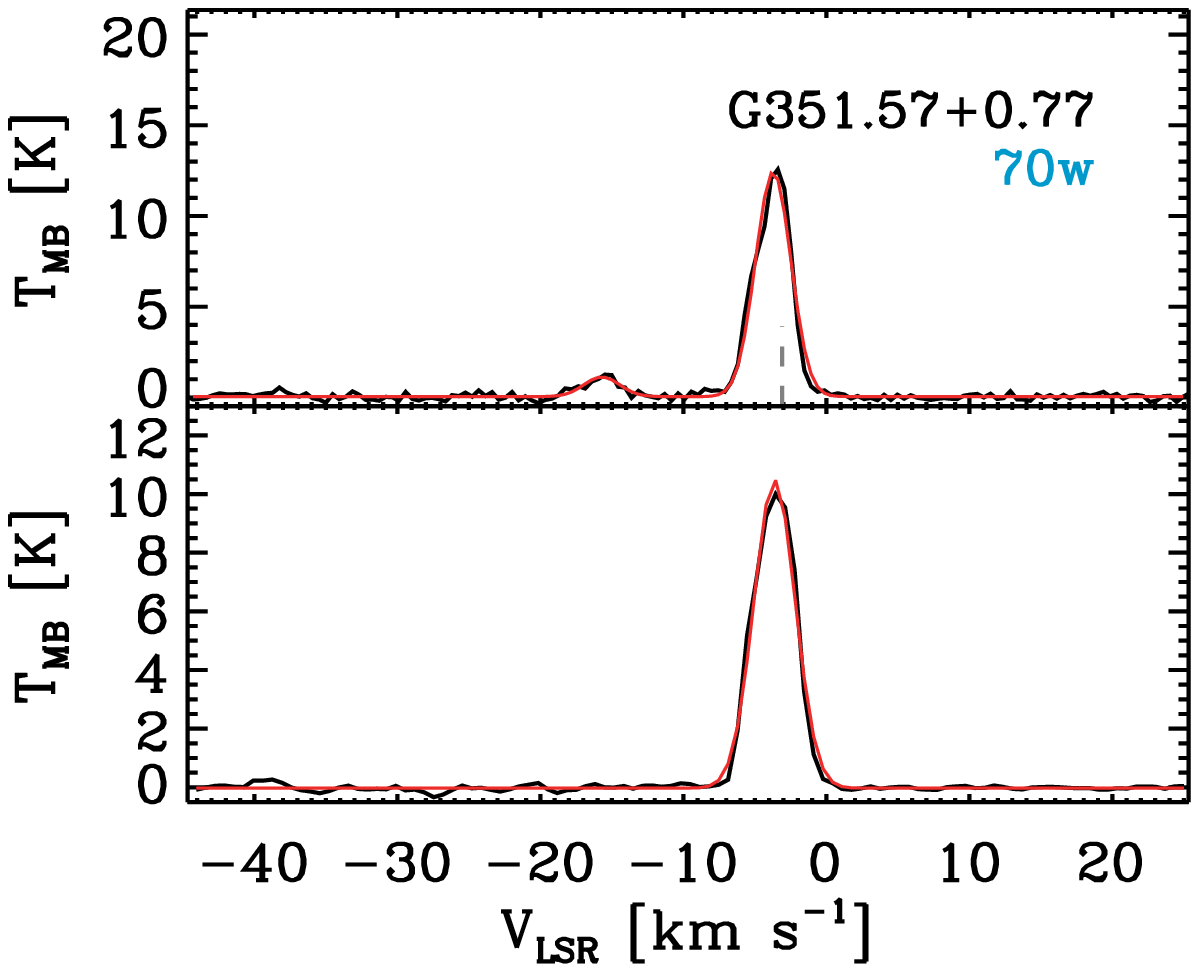}
\includegraphics[scale=0.46]{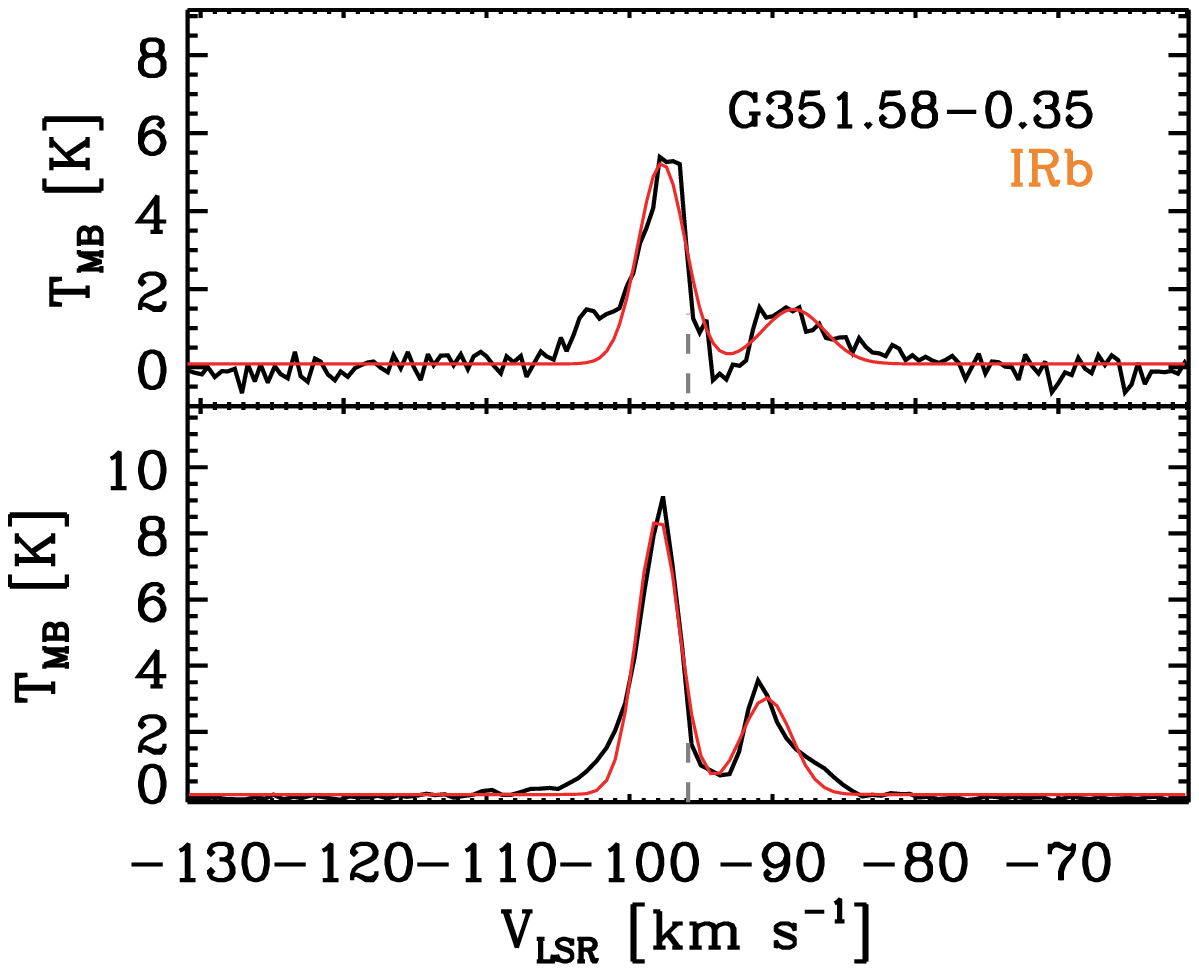}
\includegraphics[scale=0.46]{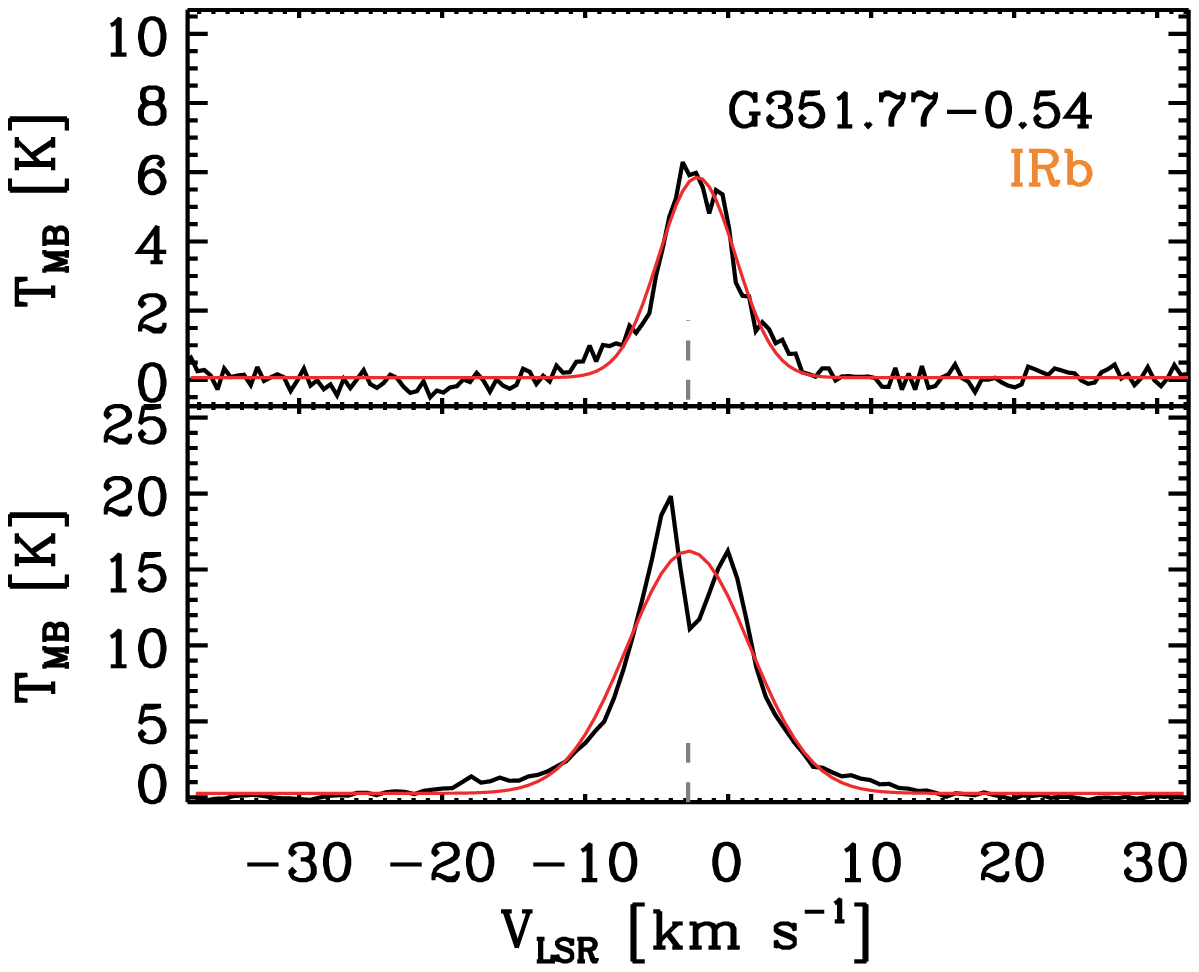}
\includegraphics[scale=0.46]{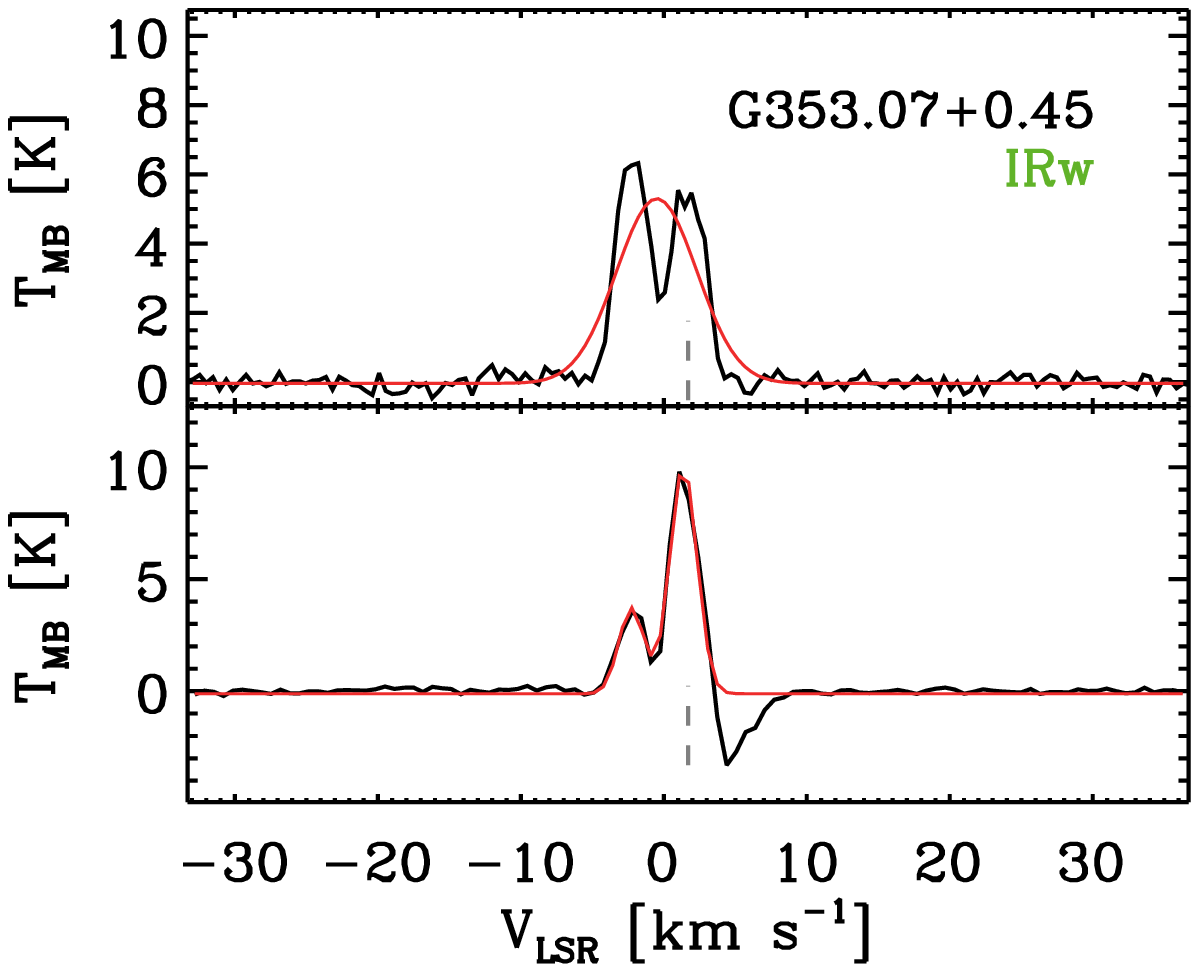}
\includegraphics[scale=0.46]{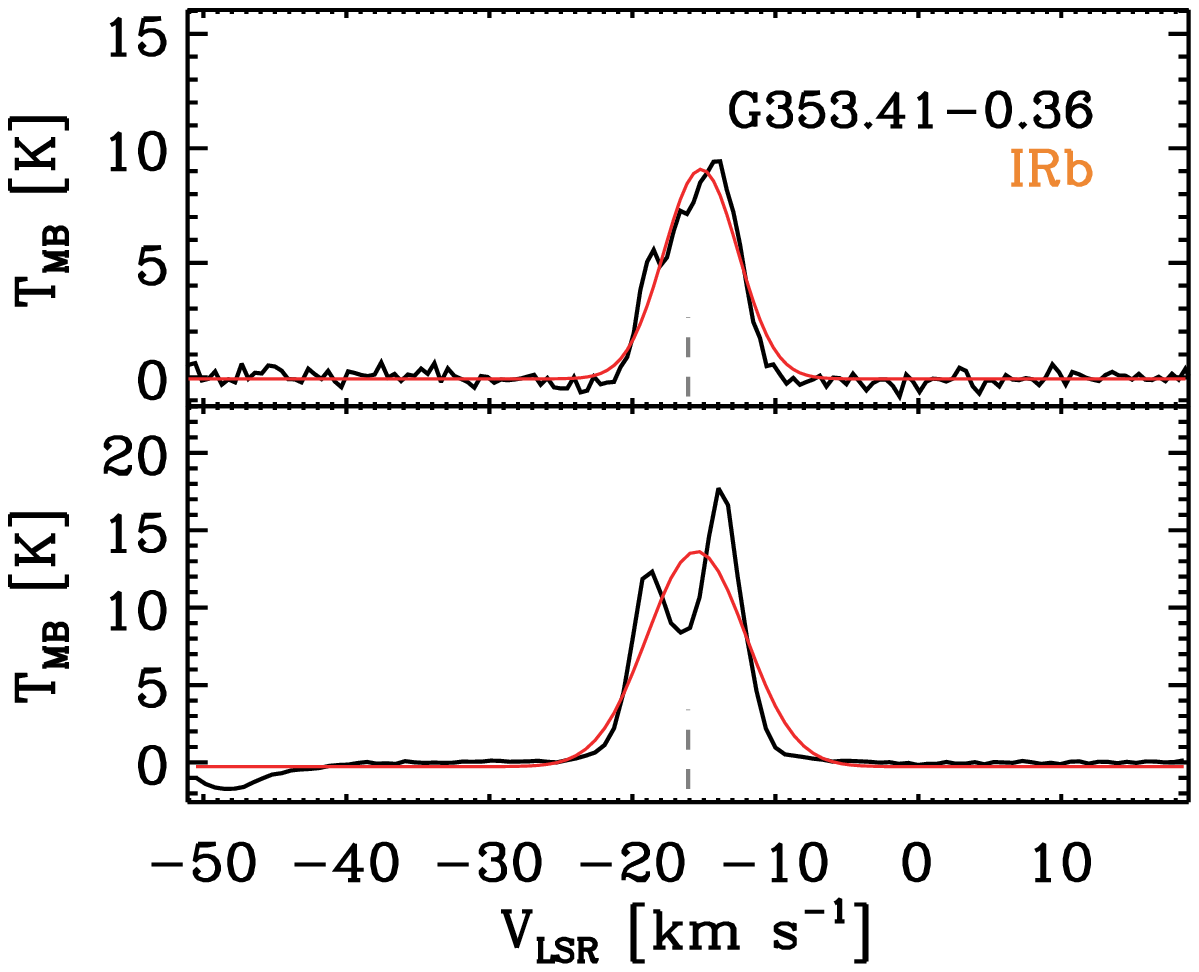}
\includegraphics[scale=0.46]{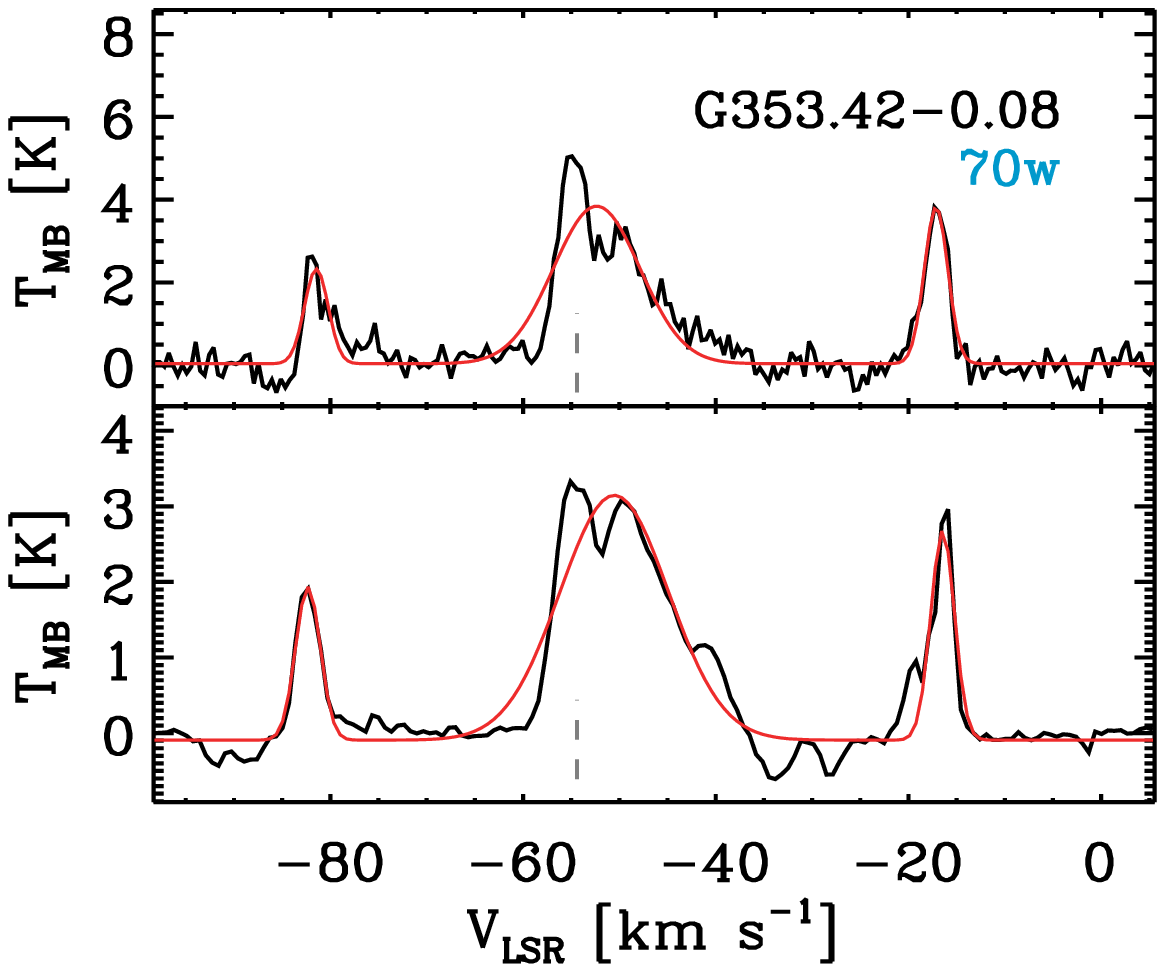}
\includegraphics[scale=0.46]{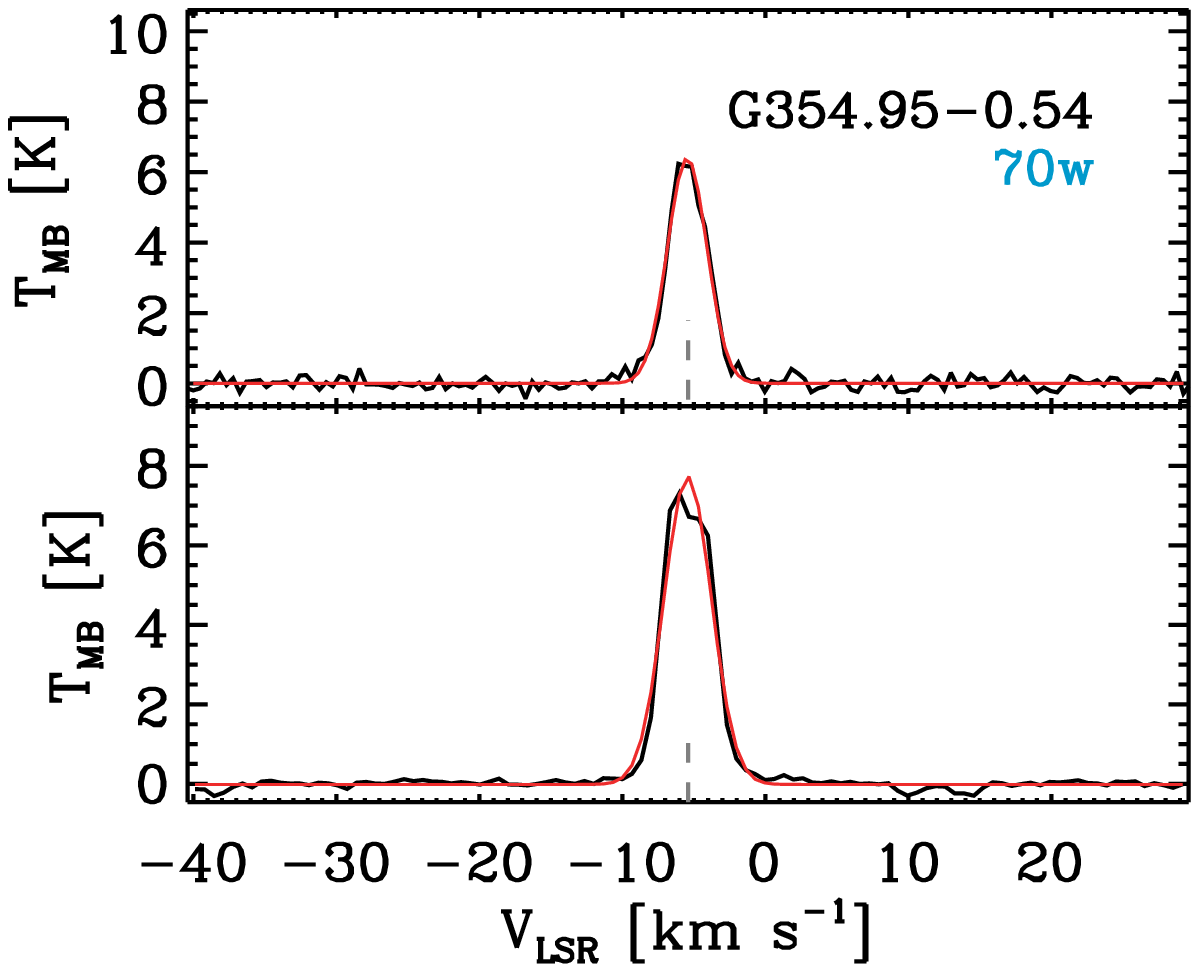}
\caption[]{(continued)}
\end{figure*}

\section{LTE calculation} 
\label{s:appendix_mc_LTE} 

To calculate the physical properties of the [C~\textsc{i}]-traced main components (Sect. \ref{s:CI_em_LTE}), 
we combined the APEX \CI 492 GHz and 809 GHz observations and followed the LTE-based procedure as below. 
First, we started with the definition of optical depth:
\begin{equation} 
\label{e:appendix_mc_LTE_eq1}
\tau_{\varv} = \frac{c^{3}}{8 \pi \nu^{3}}\left[\textrm{exp}\left(\frac{h\nu}{k_{\rm B} T_{\rm ex}}\right) - 1 \right] A_{\rm ul} \phi_{\varv} N_{\rm u}  
\end{equation}

\noindent where $c$ is the speed of light, 
$\nu$ is the frequency of the transition,
$h$ is the Planck constant, 
$k_{\rm B}$ is the Boltzmann constant, 
$A_{\rm ul}$ is Einstein's spontaneous emission coefficient,
$\phi_{\varv}$ is the normalized spectral distribution of the transition as a function of velocity, 
and $N_{\rm u}$ is the column density of atoms in the upper transition state.  
With this definition of optical depth, the $\tau_{2-1}$-to-$\tau_{1-0}$ ratio can be written as
\begin{equation} 
\label{e:appendix_mc_LTE_eq2}
\frac{\tau_{2-1}}{\tau_{1-0}} = 1.26~\textrm{exp}\left(\frac{-23.6}{T_{\rm ex}}\right)\frac{1 - \textrm{exp}(-38.9/T_{\rm ex})}{1 - \textrm{exp}(-23.6/T_{\rm ex})}
\end{equation}
\noindent This equation assumes that \CI 492 GHz and 809 GHz have the same excitation temperature and hence are in LTE. 
In addition, the atomic carbon column density can be estimated by
\begin{equation}
\label{e:appendix_mc_LTE_eq3} 
N(\textrm{C}) = \frac{8 \pi \nu^{3}}{c^{3}} \frac{Z}{A_{\rm ul}g_{\rm u}} \textrm{exp}\left(\frac{E_{\rm u}}{k_{\rm B}T_{\rm ex}}\right) 
       \left[\textrm{exp}\left(\frac{h\nu}{k_{\rm B} T_{\rm ex}}\right) - 1\right]^{-1} \int \tau_{\varv} \Delta \varv,  
\end{equation}

\noindent where $g_{\rm u}$ and $E_{\rm u}$ are the statistical weight and energy of the upper transition state. 
The partition function $Z$ is defined as
\begin{equation} 
\label{e:appendix_mc_LTE_eq4} 
\begin{split} 
Z & = \sum_{i = 0}^{2} g_{i}~\textrm{exp}\left(\frac{-E_{i}}{k_{\rm B} T_{\rm ex}}\right) \\ 
  & = 1 + 3~\textrm{exp}(-23.6/T_{\rm ex}) + 5~\textrm{exp}(-62.5/T_{\rm ex}) 
\end{split} 
\end{equation}

On the other hand, we considered the brightness temperature emitted from a region of uniform excitation temperature:
\begin{equation}
\label{e:appendix_mc_LTE_eq5} 
T_{\rm B} = \frac{h\nu/k_{\rm B}}{\textrm{exp}(h\nu/k_{\rm B}T_{\rm ex}) - 1}\left[1 - \textrm{exp}(-\tau)\right].  
\end{equation}

\noindent For \CI 492 GHz and 809 GHz, $\tau_{1-0}$ and $\tau_{2-1}$ can be written as
\begin{equation}
\label{e:appendix_mc_LTE_eq6} 
\tau_{1-0} = -\textrm{ln}\left[1 - T_{\rm B}(1-0) \frac{\textrm{exp}(23.6/T_{\rm ex}) -1}{23.6}\right] 
\end{equation} 
\begin{equation}
\label{e:appendix_mc_LTE_eq7}
\tau_{2-1} = -\textrm{ln}\left[1 - T_{\rm B}(2-1) \frac{\textrm{exp}(38.9/T_{\rm ex}) -1}{38.9}\right]. 
\end{equation}

\noindent Therefore, the $\tau_{2-1}$-to-$\tau_{1-0}$ ratio is
\begin{equation} 
\label{e:appendix_mc_LTE_eq8}
\frac{\tau_{2-1}}{\tau_{1-0}} = \frac{-\textrm{ln}\left[1 - T_{\rm B}(2-1) \frac{\textrm{exp}(38.9/T_{\rm ex}) -1}{38.9}\right]}
                                     {-\textrm{ln}\left[1 - T_{\rm B}(1-0) \frac{\textrm{exp}(23.6/T_{\rm ex}) -1}{23.6}\right]}. 
\end{equation} 

\noindent Solving Eqs. (\ref{e:appendix_mc_LTE_eq2}) and (\ref{e:appendix_mc_LTE_eq8}) with the measured brightness temperatures numerically results in 
$T_{\rm ex}$, as well as $\tau_{1-0}$ and $\tau_{2-1}$.

\end{appendix}
  
\end{document}